\documentclass[aps,rmp,twocolumn,floatfix,longbibliography,superscriptaddress]{revtex4-1}
\usepackage{graphicx}
\usepackage{braket}
\usepackage{amsmath, comment}
\usepackage{amssymb}
\usepackage{hyperref}
\usepackage{soul}

\usepackage[pdftex]{color}
\usepackage{bbm}

\newcommand{\e}{\mathrm{e}}

\newcommand{\di}{\mathrm{d}}

\newcommand{\ZZ}{\mathbb{Z}}

\newcommand{\CC}{\mathbb{C}}
\def\one{\mathbbm{1}}

\definecolor{myred}{RGB}{252,152,82}

\newcommand{\revision}[1]{{\color{black}  #1}}

\newcommand{\rp}{\mathrm{Re}}
\newcommand{\Jex}{J}
\newcommand{\Mtot}{S^z}
\newcommand{\magdens}{m_z}
\newcommand{\Onsager}{{\cal L}}
\newcommand{\CCharge}{{Q}_n}
\newcommand{\CChargem}{{Q}_m}
\newcommand{\CCharges}{{Q}_{n,s}}
\newcommand{\dens}{{q}^{(n)}}
\newcommand{\densm}{{q}^{(m)}}
\newcommand{\densk}{{q}^{(k)}}
\newcommand{\denskp}{{q}^{(k')}}
\newcommand{\curr}{{j}^{(n)}}
\newcommand{\currm}{{j}^{(m)}}

\newcommand{\sx}[1]{s^{\rm x}_{#1}}
\newcommand{\sy}[1]{s^{\rm y}_{#1}}
\newcommand{\sz}[1]{s^{\rm z}_{#1}}
\newcommand{\spm}[1]{s^{\pm}_{#1}}
\newcommand{\splus}[1]{s^{+}_{#1}}
\newcommand{\sminus}[1]{s^{-}_{#1}}
\newcommand{\jS}[1]{j^{\rm (S)}_{#1}}
\newcommand{\js}{j^{\rm (S)}}
\newcommand{\jE}[1]{j^{\rm (E)}_{#1}}
\newcommand{\JE}{{\cal J}^{(\rm E)}}
\newcommand{\JS}{{\cal J}^{(\rm S)}}
\newcommand{\J}{{\cal J}}
\newcommand{\JQ}{{\cal J}^{(\rm Q)}}
\newcommand{\JQp}{{\cal J}^{(\rm Q')}}
\newcommand{\Qcons}{{\rm Q}}
\newcommand{\Qconsp}{{\rm Q'}}
\newcommand{\sigmar}{\sigma_{\rm reg}}
\newcommand{\kappar}{\kappa_{\rm reg}}
\newcommand{\Conductivity}{\Onsager_{\rm QQ}}
\newcommand{\ConductivityOff}{\tilde{\Onsager}_{\rm QQ'}}
\newcommand{\Cbar}{\bar{C}'}
\newcommand{\Dw}{{\cal D}_{\rm w}}

\newcommand{\Dws}{{\cal D}_{\rm w}^{(\rm S)}}
\newcommand{\Dwe}{{\cal D}_{\rm w}^{(\rm E)}}
\newcommand{\Dwsbeta}{\tilde{\cal D}_{\rm w}^{(\rm S)}}
\newcommand{\DwEbeta}{\tilde{\cal D}_{\rm w}^{(\rm E)}}
\newcommand{\D}{D}
\newcommand{\DS}{D^{\rm (S)}}

\newcommand{\ii}{{\rm i}}
\newcommand{\ave}[1]{\langle {#1} \rangle}
\newcommand{\tr}[1]{\mathrm{tr}{#1}}
\newcommand{\be}{\begin{equation}}
\newcommand{\ee}{\end{equation}}

\begin{document}

\title{Finite-temperature transport in one-dimensional quantum lattice models}

\author{B. Bertini}
\affiliation{Physics Department, Faculty of Mathematics and Physics, University of Ljubljana, Ljubljana, Slovenia}

\author{F. Heidrich-Meisner}
\affiliation{Institute for Theoretical Physics, Georg-August-Universit\"at G\"ottingen, D-37077 G\"ottingen, Germany}

\author{C. Karrasch}
\affiliation{Technische Universit\"at Braunschweig, Institut f\"ur Mathematische Physik, Mendelssohnstra\ss{}e 3, D-38106 Braunschweig, Germany}

\author{T. Prosen}
\affiliation{Physics Department, Faculty of Mathematics and Physics, University of Ljubljana, Ljubljana, Slovenia}

\author{R. Steinigeweg}
\affiliation{Department of Physics,
University of Osnabr\"uck, D-49069 Osnabr\"uck, Germany}

\author{M. \v Znidari\v c}
\affiliation{Physics Department, Faculty of Mathematics and Physics, University of Ljubljana, Ljubljana, Slovenia}

\begin{abstract}

The last decade has witnessed an impressive progress in the theoretical understanding of transport properties of clean, one-dimensional quantum lattice systems. 
Many physically relevant models in one dimension are Bethe-ansatz integrable, including the anisotropic spin-1/2 Heisenberg (also called spin-1/2 XXZ chain) and the Fermi-Hubbard model. 
Nevertheless, practical computations of, for instance,  correlation functions and transport coefficients  pose hard problems from both the
conceptual and technical point of view. 
Only due to recent progress in the theory of integrable systems on the one hand  
and due to the development of numerical 
methods on the other hand has it become possible to compute  
their finite temperature and nonequilibrium transport properties quantitatively.
Most importantly, 
due to the discovery of a novel class of quasilocal conserved quantities,
there is now a qualitative understanding of the origin of ballistic finite-temperature transport, and even diffusive or super-diffusive
subleading corrections, in integrable lattice models.  
We shall review the current understanding of transport in one-dimensional lattice models, in particular, in the paradigmatic example of the spin-1/2 XXZ and Fermi-Hubbard models,
and we elaborate on state-of-the-art theoretical methods, including both analytical and computational approaches.  
Among other novel techniques,  we discuss matrix-product-states based simulation methods, dynamical typicality,  and, in particular,  generalized hydrodynamics.
We will discuss the  close and fruitful connection between theoretical models and recent experiments, with examples from both the realm of quantum magnets and  ultracold quantum gases in optical lattices. 

\end{abstract}
\date{October 28, 2020}

\maketitle

\tableofcontents

\section{Introduction}
\label{sec:intro}

The physics of strongly-correlated quantum systems in one dimension (1d) has long attracted the interest of theoreticians \cite{Giamarchi,Schoenhammer2004,Cazalilla2011,Guan2013} because of its intriguing 
properties.
For instance, quantum fluctuations can have a particularly pronounced effect in 1d, leading to the absence of finite-temperature  phase transitions and to the breakdown of 
 Landau's Fermi liquid theory, rendering  1d unique in many regards.
\revision{A particularly appealing aspect of many-body physics in one dimension is the existence of exact solutions for a subset of microscopic models, including both systems in the 
continuum such as the Gaudin-Yang model and the Lieb-Liniger gas, and lattice models such as the spin-1/2 XXZ chain and the Fermi-Hubbard chain. For the aforementioned models, 
versions of the Bethe ansatz are exploited in order to arrive at such solutions, and these are considered instances of integrable (quantum) models.\footnote{The notion of integrability in quantum systems will be commented on below.}} 

Because of the wide range of available theoretical approaches, there is the appealing ambition of developing a full theoretical understanding of these systems, both 
quantitative and qualitative. 
Moreover, many    quasi-1d materials from, e.g., quantum magnetism,  are, to a good approximation, described
by relatives of the integrable spin-1/2 Heisenberg or the Fermi-Hubbard chain. Ultracold quantum gases \cite{Bloch2008} provide another avenue
for the experimental study of 1d systems, ranging from degenerate quantum gases in the continuum [see, e.g., \cite{Langen2015,Kinoshita2004,Paredes2004,Kinoshita2006,Hofferberth2007,Liao2010}) to fermionic or bosonic lattice gases (see, e.g., \cite{Cheneau2012,Ronzheimer2013,Kaufman2016,Solomon2019,Xia2014,Vijayan2020}], including also 
realizations of Heisenberg Hamiltonians \cite{Fukuhara2013,Fukuhara2013a,Hild2014}. A renewed interest in 1d systems originates from the fields of nonequilibrium dynamics in 
closed quantum systems [for a review, see \cite{dAlessio2016,Rigol2008,Polkovnikov2011,Eisert2015,Gogolin2016,Calabrese2016}] and many-body localization [for a review, see \cite{Nandkishore2015,Altman2015,Abanin2019}], where 1d systems are the play- and testing ground for new concepts, novel phase transitions, or far-from-equilibrium dynamics.
Due to the integrability of  some 1d systems,   one can systematically  study the   transition between integrability and quantum-chaotic behavior 
[see \cite{dAlessio2016,Vidmar2016,Essler2016} and references therein].

One of the most generic nonequilibrium situations is  steady-state transport. This question has a very rich history. It was Joseph Fourier who in 1807 presented his manuscript to the French Academy, describing heat transport in terms of the diffusion equation~\cite{Fourier1822}. The work was groundbreaking in several ways~\cite{Narasimhan99}. 
Prior to  that, physicists were trying to understand heat conduction in terms of the complicated motion of the constituent particles but  Fourier changed the mindset by suggesting an effective continuum description in terms of a partial differential equation. 
Fourier's law (or its extensions to other conserved quantities, such as Fick's law, Ohm's law, etc.) states that the energy current $\jE{}$ is proportional to the temperature gradient and to the inverse of the system's length\footnote{As we will deal with lattice Hamiltonians, we will use $L$ for denoting the number of sites as well, with the 
understanding that the lattice spacing is set to unity.} $L$. Empirically, it holds in real materials.
However, the microscopic origin of such normal, i.e., diffusive transport is, even today, not entirely understood. Particularly in low-dimensional systems, one often finds that simple Hamiltonian systems do not obey Fourier's law -- instead, transport is anomalous with a nontrivial power-law scaling of the current, $\jE{} \sim 1/L^\gamma$. Understanding under what conditions one gets normal transport is one of the main challenges of theoretical physics~\cite{Lebowitz00,buchanan2005}.

In classical systems, this question has been studied since Fermi, Pasta, Ulam and Tsingou's work  on equilibration in anharmonic chains~\cite{FPU,FPUPT}, which eventually  led to the birth of the theory of classical Hamiltonian chaos. 
One would naively expect integrable systems to be ballistic conductors, i.e.,  exhibiting a zero bulk resistivity, while chaotic ones should display diffusion\revision{; this is rooted in the existence of extra conservation laws, which may prevent currents from decaying.}
Such a distinction, however, is not as clear-cut as one might think. While no rigorous conclusions have been reached yet \revision{[for reviews, see~\cite{lepri2003,dhar2008,Benenti2020}]}, explicit examples  demonstrate that even systems without classical chaos can display a wide spectrum of transport types.

In the quantum domain, the situation is even more interesting. There has been  significant progress over the last years in understanding transport in 1d quantum lattice systems, thanks to both  analytical and  numerical work. 
Due to the large number of studies since the latest overview articles appeared \cite{Zotos-review,Heidrich-Meisner2007,Zotos2002,Zotos2005},
there is a clear need for a comprehensive survey of the state-of-the-art of this field.
The aim of the present review is to give an  overview over the  transport properties of 1d  quantum lattice models at finite  temperatures, to 
describe the established results, to identify open questions, and to point out future directions.
Specifically, we are interested in lattice systems in the thermodynamic limit, including examples of  integrable and nonintegrable cases. 

We stress that the field was by no means only driven by theoretical questions, but equally importantly, also by experiments on quantum magnets \cite{Hess2019,Hess2007,Sologubenko2007}, which
show that low-dimensional quantum magnets typically feature significant contributions from magnetic excitations to the thermal conductivity.
Moreover, experiments with ultracold atomic gases in optical lattices can investigate transport properties of well \cite{Guardado-Sanchez2020,Xia2014,Schneider2012,Ronzheimer2013,Brown2019,Nichols2019}.

The universal features of 1d quantum systems at low temperatures are well captured by a universal Tomonaga-Luttinger low-energy theory, which can be solved using bosonization (see \textcite{Giamarchi,Schoenhammer2004} for a review). This reflects the general failure of the Landau quasiparticle description and accounts for the phenomenon of spin-charge separation.
Moreover, many numerical tools
work particularly well in the one-dimensional case, 
such as the density-matrix-renormalization group (DMRG)
technique and its relatives \cite{white1992,Schollwoeck2011,Schollwoeck2005}.   
As a consequence, many of the equilibrium properties of one-dimensional quantum systems are well understood.
Despite the power of such methods, there are, nevertheless, open questions and limitations.
 When deriving the universal low-energy theory, it is not straightforward
to capture nontrivial conservation laws inherited from the microscopic lattice models, and a description of the 
transport properties therefore remains a challenging task.
Numerical methods often suffer from limitations in the accessible time scales and system sizes, rendering the calculation
of dc transport coefficients a particulary difficult problem.

A number of specific 1d Hamiltonians  allow for exact solutions via Bethe-ansatz techniques \cite{Bethe}. These include the anisotropic spin-1/2 Heisenberg, its anisotroic extension, the spin-1/2  XXZ chain  
\cite{Takahashi:1999aa}, and the Fermi-Hubbard chain \cite{esslerbook},  which serve as paradigmatic models of 1d quantum physics.
For concreteness and because of its significance within the scope of the review,  let us detail the Hamiltonian of the anisotropic Heisenberg chain. It  can be written as 
$H=\sum_r
h_{r,r+1}$ with
\begin{equation}
h_{r,r+1}= \Jex (\sx{r}\sx{r+1} + \sy{r}\sy{r+1} + \Delta \sz{r}\sz{r+1}) \, .
\label{eq:XXZ-intro}
\end{equation}
Here, ${s}_r^{\textnormal{x,y,z}}$ are spin-$1/2$
operators at site $r$ ($\hbar = 1$), $\Jex$ is the exchange coupling
constant, and $\Delta$ parametrizes the exchange anisotropy. We choose $\Jex>0$, i.e., an antiferromagnetic coupling, unless stated otherwise. \revision{The spin-$1/2$ XXZ chain is gapless for $|\Delta|\leq1$ and features a gapped charge-density wave phase for $\Delta>1$. By using a
Jordan-Wigner transformation \cite{Giamarchi}, the model can be mapped to a system of spinless lattice fermions
$c_r^{(\dagger)}$:
\begin{equation} \label{eq:xxz_fermions-intro}
\begin{split}
h_{r,r+1} = &\frac{J}{2} \, c_r^\dagger c_{r+1} + \textnormal{h.c.} +
 J \Delta \left ( n_r-\frac{1}{2} \right) \left ( n_{r+1} -\frac{1}{2} \right).
\end{split}\end{equation}
The limit $\Delta=0$ corresponds to free fermions and can thus be solved
analytically by a simple Fourier transform from real to (quasi)momentum space.
Because of this mapping, the spin-$1/2$ XXZ chain is often considered to be one of the simplest models of interacting (spinless) fermions.
}

While the aforementioned  Bethe-ansatz methods provide access to the eigenenergies, excitations \cite{Orbach,esslerbook}, thermodynamics [see, e.g., \cite{Gaudin1971,Taka1971,Takahashi73,Takahashi:1999aa, Kluemper93,Kluemper2000}], and even response functions [see, e.g., \cite{Klauser2011,Caux2005}] of such Hamiltonians \cite{qmag-book}, the exact calculation of transport coefficients  
is a very difficult task and has remained controversial for decades.

The notion of integrability is not unambiguously defined in quantum physics 
\cite{Caux2011}.
Within the scope of this review, we will exclusively deal with examples of Bethe-ansatz integrable
models that possess an infinite number of local conservation laws. These are primarily the spin-1/2
XXZ chain and the 1d Fermi-Hubbard model.
The nonintegrable models that are covered here emerge from these integrable models via adding perturbations
that are expected to break all nontrivial conservation laws, such as (generic) spin-1/2 ladders, chains with a staggered magnetic
field, frustrated spin chains, or dimerized spin chains.

The following discussion is based on the description of transport within linear-response theory, which relates transport coefficients
to current autocorrelation functions via Kubo formulae. At zero temperature $T=0$, the transport coefficients
of clean systems are well understood \cite{kohn64,Scalapino1993,Shastry90}: in gapless phases, we deal with ideal metals
and hence a divergent dc conductivity. This divergence is captured via the so-called Drude weight, the prefactor of a $\delta$-singularity in the real part of the conductivity.
At $T=0$, the presence or absence of such a singularity simply distinguishes metallic behavior from insulators, respectively, and therefore, in this limit, integrability
of the microscopic model is not relevant for the existence of nonzero Drude weights.
 
An intriguing property of integrable models with regard to their transport properties 
is that they can be {\em ideal finite-temperature} conductors despite the presence of two-body interactions.
This connection was comprehensively worked out in seminal papers \cite{Castella1995,Zotos1996,Zotos1997} and is 
explained by the presence of nontrivial conservation laws preventing current autocorrelation functions from decaying to zero.
This is reflected by a nonzero finite-temperature Drude weight in the corresponding transport coefficient.\footnote{
We note that in this review, the term `transport coefficient' refers to the entire frequency-dependent object, including potential
zero-frequency singularities such as the Drude weight.
Note further that a nonzero Drude weight does not exclude the existence  of nonzero and nondivergent zero-frequency contributions stemming from the regular part (see \cite{Spohn2012large} for a review and referencs therein).
This is, in fact, a generic situation in normal fluids in the continuum.}
Similarly, one can view this as a quantum-quench problem: Imagine a current is induced in a ring at finite temperature by applying and then turning
off a force. If there is an overlap with conserved quantities, then the induced current will never decay, not even in the thermodynamic limit \cite{Mierzejewski2014}. 
Therefore, there is an intimate connection to the intensely debated topic of thermalization and relaxation in closed quantum many-body systems \cite{Polkovnikov2011,Eisert2015,Gogolin2016, dAlessio2016,Vidmar2016,Essler2016}.

\revision{The existence of a finite-temperature Drude weight is trivial in a system of free fermions (or bosons) such as the spin-1/2 XX chain.
In an ordinary metal and in the Drude model, a finite Drude weight arises in the limit of a diverging relaxation time 
In a Fermi liquid, this occurs in the limit
of $T\to 0$, where the quasi-particle lifetime becomes infinite, provided there are no impurities.
}

In some famous cases of \revision{integrable interacting models}, the conservation laws relevant for ballistic transport properties are easy to identify~\cite{Grabowski95}: For thermal transport in the spin-1/2 XXZ chain,
the total energy current $\JE$ itself is conserved, rendering both the transport coefficients for energy and thermal transport divergent.
The conservation of $\JE$ is also sufficient to prove that {\it spin transport} is ballistic at any {\it finite} magnetization   
$\magdens=2 \langle \Mtot\rangle/L\not=0 $ where $\Mtot = \sum_r \sz{r}$ \cite{Zotos1997}.
For thermal transport in spin-1/2 XXZ chains at zero magnetization, the energy Drude weight\footnote{Throughout this review, we use the term energy Drude weight instead of thermal Drude weight.} was  computed from Bethe-ansatz methods \cite{Kluemper2002,Sakai2003,Zotos2016}.

For spin transport and at {\it zero magnetization} (either in the canonical or grand-canonical ensemble), the problem turned out to be much harder and has evolved into one of the key open questions in the theory of low-dimensional quantum systems. 
While a first Bethe-ansatz calculation \cite{Zotos1999} indicated nonzero spin Drude weights 
in a wide parameter range, consistent with exact diagonalization  \cite{Zotos1996,Narozhny1998,Heidrich-Meisner2003}, the actual relevant conservation laws were not known until 2011.
Exact diagonalization was often argued to be inconclusive due to the small accessible system sizes \cite{Sirker2009,Sirker2011} while the Bethe-ansatz results
from \cite{Zotos1999} were challenged as well: The calculation of the spin Drude 
weight cannot be done in the same rigorous manner as for the energy Drude 
weight, and
qualitatively different results were obtained from another Bethe-ansatz calculation using different assumptions \cite{Benz2005}.
Therefore, the questions of whether or not the spin Drude weight was finite in the spin-1/2 XXZ chain at $\magdens=0$ and how to compute it quantitatively  attracted the attention of theoreticians using a wide range of methods such as Quantum Monte Carlo \cite{Alvarez2002,Heidarian2007,Grossjohann2010}, field theory \cite{Fujimoto2003,Sirker2009,Sirker2011}, density-matrix-renormalization-group simulations at finite temperatures \cite{Karrasch2012,Karrasch2013}, dynamical typicality \cite{Steinigeweg2014}, DMRG simulations of open quantum systems \cite{Prosen2009,Znidaric2011}, and more recently, generalized hydrodynamics (GHD) \cite{Ilievski2017a,Bulchandani2017a}.
GHD is a hydrodynamic description valid for general Bethe-ansatz integrable 
models developed in \cite{Bertini2016,Castro-Alvaredo2016} [see also the recent review \cite{Doyon2019b}].

The question of finiteness of the finite-temperature spin Drude weight in the gapless regime ($|\Delta|<1$) of the spin-1/2 XXZ chain has been resolved in 2011 \cite{Prosen2011,Prosen2013} by the
discovery of the so-called quasilocal charges which were derived, quite unexpectedly, from an exact solution of a boundary-driven many-body Lindblad master equation. These conserved quantities are fundamentally different from the previously known local conserved charges derived  from the algebraic Bethe ansatz since they break spin-reversal symmetry. This can be interpreted as a consequence of the dissipative, non-time-reversal invariant setup that they are derived from. Soon after, the quasilocal charges have been extended to periodic (or more generally, twisted) boundary conditions \cite{Prosen2014,Pereira2014}, and generalized to a one-parameter family \cite{Prosen2013}.
The existence of these hitherto unknown quasilocal charges  quantitatively explained the results of numerical simulations and qualitatively confirms the TBA result \cite{Zotos1999}. 
Remarkably, the lower bound to the spin Drude weight  agrees exactly with recent
analytical results for the spin Drude weight based on GHD \cite{Ilievski2017a} and the thermodynamic Bethe ansatz \cite{Zotos1999,Urichuk2018}.
\revision{Table~\ref{tab:drudes} summarizes the Drude weights that will be covered in this review for the spin-1/2 XXZ chain.}

\begin{table}[t]
\begin{tabular}{c| cccc}
Transport channel & $\magdens$ &$0\leq \Delta <1$ & $\Delta=1$ & $\Delta >1$ \\\hline\hline
Energy Drude weight $\Dwe$ & 0, $\not =$0 & $>0$ & $>0$ & $>0$ \\\hline 
Spin Drude weight $\Dws$ & $0$ &  $>0$ & 0 & 0 \\ \hline
Spin Drude weight $ \Dws$ & $\not =$0 & $>0$ & $>0$ & $>0$ \\\hline
\end{tabular}
\label{tab:drudes}
\caption{\revision{Overview  of the different finite-temperature Drude weights of the antiferromagnetic spin-1/2 XXZ chain whose different behaviors will be covered in this
review: the spin Drude weight $\Dws$ and the energy Drude weight $\Dwe$ as a function of magnetization $\magdens = 2\sum_r \langle \sz{r}\rangle/L $ and model parameters, where $\Delta$ is the exchange anisotropy. The actual definitions for the Drude weights will be given 
in Sec.~\ref{sec:theory} and the theoretical predictions are covered in Secs.~\ref{sec:mazur} and \ref{sec:xxz}.}}
\end{table}

Apart from the issue of Drude weights, there are equally interesting questions concerning diffusion and finite-frequency behavior.\footnote{\revision{The range of possible transport types -- ballistic, diffusive, superdiffusive, subdiffusive -- will be introduced in Sec.~\ref{sec:theory.ballistic}, see also Fig.~\ref{sketch_sigma}.}} In the gapless regime of the spin-1/2 XXZ chain ($|\Delta| <1$), a regular
diffusive subleading contribution to transport  was advocated for by \cite{Sirker2009,Sirker2011} while a pseudogap structure in the low-frequency window was suggested in \cite{Herbrych2012}.
In the regime $|\Delta|>1 $, anomalous low-frequency properties were
observed on finite systems \cite{Prelovsek2004}, while most studies indicate a nonzero dc spin conductivity and thus a finite diffusion constant \cite{Prosen2009,Znidaric2011,Karrasch2014,Steinigeweg2009,Steinigeweg2011}. Remarkably, diffusion in integrable systems has
been recently explained within the GHD framework, also yielding a  quantitative prediction for the diffusion  constant \cite{DeNardisDiffusion,Gopalakrishnan2019}.
Moreover, numerical evidence for superdiffusive spin transport with a dynamical exponent of $z=3/2$  at the Heisenberg point $\Delta=1$ has been found in 
\cite{Ljubotina2017,Ljubotina2019} and self-consistently explained within GHD \cite{Gopalakrishnan2019,Bulchandani2019b,DeNardis2019,DeNardis2020}.
This is the same exponent as in the Kardar-Parisi-Zhang universality class \cite{Kardar1986} leading to the actively investigated question of whether
this scenario is realized in the spin-1/2 Heisenberg chain and possibly other systems with SU(2)-symmetric exchange \cite{Ljubotina2019,Ljubotina2017,Weiner2020,Spohn2019,DeNardis2019,Dupont2019}.

While much of the research concentrated on the linear-response regime of the spin-1/2 XXZ chain, current activities have evolved into a number of interesting directions.
An immediate goal \cite{Karrasch2014a,Jin2015,Karrasch2016,Karrasch2017a,Karrasch2017b}
 is to establish a  complete
picture for the linear-response transport in the Fermi-Hubbard chain, which is perhaps the second equally important
integrable lattice model with regards to experimental realizations. 

Next, also having real materials in mind, another important question is how robust transport properties are against perturbations. This has  triggered  much research into nonintegrable models
[see, e.g., \cite{Rabson2004,Zotos1996,Saito1996,Alvarez2002a,Heidrich-Meisner2002,Huang2013,Steinigeweg2015,Heidrich-Meisner2004,Heidrich-Meisner2003,Zotos2004,Prosen1999,Jung2006,Jung2007,Steinigeweg2016a} and further
references mentioned in Sec.~\ref{sec:nonint}].
In this regime, numerical methods play a crucial role. While the expectation is that nonintegrable models should exhibit diffusive transport at finite temperature, demonstrating this in an exact manner or
in numerical simulations is a challenging task. Significant progress has been made with modern computational methods that allow one to obtain diffusion constants at least at high temperatures \cite{Steinigeweg2015,Steinigeweg2016a,Karrasch2014,Znidaric2011}.
The generic description of nonintegrable models at low temperatures results from extensions of Tomonaga-Luttinger low-energy theories for gapless systems \cite{Sirker2009,Sirker2011}
or field theories for gapped situations \cite{Sachdev1997,Damle2005}.
Moreover, nonintegrable models in 1d may still possess long-lived dynamics and hydrodynamic tails and it is by no means obvious that diffusion is the only possible scenario [see, e.g., \cite{Medenjak2019,DeNardis2020} for recent work].

In the discussion of nonintegrable models, we exclude systems with  disorder  \cite{Abanin2019,Altman2015,Nandkishore2015,luitz2017,Gopalakrishnan2019c}. Many-body lattice systems with disorder
are believed to host both ergodic and many-body localized phases [see also the recent discussion in
\cite{Suntajs2019,Abanin2019a,Sierant2019,Panda2019}]. The transport properties of the
ergodic phase are quite interesting and there is a number of studies \cite{Agarwal2015,Scardicchio16} that claim the existence of
a subdiffusive regime within the ergodic phase. This result, however, is still controversial
\cite{Barisic2016,Steinigeweg2016c,Bera2017}. Nevertheless, the ergodic phase of disordered models is often considered a generic
example of a thermalizing phase with diffusive transport (then obviously excluding the putative subdiffusive regime).

Moreover, there has been a fervent activity  concerning the studies of more general forms of transport.  
For instance, manifestly nonequilibrium situations with inhomogeneous density profiles are intensely investigated \cite{Karrasch2013a,Bertini2016,Castro-Alvaredo2016,Ruelle2000,Aschbacher2003,Gobert2005,Langer2009,Langer2011,Jesenko2011,Lancaster2010,Steinigeweg2017,Ljubotina2017}, partially also because such initial conditions can be realized with both quantum magnets \cite{Otter2009,Montagnese2013} and quantum gases 
\cite{Schneider2012,Ronzheimer2013,Fukuhara2013,Fukuhara2013a}. In addition, there is a growing interest in using insights from CFT and AdS/CFT correspondence 
for the description of such nonequilibrium situations \cite{Bernard2012,Bhaeseen2015,Dubail:2017aa}.

For both the description of transport in the linear-response regime and for nonequilibrium situations,
GHD has been established as a powerful theoretical framework for Bethe-ansatz integrable quantum lattice models \cite{Bertini2016,Castro-Alvaredo2016}.
The approach allows to compute Drude weights \cite{Ilievski2017a}, diffusion constants \cite{DeNardisDiffusion} and can provide the full temperature dependence of 
both quantities. Moreover, subleading corrections to transport coefficients can be extracted such as  diffusive or superdiffusive corrections 
in the presence of a Drude weight \cite{Agrawal2019}.
 Most importantly,
GHD often allows for developing an intuition and interpretation as it is based on a kinetic theory of the characteristic excitations
of integrable models. While GHD is a recent development, it will be prominently featured throughout the review.
 
Furthermore, we will complement the picture emerging from linear-response theory or closed quantum system simulations with insights 
 from studies of open-quantum systems. In our context, these are long pieces of spin or Fermi-Hubbard chains coupled to an environment
via boundary driving. The theoretical description is based on quantum master equations, and the Lindblad equation is the most commonly employed
starting point. The boundary-driving terms can be used to induce a temperature 
or magnetization difference across the region of interest.
The focus is on the steady state that can be close or far away from equilibrium and is referred to as a nonequilibrium steady state (NESS).
While there are methods to solve such set-ups exactly for free systems \cite{3rdQuant,ProsenSpectral} and statements about the existence and uniqueness of the steady state \cite{spohn77,frigerio77,evans77},
one frequently needs to resort to numerical methods, in particular when dealing with interacting systems. Time-dependent DMRG
has emerged as a useful solver and comparably large systems sizes are studied \cite{Prosen2009}. The scaling behavior of the NESS current with system size allows to characterize
transport as diffusive, ballistic or super(sub)-diffusive and is therefore a very valuable complementary approach. For instance, the notion of superdiffusive
dynamics in the spin-1/2 Heisenberg chain was first established from open-quantum system simulations \cite{Znidaric2011}.
One can also extract diffusion constants which in certain limiting cases should agree with the results from linear-response theory \cite{Znidaric2019}.
Open-quantum system simulations were extensively used to investigate transport in spin-1/2 XXZ chains, the Fermi-Hubbard chain, and spin-ladders, to name but a few examples [see, e.g., \revision{\cite{michel03,Poletti19,michel08,Saito1996,Znidaric2013b,Mendoza2015,Prosen2012,mendoza-arenas13a,Mejia2007,Katzer2020}].}

As with any review article, choices regarding the scope, topics, and focus need to be made. 
This review  will not discuss transport in mesoscopic systems,
transport in  systems with disorder, or in continuum models. Out of the wide range of transport theory in lattice models, here, we emphasize certain Hamiltonians, 
results from Bethe ansatz, the role of the newly discovered quasilocal charges, results from GHD, from a  range of numerical
methods, and a comparison between linear-response theory and open-quantum systems. 
Field-theoretical approaches are very important in the field, yet a full coverage of the technical aspects and its predictions are 
beyond the scope of this work and the reader is  referred to recent reviews \cite{Sirker2020} and the original literature for more details.
The same goes for a wide range of results for nonintegrable models, \revision{Floquet systems [see, e.g., \cite{Lenarcic2018,Lenarcic2018a,Lange2018a}],} transport in disordered systems,  and many nonequilibrium studies  that will not be covered in full
detail.
 
This review is organized as follows. 
First, we introduce the calculations of transport coefficients within linear-response theory in Sec.~\ref{sec:theory}. Then, we discuss how nontrivial
conservation laws can constrain the dynamics of current correlations, approaches based on Bethe ansatz, and generalized hydrodynamics in Sec.~\ref{sec:mazur}. 
In Sec.~\ref{sec:methods}, we cover recent developments in theoretical and numerical methods, which are intimately intertwined with the progress in the theory of finite-temperature transport.  
The introductory sections are concluded by Sec.~\ref{sec:open_systems} that discusses  open-quantum systems.
The readers who are  familiar with the theoretical background and the methods can immediately jump to Secs.~\ref{sec:xxz} --~\ref{sec:experiments},
which cover specific models and results.

We will extensively discuss the properties of the spin-1/2 XXZ chain and stress the importance of local and quasilocal conservation
laws in Sec.~\ref{sec:xxz}. 
Moreover, we will provide an overview over the established results and the open questions for the Hubbard chain in Sec.~\ref{sec:TransportHubbard}, while Sec.~\ref{sec:nonint} 
is devoted to transport in nonintegrable systems. Section \ref{sec:noneq} covers examples of far-from-equilibrium transport.

Finally, we  will provide a brief overview over key experimental results in Sec.~\ref{sec:experiments}. Besides experiments investigating the steady-state thermal conductivity in quantum magnets, these also include measuring spin diffusion using NMR methods and a more recent approach, namely
the driving of spin currents in quantum magnets via the Seebeck effect \cite{Hirobe2017}.
In parallel, ultracold quantum gases have emerged as an additional platform to investigate transport in one-dimensional lattice models [see, e.g., \cite{Vijayan2020,Ronzheimer2013,Hild2014,Xia2014}]. A major result is the
first  observation of ballistic nonequilibrium mass transport in a 1d integrable model of strongly interacting bosons \cite{Ronzheimer2013}.

\revision{The theoretical progress in characterizing the different spin-transport regimes in the spin-1/2 XXZ chain that include ballistic transport (i.e., finite Drude weights), diffusive and
superdiffusive dynamics have stimulated very recent experiments with both quantum magnets and quantum gases. A neutron-scattering study carried out in the high-temperature regime on KCuF$_3$ 
reports evidence for superdiffusive spin dynamics that is consistent with the Kardar-Parisi-Zhang  behavior \cite{Scheie2020}. 
A nonequilibrium optical-lattice experiment using ${}^7$Li atoms has investigated the crossover from ballistic transport to superdiffusion and diffusion in the same model
as a function of $\Delta$ \cite{Jepsen2020}}.

\section{Linear-response theory}
\label{sec:theory}

In most studies of transport in interacting 1d lattice quantum systems, the 
linear response is the  dominant approach. In the context of this review, one reason is that much of 
the focus has been on ballistic transport in integrable models which can be 
characterized by the so-called Drude weight, naturally appearing in linear 
response theory. One appealing aspect of linear-response theory is that correlation 
functions, in terms of which transport coefficients are expressed, and specifically 
their Fourier transformations (i.e., spectral functions) are readily accessible in 
various scattering experiments.

\subsection{Framework}

We are interested in the transport of conserved quantities. Specifically, we 
consider extensive quantities $Q$ which (i) are conserved, $[Q,H]=0$, and (ii) are 
expressed as a sum of local terms $q_r$ whose support is localized around the 
site $r$, $Q = \sum_r q_r$. These quantities are often referred to as 
``conserved charges''. If $Q$ is not conserved, one cannot, in the strict sense,
speak about transport because $Q$ is not just {\em transported} from one 
place to another, but is also locally generated. 
To be 
concrete, 
we will often refer to 
a typical local Hamiltonian $H=\sum_r 
h_{r,r+1}$, 
with $h_{r,r+1}$ given in Eq.~\eqref{eq:XXZ-intro}, i.e., the spin-$1/2$ XXZ 
chain. We shall focus on the two most 
local conserved quantities that are connected to global symmetries of the model: 
energy $q_r=h_{r,r+1}$ stems from the invariance under time translations, while 
conservation of magnetization or spin $q_r=\sz{r}$ is due to the $U(1)$ symmetry 
associated with rotations around the $z$ axis. 
For spin and energy, we have $Q=\Mtot=\sum_r\sz{r}$ and $Q=H$, respectively.

The definition of the 
corresponding local current $j^{(\Qcons)}_r$, where the superscript labels the conserved 
quantity $Q$,\footnote{For simplicity and in order to be consistent with the bulk of the literature in the field, we use the labels S and E for spin and energy, respectively.} follows from requiring the validity of a continuity equation and 
Heisenberg's equation of motion. For instance, take the total magnetization 
$\Mtot_{[l,l']}:=\sum_{r=l}^{l'} \sz{r}$ of a chain subsection with indices 
$\{ l,l+1,\ldots,l' \}$. The time derivative of $\Mtot_{[l,l']}$ should be 
given by the difference of local spin currents $\jS{r}$ flowing at the 
section's edge,
\begin{equation}
\frac{\di \Mtot_{[l,l']}}{\di t} + \jS{l'} - \jS{l-1} = 0 \, ,
\label{eq:cont}
\end{equation}
which together with Heisenberg's equation of motion $\dot{\Mtot}_{[l,l']} = \ii 
[H,\Mtot_{[l,l']}]$ naturally leads to the identification
\begin{eqnarray}
\label{eq:jS}
\jS{r} &:=& \ii [\sz{r},h_{r,r+1}] \\
&=& \Jex (\sx{r}\sy{r+1} - \sy{r}\sx{r+1}) \nonumber \, .
\end{eqnarray}
Similarly, energy conservation leads to the energy current $\jE{r}$
defined as
\begin{eqnarray}
\label{eq:jE}
\jE{r} &:=& \ii [h_{r-1,r},h_{r,r+1}] \\
&=& \Jex [\Delta( \jS{r-1}\sz{r+1} + \sz{r-1}\jS{r}) - \jS{r-1,r+1}\sz{r}] \, , 
\nonumber
\end{eqnarray}
where the explicit expression is again written for the XXZ model  
(\ref{eq:XXZ-intro}), and a two-index spin current is $\jS{r-1,r+1} := \Jex 
(\sx{r-1}\sy{r+1} - \sy{r-1}\sx{r+1})$. We note that the continuity equation 
(\ref{eq:cont}) does not uniquely define the current; one can always add a 
divergence-free operator (e.g., a constant). \revision{This  ambiguity does not 
affect the dc-conductivity, yet it may affect the finite-frequency behavior.} While energy and spin currents can be defined 
microscopically, a definition of ``heat'' requires an excursion into 
thermodynamics [see, e.g., \cite{ashcroft}], which is beyond the scope of this review.

Before writing down the linear-response expressions, let us give a simple {\it
classical}
example that illustrates their general form. Let us assume that we 
are following a particle with a coordinate $x(t)$ and are interested in the 
variance $\Sigma^2 := \ave{x^2(t)}$, where the average can be taken over 
different realizations of the stochastic trajectory $x(t)$ (or, e.g., the 
distribution of positions). Kinematics gives $x(t)=\int_0^t v(t_1)\di t_1$ and therefore,   
the variance becomes $\int_0^t\int_0^t \ave{v(t_1)v(t_2)}\di t_1 \di t_2$. 
Provided the process becomes stationary at long times and $\ave{v(t)} \to 0$, 
the correlation function will depend only on the time 
difference, $\ave{v(t_1)v(t_2)}=\ave{v(t_2-t_1) v(0)}$, \revision{leading to 
$\Sigma^2 \longrightarrow \int_0^t 2(t-\tau)\ave{v(\tau)v(0)}\di\tau$ in the 
long-time limit.} If in addition the correlation function decays to zero for 
large $\tau$ \revision{(which is assumed at this point but may not necessarily 
happen for a specific model)}, one finally gets
\begin{equation}
\Sigma^2 \stackrel{t\to\infty}{\longrightarrow} 2 \D t \, ,\quad \D := \int_0^\infty \ave{v(\tau)v(0)} 
\, \di\tau \, .
\label{eq:Dexact}
\end{equation}
The interpretation is very simple: the diffusion constant of the coordinate is 
given by an integral of an autocorrelation function of a ``coordinate current'' 
-- the velocity. This is the spirit of all linear-response formulae for 
transport coefficients and rests on simple kinematics or, equivalently, on 
the continuity equation for a conserved quantity. As we shall see, the same 
type of kinematic relation (an equality of the 2nd moment of the spatial
autocorrelation function and the integral of the current autocorrelation 
function) holds also in lattice systems (see Sec.\ \ref{sec:transport_inhom}). 
One remark is that the above derivation is exact because it involves the full 
non-equilibrium process $v(t)$, while in linear response, the validity is 
limited to small (gradients of) driving fields. 

Linear-response theory deals with the response of a system to an additional 
perturbation in the Hamiltonian. It sprouted up from studies conducted in the 1950s 
that connected equilibrium correlation functions and nonequilibrium properties, 
leading to the fluctuation-dissipation relation obtained by \cite{Callen51} and 
to Green-Kubo type formulae for transport coefficients obtained in 
\cite{Green52, Green54} and \cite{Kubo57} [for an early review, see 
\cite{Zwanzig65}].

The frequency-dependent conductivity $\Conductivity(\omega)$ is 
defined via a Fourier-space proportionality $\JQ(\omega)=\Conductivity(\omega) 
{\cal F}_\Qcons(\omega)$, where ${\cal F}_\Qcons(t)= 1/(2 \pi) \int_{-\infty}^\infty {\cal F}_\Qcons (\omega) \e^{-\ii 
\omega t} \di t$ is the driving field and $\JQ(\omega)$ is the extensive 
current, which in a lattice model is $\JQ(\omega) := \int_{-\infty}^\infty \JQ(t) 
\e^{\ii \omega t} \di t$ with
\begin{equation}
\JQ(t):=\sum_r j^{(\Qcons)}_r(t)
\end{equation} 
being a sum of local currents at lattice sites $r$. Note that here and in the following, 
we use the Heisenberg 
picture, i.e.,  $\JQ(t):=\e^{\ii H t} \JQ \e^{-\ii H t}$. One can think of the spin conductivity in the XXZ chain as a 
concrete example. In this case, $Q=\Mtot=\sum_r\sz{r}$, and the role of the driving field is played by 
\revision{the gradient of the} magnetic field. For the spin conductivity, we will use the following notation throughout this review:
\begin{equation}
\sigma(\omega) := \Onsager_{\rm SS}(\omega).
\end{equation}

Calculating the lowest-order response of the current operator to a Hamiltonian 
perturbation that consists of a linearly increasing potential corresponding to a homogeneous 
field $F$, or, equivalently, the linear perturbation of an equilibrium initial 
density operator, one gets the conductivity\footnote{For a concise derivation, see 
\cite{Kubo57} and for a more pedagogical exposition, see \cite{KuboII,Pottier10}.}
\begin{equation}
\label{eq:Kubo}
\Conductivity(\omega) = \beta \lim_{t\to \infty} \lim_{L \to 
\infty} \int_0^t \! \e^{\ii \omega \tau} \, \frac{K_{\JQ \JQ}(\tau)}{L} \, \di 
\tau \, ,
\end{equation}
\begin{equation}
K_{AB}(t) := \frac{1}{\beta} \int_0^\beta \! \ave{B \, A(t + \ii \lambda)} \, 
\di \lambda \nonumber \, ,
\end{equation}
where $K_{AB}(t)$ is the so-called Kubo (or canonical) correlation function  
with the bracket denoting the canonical average, $\ave{\bullet} := 
\tr{(\e^{-\beta H}\bullet)}/Z$, $Z:=\tr{(\e^{-\beta H})}$, and $\beta = 
1/T$ $(k_\mathrm{B} = 1)$. The conductivity $\Conductivity(\omega)$ has a standard 
form, being a Fourier transformation of the correlation function in 
Eq.~(\ref{eq:Kubo}).

The Kubo correlation function $K_{AB}(t)$ is real~\cite{Kubo57} for Hermitian  
$A$ and $B$ and therefore, $\Conductivity(\omega)$ is complex, $\Conductivity(\omega) := 
\Conductivity'(\omega) + \ii \, \Conductivity''(\omega)$, where $\Conductivity'(\omega) = 
\Conductivity'(-\omega)$ and $\Conductivity''(\omega) = -\Conductivity''(-\omega)$ (as well as 
$\Conductivity''(\omega > 0) \ge 0$). In the context of the electrical conductivity, where $Q$ is the electrical charge, $\Conductivity(\omega)$ is often called the optical 
conductivity because it can be probed with light-reflectivity measurements.\footnote{Energy scales of correlated electrons in most materials are of the order of 
electron volts (coinciding with visible light), the magnetic-field strength is 
negligible, and the penetration depth of light in a conductor $\sim 
1/\sqrt{\omega \mu_0 \Conductivity}$ ($\approx 2-20\,$nm) is larger than the lattice 
spacing ($\approx 0.5\,$nm) such  that one probes the zero-wavevector limit of $F(k 
\to 0)$ described by $\Conductivity(\omega)$.} The order of limits in  
Eq.~(\ref{eq:Kubo}) is important: if one takes the wrong order, taking 
the limit $t \to \infty$ first, one will probe the edge/finite-size effects 
instead of bulk physics.

In the classical limit $\hbar \to 0$, or in the high-temperature limit $\beta 
\to 0$, the Kubo correlation function goes to a classical correlation function, 
$K_{AB}(t) \to \ave{B\, A(t)}$ and therefore,   one gets a classical expression for the 
conductivity $\Conductivity(\omega)=\lim_{t \to \infty} \lim_{L \to \infty} 
\frac{\beta}{L} \int_0^t \e^{\ii \omega \tau} \ave{\JQ \,\JQ(\tau)}\di \tau$. The 
zero-frequency conductivity at infinite temperature $T\to \infty$ is therefore
\begin{equation}
\lim_{\beta \to 0} \frac{\Conductivity(0)}{\beta} = \lim_{t \to \infty} \lim_{L \to 
\infty} \frac{1}{L} \int_0^t \! \langle\JQ \JQ(\tau)\rangle \, \di \tau \, ,
\label{eq:sigma-Inf}
\end{equation}
This infinite-temperature limit will frequently be referred to          
in this review.
Instead of the Kubo correlation $K_{\JQ\JQ}(t)$, one can also express 
Eq.~(\ref{eq:Kubo}) in terms of other types of correlation functions. For instance, 
one has the relation~\cite{Pottier10} $K_{AB}(\omega) = 2/(\beta 
\omega) \, \xi_{AB}(\omega)$ with the spectral function $\xi_{AB}(t) := 
1/2 \, \ave{[A(t),B]}$. Because $K_{\JQ \JQ}(t)$ is real and even, 
$K_{\JQ \JQ}(\omega)$ is real as well and can be written as 
$K_{\JQ \JQ}(\omega) = 2 \int_0^\infty \cos{(\omega t) K_{\JQ\JQ}(t) \di 
t}$. Such a ``one-sided'' Fourier transformation is exactly what is needed for 
$\Conductivity'(\omega)$ in Eq.~(\ref{eq:Kubo}), resulting in the real part of 
the conductivity
\begin{equation}
\Conductivity'(\omega) = \frac{\ii}{\omega} \int_0^\infty \! \lim_{L \to 
\infty} \frac{\sin{(\omega \tau)}}{L} \ave{[\JQ(\tau),\JQ]} \, \di \tau \, ,
\label{eq:KuboCom}
\end{equation}
where we have used that $\xi_{\JQ\JQ}(t)$ is odd and performed the limit $t \to 
\infty$. Similarly, $K_{AB}(\omega) = (1-\e^{-\beta \omega})/(
\beta \omega) \, C_{AB}(\omega)$, where $C_{AB}(t) := \ave{A(t) \, B}$, 
leading to equivalent expressions
\begin{equation}\begin{split}
\label{eq:KuboJJ}
\Conductivity'(\omega) &= \frac{1-\e^{-\beta \omega}}{\omega} 
\int_{0}^\infty \! \lim_{L \to \infty} \, \frac{\rp \! \left( 
\e^{\ii \omega \tau} \ave{\JQ(\tau)\JQ} \right)}{L} \, \di \tau \\
&= \frac{2\,\mathrm{th}{(\frac{\beta \omega}{2})}}{\omega} \int_{0}^\infty 
\! \lim_{L \to \infty} \frac{\cos{(\omega \tau)}}{L} \, \rp
\ave{\JQ(\tau)\JQ} \, \di \tau \, . 
\end{split}\end{equation}
The imaginary part $\Conductivity''(\omega)$ can be obtained using Kramers-Kronig  
(Plemelj-Sokhotski) relations~\cite{Goldbart} or the fluctuation-dissipation 
theorem.

If $H$ conserves the total number of particles, so does the current $\JQ$, and 
therefore, the same expression holds also for a grandcanonical average with the 
density operator $\rho \sim \e^{-\beta(H-\mu N)}$. In case the average current 
is not zero, $\ave{\JQ}\neq 0$, which, for instance, happens if the total momentum is conserved, 
one has to take the connected correlation function or work in an ensemble with 
zero total momentum. \revision{For a detailed discussion and definition of corresponding
connected correlation functions, we refer to \cite{Lebowitz00,lepri2003}.}

The linear-response formulae for the specific case of energy transport are 
somewhat trickier to derive as there is no obvious microscopic driving 
potential \cite{Zwanzig65} [see also, e.g., \cite{gemmer06} for studies in 
concrete systems], 
such as, e.g., the  magnetic or electric field for magnetization or particle 
transport. The driving force is the gradient of the inverse temperature which
is a thermodynamic quantity and not a microscopic one. This is connected to the 
fact that the Hamiltonian, whose expectation value is the energy, is itself the 
generator of dynamics and therefore plays a special role in thermodynamics. 
Nevertheless, one can, for instance, identify a perturbation ``Hamiltonian'' 
that is equivalent to a thermal perturbation, ultimately leading to the same 
Green-Kubo type expression~\cite{luttinger64, Pottier10} as for the generic conductivity $\Onsager_{QQ}(\omega)$  discussed above. 
Defining the energy-transport coefficient $\kappa(\omega)=\beta\Onsager_{\rm EE}(\omega)$ as the 
proportionality factor of the energy current, $\JE(\omega) = -\kappa(\omega) 
\nabla T(\omega)$ (at vanishing expectation value of the particle 
current), one gets
\begin{equation}
\kappa(\omega)=\beta\Onsager_{\rm EE}(\omega)=\!\!\lim_{t\to \infty}\lim_{L \to \infty} \beta^2 
\int_0^t \! \e^{\ii \omega \tau} \frac{K_{\JE \JE}(\tau)}{L} \, \di \tau  \, .
\label{eq:KuboE}
\end{equation}
The difference compared to the  conductivity  given in Eq.~(\ref{eq:Kubo}) is an additional factor of 
$\beta = 1/T$ stemming from the fact that $\kappa$ is the proportionality 
factor between current and $\nabla T$ instead of $\nabla T/T$.

In general, one can also have nonzero cross-transport coefficients, in 
which case one has to deal with the whole Onsager matrix\footnote{Note 
that $\Onsager$ and $\tilde \Onsager$ differ by a factor of $\beta$.} $
\ConductivityOff$. In order to ensure that the  matrix $\tilde \Onsager$ has the correct symmetry,\footnote{For a time-reversal invariant system and observables with the same parity 
under time reversal, $\tilde \Onsager$ is symmetric.} one has to be 
careful~\cite{Pottier10,Mahan} with the choice of driving forces
$\cal{F}_\Qcons$ which are equal to gradients of intensive quantities obtained 
by entropy derivatives. One way is to start from  the entropy production rate $\di 
s/\di t=\sum_Q \JQ {\cal F}_\Qcons/L$ from which one can identify 
currents $\JQ$ and corresponding forces ${\cal F}_\Qcons$. To linear order, the 
relations between currents and forces take the form
\begin{equation}
\JQ = \sum_{Q'} \ConductivityOff \, {\cal F}_{\Qcons'} \, .
\label{eq:Onsager}
\end{equation}
Since the entropy production rate is $\sum_{Q,Q'} \ConductivityOff {\cal F}_\Qcons 
{\cal F}_\Qconsp$, the Onsager matrix has to be positive semidefinite, 
$\tilde \Onsager \ge 0$. Using Hamiltonian linear-response theory, 
$\ConductivityOff $ are given by the Kubo correlation function 
$K_{\JQ  \dot{A}_{\Qcons}}$, where $A_\Qcons$ is the operator 
coupled to ${\cal F}_\Qcons$. For instance, one has $\dot{A}_{\rm E}=T \JE$ and 
${\cal F}_{\rm 
E}=\nabla(\frac{1}{T})$ for energy transport and $\dot{A}_{\rm S}=T\JS$ and 
${\cal F}_{\rm S}=-\nabla(\frac{b}{T})$ for spin  transport ($b$ is the 
magnetic field), so that zero-frequency transport coefficients can be 
written as
\begin{equation}
\ConductivityOff = \int_0^\infty \lim_{L \to \infty} 
\frac{K_{{\JQ}\JQp}(t)}{L} \, \di \tau \, .
\label{eq:Lad}
\end{equation} 
In the uncoupled case, \revision{i.e.,\ $\tilde \Onsager_{\rm ES}\equiv 0$}, one 
has $\kappa = 
\tilde \Onsager_{\rm EE}/T^2$ and $\sigma = \tilde\Onsager_{\rm SS}/T$, recovering the 
previous expressions (\ref{eq:KuboE}) and (\ref{eq:Kubo}). 

The conductivity satisfies various sum rules -- formulae expressing moments of 
$\Conductivity(\omega)$ in terms of correlation functions (or derivatives thereof) at 
$t=0$. They are mostly useful in phenomenological theories as well as in 
experiments because they represent rigorous constraints on $\Conductivity(\omega)$, for 
instance, on the large-frequency behavior. For their form see, e.g., 
\cite{Pottier10}. \revision{A particularly simple example is
\begin{equation}
\int_{-\infty}^\infty \di \omega \lim_{\beta \to 0} 
\frac{\Conductivity(\omega)}{\beta} = \pi \lim_{L \to \infty} \frac{\tr{[\JQ 
\JQ(t = 0)]}}{L\,Z_0}  \, .
\end{equation}}
For sum rules for the thermal conductivity $\kappa(\omega)$ 
see, e.g., \cite{shastry06}.

Linear response is limited to sufficiently small driving 
fields. While the range of validity of linear response is system-specific, let 
us briefly comment on the validity of perturbation theory used in its 
derivation. One can argue~\cite{KuboII} that
linear response should not work since the microscopic evolution 
is, in general, unstable against perturbations. This applies, in particular,  to the limit $t \to 
\infty$ needed to evaluate the conductivity.  
The point is rather subtle: it is true that for generic observables and initial 
(pure) states, perturbation theory will fail, yet nevertheless, in the 
linear-response regime we are interested in smooth observables and very specific 
states -- the equilibrium density matrices. A perturbation will change 
microscopic dynamics and potentially even make it chaotic, but this very same 
chaoticity also guarantees that at long times, the system will locally 
self-thermalize such that the density matrix will change little. In short, a 
generic system with good thermalization properties is microscopically 
unstable but macroscopically stable \cite{Dorfman}.

\subsection{Ballistic versus diffusive transport in the context of current 
correlations}
\label{sec:theory.ballistic}

In this section, we discuss the small-frequency behavior of transport coefficients. This is of special importance 
because the limit $\omega \to 0$ probes the slowest long-wavelength modes that 
are often of a hydrodynamic nature (note that we also implicitly take momentum
$k\to 0$, preceeding frequency $\omega \to 0$). Here and in Sec.~\ref{sec:transport_inhom}, we exclusively focus on the case of spin transport, $\sigma(\omega)=\Onsager_{\rm SS}(\omega)$.

Of particular interest is the real part of the conductivity $\sigma(\omega)$, 
\revision{the imaginary part being zero, $\sigma''(0)=0$, due to the symmetry
$\sigma''(\omega)=-\sigma''(-\omega)$.}
It can happen that $\sigma'(\omega \to 0)$ diverges. To 
this end, it is useful to decompose $\sigma'(\omega)$ into a singular and a 
regular part, 
\begin{equation}
\sigma'(\omega) := 2 \pi \, \Dws \, \delta(\omega) + \sigmar(\omega) \, ,
\label{eq:sigmar}
\end{equation}
where the
prefactor $\Dws$ is called the Drude weight~\cite{kohn64, 
Scalapino92}. We use the symbol $\Dws$ to distinguish it from the diffusion constant $\DS$. 
In older literature, it is often called spin stiffness \cite{Shastry90}. 
Alternatively, using Kramers-Kronig relations, one can see that 
$\sigma''(\omega \to 0) = \lim_{\epsilon \to 0} \frac{2 \omega}{\omega^2 + 
\epsilon^2} \Dws$ and therefore, $\Dws = \lim_{\omega \to 0^+} \frac{\omega}{2} 
\sigma''(\omega)$.

To get an idea of  the typical behavior of $\sigma(\omega)$, it is instructive 
to have a look at the simple Drude model of conduction~\cite{ashcroft}.\footnote{To this end, we make use of the mapping of spin-1/2 degrees of freedom to spinless fermions via the Jordan-Wigner transformation.}
The original Drude model 
consists of classical charged particles that are accelerated by the electric 
field and damped by a force proportional to their velocity. One gets 
$\sigma(\omega)=\frac{\sigma_0}{1-\ii \omega \tau}$, where $\tau$ is the 
relaxation (damping) time and $\sigma_0:=ne^2\tau/m$, with $m$ being the mass 
and $n$ the carrier density.\footnote{In good conductors at room temperature, $\tau \sim 
10^{-14} \, {\rm s}$, corresponding to a mean-free path of a few lattice 
spacings.} The real part is therefore $\sigma'(\omega) = \frac{\sigma_0}{1 + 
\omega^2 \tau^2}$, while the imaginary part is $\sigma''(\omega) = 
\frac{\sigma_0 \omega \tau}{1 + \omega^2 \tau^2}$. At finite $\tau$, one has 
diffusive transport with a Lorentzian $\sigma'(\omega)$ corresponding to an 
exponential decay of the autocorrelation function $C_{\JS\JS}(t)\sim 
\e^{-t/\tau}$. In the limit of no relaxation, $\tau \to \infty$, $\sigma'$ 
diverges as $\sim \tau$ at its peak at $\omega = 0$, resulting in a nonzero 
Drude singularity $\sigma'(\omega \to 0) \to 2\pi \Dws \delta(\omega)$, with 
$\Dws=n e^2 / (2m)$. In the opposite limit of fast relaxation, $\tau \to 0$, 
where the autocorrelation function is $C_{\JS\JS}(t) \sim \delta(t)$, one gets a 
broad ``white noise'' conductivity $\sigma(\omega) = {\rm const}$.

\begin{figure}[t]
\begin{center}
\includegraphics[width=0.90\linewidth]{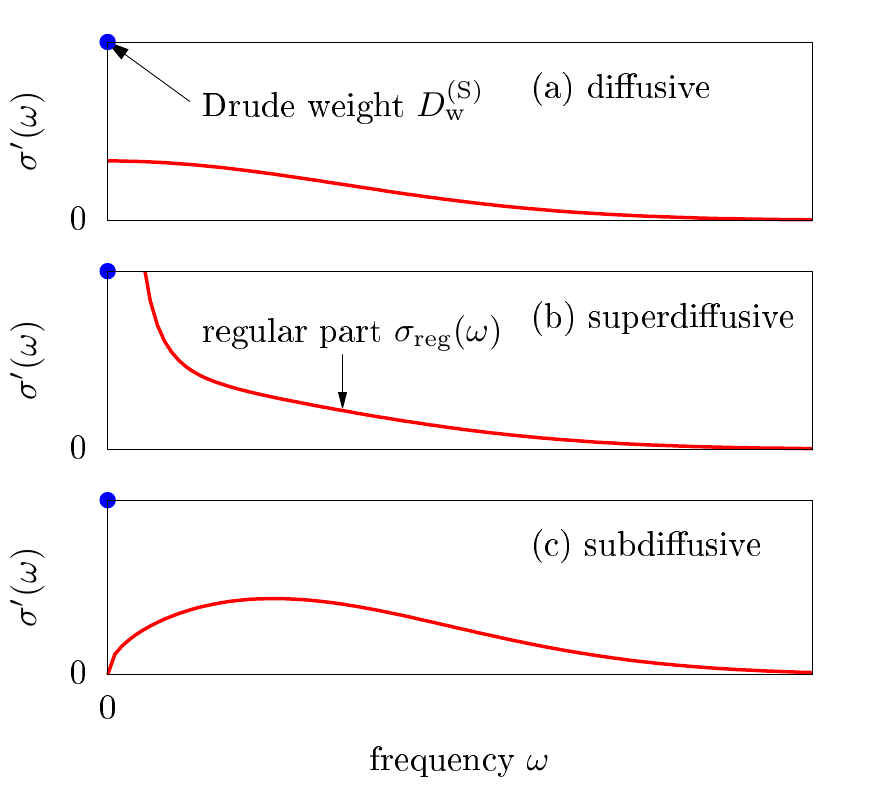}
\caption{(Color online) Sketch of the three different scenarios that one can 
envision for the behavior of the 
regular part of optical conductivity $\sigma_\textnormal{reg}(\omega)$ (solid lines) at finite 
temperature. The point at $\omega=0$ indicates the Drude weight, which may coexist with a nonzero regular part.
}
\label{sketch_sigma} 
\end{center}
\end{figure}

The definition of the Drude weight by Eq.~(\ref{eq:sigmar}) is per se not 
unique. namely, for a physicist the Dirac delta function means just a 
singularity without specifying its type, with different possible  representations. 
The singularity can be characterized with a scaling exponent $\alpha$ 
as,
\begin{equation}
\sigma'(\omega \to 0) \sim |\omega|^\alpha \, .
\label{eq:alpha}
\end{equation}
We shall use a self-consistent convention where the singularity with $\alpha = 
-1$ (like in the Drude model in the limit of zero relaxation) is put into the 
Dirac delta, while weaker (integrable) singularities with $-1 < \alpha < 0$ are 
retained in $\sigmar$. Note that in systems with a bounded local Hilbert space 
(or in 
an unbounded one at finite energy density) the singularity cannot be stronger 
than $1/|\omega|$. That is, if one splits the correlation function
\begin{equation}\begin{split}
C_{\JS\JS}'(t) :& = \frac{\rp \langle \JS(t) \JS \rangle}{L}\\ & = \frac{\langle \JS(t) \JS 
\rangle + \langle \JS(-t) \JS \rangle}{2L}
\end{split}\end{equation}
as $C'_{\JS\JS}(t):= \Cbar_{\JS\JS} + \tilde C'_{\JS\JS}(t)$ into the average $\Cbar_{\JS\JS} := \lim_{t 
\to \infty}\frac{1}{t} \int_0^t\! C'_{\JS\JS}(\tau)\di \tau$ and an oscillating part 
$\tilde C'_{\JS\JS}(t)$, the Green-Kubo formula (\ref{eq:KuboCom}) gives
\begin{equation}\begin{split}
\sigma'(\omega) & = \beta \pi \Cbar_{\JS\JS} \delta(\omega)\\& + 
\frac{2\,\mathrm{th}{(\frac{\beta \omega}{2})}}{\omega} \int_0^\infty \!
\cos{\omega \tau} \, \tilde C'_{\JS\JS}(\tau)  \, \di \tau  \, .
\label{eq:DrudeC}
\end{split}\end{equation}
Comparing with Eq.~(\ref{eq:sigmar}), we see that
\begin{equation}
\Dws = \frac{\beta}{2}\,  \Cbar_{\JS\JS} \, .
\label{eq:DwC}
\end{equation}
$\sigma'(\omega)$ can now be used to classify transport, originally used at 
zero temperature \cite{Shastry90, Scalapino92, Scalapino1993}. Since the Drude weight
$\Dws$ \revision{in Eq.~\eqref{eq:DwC}} trivially vanishes in the 
high-temperature 
limit $\beta \to 0$, a suitable quantity for the classification of transport is 
not $\Dws$ itself but rather the quantity
\begin{equation}
\Dwsbeta := \frac{\Dws}{\beta} \, .
\end{equation}
If $\Dwsbeta \neq 0$, i.e., $\alpha = -1$, one speaks of an ideal conductor, which we will refer to as ballistic transport. 
If $\Dwsbeta = 0$, one can distinguish three situations, 
see Fig.\ \ref{sketch_sigma}: 
(i) if $0 < \sigmar(0) / \beta < \infty$, 
i.e., $\alpha = 0$, the system is a normal, diffusive conductor; \revision{(ii) if 
$\sigmar(\omega \to 0) / \beta \to \infty$, i.e., $-1 < \alpha < 0$, one has 
superdiffusion; (iii) if $\sigmar(0) / \beta=0$, i.e., $\alpha>0$, one has 
subdiffusive transport (including the extreme case of localization).}
\revision{If $\Dwsbeta \neq 0$, the transport types (i)-(iii) must be 
understood as subleading corrections to ballistic transport.}

In the case (i) above one obtains  a finite diffusion constant. While 
$\sigmar(\omega)$ is a microscopic quantity, this is not the case for the 
diffusion constant and one has to define it in terms of an appropriate phenomenological 
macroscopic relation. A common way is via Fick's law, 
\begin{equation}
\JS = -\DS \, \nabla \Mtot \, ,
\label{eq:Fick}
\end{equation}
where $\DS$ is the spin-diffusion constant. We can express it with $\sigmar(\omega 
\to 0)$ using Eqs.\ (\ref{eq:Onsager}) and (\ref{eq:Lad}). At fixed $T$, we 
namely also have $\JS=-\Onsager_{\rm SS} \nabla(b)/T$, which, after equating it 
with Fick's 
law, gives the spin-diffusion constant
\begin{equation}
\DS = \frac{\sigma_{\rm reg}(0)}{\partial \Mtot/\partial b} \, ,
\label{eq:Ds}
\end{equation}
where $b$ is the magnetic field. The denominator is equal to the static
spin susceptibility, $\frac{\partial \Mtot}{\partial b} := \chi = \frac{\beta}{L} 
[\ave{(\Mtot)^2} - \ave{\Mtot}^2]$, which is, at infinite temperature, equal to 
$\chi=\beta/4$, and in turn, the diffusion constant at infinite $T$ is
\begin{equation}
\DS = 4 \lim_{t \to \infty} \lim_{L \to \infty} \frac{1}{L} \int_0^t 
\! \langle\JS \JS(\tau)\rangle \,\di \tau \, .
\label{eq:D-Inf}
\end{equation}
We stress that Eq.\ (\ref{eq:D-Inf}) holds in the case of a vanishing Drude 
weight only.

The Drude weight can also be connected to the sensitivity 
of the spectrum to a threading flux $\phi$, in essence probing the sensitivity 
to boundary conditions. This was originally used for the ground state~\cite{kohn64} and 
extended to finite $T$ in \cite{Castella1995}, leading to
\begin{equation}
\Dws = \frac{1}{L} \sum_\alpha p_\alpha \frac{1}{2} \left. \frac{\di^2 
E_\alpha}{\di \phi^2} \right \rvert_{\phi=0} \, ,
\label{eq:Kohn}
\end{equation}
where $E_\alpha$ are eigenenergies and $p_\alpha:=\e^{-\beta E_\alpha}/Z$ are 
the Boltzmann weights. Completely analogous Drude weights can be defined 
for transport of other quantities as well.

A finite Drude weight implies that the current autocorrelation 
function exhibits a plateau at long times. Such a nonzero plateau is typically an indication 
of a conserved quantity. Indeed, it is intuitively clear that a conserved 
operator that has a nonzero overlap with the current operator causes a plateau 
in the current autocorrelation function. The argument can be formalized in the 
form of the so-called Mazur (in)equality, first studied 
in~\cite{mazur69,suzuki71}, that bounds the time-averaged autocorrelation 
$\Cbar$ by constants of motion. One has
\begin{equation}
 \Cbar_{\JS\JS} \ge \sum_n \frac{1}{L}\frac{\ave{\JS 
\CCharge}^2}{\ave{\CCharge^2}},
\label{eq:MazurC}
\end{equation}
where the sum runs over Hermitian constants of motion $\CCharge$, 
$[\CCharge,H]=0$, that are chosen to be orthogonal, $\ave{\CCharge 
\CChargem}\propto \delta_{nm}$. The equality in \eqref{eq:MazurC} holds if the 
sum is over \revision{(a complete set of)} all $\CCharge$. The bracket is a standard canonical average. 
However, if one wants to bound the Kubo autocorrelation function, one uses the 
Kubo-Mori~\cite{Mori65} (also called Bogoliubov) inner 
product \revision{$K_{\JS\CCharge}(0)$ as defined in Eq.~(\ref{eq:Kubo}).}
Mazur's inequality (\ref{eq:MazurC}), together with Eq.~(\ref{eq:DrudeC}), can 
be used to bound the Drude weight from below~\cite{Zotos1997},
\begin{equation}
\Dws \ge \frac{\beta}{2}\, \lim_{L\to \infty} \sum_n \frac{1}{L}\frac{\ave{\JS  
\CCharge}^2}{\ave{\CCharge^2}}.
\label{eq:MazurD}
\end{equation}
We remark that simply using a complete set of eigenstate projectors as 
$\CCharge$ in Eq.\ (\ref{eq:MazurD}) of course does not work because the right 
hand side is zero since the sum is exponentially small in $L$. The important conserved quantities 
are (quasi)local conserved $\CCharge$ for which overlaps are not necessarily 
exponentially small.

For anomalous superdiffusive transport, the Drude weight is zero 
but the decay of the autocorrelation function is slow, resulting in a diverging 
diffusion constant $\D^{(S)}$. We note that in such anomalous cases, the application 
of the linear-response formula is in practice not straightforward 
\cite{berciu10,dhar09}.

Above, we discussed the effect of exact conservation laws, captured via Mazur's inequality.
Weakly violated or approximately conserved quantities may also affect the long-time decay of 
current autocorrelation functions [see, e.g.,  \cite{Rosch2006} for a discussion].

\subsection{Time evolution of inhomogeneous densities}

\subsubsection{Generalized Einstein relations}
\label{sec:transport_inhom}

Another widely used approach to study transport (we again focus exclusively on the spin case) is to prepare a non-equilibrium 
initial state
\begin{equation}
\rho \neq \rho_\text{eq}
\end{equation}
and to follow the dynamics of expectation values
\begin{equation}
\langle \delta \sz{r}(t) \rangle = \text{tr} [ \rho(t) \, \delta \sz{r} ] \, ,
\end{equation}
where $\rho(t) = e^{-i H t} \, \rho \, \e^{i H t}$ is the unitary 
time evolution in an isolated quantum system governed by $H$ and $\delta \sz{r} = 
\sz{r} - \langle \sz{r} \rangle_\text{eq}$ measures the deviation of the local 
density $\sz{r}$ from its value $\langle \sz{r} \rangle_\text{eq}$ at equilibrium. In 
such a situation, a large variety of different initial states can be prepared: 
They can be mixed or pure, entangled or non-entangled, close to or far away 
from equilibrium, e.g., as resulting from sudden quenches or from joining two 
semi-infinite chains at different equilibrium [see Sec.\ 
\ref{sec:bipartitioning}]. Various initial profiles can be realized as well: They 
can be spatially localized, domain walls, staggered, etc. We stress that the 
situations considered in this subsection are not necessarily limited to the linear-response regime
and are therefore more general.

A general strategy to analyze the dynamical behavior is given by the spatial 
variance
\begin{equation}
\Sigma^2(t) = \sum_{r} \frac{\langle \delta \sz{r}(t) \rangle }{\langle \delta \Mtot 
\rangle} \, r^2 - \Big [\frac{\langle \delta \sz{r}(t) \rangle}{\langle \delta \Mtot 
\rangle} \, r \Big ]^2 
\label{variance}
\end{equation}
with the time-independent sum $\langle \delta \Mtot \rangle = \sum_r \langle \delta 
\sz{r}(t) \rangle$, i.e., $\sum_r \langle \delta \sz{r}(t) \rangle/\langle \delta \Mtot 
\rangle = 1$ is properly normalized, and we assume $\langle \delta \sz{r}(t) 
\rangle > 0$. Thus, the spatial variance yields information on the overall 
width of the profile. In the case that  diffusive dynamics is realized at {\em all} times,
\begin{equation}
\frac{\text{d}}{\text{d} t} \Sigma^2(t) = 2 \, \DS \, . 
\label{variance_diffusion}
\end{equation}
Here, the quantity $\DS$ is a time- and space-independent diffusion constant.

In general, the spatial variance in Eq.\ (\ref{variance}) is unrelated to the 
linear-response functions discussed in the previous sections. However, a 
relation can be derived if the initial state $\rho$ is close enough to the
equilibrium state $\rho_\text{eq}$. To this end, consider the specific 
non-equilibrium state
\begin{equation}
\rho \propto \exp \Big [-\beta \Big (H - \varepsilon \sum_r p_r \, \sz{r} \Big) 
\Big ] \, , 
\label{Kubo_state}
\end{equation}
i.e., a thermal state of the Hamiltonian $H$ but now with an additional 
potential $\sum_r p_r \, \sz{r}$ of strength $\varepsilon$. As shown by 
Kubo \cite{KuboII}, Eq.\ (\ref{Kubo_state}) can be expanded in $\varepsilon$ as
\begin{equation}
\rho = \rho_\text{eq} \Big [1 + \varepsilon \int_0^\beta \text{d} \beta' 
\sum_r 
p_r \, e^{\beta' H} \, \delta \sz{r} \, e^{-\beta' H}  + {\cal O}(\varepsilon^2 )
\Big ] \, .
\end{equation}
If $\varepsilon$ is a sufficiently small parameter, the expansion can be 
truncated to linear order. Using this truncation, the expectation values 
$\langle \delta \sz{r} 
(t) \rangle$ become
\begin{equation}
\langle \delta \sz{r} (t) \rangle = \varepsilon \, \beta \, \sum_{r'} p_{r'} \, 
K_{\delta \sz{r'} \delta \sz{r}}(t) \, .
\end{equation}
Assuming that $\langle \delta \sz{r} (t) \rangle$ remains negligibly small at 
the boundary of the lattice, the time derivative of the spatial variance can be 
written in the form \cite{bohm1992, Steinigeweg2009a, yan2015}
\begin{equation}
\frac{\text{d}}{\text{d} t} \Sigma^2(t) = 2 \, \DS(t) \, ,
\end{equation}
where the time-dependent diffusion constant is given by the relation
\begin{equation}
\DS(t) = \frac{\beta}{\chi} \int_0^t \text{d}t' \, K_{\JS \JS}(t') 
\label{Einstein_relation}
\end{equation}
with the static susceptibility 
\begin{equation}
\chi = \beta \, K_{\delta \Mtot \, \delta \Mtot} \, .
\end{equation}
As mentioned above, one has $\chi/L =\beta/4$ for the specific case of high 
temperatures.

Equation~(\ref{Einstein_relation}) is a generalized Einstein relation 
as it holds for any time $t$. In particular, in the long-time limit $t \to \infty$, it 
simplifies to the usual Einstein relation, if the current autocorrelation 
function decays sufficiently fast to zero,
\begin{equation}
\lim_{t \to \infty} \DS(t) = \DS = \frac{\sigma_\text{dc}}{\chi/L} \, ,
\label{eq:Ds_2}
\end{equation}
where $\sigma_\text{dc}$ is the dc conductivity as obtained from linear 
response theory, \revision{i.e., Eq.\ (\ref{eq:Ds_2}) is identical to Eq.\ 
(\ref{eq:Ds}).} Therefore, the existence of $\sigma_\text{dc}$ implies a 
diffusive scaling of the spatial variance in time, at least for the specific 
initial state $\rho$ in Eq.\ (\ref{Kubo_state}) with a small parameter 
$\varepsilon$. However, it is worth pointing out that the requirement of a 
strictly mixed state can be relaxed by employing the concept of typicality 
(see Sec.\ \ref{sec:typicality}).

Since the generalized Einstein relation is neither restricted to the limit of 
large times nor to the case of diffusion, it allows one to investigate both 
different time scales and different types of transport. For example, it 
predicts a ballistic scaling $\DS(t) \propto t$ and $\Sigma^2(t) \propto t^2$ at 
short times $t \ll \tau$, before a diffusive scaling $\DS(t) = \DS$ and 
$\Sigma^2(t) \propto t$ may finally set in at intermediate times $t > \tau$. 
Remarkably, it also captures the influence of a Drude weight $\Dws > 0$. A 
finite Drude weight $\Dws > 0$ implies a ballistic scaling
\begin{equation}
\DS(t) \propto \frac{\Dws}{\chi/L} \, t
\end{equation}
and $\Sigma^2(t) \propto t^2$ at large times.

\begin{figure}[tb]
\includegraphics[width=0.9\columnwidth]{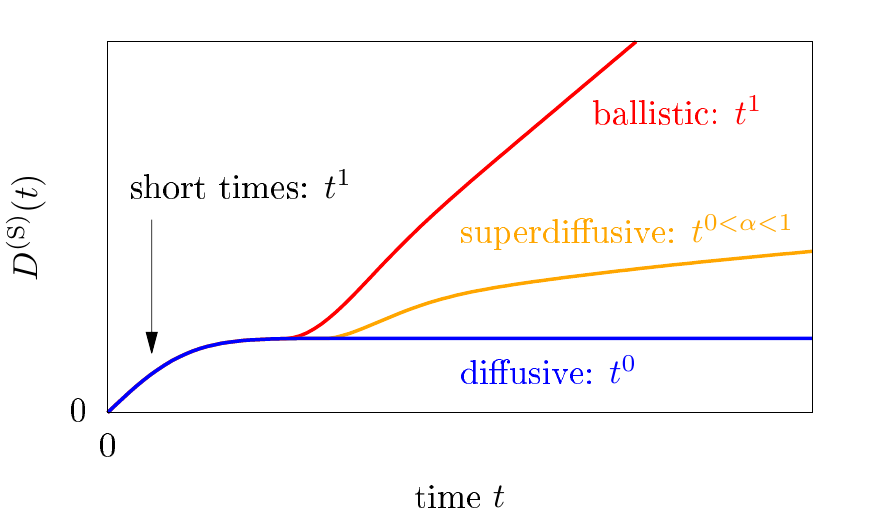}
\caption{(Color online) Sketch of different scenarios for the time-dependent 
diffusion constant $\DS(t)$: ballistic,  superdiffusive, and diffusive (top to bottom). The behavior 
in the short-time limit is always ballistic, the typical exponents in the long-time dynamics are indicated as $t^1$, $t^\alpha$ ($0<\alpha<1$), and $t^0$, respectively.}
\label{sketch_D}
\end{figure}

Finally, we remark that a power-law scaling of
\begin{equation} \label{eq:alpha2}
\Sigma^2(t) \propto t^{\alpha'}
\end{equation}
indicates subdiffusion for $0 < \alpha' < 1$ and superdiffusion
for $1 < \alpha' < 2$ (see Fig.\ \ref{sketch_D}). Due to the generalized 
Einstein relation in Eq.\ (\ref{Einstein_relation}), such a power-law scaling 
in time also implies that the frequency dependence of the conductivity 
$\sigma'(\omega)$ is given by the power law \cite{maass1991, dyre2009, 
Stachura2015, luitz2017}
\begin{equation} 
\sigma'(\omega) \propto \int_0^\infty \text{d}t \, e^{\ii \omega t} \, 
t^{\alpha'-2} \propto |\omega|^{1-\alpha'} \, ,
\end{equation}
i.e., Eq.\ (\ref{eq:alpha}) with $\alpha = 1 - \alpha'$.

\subsubsection{Diffusion}

While the spatial variance in Eq.\ (\ref{variance}) is a useful quantity to 
study transport, it yields no information beyond the overall width of a 
profile. In particular, in order to draw reliable conclusions on the existence of 
diffusion, it is necessary to require the full spatial dependence of a profile 
to be described by the diffusion equation. In one dimension and for a 
discrete lattice, the diffusion equation reads
\begin{equation}
\frac{\text{d} \langle \delta \sz{r}(t) \rangle}{\text{d}t} \!=\! \DS \Big
[ \! \langle \delta \sz{r-1}(t) \rangle - 2 \langle \delta \sz{r}(t) \rangle + 
\langle \delta \sz{r+1}(t) \rangle \! \Big ] \, , \label{diffusion_equation}
\end{equation}
where $\D^{(S)}$ again denotes a time- and space-independent diffusion constant  
and the right-hand side can be viewed as a discretized version of the Laplacian 
$\partial^2/\partial r^2$. It is important to note that Eq.\ 
(\ref{diffusion_equation}) is a phenomenological description for the expectation 
values $\langle \sz{r} (t) \rangle$ and their irreversible relaxation towards 
equilibrium. A rigorous justification is still  a challenge to theory 
\cite{casati1984, Lebowitz00, lepri2003, michel2005, buchanan2005, dhar2008,
Ljubotina2017, Steinigeweg2017, Steinigeweg2017b}. Within such a description 
and in the following discussion, one does not need to specify the initial state in 
detail, however, note that this statistical description is often discussed in 
the context of correlation functions \cite{kadanoff1963, steiner1976, 
forster1990}. \revision{We stress that the diffusion in Eq.\
(\ref{diffusion_equation}) is a statistical process starting at time $t = 0$
and occurring between individual lattices sites, i.e., it implicitly assumes
a mean-free time $\tau = 0$ and a mean-free path $l=0$. Since $\tau > 0$ and
$l>0$ in specific models, it can only hold when the density profile has become
sufficiently broad. In terms of the density modes discussed below, statistical
behavior is thus restricted to sufficiently small momenta.}

For a local injection at some site $r'$, i.e., $\langle \delta \sz{r = 
r'}(0) \rangle \neq 0$ and $\langle \delta \sz{r \neq r'}(0) \rangle = 0$, the 
solution of Eq.\ (\ref{diffusion_equation}) reads
\begin{equation}
\frac{\langle \delta \sz{r}(t) \rangle}{\langle \delta \Mtot \rangle} = \exp(-2 \DS t) 
\, I_{r-r'}(2 \DS t)\ , 
\label{Bessel}
\end{equation}
where $I_r(t)$ is the modified Bessel function of the first kind and order $r$. 
This lattice solution can be well approximated by the corresponding continuum 
solution
\begin{equation}
\frac{\langle \delta \sz{r}(t) \rangle}{\langle \delta \Mtot \rangle} = 
\frac{1}{\sqrt{2\pi} \, \Sigma(t)} \exp \Big [ - \frac{(r -r')^2}{2 
\, \Sigma^2(t)} \Big] \, , \label{Gaussian}
\end{equation}
where the spatial variance $\Sigma^2(t) = 2 \, \DS \, t$ has been introduced in 
the previous section. Thus,
\begin{equation}
\langle \delta \sz{r = r'}(t) \rangle \propto \frac{1}{\Sigma(t)} \propto 
\frac{1}{\sqrt{t}} \, .
\end{equation}
Obviously, since the diffusion equation is a linear differential 
equation, the general solution can be constructed as a superposition of 
$\delta$ injections at different positions.

At this point, it is certainly instructive to provide a link to correlation 
functions. To this end, consider the specific initial state $\rho$ in Eq.\ 
(\ref{Kubo_state}) with coefficients $p_{r = r'} \neq 0$ and $p_{r \neq r'} 
= 0$.
For 
$\varepsilon$ sufficiently small, the expectation values $\langle \delta \sz{r} 
(t) \rangle$ become
\begin{equation}
\langle \delta \sz{r} (t) \rangle = \varepsilon \, \beta \, p_{r'} \, K_{\delta 
\sz{r'} \delta \sz{r}}(t) \, . \label{local_correlation}
\end{equation}
For high temperatures, $K_{\delta \sz{r'} \delta \sz{r}}(0) \propto 
\delta_{r,r'}$, and in the case of diffusion, $\langle \delta \sz{r} (t) \rangle$ satisfies Eqs.\ (\ref{Bessel}) and (\ref{Gaussian}) 
\cite{Steinigeweg2017}.

Coming back to the general case, it is often convenient to study diffusion not 
only in real space but also in the space of lattice momenta (reciprocal space)
\begin{equation}
q = \frac{2 \pi k}{L} \, , \quad k = 0, \ldots, L-1 \, .
\end{equation}
Note that the lattice spacing is set to one. The quasimomentum representation is particularly 
useful, since a discrete Fourier transform
\begin{equation}
\langle \delta \sz{q}(t) \rangle = \frac{1}{\sqrt{L}} \sum_r e^{\ii q r} \, 
\langle \delta \sz{r}(t) \rangle \label{ISF}
\end{equation}
decouples the diffusion equation in Eq.\ (\ref{diffusion_equation}). Hence, 
after this transformation, it becomes the simple rate equation
\begin{equation}
\frac{\text{d} \langle \delta \sz{q}(t)  \rangle}{\text{d} t} = -\tilde{q}^2 
\, \DS \, \langle \delta \sz{q}(t)  \rangle \, , \label{rate_equation}
\end{equation}
where the momentum dependence $\tilde{q}^2 = 2 (1 - \cos q)$ may be 
approximated as $\tilde{q}^2 \approx q^2$ for sufficiently small $q$. The 
solution of Eq.\ (\ref{rate_equation}) is obviously an exponential decay of the 
form \cite{steiner1976}
\begin{equation}
\frac{\langle \delta \sz{q}(t)  \rangle}{\langle \delta \sz{q}(t=0)  \rangle} 
= e^{-\tilde{q}^2 \DS t} \, .
\end{equation}
Thus, the general solution of the diffusion equation can also be written as 
a superposition of exponential decays at different momenta. For instance, the 
Bessel solution in Eq.\ (\ref{Bessel}) can be written in the form
\begin{equation}
\frac{\langle \delta \sz{r}(t) \rangle}{\langle \delta \Mtot \rangle} = 
\frac{1}{L} \sum_q e^{-\ii q (r-r')} \, e^{-\tilde{q}^2 \DS t} \, .
\end{equation}
This form makes it particularly clear when the Gaussian in Eq.\ (\ref{Gaussian}) 
is a good approximation: Quasimomentum $q$ must be sufficiently dense, i.e.,  $L$ must be sufficiently
large and in addition, time $t$ must be sufficiently long.

As Fourier modes $\langle \delta \sz{q}(t)  \rangle$ decay exponentially in 
the case of diffusion, their spectral representation
\begin{equation}
\langle \delta \sz{q}(\omega) \rangle = \int_{0}^{\infty} \text{d} t \, 
e^{\ii \omega t} \, \langle \delta \sz{q}(t)  \rangle
\end{equation}
becomes a Lorentzian of the form \cite{kadanoff1963}
\begin{equation}
\rp \left [ \frac{\langle \delta \sz{q}(\omega) \rangle}{\langle \delta 
\sz{q}(t=0)  \rangle} \right ] = \frac{\tilde{q}^2 \DS}{(\tilde{q}^2 \DS)^2 + 
\omega^2} \label{Lorentzian}
\end{equation}
with the sum rule
\begin{equation}
\int_{-\infty}^\infty \text{d} \omega \, \rp \left [ \frac{\langle \delta 
\sz{q}(\omega) \rangle}{\langle \delta \sz{q}(t=0)  \rangle} \right ] = 
\pi \, .
\end{equation}
This Lorentzian line shape occurs for all momenta (and 
frequencies), which reflects the fact that the diffusion equation in Eq.\ 
(\ref{diffusion_equation}) assumes a mean-free path $l = 0$ (and mean free 
time $\tau = 0$). However, if $l$ and $\tau$ are finite, a Lorentzian 
line shape can only occur in the hydrodynamic limit where momentum and 
frequency are both sufficiently small.

Eventually, it is instructive to discuss correlation functions again. 
Focusing on the specific initial state $\rho$ in Eq.\ (\ref{Kubo_state}), 
starting from Eq.\ (\ref{local_correlation}), and assuming translation 
invariance of $H$, it is straightforward to show that
\begin{equation}
\langle \sz{q}(t)  \rangle = \varepsilon \, p_{r'} \, \chi_q(t)
\end{equation}
with the correlation function
\begin{equation}
\chi_q(t) = \beta \, K_{\delta \sz{q}  \, \delta \sz{-q} }(t)
\, .
\label{eq:chi_time}
\end{equation}
Therefore, in the case of diffusion, the correlation function $\chi_q(t)$ is 
an exponential and the real part of its Fourier transform 
$\chi_q(\omega)$ is a Lorentzian.

Remarkably, the continuity equation in momentum space,
\begin{equation}
\frac{\text{d} \sz{q}(t)}{\text{d} t} = (e^{\ii q} - 1) \, \JS_q(t) \, ,
\end{equation}
allows to relate $\chi_q(t)$ to the correlation function
\begin{equation}
\sigma_q(t) = \beta \, K_{\JS_q \, \JS_{-q}} (t) \, .
\end{equation}
In the time domain, this relation reads
\begin{equation}
\sigma_q(t) = -\frac{1}{\tilde{q}^2} \, \frac{\text{d}^2 
\chi_q(t)}{\text{d}t^2} 
\end{equation}
and, as a function of  frequency, it becomes
\begin{equation} \label{eq:sigma_chi}
\rp \, \sigma_q(\omega) = \frac{\omega^2}{\tilde{q}^2} \, 
\rp \, \chi_q(\omega) \, .
\end{equation}
Therefore, if the dynamics is diffusive, the Lorentzian in Eq.\ 
(\ref{Lorentzian}) implies
\begin{equation}
\rp \left [ \frac{\sigma_q(\omega)}{\chi_q(t=0)} \right ] = 
\frac{\DS \, \omega^2}{(\tilde{q}^2 \DS)^2 + \omega^2} \, .
\end{equation}
In the limit of small momentum, one thus obtains the Einstein relation
\begin{equation}
\lim_{q \to 0} \, \rp \left [ 
\frac{\sigma_q(\omega)}{\chi_q(t=0)} \right ] = \DS \, .
\end{equation}
Note that no frequency dependence is left as the mean-free time is assumed to 
be $\tau = 0$. This broad conductivity also results in the Drude 
model of conduction discussed before.

\begin{figure}[tb]
\includegraphics[width=0.9\columnwidth]{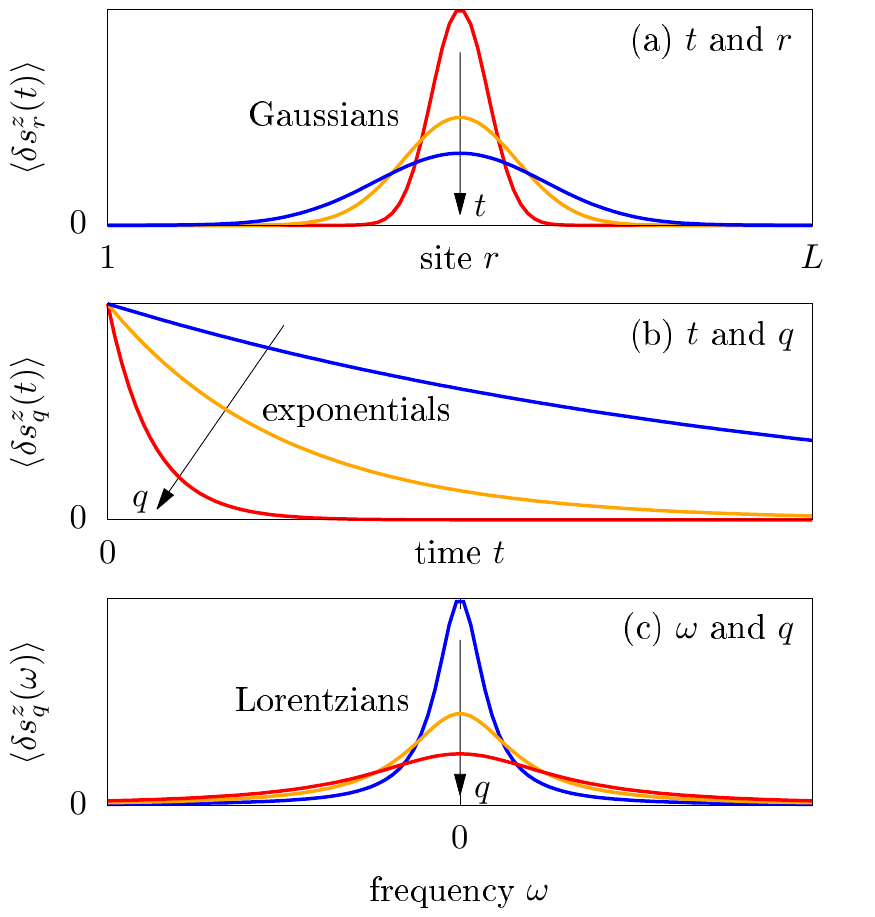}
\caption{(Color online) Sketch of the diffusive spreading of a spin-density perturbation  as a function of (a) time $t$ and real space $r$, 
(b) time $t$ and momentum space $q$, and (c) frequency $\omega$ and momentum 
space $q$.}
\label{sketch_diffusion}
\end{figure}

To summarize this section, Fig.\ \ref{sketch_diffusion} sketches diffusion in
(a) time $t$ and real space $r$, (b) time $t$ and momentum space $q$, and 
(c) frequency $\omega$ and momentum space $q$.

\section{Exploiting integrability}
\label{sec:mazur}

In this section, we will see how integrability affects the finite-temperature transport properties. 
We will stress the important role played by \emph{local} and \emph{quasilocal} conservation laws, showing that they can lead to \emph{ballistic} transport. Specifically, in Sec.~\ref{sec:charges}, we will show that a systematic construction of quasilocal charges provides lower bounds for Drude weights and diffusion constants. In Secs.~\ref{sec:betheansatz} and \ref{sec:GHD}, we will  describe methods to obtain closed-form analytical predictions for these quantities. In particular, Sec.~\ref{sec:DrudeTBA} reports on the predictions for 
spin and energy Drude weights obtained using the Thermodynamic Bethe Ansatz (TBA) formalism, whereas Sec.~\ref{sec:GHD} gives an introduction to GHD and describes its predictions for the Drude weights and diffusion constants of \emph{all} conserved charges. Most of the ideas will be exemplified in the paradigmatic case of the spin-1/2 XXZ chain.

We remark that Secs.~\ref{subsec:BA} and \ref{sec:tba} give a rather detailed introduction to Bethe ansatz and TBA and serve to establish a coherent formalism to express both the TBA results for  transport coefficients and GHD. The reader not interested in technical aspects should skip these subsections and go directly to Secs.~\ref{sec:DrudeTBA} and \ref{sec:GHD}.

\subsection{Role of local and quasilocal conserved charges}
\label{sec:charges}

Quantum integrability is based on the existence of two key objects \cite{korepinbook,faddeevreview}. The first one is the $R$-matrix, which can be understood as an abstract unitary scattering operator $\check{R}_{j,l}(\lambda)$ acting over a pair of local finite-dimensional physical Hilbert spaces, ${\cal H}_{j}\simeq {\cal H}_l \simeq \CC^d$. The $R$-matrix depends on a free (complex) spectral parameter $\lambda$ and satisfies the celebrated Yang-Baxter equation. The second key object is the Lax operator $L_{j,a}(\lambda)$, which acts on a pair of Hilbert spaces that are in principle different: the local Hilbert space ${\cal H}_j$ and the so-called ``auxiliary" space $V_a$ of dimension $N_a$, which can be finite or infinite.
These two spaces carry the  physical and auxiliary representation of the quantum symmetry of the problem, respectively. \revision{This symmetry is concisely expressed via the so-called RLL relation},\footnote{RLL stands for $R$-matrix -- Lax Matrix -- Lax Matrix} 
\begin{eqnarray}
&\check{R}_{j,l}(\mu) L_{j,a}(\lambda+\frac{\mu}{2})L_{l,a}(\lambda-\frac{\mu}{2}) \nonumber \\
&\;\;\quad\qquad = L_{j,a}(\lambda-\frac{\mu}{2})L_{l,a}(\lambda+\frac{\mu}{2}) \check{R}_{j,l}(\mu). \label{RLL}
\end{eqnarray}
The RLL equation is  another form of the Yang-Baxter relation. \revision{For a given $\check{R}_{j,l}(\mu)$, one can construct the two-site local Hermitian operator
\begin{equation}
h_{j,j+1}=\frac{{\rm d}\,\,}{{\rm d}\lambda}\check{R}_{j,j+1}(\lambda)\big |_{\lambda=0}\,,
\end{equation} 
which gives the Hamiltonian density ($H=\sum_{j=1}^{L} h_{j,j+1}$) of the corresponding integrable model, where periodic boundary conditions can be assumed for simplicity.}

A critical consequence of integrability is the existence of an extensive number of local conserved quantities, which are generated via
logarithmic derivatives 
\begin{equation}
\CCharge = \frac{{\rm d}^n\,\,\,}{{\rm d}\lambda^n} \log \tau(\lambda)\big |_{\lambda=\lambda_0}
\label{LocalCharges}
\end{equation}
of the ``fundamental transfer matrix", an operator over $\bigotimes_j \!\mathcal H_j \simeq (\CC^d)^{\otimes L}$ defined as follows
\begin{equation}
\tau(\lambda) = {\rm tr}_0 L_{1,0}(\lambda) L_{2,0}(\lambda)\cdots L_{L,0}(\lambda). 
\label{fundT}
\end{equation}
Here, $L_{j,0}(\lambda)$ is the Lax operator in the fundamental representation, where the auxiliary space is isomorphic to the local physical space. At the special point $\lambda=\lambda_0$, the Lax operator $L_{j,0}(\lambda)$ degenerates to a permutation operator $L_{j,0}(\lambda_0) = P_{j,0}$, acting as $P \ket{\psi}\otimes \ket{\phi} = \ket{\phi}\otimes\ket{\psi}$. This property is instrumental for showing that $\CCharge=\sum_{l=1}^L \dens_l$ are in fact extensive sums of local densities $\dens_l$. The conservation law property $[H,\tau(\lambda)]\equiv [H,Q_k]\equiv 0$ is then a simple consequence of the RLL relation Eq.~(\ref{RLL}), and similarly, the involution property $[\tau(\lambda),\tau(\mu)]\equiv [Q_j,Q_k]\equiv 0$ follows from another form of Yang-Baxter equation. In fact, one can fix normalization such that $H=Q_1$.

This construction applies, for example, to the paradigmatic example of the spin-1/2 XXZ chain. In this case, the local Hilbert space is ${\cal H}_j = {\mathbb C}^2$ and $\check{R}_{j,l}(\lambda)$ is the standard 6-vertex $R$-matrix \cite{baxter}. Using the parametrization 
\be
\label{eq:standardparametrisationDelta}
\Delta=\cos(\eta),
\ee
the general Lax operator \revision{can be written as} 
\begin{eqnarray}
\label{eq:Lmatrix}
L_{j,a}(\lambda,s) &=& \frac{2\sin\eta}{\sin\lambda}\left(S_a^+ s^-_j + S_a^- s^+_j\right) \nonumber\\ 
&+&  \cos(\eta S_a^z) \one + 2(\cot\lambda)\sin(\eta S_a^z)s^z_j,  \label{Lax}
\end{eqnarray}
where the local spin operators $s^\alpha_j=(1/2)\sigma^\alpha_j$, $\alpha\in\{+,-,z\}$, act over the local physical space while $S_a^{+,-,z}$ span an irreducible highest-weight representation of the $q$-deformed angular momentum algebra ($q=e^{i \eta}$) \revision{$SU_q(2)$}. This representation depends on a free (complex) parameter
$s\in\CC$ and is generically infinite-dimensional 
\begin{equation}
\begin{split}
S_a^{z} &= \sum_{n=0}^{\infty} (s-n)\ket{n}\bra{n}, \\
S_a^{+} &= \sum_{n=1}^{\infty} \frac{\sin n \eta}{\sin\eta}\ket{n-1}\bra{n},\\
S_a^{-} &= \sum_{n=1}^{\infty} \frac{\sin  (2s-n+1)\eta}{\sin\eta}\ket{n}\bra{n-1}.
\end{split}
\label{s_representation}
\end{equation}
However, either (i) for half-integer spin $s\in \frac{1}{2}\ZZ$ or (ii) for any $s\in\CC$ but root-of-unity anisotropies $\eta = \pi \ell/m$ ($\ell,m$ coprime integers) the above irrep truncates to a finite dimension: $N_a=2s+1$ or $N_a=m$, respectively. In this case, the sums above run up to $n=N_a-1$. One can thus define a general family of commuting transfer matrices
\begin{equation}
\tau(\lambda,s) = {\rm tr}_a L_{1,a}(\lambda,s) L_{2,a}(\lambda,s)\cdots L_{L,a}(\lambda,s),
\end{equation}
satisfying $[H,\tau(\lambda,s)]=0$ for all $\lambda,s$, again as a consequence of (\ref{RLL}), while clearly $\tau(\lambda,0) \equiv \tau(\lambda)$. 

\revision{For every fixed $s$, the transfer matrix $\tau(\lambda,s)$ generates the following sequence of additional conserved charges 
\begin{equation}
\CCharges = \frac{{\rm d}^n\,\,\,}{{\rm d}\lambda^n} \log \tau(\lambda,s)\big |_{\lambda=\eta/2}\,.
\label{eq:higherconservations}
\end{equation}
Therefore, one can argue that the sequence of local charges $Q_n$ stemming from the fundamental transfer matrix Eq.~(\ref{fundT}) is ``not complete" and is not sufficient to describe the statistical mechanics of integrable models. Indeed,  \cite{Ilievski15a} showed that, for $s>1/2$, the charges \eqref{eq:higherconservations} are linearly independent from the family of local charges $\CCharges \equiv Q_{n,1/2}$ and are ``essentially local''. More formally, for any size $L$, a generic charge $Q=\CCharges$ in the family \eqref{eq:higherconservations} can be written as an extensive series $Q=\sum_r \sum_{l=1}^L {q_{l,r}}$
of $r$-site local densities $q_{l,r}$ with exponentially decaying vector norm  (i.e., 
$\ave{[q_{l,r}]^2} < C e^{-r/\xi}$ for some fixed $C,\xi > 0$). This property, called quasilocality, implies extensivity in the sense
$0 < \lim_{L\to\infty} \ave{Q^2}/L < \infty$. 
Note that Eq.~(\ref{eq:higherconservations}) provides a full set of charges for $|\Delta| \geq 1$, while for $|\Delta| < 1$ one can establish 
a one-to-one correspondence between the known (quasi)local charges and the string excitations using the so-called string-charge duality \cite{ilievski2016string}.}

All the charges $\CCharges$ generated by unitary representations of \revision{$SU_q(2)$} are {\em even} under a generic $\ZZ_2$ `particle-hole' symmetry of the model, e.g., in the case of the spin-1/2 XXZ chain, under the spin-reversal (spin-flip) transformation $F=\prod_{l=1}^L \sigma^x_l$,
$F \CCharges = \CCharges F$. However, the spin current ${\cal J}^{({\rm S})}$ is odd, ${\cal J}^{({\rm S})} F = -F {\cal J}^{({\rm S})}$,  and hence $
\ave{{\cal J}^{({\rm S})}\CCharges} = 0$. In other words, irrespective of the temperature, these charges cannot contribute to the Mazur bound Eq.~(\ref{eq:MazurD}) at vanishing magnetization.

Nevertheless, one can explore non-unitary representations of the symmetry algebra \revision{$SU_q(2)$} to search for charges that are not invariant under spin-reversal using the general relation $F \tau(\lambda,s) F^{-1} = \tau(\pi-\lambda,s)^{\rm T}$. For root-of-unity anisotropies $\eta=\pi \ell/m$  ($\ell,m$ coprime integers), this procedure leads to an additional family of quasilocal conserved charges that are non-Hermitian and odd under spin reversal \cite{Prosen2014}. They can be expressed as 
\begin{equation}
\!\! Z(\lambda) =  \frac{\sin{(\lambda})^{2}}{2\eta\sin{(\eta)}}\,\partial_{s} \tau(\lambda,s)\vert_{s=0}-
\frac{\sin{(\lambda)}\cos{(\lambda)}}{\sin{(\eta)}}\, \Mtot\!,
\label{eqn:Zdef0}
\end{equation}
where $\lambda$ lies inside the 
analyticity strip ${\cal S}=\{ \lambda\in\CC; |{\rm Re}\,\lambda - 
\frac{\pi}{2}| < \frac{\pi}{2m}\}$ and  $\Mtot=\sum_{r=1}^L \sz{r}$ denotes the 
total magnetization in the $z$ direction. 

\subsubsection{Lower bound on spin Drude weight at high temperature}
\label{sec:lowerbounddrude}

\begin{figure}[t]
\begin{center}
\includegraphics[width=0.90\linewidth]{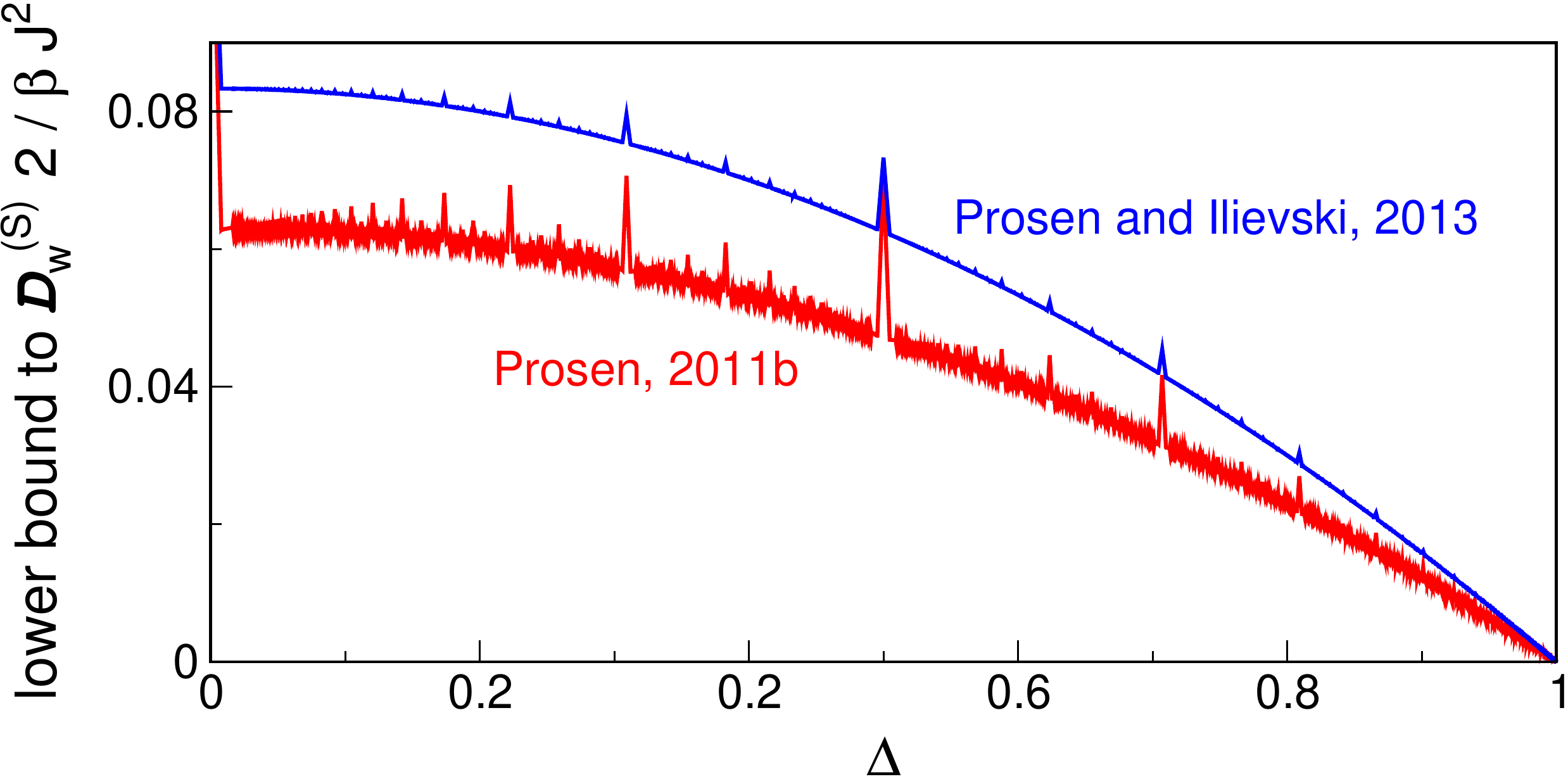}
\caption{(Color online) Lower bound for the spin Drude weight $\Dws$ of the spin-1/2 XXZ chain according
to Eq.\ (\ref{eq:prosen_bound}), as obtained in \cite{Prosen2013}, and
another lower bound for $\Dws$, as obtained earlier in \cite{Prosen2011}.
Both bounds exhibit a pronounced fractal-like (i.e., nowhere continuous)
dependence on the anisotropy parameter $\Delta$.}
\label{fig:prosen_bound}
\end{center}
\end{figure}

Since the quasilocal charges generated from non-unitary representations are not spin-reversal invariant, they have a non-vanishing overlap with the spin current and may contribute to the Mazur bound. 
For example, in the high-temperature regime ($\beta\to 0$), the overlap is also extensive, $\ave{ Z(\lambda) {\cal J}^{({\rm S})}} = i L/4$, yielding a finite
contribution to Eq.~(\ref{eq:MazurD}). However, the $Z(\lambda)$ are not mutually orthogonal and  their overlaps are given by the following analytic kernel
\begin{eqnarray*}
K(\lambda,\mu) &=& \lim_{L\to\infty}\frac{\ave{Z(\bar{\lambda})^\dagger Z(\mu)}}{L} \nonumber \\
&=& -\frac{\sin{(\lambda)}\sin{(\mu)}\,\sin((m-1)(\lambda+\mu))}{2\sin^2{(\eta)}\sin{(m(\lambda+\mu))}}
 \end{eqnarray*}
 while
 ${\ave{Z(\lambda)Z(\mu)} \equiv 0}$. The Mazur bound for the spin-Drude weight generally follows \cite{Ilievski2013} from finding an extremum of  the nonnegative action
 \begin{equation}
 {\cal S}[f]:= \lim_{t\to\infty}\lim_{L\to\infty} \frac{1}{L} \ave{(B_{L,t}[f])^\dagger B_{L,t}[f]} \ge 0,
 \end{equation} 
  with respect to an unknown function $f(\lambda)$. Here, we introduced 
 \begin{equation}
 B_{L,t}[f(\lambda)] := \frac{1}{t}\int_0^t {\rm d} s\, {\cal J}^{(S)}(s) - \int_{{\cal S}} {\rm d}^2\lambda\, f(\lambda)Z(\lambda).
 \end{equation} 
The variation $\delta S/\delta f(\lambda)=0$ results in the Fredholm equation of the first kind on a two-dimensional (complex) domain ${\cal S}$
 \begin{equation}
 \int_{\cal S} d^2 \mu K(\lambda,\mu)f(\mu) = \ave{Z(\lambda)j^{({\rm S})}}\,, 
\end{equation}
 which for the spin-1/2 XXZ chain with $\Delta=\cos(\pi \ell/m)$ yields 
 \begin{equation}
 f(\lambda) = \frac{m \sin^2(\pi/m)}{i \pi |\sin\lambda|^4}.
 \end{equation} 
 This in turn results in the following rigorous lower bound for the leading coefficient in the high-temperature expansion of the Drude weight ${\tilde {\cal D}^{(S)}_{\rm w}}$ in $\beta$, defined as
 \begin{eqnarray} \label{eq:prosen_bound}
{\tilde {\cal D}^{\rm (S)}_{\rm w}} &=& \lim_{\beta\to 0} \frac{\Dws}{\beta} \ge \frac{1}{2}\int_{\cal S} d\lambda^2 f(\lambda)\ave{Z(\lambda)j_0^{({\rm S})}} \nonumber \\
&=& \frac{1}{16} \frac{\sin^{2}{(\pi \ell/m)}}{\sin^{2}{(\pi/m)}}
\left(1 - \frac{m}{2\pi}\sin{\left(\frac{2\pi}{m}\right)}\right).
 \end{eqnarray}
Note that the r.h.s. of~\eqref{eq:prosen_bound} is a nowhere continuous function of $\Delta$ whose graph is a fractal set.
The dependence on $\Delta$ is illustrated in Fig.~\ref{fig:prosen_bound}.
 
We refer the reader to  Sec.~\ref{sec:xxz} for a detailed discussion of the saturation of this bound 
and to \cite{Matsui2020} for an explanation of why the natural non-quasilocal extension of the quasilocal charges given in Eq.~\eqref{eq:higherconservations}  cannot improve the bound. 
A more comprehensive review on quasilocal charges can be found in \cite{Ilievski2016}, whereas the extension of Drude weights and quasilocal charges to integrable periodically driven (Floquet) systems is given in \cite{Ljubotina2019a}.

 \subsubsection{Lower bounds on spin diffusion constant at high temperature}
\label{sec:lowerbound_diff}

In typical integrable models, e.g., the spin-1/2  XXZ chain for $|\Delta| \ge 1$ 
or the 1d Fermi-Hubbard model, the spin or charge Drude weight vanishes at zero 
magnetization $\magdens= 2 \langle \Mtot\rangle/L=0$  
or in the half-filled sector $\rho = N/L=1/2$, respectively. However, moving slightly away from half filling, one typically obtains a finite Drude weight. More precisely, calling $\delta$ the small deviation from either zero magnetization or half filling, one observes a Drude-weight scaling as $\Dw^{(Q)} \propto \delta^2$. At first sight, this seems to exclude the onset of spin diffusion: a finite Drude weight implies a diverging diffusion constant. Nevertheless, for large $L$, the Hilbert-space sector at $\delta=0$ dominates over all sectors with $\delta\ne 0$. Therefore, one may argue that, after performing a careful (grand)canonical average, the two effects compensate each other giving rise to a finite spin- or charge-diffusion constant in the thermodynamic limit.

In fact, this argument can be  justified rigorously by studying the Mazur bound for the dynamical susceptibility in a double-scaling limit, $L\to \infty$ and $t \to \infty$ with $L/t > v_{\rm LR}$, giving rise to a universal lower bound on the 
diffusion constant $D^{(Q)}$ in terms of the curvature of the Drude weight ${\cal D}^{(Q)}_{\rm w}(\beta,\delta)$ around $\delta=0$: \cite{Medenjak2017} [see also \cite{Spohn2018}].
For spin transport, one obtains: 
\begin{equation}
	\DS(\beta)\geq \frac{1}{8\beta v_{\rm LR}\chi(\beta) f_1(\beta)} \frac{\partial^2}{\partial \delta^2}\Dws(\beta,\delta)\Bigr|_{\delta=0},
	\label{diflb}
\end{equation}
where $ v_{\rm LR} $ is the Lieb-Robinson velocity \cite{liebrobinson}, and 
\be
f_1(\beta)=\lim_{L\to\infty}\frac{1}{2L} \frac{\partial^2}{\partial \delta^2} F_{L}(\beta,\delta)|_{\delta=0}
\ee
is a second derivative of the free-energy density 
at zero magnetization, while $\chi(\beta)$ is the static susceptibility 
         $
	{\chi(\beta)}/{\beta}=\lim_{L\to\infty}{\langle (\Mtot)^2 
\rangle-(\langle \Mtot \rangle)^2}/L
	$.
The inequality holds in general, even for a nonintegrable system, if there exist conserved quantities which would make Drude weights nonvanishing away from the symmetric Hilbert-space sector $\delta=0$. However, for integrable systems with a well-understood quasiparticle content, such as the spin-1/2 XXZ chain, the inequality can be further refined by decomposing the contribution to the diffusion constant in terms of the curvatures of the Drude weight contributions
associated to independent Bethe-ansatz quasiparticle species (see Sec.~\ref{sec:GHDDrude}). In this case, the velocity $v_{\rm LR}$ can be replaced by the corresponding dressed quasiparticle  velocity \cite{Ilievski2018}.

One can approach lower bounds on diffusion constants from another angle. In the same way as for the Mazur bound Eq.~(\ref{eq:MazurD}) suggests that a non-vanishing high-temperature Drude weight is connected to the existence of linearly extensive --- i.e., proportional to the volume --- (quasi)local charges, one might argue that a non-vanishing high-temperature diffusion constant suggests the existence of 
conserved charges which are quadratically extensive. 
\revision{Indeed, for any locally interacting lattice system, the existence of an almost conserved operator $Q$ which has an overlap with any current operator $j_r^{(Q')}$, associated with some charge $Q'$, leads to a rigorous bound on high-temperature diffusion constants \cite{TP2014QuadCharges} associated with that current. In other words,  
\be
\D^{(Q')}(\beta\to 0) \ge \frac{ |Q^j|^2} {8 v_{\rm LR} q},
\ee
where we used that the commutator $[H,Q]$ only contains boundary terms,  $0 < q:=\lim_{L\to\infty} \langle Q^2 \rangle/L^2 <\infty$, and ${Q^j:=\lim_{L\to\infty}\langle j_r^{(Q')}  Q\rangle}$ is finite.}

This gives, e.g., nontrivial lower bounds for spin-diffusion constants in the spin-1/2 Heisenberg chain as well as for  spin- and charge-diffusion constants for the  1d Fermi-Hubbard model. The bound has been recently generalized and formalized within the method of ``hydrodynamic projections" in \cite{Doyon2019QuadCharges} (cf.\ Sec.~\ref{sec:GHDDrude}), where similar ideas were also used to provide bounds on anomalous (e.g., superdiffusive) transport, i.e., to estimate the dynamical exponents.

\subsection{Bethe Ansatz}
\label{sec:betheansatz}

Here, we consider an important subclass of integrable models: those treatable by the collection of techniques grouped under the name of Bethe ansatz. The key property of these models is that their energy eigenstates can be expressed as scattering states of stable quasiparticles~\cite{korepinbook, faddeevreview, esslerbook}. This gives direct access to their energy spectrum and, more generally, to their thermodynamic properties. Although the stable quasiparticles of integrable models generically undergo nontrivial scattering processes, integrability ensures that every scattering process can always be decomposed into a sequence of two-particle scatterings.

Focussing on the paradigmatic example of the spin-1/2 XXZ chain, we will 
introduce the central equations of Bethe ansatz --- the Bethe equations --- 
which give access to all possible eigenstates of the systems. Then, we will 
explain how to take their thermodynamic limit, arriving at the so called 
thermodynamic Bethe ansatz (TBA) description~\cite{Takahashi:1999aa}, where one 
characterizes the eigenstates in terms of ``densities" of quasiparticles. 
Finally, we will recall some results for the energy and spin Drude weight 
obtained using TBA.    

\subsubsection{Bethe Equations}
\label{subsec:BA}

There are two known routes to \revision{diagonalize the Hamiltonian using Bethe Ansatz}. The first one consists in writing an ansatz many-body wave-function in real (coordinate) space. This is the original method introduced in \cite{Bethe} and is now known as coordinate Bethe ansatz. The second, more recent, route consists of constructing a basis of eigenstates of the fundamental transfer matrix \eqref{fundT} for all values of the spectral parameter $\lambda$ (cf. Sec.~\ref{sec:charges}). This is always possible since transfer matrices with different spectral parameters commute. Since the Hamiltonian is proportional to the logarithmic derivative of the transfer matrix \revision{(cf. the discussion after \eqref{fundT})}, these states are also eigenstates of $H$.  \revision{The latter route}, called algebraic Bethe ansatz, is more powerful: it gives direct insights into the conservation laws of the system and correlation functions~\cite{korepinbook, faddeevreview, esslerbook}. \revision{For the sake of brevity, we do not describe such approaches in detail but only report the final results (we refer the reader interested in the derivations to the aforementioned references). 

The Bethe-ansatz procedure yields the eigenstates of the system parametrized by a set of (generically complex) numbers $\{\lambda_j\}$ called rapidities and obtained by solving a set of non-linear algebraic equations. For example, in the case of the spin-1/2 XXZ chain, the eigenstates with magnetization $L/2-N$ are parametrized by the solutions $\{\lambda_j\}_{j=1}^N$ of} 
\be
\!\!\!\left[\frac{\sinh(\lambda_j+i\eta/2)}{\sinh(\lambda_j-i\eta/2)}\right]^L \!\!= -\prod_{k= 1}^N \left[\frac{\sinh(\lambda_j-\lambda_k+i\eta)}{\sinh(\lambda_j-\lambda_k-i\eta)}\right],
\label{eq:BetheEquations}
\ee
for $j=1,\ldots,N$. These are the illustrious Bethe equations, first found in \cite{Bethe} for $\Delta=1$ and then in \cite{Orbach} for generic $\Delta$. 

All Bethe-ansatz integrable models produce sets of nonlinear, coupled, algebraic equations of this form. In some cases, however, one needs to repeat the procedure multiple times before finding the eigenstates of the Hamiltonian. This produces multiple sets of equations similar to Eq.~\eqref{eq:BetheEquations}, involving different sets of rapidities, which are coupled together. This procedure is known as nested Bethe ansatz and is necessary, e.g., for the Fermi-Hubbard model. For simplicity, we restrict the discussion  to the non-nested case in our presentation. 

The eigenvalues of quasimomentum\footnote{On the chain, the quasimomentum operator is defined as $-i\log \Pi$ ($\Pi$ acts as the one-site-shift operator).} and the Hamiltonian \revision{in the eigenstate parametrized by $\{\lambda_j\}_{j=1}^N$} 
\be
\!\!P \!=\!\! \left[\sum_{k=1}^{N} p(\lambda_k, \tfrac{1}{2})\!\right]{\rm mod}\,2\pi,\quad E\!=\!\!\sum_{k=1}^{N} e(\lambda_k)+e_0 L,
\label{eq:EVEP}
\ee
where we set ${p(\lambda, s)\!\!=\!i\log\left[{\sinh(\lambda-i\eta s)}/\sinh(\lambda+i\eta s)\right]}$, ${e(\lambda)=\!-\sin\eta/2\, \partial_\lambda p(\lambda, \tfrac{1}{2})}$ and ${e_0={\Delta}/{4}}$. An expression similar to the one for the energy holds for higher local (and quasilocal) conservation laws \eqref{eq:higherconservations}. In particular, in the eigenvalue of $\CCharges$ the function $e(\lambda)$ is replaced by ${q_n(\lambda,s)=(-\sin\eta/2)^n \partial_\lambda^n p(\lambda, s)}$ while the constant shift $e_0$ is replaced by 0.  

The Bethe equations might be viewed as convoluted quantization conditions for the momenta (or better the ``rapidities'') of a gas of quasiparticles confined in a finite volume $L$. However, one should be careful  with such an interpretation as the solutions to these equations are generically \emph{complex}: this is a common feature of many Bethe-ansatz integrable models.

To understand the distribution of the roots in the complex plane, it is useful to look at the solutions for $L\to\infty$ and fixed $N$~\cite{Takahashi:1999aa, esslerbook}. In this case, any ${\rm Im}[\lambda_j]\neq0$ causes the l.h.s. to either go to infinity or to 0. Requiring the r.h.s. to do the same forces the solutions to follow ordered patterns in the complex plane known as ``strings". 
Strings can be interpreted as stable bound states formed by the elementary particles \cite{esslerbook} and appear in all Bethe-ansatz integrable models with complex rapidities, but their specific form depends on the model and on the values of its parameters. 
Specifically, in the spin-1/2 XXZ chain, the string-structure depends on whether $\eta$ is real ($|\Delta|<1$) or imaginary ($|\Delta|>1$). For instance, for $\eta\in\mathbb R$, we have strings of the form~\cite{Takahashi:1999aa} 
\be
\lambda_{\alpha, a}^k = \lambda_\alpha^k+i\frac{\eta}{2}(n_k+1-2a)+i \frac{\pi}{4}(1-\upsilon_k)+\delta_{\alpha,a}^k,
\label{eq:strings}
\ee
where $\lambda_\alpha^k\in\mathbb R$ is called ``string center", $k=1,\ldots,N_s$ is called ``string type", $\alpha=1,\ldots,M_k$ labels different strings of the same type, and $a=1,\ldots,n_k$ labels rapidities in the same string. Finally, the ``string deviations" $\delta_{\alpha,a}^k$ are exponentially small in $L$. 

The number $N_s$ of type of strings, the ``length" $n_k$ of the $k$-th string, and its ``parity" $\upsilon_k$ depend on $\eta$ in a discontinuous way: they change drastically depending on whether $\eta/\pi$ is rational. For example, for $\eta=\pi/m$, we have $N_s=m$, ${n_k=(k-1)(1-\delta_{k,m})+1}$, and $\upsilon_k=1-2\delta_{k,m}$. A similar parameterisation of strings can be performed also for $i\eta\in\mathbb R$ and more generally, for other Bethe-ansatz integrable models~\cite{Takahashi:1999aa}.  

\subsubsection{Thermodynamic Bethe Ansatz}
\label{sec:tba}

For small numbers $N$ of rapidities, the Bethe equations can be easily solved on a computer [see, e.g., \cite{hagemansthesis, Shevchukthesis}]. For a full classification of the solutions of Eq.~\eqref{eq:BetheEquations}, this is feasible for  for $N\leq L=10$.  However, this procedure becomes quickly impractical when $N$ and $L$ increase. In particular, to study the thermodynamic limit --- $N,L\to\infty$ with finite $N/L$ --- a brute force numerical solution of the equations is unfeasible and some analytical treatment becomes unavoidable. The standard approach --- known as thermodynamic Bethe ansatz (TBA) --- is based on the crucial assumption that the solutions to Eq.~\eqref{eq:BetheEquations} continue to follow the string patterns \emph{even at finite density} \cite{Bethe,Taka1971}, i.e., when $N$ is not fixed but goes to infinity with $L$. Although this assumption --- usually called \emph{string hypothesis} --- does not strictly hold for all states in large but finite systems, it is believed to describe exactly the thermodynamic properties of all Bethe-ansatz integrable models. In particular, \cite{Tsvelik83} proved the self consistency of the string hypothesis for the spin-1/2 XXZ chain at finite temperature. A more rigorous alternative to the string hypothesis exists \cite{Kluemper92,Kluemper93,SuzukiInoue87} and is often referred to as quantum transfer-matrix approach. 
Even though this approach is very powerful, it is generically less versatile than TBA (currently most of the results have been found for thermal states). Importantly, the two approaches have been proven to give an equivalent description of the thermodynamic properties of the spin-1/2 XXZ chain at finite temperatures \cite{Kluemper92, Kuniba1998}. 

Embracing the string hypothesis and multiplying together all Bethe equations referring to particles in the same string one arrives at a set of equations --- known as Bethe-Takahashi equations --- for the real string centers [cf.~Eq.~\eqref{eq:strings}]. These equations can readily be viewed as quantization conditions for the momenta of the original particles and all their bound states and are most commonly expressed in ``logarithmic form" (taking $-i\log[\cdot]$ of both sides). In particular, the Bethe-Takahashi equations for the spin-1/2 XXZ chain read as \cite{Takahashi:1999aa} 
\be
L \theta_j(\lambda_\alpha^j)-
{\sum_{k=1}^{N_s}\sum_{\gamma=1}^{M_k}}\Theta_{jk}(\lambda_\alpha^j-\lambda_\gamma^k)=2\pi I_{\alpha}^{(j)}\,,
\label{eq:BetheTaka}
\ee  
where the ``quantum numbers" $I_{\alpha}^{(j)}$ are integer (half-odd integers) for odd (even) $L-M_j$ (also their allowed ranges depend on $\{M_j\}$) and the string centres $\lambda_\alpha^j$ lie in the symmetric interval $[-\Lambda,\Lambda]\subset \mathbb R$ while the smooth functions $\theta_j(x)$ and $\Theta_{ij}(x)$ can be expressed as 
\begin{align}
\theta_j(x)=&f(x;n_j,\upsilon_j),\\ 
\Theta_{ij}(x)=&f(x;|n_i-n_j|,\upsilon_i\upsilon_j)+f(x;n_i+n_j,\upsilon_i\upsilon_j)\notag\\
&+2 \!\!\!\!\!\!\!\!\!\!\! \sum_{k=1}^{\min(n_i,n_j)-1} \!\!\!\!\!\!\!\!\!\!\!f(x;|n_i-n_j|+2k,\upsilon_i\upsilon_j).
\end{align}
Both $\Lambda$ and the form of the auxiliary function $f(x; n; \upsilon)$ depend on whether  $|\Delta|<1$ or  $|\Delta|\geq1$; their form is reported in Table~\ref{table:XXZ}.

\begin{table}
\begin{tabular}{|c|c|c|}
 \hline
 & $|\Delta|<1$ &  $|\Delta|\geq1$  \\
 \hline
$\Lambda$ & $\infty$ & $\pi/2$ \\
\hline
$\frac{f(x; n; \upsilon)}{2}$ & $
\begin{cases}
\upsilon\,{\rm atan}\!\!\left[\frac{\tanh x}{[\tan\frac{n \gamma}{2}]^{\upsilon}}\right] &n\gamma\notin\mathbb Z\\
0 &n\gamma\in\mathbb Z\\
\end{cases}$ &${\rm atan}\!\!\left[\frac{\tan x}{\tanh\frac{n \gamma}{2}}\right]\!\!+\!\pi \lfloor \frac{x}{\pi}+\frac{1}{2}\rfloor$ \\[1.5pt]
   \hline
\end{tabular}
\caption{Auxiliary function $f(x; n; \upsilon)$ for the spin-1/2 XXZ chain for $\Delta=\cos\eta$. We defined $\gamma\equiv|\eta|$. \label{table:XXZ}}
\end{table}

Furthermore, by substituting the string hypothesis in the expectation value of the energy density [see Eq.~\eqref{eq:EVEP}], we have 
\be
E\!=\!\!\sum_{k=1}^{N_s}\sum_{\gamma=1}^{M_k} e_k(\lambda_\gamma^k)+e_0 L,
\label{eq:energySH}
\ee
\revision{where ${e_k(\lambda)\!\equiv\! -{\rm sgn}(\Delta+1)} [\sqrt{|\Delta^2-1|}/2]\partial_\lambda \theta_k(\lambda)$ are known as ``bare energies''. We see that the energy of an eigenstate is obtained by summing up the bare energies of all quasiparticles characterizing it. A similar expression holds for higher conservation laws.}

The set $\{I_{\alpha}^{(j)}\}$ is in one-to-one correspondence with the set of string centres (or ``particle rapidities") $\{\lambda_{\alpha}^{(j)}\}$ and can be used to specify the state of the system, much like momentum occupation numbers in free systems. We note in passing that the correspondence between $\{I_{\alpha}^{(j)}\}$ and the solutions of the Bethe equations has been used to prove the combinatorial completeness of Bethe ansatz for the XXZ chain~\cite{KirillovCompleteness, KirillovCompleteness2} and for the Fermi-Hubbard model \cite{CompletenessHubbbard}. The correspondence is explicitly established by introducing the ``counting functions" 
\be
{z_j(x|\{\lambda_{\alpha}^{(j)}\}) \equiv \theta_j(x)-\frac{1}{L}{\sum_{k=1}^{N_s}\sum_{\gamma=1}^{M_k}}\Theta_{jk}(x-\lambda_\gamma^k)}.
\ee
These functions are monotonic in $x$ and, by definition, satisfy ${z_j(\lambda_{\gamma}^{(k)} |\{\lambda_{\alpha}^{(j)}\})=2\pi I^{(k)}_\gamma /L}$ (this is just a rewriting of \eqref{eq:BetheTaka}). There exist, however, some holes ${\{\bar \lambda_\gamma^j\}\not\subset\{\lambda_{\alpha}^{(j)}\}}$ such that ${z_j(\bar \lambda_\gamma^j |\{\lambda_{\alpha}^{(j)}\})=2\pi J^{(j)}_\gamma /L}$ with $\{J_{\alpha}^{(j)}\}$ integers (or half-odd integers)  in the allowed range to be quantum numbers but not appearing in $\{I_{\alpha}^{(j)}\}$ (they can be thought of as empty slots).  

In the thermodynamic limit, both particle and hole rapidities become dense in $[-\Lambda,\Lambda]$ (differences of neighboring rapidities scale like $L^{-1}$) and it is convenient to switch  to a coarse grained description of the system in terms of their densities $\{\rho_j(\lambda)\}$ and $\{\rho^h_j(\lambda)\}$ ($h$ stands for ``hole"). It is easy to verify that $2 \pi \sigma_j(\rho_j(\lambda) +\rho^{h}_j(\lambda))= \lim_{L\to\infty} \partial_\lambda z_j(\lambda|\{\lambda_{\alpha}^{(j)}\})$, where the sign ${\sigma_j=\{\pm1\}}$ accounts for strings where ${z_j(x|\{\lambda_{\alpha}^{(j)}\})}$ is monotonically decreasing (they occur for ${|\Delta|<1}$) \cite{Takahashi:1999aa}. Computing the derivative explicitly, we find the so called thermodynamic Bethe-Takahashi equations
\be
\rho_j^t(\lambda) \equiv \rho_j(\lambda) +\rho^{h}_j(\lambda)= \sigma_j a_j(\lambda) - \sum_{k=1}^{N_s} \sigma_j T_{jk}\ast \rho_k (\lambda).
\label{eq:TBT}
\ee
Here, we introduced the driving ${a_j(\lambda)=(1/2\pi)\partial_\lambda \theta_j(\lambda)}$, and the kernel ${T_{jk}(\lambda)=(1/2\pi)\partial_\lambda \Theta_{jk}(\lambda)}$ (encoding all information about the interactions), while $\ast$ denotes the convolution $f\ast g(x)=\int_{-\Lambda}^\Lambda {\rm d}y f(x-y)g(y)$.

Equations \eqref{eq:TBT} fix the densities of holes in terms of the densities of particles. In other words, for each state, they provide the densities $\rho_j^t(\lambda)$ of ``rapidity slots" (called ``vacancies")  that can be occupied by a particle. Due to the interactions, the density of slots depends on the state [cf. the second term on the r.h.s.\ of Eq.~\eqref{eq:TBT}]. Integral equations of this form are very common in TBA. In the following, we will find many instances of these equations with the same kernel but different driving functions. 

\revision{We remark that,  even though each eigenstate of the Hamiltonian corresponds to a set of densities $\{\rho_j(\lambda)\}$, the correspondence is generically not one-to-one: in a large finite volume $L$, there are approximately $\exp[L s[\rho]]$ eigenstates of the Hamiltonian corresponding to the same set of densities $\{\rho_j(\lambda)\}$, where the functional ${s[\rho]=\sum_{k} \!\int \!{\rm d}\lambda (\rho^t_k\log\rho^t_k-\rho_k\log\rho_k-\rho^h_k\log\rho^h_k})$ is known as the Yang-Yang entropy density. This fact is often referred to by saying that the densities of rapidities specify a ``macrostate" of the system, as opposed to a single eigenstate of the Hamiltonian that is called ``microstate".}

The densities of rapidities in principle allow one to compute the expectation values of all local operators in the thermodynamic limit. In practice, however, explicit expressions are known only for few classes of observables (see also Sec.~\ref{sec:GHD}). A relevant example is that of  local (and quasilocal) conserved-charge densities. Specifically, considering the density of the generic charge $Q$, we have 
\begin{align}
q[\rho]&=\sum_{k=1}^{N_s} \!\int \!\!{\rm d}\lambda\, q_k(\lambda)\rho_k(\lambda),
\label{eq:charge}
\end{align}
where the set of functions $q_k(\lambda)$ specifies the charge and it is often called ``bare charge". For example, the energy density is obtained by replacing $q_k(\lambda)$ with $e_k(\lambda)$ and adding the constant shift $e_0$. Moreover, setting $q_k(\lambda)=q_{n,k}(s,\lambda)$ for some appropriate $q_{n,k}(s,\lambda)$, one reproduces the density of  higher conservation laws \eqref{eq:higherconservations}. In particular, for the densities $q_n[\rho]$ of the higher local conserved charges \eqref{LocalCharges}, we have $q_{k}(\lambda)=q_{n,k}(\lambda)=(-{\rm sgn}(\Delta+1) [\sqrt{|\Delta^2-1|}/2] \partial_\lambda)^{n+1} \theta_k(\lambda)$.

TBA can also be used to analyze excitations over macrostates.  Let us take a large finite volume $L$ and consider the system in one of the microstates corresponding to the densities $\{\rho_j(\lambda)\}$. Injecting an extra string of type $j$ and rapidity $\lambda$ induces a change in the expectation values of the conserved charges
\be
L q[\rho] \mapsto L q[\rho]+ q_j^d(\lambda).
\ee
\revision{Due to the presence of interactions, $\{q_j^d(\lambda)\}$ differ from the bare charges of Eq.~\eqref{eq:charge} and are commonly referred to as ``dressed charges". Specifically, given a set of bare charges $\{q_j(\lambda)\}$, one can find the corresponding dressed charges through the following integral equation}
\be
\partial_\lambda q^d_j (\lambda)= \partial_\lambda q_j(\lambda) - \sum_{k=1}^{N_s}\sigma_k \left[T_{jk}\ast  \vartheta_k \partial_\lambda q^d_k \right](\lambda),
\label{eq:dressing1}
\ee
where we introduced the ``filling" function 
\be
\vartheta_j(\lambda)\equiv \frac{\rho_j(\lambda)}{\rho^t_j(\lambda)}.
\label{eq:thetafun}
\ee
Even though the momentum is only conserved modulo $2\pi$, a dressed momentum is well defined as long as $p_j^d(\lambda)<2\pi$. In particular, since the bare charge related to the momentum is  $\theta_j(\lambda)$ [cf.\  Eq.~\eqref{eq:EVEP}], the dressed momentum fulfills $ \partial_\lambda p_j^d(\lambda)= 2\pi \sigma_j \rho^t_j(\lambda)$. This can be established comparing the equation for the dressed momentum with Eq.~\eqref{eq:TBT} and allows us to express the group velocity of the excitation $(\lambda,j)$ as 
\be
v^d_{j}(\lambda)=\frac{\partial_\lambda e^d_{j}(\lambda)}{\partial_\lambda p^d_{j}(\lambda)}=\frac{\partial_\lambda e^d_{j}(\lambda)}{2\pi \sigma_j \rho^t_j(\lambda)}\,.
\label{eq:velocityrhot}
\ee
In other words, $2\pi \sigma_j \rho^t_j(\lambda) v^d_{j}(\lambda)$ fulfills Eq.~\eqref{eq:dressing1} with $\partial_\lambda e_{j}(\lambda)$ as a driving. In addition to the dressed charge, we can associate  another ``dressed" quantity, sometimes called the ``effective charge",
to each quasilocal conservation law (and also to the momentum).
For a given bare charge $q_j(\lambda)$, we define the associated effective charge $q_j^{\rm eff}(\lambda)$ as the solution of the following Eq.~\eqref{eq:dressing}, which has $q_j(\lambda)$ as its driving term, 
\be
q^{\rm eff}_j (\lambda)= q_j(\lambda) - \sum_{k=1}^{N_s}\sigma_k \left[T_{jk}\ast  \vartheta_k  q^{\rm eff}_k \right](\lambda).
\label{eq:dressing}
\ee
Note that in this case, one directly ``dresses" the charge and not its derivative and hence dressed and effective charges do not coincide. We have, however, ${\partial_\lambda (q_j^{\rm d}(\lambda))=(\partial_\lambda q_j(\lambda))^{\rm eff}}$ such  that we can equivalently express \eqref{eq:velocityrhot} as $v^d_{j}(\lambda)={(\partial_\lambda e_{j}(\lambda))^{\rm eff}}/{(\partial_\lambda p_{j}(\lambda))^{\rm eff}}$. This formulation is used in a large portion of the GHD literature.

In closing, we remark that, even though here we assumed the system to be in an eigenstate of the Hamiltonian, the TBA description can be used also for some (stationary) mixed states. This is true every time a generalized microcanonical representation applies \cite{Essler2016, Vidmar2016}. In essence, this means that the expectation values of all local observables in the mixed state can be reproduced, in the thermodynamic limit, by expectation values in a single, appropriately  chosen eigenstate of the Hamiltonian. For example, the densities corresponding to a generalized Gibbs state $\rho_{\rm GGE}\propto \exp[\sum_n \beta_{n} \CCharge]$ can be found minimising the ``generalized free energy" $f[\rho]= \sum_n \beta_{n} q_n[\rho]- s[\rho]$, which yields the following integral equations \cite{Yang1969}
\be
\log\eta_j(\lambda) \!\!= \!\!\sum_n \beta_n q_{n,j}(\lambda) \!+\! \sum_{k=1}^{N_s}\! \sigma_k T_{kj}\ast\log[1+\eta_k^{-1}](\lambda)\,,
\label{eq:GTBA}
\ee 
where we introduced the function
\be
\eta_j(\lambda)\equiv \frac{\rho^h_j(\lambda)}{\rho_j(\lambda)}=\frac{1}{\vartheta_j(\lambda)}-1.
\label{eq:etafun}
\ee
These equations, together with Eq.~\eqref{eq:TBT}, completely fix the densities of the generalized Gibbs state. Note that, if $\{\rho_j(\lambda)\}$ and $\{\rho^h_j(\lambda)\}$ solve  Eqs.~\eqref{eq:TBT} and  \eqref{eq:GTBA}, the generalized free energy can be written compactly as
\be
f = \frac{e_0}{T} - \sum_{k=1}^{N_s} \sigma_k \!\!\int_{-\Lambda}^{\Lambda} \!\!\!{\rm d}\lambda\,\, a_k(\lambda) \log\!\left[1+\frac{\rho_k(\lambda)}{\rho_k^h(\lambda)}\right]\!.
\label{eq:thermalfreeenergy}
\ee 
We also remark that the derivatives of $\log\eta_k(\lambda)$ with respect to the ``chemical potentials" $\beta_n$ are related to the dressed quantities. Indeed, comparing Eqs.~\eqref{eq:dressing1} and \eqref{eq:GTBA} we find ${\partial_{\beta_n}\!\!\log\eta_k(\lambda)= -{\rm sgn}(\Delta+1) [\sqrt{|\Delta^2-1|}/2] \partial_\lambda q^d_{n-1,k}(s, \lambda)}$.
In order to find the explicit form of $q_{n,k}(\lambda)$, we use the explicit form of $q_{n,k}(\lambda)$ reported after Eq.~\eqref{eq:charge}.

\subsubsection{Drude weights from TBA}
\label{sec:DrudeTBA}

As an application of the TBA formalism, here we present the calculation of certain Drude weights. \revision{We remark that the calculation of generic Drude weights remained unfeasible for a long time even in Bethe-ansatz integrable models. Indeed, Drude weights are expressed in terms of dynamical correlations and the calculation of the latter falls outside the compass of standard Bethe-ansatz techniques.} In some cases, however, it has been possible to relate Drude weights to simple spectral or thermodynamic properties that can be efficiently determined using TBA. 
In particular, here we briefly review Zotos's calculations of the energy \cite{Zotos2016} and spin 
\cite{Zotos1999} Drude weights for the spin-1/2 XXZ chain with 
$\Delta=\cos(\pi/m)$ at finite temperature $T$. The results for the energy 
Drude weight are directly generalized to any $\Delta$ while those for the spin 
Drude have been extended to $\Delta=\cos(\pi \ell/m)$ with coprime integers $\ell$ and 
$m$ \cite{Urichuk2018} [see Sec.~\ref{sec:GHDDrude}  for a discussion].      

\begin{figure}[t]
\begin{center}
\includegraphics[width=0.90\linewidth]{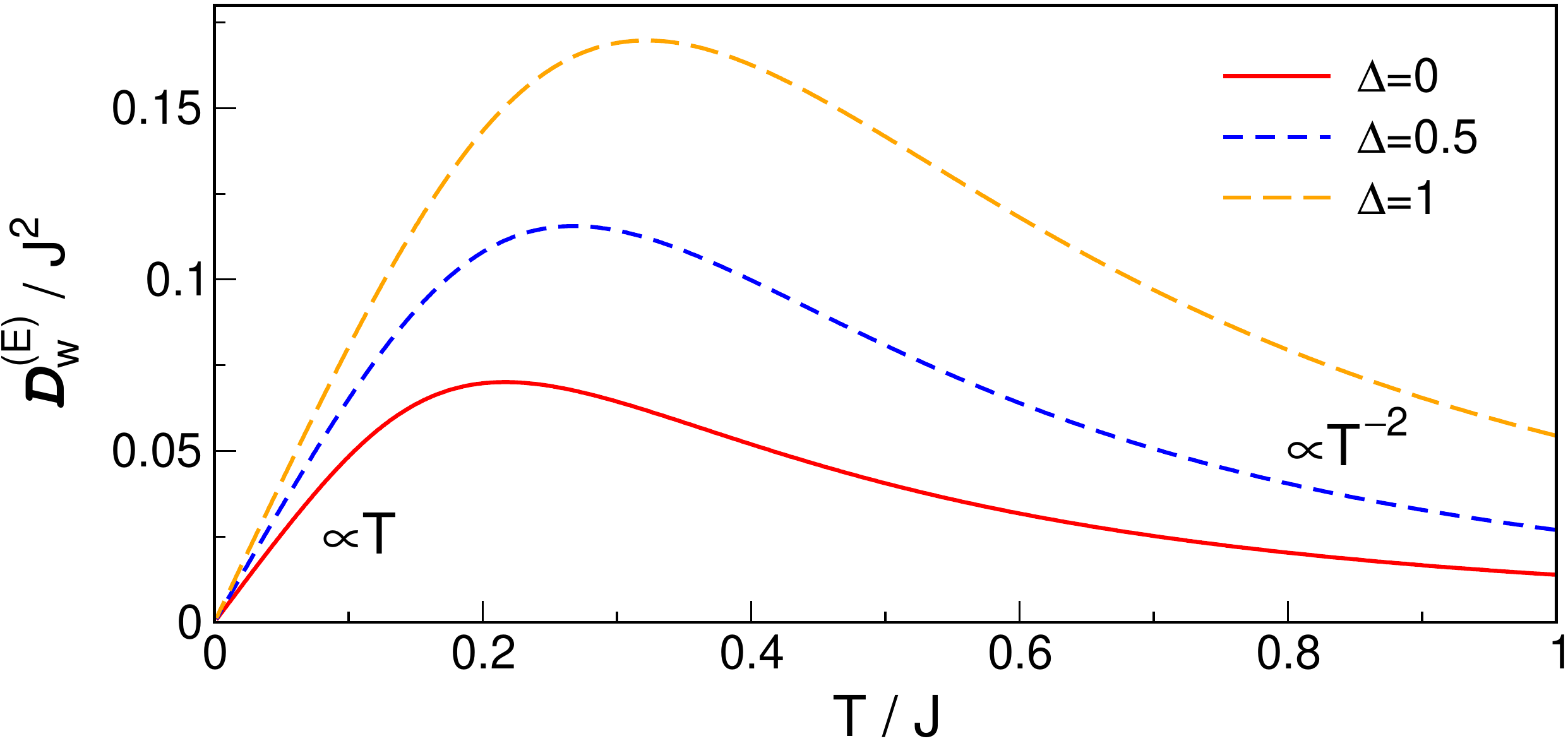}
\caption{(Color online) Exact results for the energy Drude weight of the spin-1/2 XXZ
chain given in Eq.~(\ref{eq:XXZ-intro}) at zero magnetization. The data is taken from
\cite{Kluemper2002}.}
\label{fig:drude_bethe}
\end{center}
\end{figure}

Let us begin by considering the case of the energy Drude weight, which, as we 
shall see, is considerably simpler. The crucial observation \cite{Zotos1997} is 
that in the spin-1/2 XXZ chain, the total energy current ${\JE=\sum_r 
j^{\rm (E)}_r}$ [see Eq.~\eqref{eq:jE}] is itself a conserved quantity. In 
particular, in our notation, $\JE$ coincides with $Q_2$ [see 
Eq.~\eqref{LocalCharges}]. This means that one can define a generalized Gibbs 
ensemble including such a current as a charge, i.e., $\rho_{GGE}\propto 
e^{-\beta H - \xi \JE}$, and compute its root densities following the 
last paragraph of the Sec.~\ref{sec:tba}. In particular, the free-energy 
density $f_\xi$ of this state takes the form \eqref{eq:thermalfreeenergy} where 
the densities of rapidities fulfill Eqs.~\eqref{eq:TBT} and \eqref{eq:GTBA} with  
$\beta_1=\beta$, $\beta_2=\xi$, and $\beta_{n\geq3}=0$. The Drude weight is then 
straightforwardly evaluated as [see  Eq.~\eqref{eq:DwC}]
\be
\Dwe=\frac{\beta^2}{2 L} \braket{(\JE)^2} = -\frac{\beta^3}{2}  \partial_\xi^2 f_\xi \big |_{\xi=0}\,.
\ee
\revision{Note that this identity} has been first used in \cite{Kluemper2002} to compute the energy 
Drude weight within the quantum transfer-matrix approach. 
The results are shown in Fig.~\ref{fig:drude_bethe}. 
Subsequently, Zotos found the 
explicit result from  TBA by combining Eqs.~\eqref{eq:TBT} and \eqref{eq:GTBA}. 
Using some straightforward identities among TBA functions [see, e.g., 
\cite{Urichuk2018}], the final expression can be written as 
\be
\Dwe=\frac{\beta^2 }{2} \sum_{k=1}^{N_s} \int_{-\Lambda}^{\Lambda} \!\!\!{\rm d}\lambda\,\ \!\!\! 
\frac{\rho_k^t(\lambda) (e_k^{\rm eff}(\lambda))^2 (v_k^d(\lambda))^2}{(1+\eta_k(\lambda))(1+\eta_k^{-1}(\lambda))}\,,
\label{eq:Drudeenergy}
\ee
where $e_k^{\rm eff}(\lambda)= -2\pi {\rm sgn}(\Delta+1) [\sqrt{|\Delta^2-1|}/2] \sigma_k \rho^t_k(\lambda)$ is the effective energy and $v_k^d(\lambda)$ the group velocity of the dressed excitations [cf. \eqref{eq:velocityrhot}]. The same method can be used to find higher cumulants of ${\cal J}^{(E)}$ \cite{Zotos2016, Urichuk2018}. 
   
\begin{figure}[t]
\begin{center}
\includegraphics[width=0.90\linewidth]{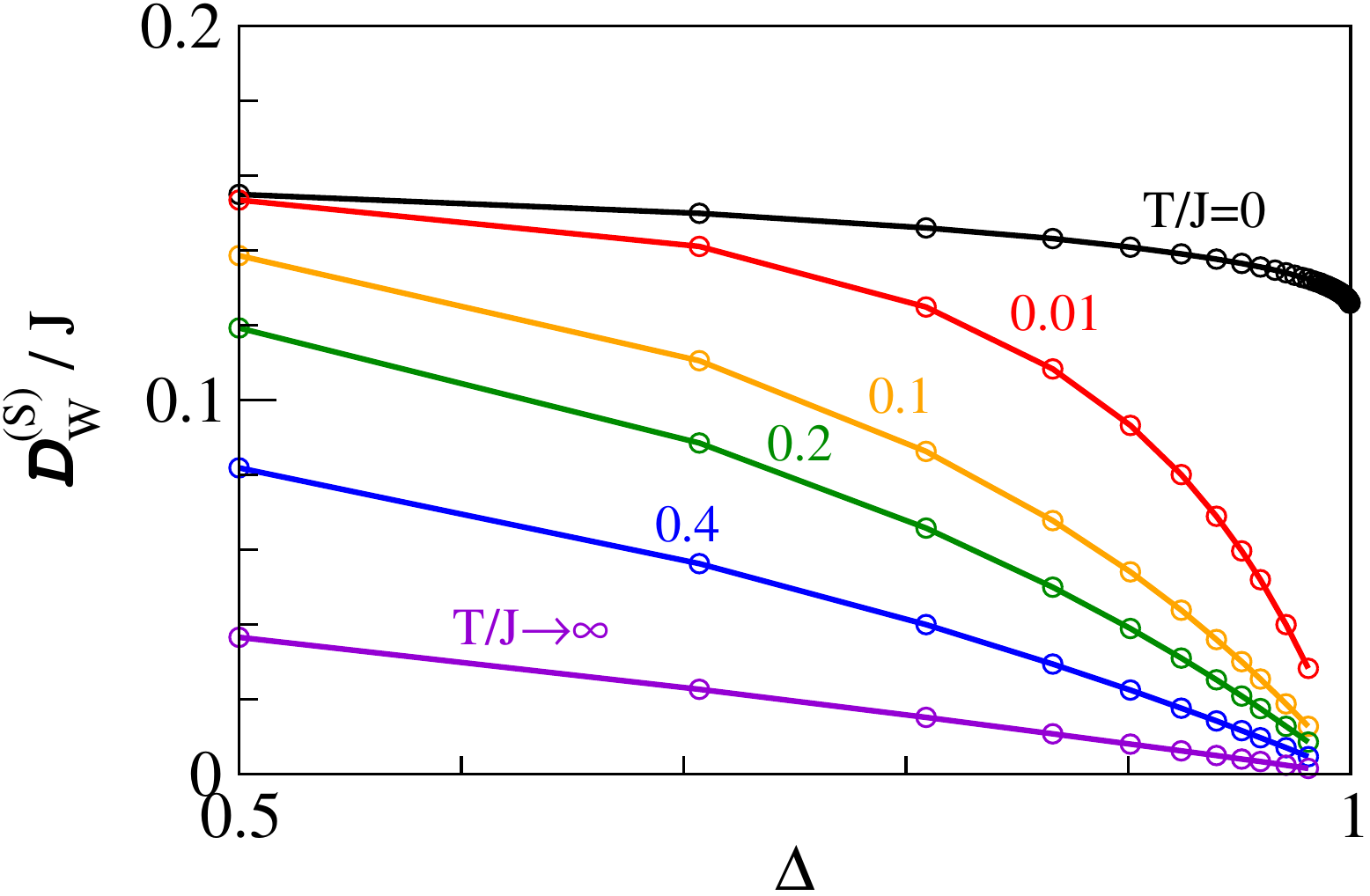}
\caption{(Color online)
TBA results for the spin Drude weight $\Dws$ of the spin-1/2 XXZ chain versus $\Delta$ for different temperatures $T$ (measured in units of $J$) from \cite{Zotos1999}.
\label{fig:zotos99}}
\end{center}
\end{figure}

The complication arising when considering the spin Drude weight is that the total spin current, as opposed to the total energy current, is not conserved. The calculation, however, can still be performed avoiding the explicit evaluation of correlation functions. The idea is to consider the system in a large finite volume $L$, to introduce a finite magnetic flux $L \phi$ through the chain, and to compute the Drude weight using the finite-$T$ Kohn formula \eqref{eq:Kohn}, i.e., in terms of the second derivative of the energy density with respect to the magnetic flux. The insertion of a magnetic flux can be easily treated in Bethe ansatz and results in a phase (``twist") $e^{i \phi L}$ multiplying the r.h.s.\ of \eqref{eq:BetheEquations}. For $L \phi$ finite in the thermodynamic limit (i.e., when $L \phi$ does not scale with the volume), the twist modifies the position of the rapidities of the strings only at sub-leading orders. This leads to 
\be
\lambda_{\alpha,L}^j(\phi)=\lambda_{\alpha,\infty}^j+\frac{g_{1,j}(\lambda_{\alpha,\infty}^j,\phi)}{L}+\frac{g_{2,j}(\lambda_{\alpha,\infty}^j,\phi)}{L^2}\,,
\label{eq:shiftedrapidities}
\ee
where we neglected $O(L^{-3})$ and introduced the subscripts $L/\infty$ to label rapidities in finite and infinite volume, respectively. The $\phi$-dependent functions $g_{1,j}(x,\phi)$ and $g_{2,j}(x,\phi)$ fulfill some integral equations determined through a $1/L$ expansion of the Bethe-Takahashi equations \eqref{eq:BetheTaka}. Plugging \eqref{eq:shiftedrapidities} into Eq.~\eqref{eq:energySH}, one can determine the second derivative of the energy density with respect to the twist in the thermodynamic limit and hence the Drude weight. This method has been introduced in \cite{Fujimoto1998} for the calculation of the charge Drude weight in the Fermi-Hubbard model and has been applied in \cite{Zotos1999} for the spin-1/2 XXZ chain. In the case of the XXZ chain, the result can be cast  into the following form 
\be
\Dws=\frac{\beta}{2} \sum_{k=1}^{N_s} \int_{-\Lambda}^{\Lambda} \!\!\!{\rm d}\lambda\,\ \!\!\! 
\frac{  \rho_k^t(\lambda) (n_k^{\rm eff}(\lambda))^2 (v_k^d(\lambda))^2}{(1+\eta_k(\lambda))(1+\eta_k^{-1}(\lambda))}\,,
\label{eq:Drudespin}
\ee  
where $n_k^{\rm eff}(\lambda)=2 \pi \sigma_k \rho_k^t(\lambda) \partial_\phi g_{1,k}(\lambda)$ fulfills the dressing equation \eqref{eq:dressing} by replacing $n_k^{\rm eff}(\lambda) \to q_k^{\rm eff}(\lambda)$ with driving $n_k$,  thus replacing $q_k(\lambda) $ by $n_k$ in Eq.~\eqref{eq:dressing} [cf. Eq.~\eqref{eq:strings} for the definition of $n_k$]. 
The temperature dependence of $\Dws$ is illustrated in Fig.~\ref{fig:zotos99}.

As we will see in Sec.~\ref{sec:GHD},  GHD provides a general framework for computing Drude weights in TBA formalism. In particular, both Eq.~\eqref{eq:Drudeenergy} and Eq.~\eqref{eq:Drudespin} are
special cases of the generic GHD result Eq.~\eqref{eq:DrudeGHD}, which describes the Drude weights of all conserved charges.

\subsection{Generalized hydrodynamics}
\label{sec:GHD}

The theory of generalized hydrodynamics concerns the evolution of integrable systems initially prepared in a state ${\rho}_0$ that is spatially inhomogeneous and then let to evolve unitarily with a homogeneous Hamiltonian. The main idea is that at large times, the expectation values of local observables become slowly varying functions of $x$ and $t$. This is much like the situation observed in the case of homogeneous quantum quenches [see, e.g., \cite{Essler2016, Eisert2015, Gogolin2016}]: initially the expectation values of local observables display fast oscillations but the latter dephase away for large times and expectation values become stationary even in the presence of a coherent unitary evolution. In the slow, late-time, regime, it is reasonable to expect that the expectation values can be described by a quasistationary state, namely 
\be\label{eq:hydrogen}
{\rm tr }\left[{\mathcal O}_x  e^{-i { H} t}   {\rho}_0 e^{i { H} t}\right]\overset{t\gg \tau_0}{\sim} {\rm tr }\left[{\mathcal O}_x   {\rho}_{\rm st}({x,t})\right],
\ee
where $H$ is the Hamiltonian of the system, ${\mathcal O}_x$ is a generic observable localized around the point $x$, ${\rho}_{\rm st}(x,t)$ is the density matrix describing the quasi-stationary state (retaining a slow space-time dependence), and $\tau_0$ is the time-scale for local relaxation. In general, the $x$-dependence in Eq.~\eqref{eq:hydrogen} is nontrivial for large but \emph{finite} times, while it is typically washed away at infinite times. 
Think, for example, of the free expansion of a gas released from a trap: the density of the gas vanishes \revision{for all $x$ at infinite times, corresponding to an $x$-independent $\lim_{t\to\infty}{\rho}_{\rm st}(x,t)$}. There are some cases, however, where nontrivial effects of the problem's inhomogeneity persist even at infinite times. In that case, one can explicitly take the infinite-time limit of Eq.~\eqref{eq:hydrogen} turning it into an exact statement. An example is the so-called bipartitioning protocol where one suddenly joins together two systems that are initially in different stationary states (see Sec.~\ref{sec:bipartitioning}). 

\revision{The state ${\rho}_{\rm st}(x,t)$ in Eq.~\eqref{eq:hydrogen} has been termed \emph{locally quasi-stationary state} (LQSS) in \cite{BertiniFagotti2016}. Specifically, 
it was argued that, at the leading order in time, ${\rho}_{\rm st}({x,t})$ is a generalized Gibbs state constructed with the charges of the Hamiltonian that controls the unitary time evolution and $(x,t)$-dependent chemical potentials. Note that the time scale at which the simplification \eqref{eq:hydrogen} arises --- often referred to as Euler time scale --- is much larger than the local relaxation time scale $\tau_0$. 
This means that, at fixed $(x,t)$, ${\rho}_{\rm st}({x,t})$ is homogeneous, stationary, and admits a ``microcanonical'' representation in terms of a TBA representative eigenstate, or, equivalently, of a set of densities of rapidities $\{\rho_{k}(\lambda,x,t)\}$. Determining such space-time dependent functions is the central problem of the theory.} 

A macroscopic number of constraints on these functions are obtained by considering the expectation values of the continuity equations of all local and quasilocal conserved charges from Eq.~\eqref{eq:higherconservations}, namely 
\be
\!\!\partial_t \dens_{x}(t) + \curr_{x}(t)-\curr_{x-1}(t)=0,\quad  x=1,\ldots,L,
\label{eq:continuity}
\ee
where $\dens_x$ is the density of charge $Q_n$ and $\curr_{x}$ its current.\footnote{We here simplify the notation introduced in Sec.~\ref{sec:theory} by writing $\curr_{x}$ instead of $j_x^{(Q_n)}$.} 
Here and in the following, we suppress the additional index $s$, keeping only the generic index $n$ for conserved charges. 
Assuming the validity of Eq.~\eqref{eq:hydrogen},  one obtains that, to leading order in time,  the expectation value of \eqref{eq:continuity} reads as  
\be
\partial_t {\rm tr }\left[\dens_{0}  {\rho}_{\rm st}({x,t})\right]+ \partial_x {\rm tr }\left[ \curr_{0}   {\rho}_{\rm st}({x,t})\right]=0\,.
\label{eq:evcontinuityleading}
\ee
We remark that this equation is already in the thermodynamic 
limit and, moreover, on its r.h.s., there are sub-leading corrections 
of $\mathcal{O}(t^{-b})$ with ${b>0}$. As shown in~\cite{Bertini2016,Castro-Alvaredo2016}, 
the constraint \eqref{eq:evcontinuityleading} is  sufficient  to fix the densities of 
rapidities to leading order in time. Specifically, 
Eq.~\eqref{eq:evcontinuityleading} is equivalent to the following continuity 
equation for the densities of rapidities
 \begin{equation}
  \partial_{t} \rho_{k}(\lambda, x,t)+  \partial_{x}(v^d_{k}(\lambda,x,t) \rho_{k}(\lambda, x,t)) = 0.
\label{eq:continuityrhoxt}  
\end{equation}
Here, $v^d_{k}(\lambda,x,t)$ is the group velocity of dressed excitations on the state ${\rho}_{\rm st}({x,t})$. The physical interpretation of this equation is straightforward: to leading order in time, the dynamics of $\{\rho_{k}(\lambda,x,t)\}$ can be described as if they were quasimomentum distributions for $N_s$ species of free classical particles moving in a density-dependent background. Indeed, the only effect of the interaction is a dressing of the group
velocity. These classical particles can be thought of as an ``asymptotic" version of the stable modes characterizing Bethe-ansatz integrable models. Indeed, for very large times and distances, the modes loose all phase information and behave like classical particles.

The crucial step in passing from Eq.~\eqref{eq:evcontinuityleading} to Eq.~\eqref{eq:continuityrhoxt} makes use of the following expression for the expectation value of generic currents on the macrostate $\{\rho_{n}(\lambda)\}$  
\be
j^Q[\rho]=\sum_{k=1}^{N_s} \!\int \!\!{\rm d}\lambda\, q_k(\lambda) v^d_{k}(\lambda) \rho_k(\lambda),
\label{eq:current}
\ee
where $q_k(\lambda) $ is the bare charge of the associated density (c.f.\ Eq.~\eqref{eq:charge}). This form has been originally proposed for relativistic integrable quantum-field theories with diagonal scattering \cite{Castro-Alvaredo2016} --- through a crossing-symmetry argument  --- and for the spin-1/2 XXZ chain  \cite{Bertini2016} --- through a semiclassical argument.  \revision{Initially, however, its validity could only be established numerically \cite{Bertini2016, Ilievski2017} or for some special currents \cite{Bertini2016, Urichuk2018}. The numerical accuracy of \eqref{eq:current} and its model-independent form triggered a fervent activity aimed at proving it rigorously \cite{Fagotti:2017aa, Yoshimura2018, borsi2019current, yoshimuraspohn2020} for all Bethe-ansatz integrable models. This endeavour has been concluded by \cite{Pozsgayproof}, who reports a complete proof of \eqref{eq:current} in the framework of the quantum-inverse scattering method. This proof encompasses all Yang-Baxter integrable lattice systems. In particular, this includes all Bethe-ansatz integrable lattice models (nested or not) such as the spin-1/2 XXZ chain and the one-dimensional Fermi-Hubbard model.} 
Finally, we remark that the form \eqref{eq:current} for the expectation values of currents has been shown to hold also for certain integrable classical field theories \cite{DoyonToda, Bastianello2018, Bulchandani2019, Cao_2019, spohn2020}.

The simplification introduced by Eq.~\eqref{eq:continuityrhoxt} is remarkable: to determine the late-time properties of an integrable quantum many-body system, one does not need to solve the many-body Schr\"odinger equation but an immensely simpler system of differential equations. After discretizing the rapidity, these equations can be treated by standard methods for initial-value partial differential equations~
\cite{iFluid}, by ``characteristics"~\cite{Doyon2017b, Bulchandani2017} or by ``molecular dynamics"~\cite{FleaGas}, i.e., by simulating the dynamics of the classical gas whose rapidity distributions obey Eq.~\eqref{eq:continuityrhoxt}. There is, however, a remaining nontrivial step to make before a solution can be obtained: one has to
find the right initial conditions for $\{\rho_{k}(\lambda,x,t)\}$. This problem has not yet been solved for all initial states $\rho_0$ but only for a number of particular choices \cite{Bulchandani2017a,Bulchandani2017b, Doyon:2017aa, Caux2019}. Importantly, some of these choices give a good characterization of experimentally-accessible initial configurations. \revision{This has been explicitly demonstrated in two recent cold-atom experiments~\cite{Schemmer, malvania2020generalized}, which have shown that GHD describes accurately the dynamics of nearly integrable 1D Bose gases in all accessible interaction regimes.} 

Let us now focus on the most popular initial configuration accessible with GHD: the bipartitioning protocol, i.e., the time evolution of an initial state composed of the tensor product of two different homogeneous states $\rho_0\sim \rho_{\rm L}\otimes \rho_{\rm R}$ (see Sec.~\ref{sec:bipartitioning}). As mentioned before, since in this case, we can explicitly take the infinite-time limit, Eq.~\eqref{eq:continuityrhoxt} becomes exact. The solution is a function of the scaling variable $\zeta=x/t$, usually termed ``ray", and can be ``implicitly" written as \cite{Castro-Alvaredo2016, Bertini2016} 
\be
\!\!\vartheta_{k}(\lambda, \zeta)\!=\! [\eta^{\rm L}_{k}(\lambda)\!-\!\vartheta^{\rm R}_{k}(\lambda)]\, \Theta [v^d_{k}(\lambda,\zeta)- \zeta]\!+\!\vartheta^{\rm R}_{k}(\lambda),
\label{eq:continuitythetasol}   
\ee 
where $\Theta(x)$ is the step-function and $\vartheta_{k}^{\rm L/\rm R}(\lambda)$ characterize the homogeneous GGE emerging at infinite distance from the junction on the left and on the right, respectively.\footnote{We assumed that $\rho_{\rm L/R}$ have \emph{cluster decomposition properties}. Namely, they satisfy $\lim_{|x-y|\to\infty}\braket{O_1(x)O_2(y)}_{\rm L/R}=\braket{O_1(x)}_{\rm L/R} \braket{O_2(y)}_{\rm L/R}$ where the operators $O_i(x)$ are local (i.e, they act trivially far away from the site $x_i$) and $\braket{O(x)}_{\rm L/R} \equiv {\rm tr}[O(x)\rho_{\rm L/R}]$.} This solution is implicit because $v^d_{k}(\lambda,\zeta)$ depends itself on $\vartheta_{k}(\lambda, \zeta)$. The explicit result is obtained by formulating an initial guess for $v^d_{k}(\lambda,\zeta)$ and iterating Eqs.~\eqref{eq:continuitythetasol}, \eqref{eq:TBT}, and \eqref{eq:dressing} until convergence is reached (this is typically achieved in less than ten steps). This protocol has been used for studying nonlinear transport in integrable quantum many-body systems on the lattice \cite{Bertini2016, Piroli:2017aa, BertiniPiroli2018, DeLucaspintransport, ColluraAnalytic, BertiniPRL2018, Mazza2018, Eisler2019} as well as on the continuum \cite{Castro-Alvaredo2016, Mestyan2018, Bertini2019}. Moreover, it has also been used for analysing the dynamics of entanglement in inhomogeneous situations \cite{Bertini2018, Alba2019, Alba2018, Mestyan2019}. Next, in Sec.~\ref{sec:GHDDrude}, we will discuss how this protocol can be used for computing Drude weights. 

In concluding this brief review of GHD, we must mention that Eq.~\eqref{eq:continuityrhoxt} does not represent an end point: there are currently many ongoing efforts to extend its range of applicability. First of all, the equation furnishes only a leading-order-in-time characterization, or, more precisely, describes the system for large times $t$ and lengthscales $x \sim t$. In analogy with ordinary hydrodynamics, however, one would expect GHD to describe the asymptotic behavior of the system also on other lengthscales, for example, the diffusive one where $x\sim \sqrt{t}$. This can be achieved by finding the sub-leading corrections in $t$ to Eq.~\eqref{eq:continuityrhoxt}. In particular, in Sec.~\ref{sec:GHDDiffusion}, we will briefly discuss a correction, recently identified in \cite{DeNardisDiffusion}, which is able to describe diffusive behaviors. Currently, however, a systematic method to find all sub-leading corrections to Eq.~\eqref{eq:continuityrhoxt} has been devised only in the noninteracting case \cite{FagottiHigherGHD, Fagotti2019}. Another active research strand is to extend Eq.~\eqref{eq:continuityrhoxt} to the case in which the time evolution is determined by a spatially inhomogeneous \revision{or time-dependent Hamiltonian, where space and time variations are slow}. \revision{In particular, \cite{DoyonYoshimura17} presented an extension valid in the case of a system confined to a slowly varying trapping potential, \cite{Bastianello2019} considered the case of position-dependent Hamiltonian parameters, and \cite{BastianelloDeLuca19} studied the effects of time-dependent magnetic fields.} Finally, there are ongoing efforts to describe the evolution of the initial-state correlations under \eqref{eq:continuityrhoxt}~\cite{ruggiero2019quantum}.

\subsubsection{GHD results for Drude weights}
\label{sec:GHDDrude}

Drude weights can be computed within GHD following two different approaches that, crucially, give coinciding results. Both approaches give access to the most general Drude weight 
\begin{align}
\Dw^{(n,m)}&= \frac{\beta}{2}\lim_{t\to\infty}\frac{1}{t}\sum_{r} \int_{-t/2}^{t/2}\!\!\!{\rm d}s \braket{\curr_{r}(s)\currm_{0}(0)}^{\rm c}\notag\\
&= \frac{\beta}{2}\lim_{t\to\infty}\sum_{r}{\rm Re}[\braket{\curr_{r}(t)\currm_{0}(0)}^{\rm c}],
\label{eq:GeneralDrude}
\end{align}
where $\braket{\cdot}^{\rm c}$ denotes the connected expectation value in a (grand-canonical) Gibbs Ensemble 
$\rho_{\rm GE}\propto \exp[-\beta H+ \sum_i \lambda_i O_i]$ \revision{(the sum in the exponent of $\rho_{\rm GE}$ runs over all conserved U(1) charges $O_i$ of the system like the total particle number, magnetization, etc).} 
In Eq.~\eqref{eq:GeneralDrude}, $n$ and $m$ can label two different conserved charges. In the case $n=m$, one recovers the usual ``diagonal" Drude weight of the charge $\CCharge$. All results, however, can be directly extended to the case of expectation values in more general GGEs. Note that (i) in order to treat all charges on the same footing, we divided  the energy Drude weight by $\beta$ and (ii) the correlation function in \eqref{eq:GeneralDrude} is \emph{not} the Kubo correlation used in Eq.~\eqref{eq:Kubo}. 
In the limit $t\to\infty$, the two expressions, however, can be shown to coincide under mild assumptions \cite{Ilievski2013}.\footnote{Interestingly, for integrable models, $\lim_{t\to\infty}\sum_{r} \braket{\curr_{r}(t)\curr_{0}(0)}^{\rm c}$ turns out to be real. This implies that the Drude weight can also be defined using a ``asymmetric" integration, namely $\Dw^{(n,m)}=\frac{\beta}{2}\lim_{t\to\infty}\frac{1}{t}\sum_{r} \int_{0}^{t}{\rm d}s \braket{\curr_{r}(s)\currm_{0}(0)}^{\rm c}$.}

The first approach, proposed in~\cite{Ilievski2017a, Bulchandani2017b}, evaluates the Drude weight using the following formulation. One considers a bipartitioning protocol that connects two halves of the system (left ``L" and right ``R") initially prepared in the following different GGEs 
\be
\rho_{\rm GGE, L/R}\propto \exp[-\beta H+ \sum_i \mu_i N_i \pm (\beta_m/2) \CChargem]\,,
\ee
where $\CChargem$ is the $m$-th conserved charge of the system. In this setting, one can compute $\Dw^{(n,m)}$ as follows~\cite{Vasseur2015}
\be
\Dw^{(n,m)}=\lim_{\beta_m \to0}\lim_{t\to\infty} \frac{\beta}{2 t \beta_m} {{\rm tr}[{\curr_{0} e^{-i H t} \rho_{0} e^{i H t}}]},
\ee 
where $\rho_{0}\sim \rho_{\rm GGE, L}\otimes \rho_{\rm GGE, R}$.    
Using Eq.~\eqref{eq:current} one can express this relation in terms of TBA quantities as  
\be
\!\!\Dw^{(n,m)}\!\!=\! \frac{\beta}{2}\! \sum_{k=1}^{N_s} \!\int \!\!{\rm d}\zeta\!\!\int_{-\Lambda}^{\Lambda}\! \!\!\!\!\!{\rm d}\lambda\,q_{n,k}(\lambda) \frac{\partial [v^d_{k}(\lambda,\zeta) \rho_k(\lambda,\zeta)]}{\partial  \beta_m}\bigg |_{\beta_m=0},
\label{eq:DrudeDNI}
\ee
where $q_{n,k}(\lambda)$ are the bare charges corresponding to $\CCharge$. 

The second approach, introduced in \cite{DoyonSpohnDrude}, computes the Drude weight using ``hydrodynamic projections". The idea is to write the Drude weight in the form Eq.~\eqref{eq:GeneralDrude} and expand it in the basis of conserved charges (appropriately orthogonalised). More precisely, one views 
\be
\sum_{r}\braket{\curr_{r}(t)\curr_{0}(0)}^{\rm c} \equiv (\curr|\currm),
\ee
as a scalar product in the space of local operators and assumes that the only contributions surviving at infinite times are coming from the overlap with conserved-charge densities  
\be
\!\!\!\!\!\lim_{t\to\infty} (\curr|\currm) = \sum_{k,k'} (\curr |\densk) [\frak C^{-1}]_{k k'} (\denskp |\currm)
\label{eq:projection}
\ee 
where we defined $\frak C_{nm}=(\dens|\densm)$. This reasoning is similar in spirit to that leading to the Mazur bound but it is carried out directly in the thermodynamic limit. In general, this approach can be used to compute the asymptotic behavior (large $t$ large $x$) of dynamical correlation functions in generic inhomogeneous situations \cite{DoyonCorr17}.

The quantities appearing in Eq.~\eqref{eq:projection} are all directly computed within GHD and lead to the following final result
\be
\Dw^{(n,m)}\!\!= \frac{\beta}{2} \sum_{k=1}^{N_s} \!\int_{-\Lambda}^{\Lambda} \!\!\!\!\!{\rm d}\lambda\,\ \!\!\! 
\frac{  \rho_k^t(\lambda) (v_k^d(\lambda))^2 q_{n,k}^{\rm eff}(\lambda)q_{m,k}^{\rm eff}(\lambda)}{(1+\eta_k(\lambda))(1+\eta_k^{-1}(\lambda))}\!,
\label{eq:DrudeGHD}
\ee
where $v_k^d(\lambda)$ and $q_{m,k}^{\rm eff}(\lambda)$ are the group velocity of excitations and the effective charge in the Gibbs state, respectively (i.e., with densities of rapidities obtained from solving \eqref{eq:TBT} and \eqref{eq:GTBA} with all Lagrange multipliers vanishing but $\beta$ and $\mu$). Remarkably, as shown in \cite{DoyonSpohnDrude}, this expression agrees with that obtained from Eq.~\eqref{eq:DrudeDNI} if one plugs in the implicit solution \eqref{eq:continuitythetasol} of the GHD equation for the bibartitioning protocol and takes the derivative explicitly. 

Three generic features of Eq.~\eqref{eq:DrudeGHD} are: (i) it is symmetric under the exchange of $n$ and $m$, in accord with Onsager reciprocal relations; (ii) the Drude weight is obtained by summing up ``elementary Drude weights" (the integrand of \eqref{eq:DrudeGHD}) for each quasiparticle in the system; (iii) the Drude weight of a certain quantity vanishes when the associated effective charges vanish. This happens, for example, in the case of the spin transport in the spin-1/2 XXZ chain with $|\Delta|\geq1$ at zero magnetization  and for the charge transport in the Fermi-Hubbard model at half filling. 

The expression Eq.~\eqref{eq:DrudeGHD} holds for all TBA  solvable models. Its generalization to the nested case has been first reported in \cite{Ilievski2017} and, again, corresponds to a sum of elementary Drude weights for each type of quasiparticle in the system. In particular, we see that Eq.~\eqref{eq:DrudeGHD} agrees with the special cases \eqref{eq:Drudeenergy} and \eqref{eq:Drudespin} discussed in the previous section once one restores the trivial $\beta$ factor in the energy Drude weight. Moreover, the nested generalization of Eq.~\eqref{eq:DrudeGHD} reproduces the result of  \cite{Fujimoto1998} for the charge Drude weight in the Fermi-Hubbard model. This follows by a direct comparison between Eqs.~(5) and (7) of  \cite{Ilievski2017} and Eq.~(35) of \cite{Fujimoto1998}, nonetheless, to the best of our knowledge, it has not  been noticed  in the literature. The main point is to note that $\xi_c(k), \xi_{s k}(\lambda), \xi_{b k}(\lambda)$ in \cite{Fujimoto1998} are exactly the ``effective electron charges" for the Fermi-Hubbard chain [cf.\ Eqs.~(A46) in the supplemental material of \cite{Ilievski2017}]. In other words, they fulfill the nested generalization of the dressing equations~\eqref{eq:dressing} with driving terms respectively given by $\xi^0_c(k)=1, \xi^0_{s k}(\lambda)=0, \xi^0_{b k}(\lambda)=2k$ [cf. Eqs.~(15)--(17) and Eqs.~(29)--(31) in \cite{Fujimoto1998}].

\subsubsection{GHD results for diffusion constants}
\label{sec:GHDDiffusion}
 
In order to access the diffusive regime, one needs to identify the leading 
corrections to Eq.~\eqref{eq:evcontinuityleading}, going beyond the Euler scale. 
A scheme to achieve this goal --- based on  two main assumptions --- has been 
proposed in~\cite{DeNardisDiffusion} [see also \cite{DeNardis2018, 
Gopalakrishnan2018a, Gopalakrishnan2019}]. The first assumption is that for 
large $t$, the system can be characterised using hydrodynamics also on 
lengthscales $x\sim \sqrt{t}$. Namely, one assumes that local observables are 
still described by a slowly varying quasi-stationary state $\rho_{\rm st}(x,t)$. 
This state, however, cannot be interpreted as a space-time dependent GGE 
anymore, but it has contributions proportional to the spatial derivatives of the 
Lagrange multipliers. Under this assumption, Eq.~\eqref{eq:evcontinuityleading} 
continues to hold also to  the first sub-leading order. The expectation values of 
the currents, however, are no longer given by Eq.~\eqref{eq:current} and include 
corrections written in terms of spatial derivatives of the densities of 
rapidities. Specifically, they can be written as~\cite{DeNardis2018} 
\begin{align}
&{\rm tr }\left[ j_{n,0}   {\rho}_{\rm st}({x,t})\right]=\sum_{k=1}^{N_s} \!\int \!\!{\rm d}\lambda\, q_{n,k}(\lambda) v^d_{k}(\lambda) \rho_k(\lambda,x,t)\notag\\
&\,\,\,-\frac{1}{2}\!\!\int \!\!{\rm d}\lambda{\rm d}\mu\!\! \sum_{k,k'=1}^{N_s} \! q_{n,k}(\lambda) {\frak D}_{k,k'}(\lambda,\mu) \partial_x\rho_{k'}(\mu,x,t),
\label{eq:GHDcurrentdiffusive}
\end{align}
where the kernel ${\frak D}_{k,k'}(\lambda,\mu)$ depends on $\{\rho_k(\mu,x,t)\}$. This kernel is related to the diffusion (Onsager) matrix defined as\footnote{Note that \eqref{eq:Onsagermatrix} does not coincide with the Onsager matrix given in Eq.~\eqref{eq:Lad} as in the latter we used Kubo correlation functions. Once again, however, the two matrices can be shown to coincide under mild assumptions~\cite{Ilievski2013}.} 
\be
D_{n,m}\!=\!\sum_{r} \!\!\int_{-\infty}^\infty\!\!\!{\rm d}t \left( \braket{\curr_{r}(t) \currm_{0}(0)}^c-\frac{2}{\beta}\Dw^{(n,m)}\right)
\label{eq:Onsagermatrix}
\ee
as follows~\cite{DeNardisDiffusion} 
\be
\!\!\!\!\!D_{n,m}\!= \!\!\sum_{p} \!\!\int \!\!{\rm d}\lambda{\rm d}\mu\!\!\!\!\sum_{k,k'=1}^{N_s}\!\!\! q_{n,k}(\lambda) {\frak D}_{k,k'}(\lambda,\mu) q_{p,k'}(\lambda)  {\frak C}_{p,m}
\label{eq:OnsagerD}
\ee
where the first sum is over all the conserved charges of the system and the matrix ${\frak C}_{p,m}$ has been introduced below Eq.~\eqref{eq:projection}. Note that it is always possible to add a ``derivative term" $\propto o_{x}-o_{x-1}$ to a charge density (where $o_x$ is a local operator), without modifying the total charge. This introduces an ambiguity in the definition of charge densities beyond the leading order [see \cite{Fagotti2019} for more details]. In particular, the kernel ${\frak D}_{k,k'}(\lambda,\mu)$ depends on the specific choice of the densities of charges while the Onsager matrix is invariant~\cite{DeNardis2018}. The simple relation \eqref{eq:OnsagerD} is obtained by taking charges and currents to be scalar under $\mathcal{PT}$-symmetry~\cite{DeNardis2018}. Finally, we remark that corrections similar to Eq.~\eqref{eq:GHDcurrentdiffusive}, i.e., depending on the spatial derivatives of the densities of rapidities, appear in the expectation values of all local observables with kernels that are generically unknown.  

The explicit TBA expression for Eq.~\eqref{eq:Onsagermatrix} in models with a single species of quasiparticles has been determined in~\cite{DeNardisDiffusion} through an expansion in finite-temperature form factors\footnote{\label{ff:definition}In this context, the term ``form factor" indicates the matrix element of a local operator between two Hamiltonian eigenstates.}.  In particular, it has been shown that Eq.~\eqref{eq:Onsagermatrix} is fully determined by form factors involving two particle-hole excitations. The expression for an arbitrary number of quasiparticles species has later been presented in~\cite{DeNardis2018} and reads as 
\begin{align}
D_{n,m}&\!=\!\int\!\frac{{\rm d}\mu_1{\rm d}\mu_2}{2}\!\!\!\sum_{k,k'=1}^{N_s}\!\! \biggl\{\!\frac{\rho^h_{k}(\mu_1)}{1+\eta_{k}(\mu_1)}\frac{\rho^h_{k'}(\mu_2)}{1+\eta_{k'}(\mu_2)}\notag\\
&\times\!\! \left(\frac{T^{\rm eff}_{k',k}(\mu_2,\mu_1) q^{\rm eff}_{n,k'}(\mu_2)}{\sigma_{k'} \rho^t_{k'}(\mu_2)}-\frac{T^{\rm eff}_{k,k'}(\mu_1,\mu_2) q^{\rm eff}_{n,k}(\mu_1)}{\sigma_k \rho^t_k(\mu_1)}\right)\notag\\
&\times\!\! \left(\frac{T^{\rm eff}_{k',k}(\mu_2,\mu_1) q^{\rm eff}_{m,k'}(\mu_2)}{\sigma_{k'} \rho^t_{k'}(\mu_2)}-\frac{T^{\rm eff}_{k,k'}(\mu_1,\mu_2) q^{\rm eff}_{m,k}(\mu_1)}{\sigma_k \rho^t_k(\mu_1)}\!\!\right)\notag\\
&\times  |v^d_{k}(\mu_1)- v^d_{k'}(\mu_2)|\biggr\}\,,
\label{eq:GHDDiffusion}
\end{align}
where both the effective charges $q^{\rm eff}_{n,k}(\lambda)$ and the ``effective scattering kernel" $T^{\rm eff}_{k,k'}(\lambda,\mu)$ fulfill \eqref{eq:dressing} with driving functions given by $q_{n,k}(\lambda)$ and $T_{k,k'}(\lambda-\mu)$ ($T^{\rm eff}_{k,k'}(\lambda,\mu)$ for fixed values of its ``second" arguments $k'$ and $\mu$), respectively. We note that, to obtain the result \eqref{eq:GHDDiffusion}, De Nardis, Bernard, and Doyon conjectured a general form for the kinematical poles for finite-density form factors: this represents the second main assumption of \cite{DeNardisDiffusion}.

Equation~\eqref{eq:GHDDiffusion} can be interpreted by realizing that, at the diffusive scale, the conserved modes of interacting integrable models, i.e., the quasiparticles, do not follow exactly free classical trajectories. As a consequence of the scattering, they perform a noisy motion around the classical trajectories with a variance that grows as $\sqrt{t}$. Such a noisy motion is responsible for the diffusive behavior~\cite{DeNardis2018, Gopalakrishnan2018a, Gopalakrishnan2019}. This simple argument can be refined to obtain a quantitative prediction in agreement with Eq.~\eqref{eq:GHDDiffusion} in the linear-response regime~\cite{Gopalakrishnan2018a}. Moreover, in accordance with this interpretation, Eq.~\eqref{eq:GHDDiffusion} vanishes for noninteracting models. \revision{Finally, we mention that a non-trivial check of \eqref{eq:GHDDiffusion} has recently been presented in \cite{Medenjak2019, Doyon2019QuadCharges} where the equation has been re-obtained using the hydrodynamic projection method.}

Including the ``diffusive correction" Eq.~\eqref{eq:GHDcurrentdiffusive} in the expectation value of the currents, the continuity equation for the space-time-dependent densities of rapidities takes the following Navier-Stokes form~\cite{DeNardisDiffusion} 
\begin{align}
 &\partial_{t} \rho_{k}(\lambda, x,t)+  \partial_{x}(v^d_{k}(\lambda,x,t) \rho_{k}(\lambda, x,t)) =\notag\\
 &+\frac{1}{2}\partial_x\left[\int \!\!{\rm d}\mu\! \sum_{k'=1}^{N_s} \! {\frak D}_{k,k'}(\lambda,\mu) \partial_x\rho_{k'}(\mu,x,t)\right]\!.
\label{eq:continuityrhoxtdiffusion} 
\end{align}
Of particular interest for this review is the case of the spin-1/2 XXZ chain with $|\Delta|>1$ for small perturbations around a zero-magnetization ($\magdens=0$) equilibrium state. In this case, Eq.~\eqref{eq:continuityrhoxtdiffusion} leads to the following heat-like equation for the profile $m(x,t)$ of the magnetization density~\cite{DeNardis2018}  
\begin{align}
\partial_{t} \magdens(x,t)= \DS \partial_x^2 \magdens(x,t)\,,
\end{align}
where the spin-diffusion constant is given by \revision{the following sum over the ``elementary diffusion constants" of different quasiparticles} 
\be
\DS =\sum_{k=1}^{N_s} \int_{-\pi/2}^{\pi/2}\!{{\rm d}\mu} \frac{\rho^h_{k}(\mu)}{1+\eta_{k}(\mu)} |v^d_k(\mu)| \mathcal W_k^2,
\label{eq:diffusionconstant}
\ee 
\revision{      
Here, the rapidity-independent coefficient $\mathcal W_k$ reads as \cite{DeNardis2019}
\be
\mathcal W_k = \lim_{k'\to\infty}\frac{T^{\rm eff}_{k',k}(\mu,\lambda)}{\rho^t_{k'}(\mu)}= \frac{1}{2 T \chi(\beta)} \partial_\delta n_k^{\rm eff}\,,
\label{eq:Wnk}
\ee 
where $n_k^{\rm eff}$ is the effective magnetization [cf. Eq.~\eqref{eq:strings} for the definition of $n_k$ and Eq.~\eqref{eq:dressing} for that of effective charges], $T$ is the temperature, $\chi(\beta)$ the static susceptibility, and  $\delta$ a small deviation from zero magnetization. 

As shown in \cite{DeNardis2019}, substituting \eqref{eq:Wnk} into \eqref{eq:diffusionconstant} and performing a few manipulations, one obtains an expression for the diffusion constant which has the same form as the right hand side of the bound \eqref{diflb} but involves a modified spin Drude weight. 

}

\section{Theoretical and computational methods}
\label{sec:methods}

While integrable systems as such in principle allow for analytically exact solutions, computing
the current autocorrelation functions that enter into the Kubo formalism is a formidable task and
no complete and general solution from Bethe-ansatz techniques exists so far.   Moreover, for non-integrable models,
one needs to resort to mostly numerical methods or universal low-energy descriptions such as bosonization.

\revision{We will concentrate the discussion on the specifics of the spin-1/2 XXZ chain for concreteness and will point out 
aspects that are important for the theoretical treatment of other models whenever necessary.}

\subsection{Low-energy theory}
\label{sec:fieldtheory}

\subsubsection{Field theory}

The low-energy excitations of a large class of 1d models are not fermionic  
quasiparticles but collective (bosonic) modes, forming the so-called Tomonaga-Luttinger 
liquid (TLL) \cite{Giamarchi,Schoenhammer2004}. The low-energy theory can be solved using bosonization, and the corresponding bosonic low-energy field theory takes the form (for one fermionic species)
\begin{equation}\label{eq:luttinger}
 H= \frac{v}{2}\int dx\left[\Pi^2+\left(\partial_x\phi\right)^2\right],
\end{equation}
where $\Pi$ is the conjugate momentum of the bosonic field $\phi$ with the 
commutation relation $[\phi(x),\Pi(y)]=i\delta(x-y)$. The TLL parameter $K$, which usually appears as a prefactor $1/K^2$ in front of the second term, has already been absorbed via a canonical transformation of the fields. 
\revision{For multiple species, such as is the case for the Hubbard chain, the low-energy Hamiltonian is a sum of independent Luttinger liquids. For the Hubbard chain, these describe
collective charge and spin excitations.}

For integrable 
systems, both $K$ and the spin velocity $v$ are known from Bethe 
ansatz. 
For example, for the spin-1/2 XXZ chain, one obtains [see, e.g., \cite{Essler2005}]
\begin{equation}\label{eq:xxz_luttingerpara}
K=\frac{\pi}{2}\frac{1}{\pi-\arccos(\Delta)}, 
~v=J\frac{\pi}{2}\frac{\sqrt{1-\Delta^2}}{\arccos\Delta}\,.
\end{equation}

The current operators associated with the spin density $\sim\partial_x\phi$ and 
with the energy density of the Tomonaga-Luttinger liquid Hamiltonian take the form 
\cite{Giamarchi,Heidrich-Meisner2002}
\begin{equation}
\JS = -v \sqrt{\frac{K}{\pi}}\int dx\, \Pi,~\JE=-v^2\int dx\, \Pi\partial_x\phi,
\end{equation}
and are both strictly conserved. The corresponding Drude weights read
\begin{equation}\label{eq:dw_luttinger}
\Dws=\frac{Kv}{2\pi},\quad ~\Dwe=\frac{\pi}{6}vT.
\end{equation}

If a certain microscopic model falls into the TLL universality class, the 
low-energy behavior of various correlation functions, such as the momentum 
distribution or the local density of states, is determined by 
Eq.~(\ref{eq:luttinger}). Transport properties, however, are  
nonuniversal: On the microscopic level of lattice Hamiltonians, they depend on 
integrability and the model parameters. In contrast, all gapless spin chains 
fall into the TLL universality class and at low $T$ map to 
Eq.~(\ref{eq:luttinger}), which by virtue of Eq.~(\ref{eq:dw_luttinger}) 
describes a ballistic conductor \cite{giamarchi91,Giamarchi1992}. Information 
about the microscopic origin of the integrability and on the conserved charges is thus lost by going to the continuum 
limit. \revision{The information on integrability is, in principle, contained in relations between the irrelevant operators that are discarded 
in the process. Accounting for these relations in the calculation of transport coefficients in a systematic manner is technically very hard and has not been 
accomplished yet.}

In order to describe transport beyond the purely ballistic case, one needs to resort to a more generic low-energy Hamiltonian. The RG irrelevant corrections to Eq.~(\ref{eq:luttinger}) which are most important in this context are given by 
umklapp scattering and band curvature:
\begin{equation}\label{eq:luttinger2}\begin{split}
H_\textnormal{u}&= \lambda_\textnormal{u} \int dx\cos\left(4\sqrt{\pi K}\phi\right)\\
H_\textnormal{b} & = \int dx\left[ 
\lambda_+(\partial_x\phi_L)^2(\partial_x\phi_R)^2
+\lambda_-(\partial_x\phi_L)^4+\lambda_-(\partial_x\phi_R)^4\right],
\end{split}\end{equation}
where $\phi=\phi_L+\phi_R$, and the prefactors $\lambda_{\textnormal{u},+,-}$ 
are known for integrable systems \cite{Lukyanov1998}. In an extension of 
earlier works \cite{Giamarchi1988} and \cite{Giamarchi1992}, the influence of 
these terms was studied via a finite-$T$ bosonic self-energy perturbation 
theory \cite{Sirker2009,Sirker2011}. This leads to a \textit{purely diffusive} 
form of the optical spin conductivity
\begin{equation}\label{eq:sigmaluttinger1}
 \sigma(q,\omega) = 
\frac{Kv}{\pi}\frac{i\omega}{[1+b(T)]\omega^2-[1+c(T)]v^2q^2+2i\gamma(T)\omega},
\end{equation}
whose real part takes a Lorentzian form in the long-wavelength limit $q\to0$:
\begin{equation}\label{eq:sigmaluttinger2}
\textnormal{Re }\sigma(\omega) = \frac{Kv}{\pi} 
\frac{2\gamma(T)}{[1+b(T)]^2\omega^2 + 4\gamma(T)^2}.
\end{equation}
The coefficients $b(T)$ and $c(T)$ as well as the decay rate $\gamma(T)$ are 
functions of $v,K,\lambda_{\textnormal{u},+,-}$ with $\gamma,b,c(T\to0)\to0$. 
 In the zero-temperature 
limit, Eq.~(\ref{eq:sigmaluttinger2}) recovers the expression for $\Dws$ from 
Eq.~(\ref{eq:dw_luttinger}). The Drude weight contribution to the conductivity 
at finite $T$, however, is missed and can at present only be accounted for by 
hand \cite{Sirker2011} using a memory-matrix approach [see, e.g., \cite{Rosch2000}].

Other exceptions to Luttinger-liquid universality are, e.g., real-time, 
real-space correlators, which already for free lattice fermions are governed by 
high-energy excitations. Further insights can be gained from  nonlinear 
TLL theory \cite{Imambekov2012}.

The above-mentioned bosonic self-energy 
perturbation-theory approach  \cite{Sirker2011} can also be used to compute the density 
correlation function. One finds that, at long times, the density autocorrelations 
are governed by a diffusive term $\sim\sqrt{\gamma/t}$, which is consistent with numerical tDMRG data \cite{Karrasch2015c}, but disagrees with earlier field-theory predictions \cite{Narozhny1996}. The formalism was subsequently extended to incorporate the 
effects of nonlinear Luttinger liquids at finite temperature \cite{Karrasch2015c}. While the integrability of a system drastically affects the long-time behavior of the global current autocorrelation function (i.e., the Drude weight), one does not expect a similar influence on the density-density correlations \revision{of local density operators such as $\sz{r}$}. Thus, there is no need to incorporate conserved quantities by hand, and field-theoretical approaches can be used to determine the long-time behavior of these quantities at low energies \cite{Sirker2011,Karrasch2015c}.

\subsubsection{Semiclassical approach}

\revision{

Damle and Sachdev introduced a semiclassical picture of thermally excited  
particles to compute the low-temperature behavior of the integrable, gapped, quantum $O(3)$ non-linear sigma model \cite{Sachdev1997, Damle1998} as well as of the Sine-Gordon field theory \cite{Damle2005}. The former describes the low-energy behavior of integer-$S$ (i.e., gapped) quantum spin chains in the limit of large $S$, for which the work of Damle and Sachdev predicts a zero Drude weight \cite{Sachdev2000} and diffusive dynamics with a conductivity that at low temperatures diverges as $\sigma_{\rm dc} \propto 1/\sqrt{T}$. The methodology was subsequently extended into various directions; e.g., a hybrid semiclassical-DMRG framework was developed \cite{Moca2017} and out-of-equilibrium setups were studied \cite{Werner2019,Bertini2019}.

The range of validity of the semiclassical approach was investigated both for the Sine-Gordon model as well as for integer-$S$ spin chains by comparing with DMRG or GHD results \cite{Moca2017,Werner2019,Bertini2019,DeNardis2019}. The current belief is that semiclassics give the correct qualitative prediction for the low-temperature limit.

}

\subsection{Exact diagonalization}
\label{sec:methods_ed}

Exact diagonalization (ED) has been a major work horse in the numerical 
analysis of finite-temperature transport properties \cite{Zotos1996, 
Narozhny1998, Heidrich-Meisner2003, Rabson2004, Herbrych2011, Karrasch2013}. The entire 
spectrum and all eigenstates are computed and therefore, practically any 
observable or correlation function can be extracted. However, there is the 
obvious limitation that only small system sizes can be accessed. For the 
spin-1/2 XXZ chain, routinely, the Hamiltonian can  be diagonalized for  $L\sim 
20$ sites by exploiting translational invariance (see \cite{Sandvik2010} for 
the implementation of $U(1)$ and discrete symmetries in ED). Accessing $L \sim
24$ is possible with some effort \cite{Heidrich-Meisner2006} for spin-1/2 chains.
\revision{For the Hubbard chain, the larger local Hibert space of four states further restricts the 
accessible system sizes, which can be overcome by using, e.g., dynamical typicality as described in Sec.~\ref{sec:typicality}. Technically, one 
needs to properly account for the fermionic statistics, which is important for correlation functions, yet 
a standard and well-known aspect of the numerical treatment of fermionic systems.}

\subsubsection{Formal expressions evaluated in ED}

We illustrate the main aspects for the example of the thermal and the spin conductivity 
in the spin-1/2 XXZ chain. The relevant expressions result from 
Eq.~(\ref{eq:KuboJJ}) by expanding the thermal expectation values in 
a basis  of many-body eigenstates $|n\rangle$, which we understand to be taken
from a subspace with fixed 
total magnetization  $\Mtot$.
Strictly speaking, by 
doing so, we work with a finite system and hence take $t\to \infty$ first and 
$L\to \infty$ next. We will first discuss the expressions and then comment 
on this conceptual aspect below.
Note that one can either work in a canonical ensemble, i.e., fixed $S^z$. In this case, the sums
in the following expressions run over all eigenstates from that subspace. Alternatively,
one can carry out a grandcanonical average over all values of $S^z$. Then, the sums have to be understood as:
\begin{equation}
\sum_n \rightarrow \sum_{S^z} \sum_{n(S^z)}\,,
\end{equation}
where the second sum runs over all eigenstates in the subspace with fixed $S^z$.

Foe the spin conductivity, we obtain the generic situation that both the Drude 
weight $\Dws$ and the regular part $\sigma'(\omega)$ can be nonzero in finite 
systems:
\begin{equation}
\Dws   =  \frac{ \langle - T_{\rm kin} \rangle}{2L}  - \frac{1}{T L} \sum_{n,n' \atop 
E_n\not=E_{n'}}  p_n \frac{|\langle n | \JS  | n'  \rangle|^2}{E_n - E_{n'}} \, 
\label{eq:dw0}     
\end{equation}
\begin{equation}\begin{split}
\sigma'(\omega) & =  \frac{1}{T L}   \frac{1-e^{-\beta \omega}}{\omega} \times  
\\ & \sum_{n,n' \atop  E_n\not=E_{n'}}  p_n \frac{|\langle n | \JS 
 | n'  \rangle |^2 }{E_n - E_{n'}} \delta(\omega -(E_n-E_{n'})) 
\end{split}\end{equation}
where $p_n = e^{-\beta E_n}/{Z}$ in the canonical case and $p_n = e^{-\beta 
E_n-\beta b \Mtot}/{Z}$ in the grand-canonical case, and $Z$ is the partition 
function. $T_{\rm kin}$ is the kinetic energy, which for the spin-1/2 XXZ chain from Eq.~\eqref{eq:XXZ-intro}
contains all terms but those proportional to $\sz{r} \sz{r+1}$.
    
In a 1d system, the Drude weight can also be obtained from the diagonal matrix 
elements of the current operator plus contributions from degenerate subspaces:
\begin{equation}
\Dws   = \frac{1}{2 T L} \sum_{n, n' \atop E_n =E_{n'}}  \, p_n \, |\langle n | \JS 
 | n'  \rangle|^2  \label{eq:dw1} \,, 
\end{equation}
which results from the absence of any superfluid density in a 1d system at 
finite temperatures \cite{Zotos1997}. Equations~(\ref{eq:dw0}) and 
\eqref{eq:dw1} are identical (i) at $\beta=0$ or (ii) at $\beta>0$ in the thermodynamic limit. Practically, they are
already indistinguishable at sufficiently high temperatures for the accessible system sizes $L \lesssim 20$ 
\cite{Heidrich-Meisner2003, Mukerjee2008}.

\begin{figure}[t]
\begin{center}
\includegraphics[width=0.90\linewidth]{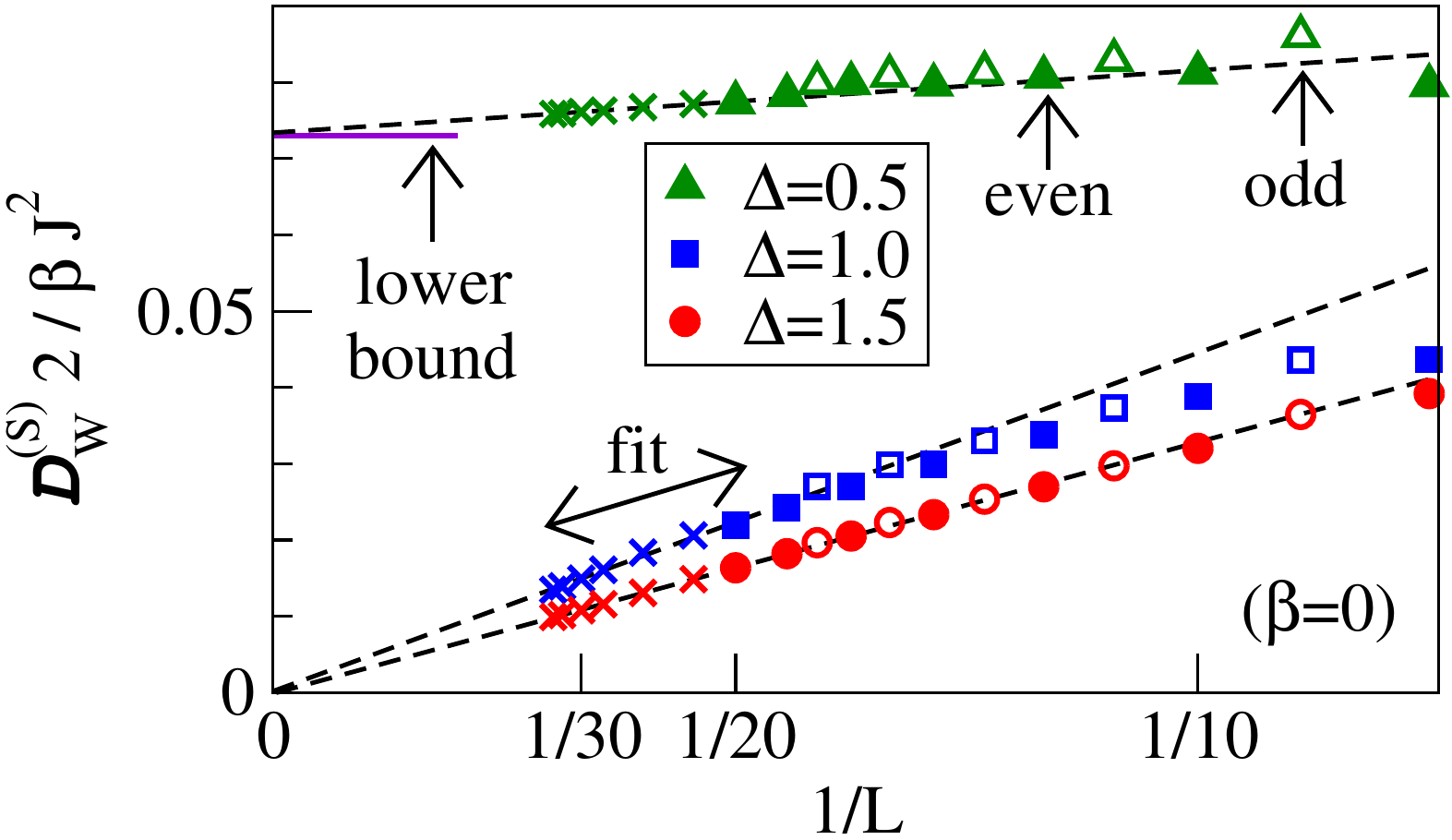}
\caption{(Color online) Finite-size scaling of the spin Drude weight $\Dws$ in
the high-temperature limit $\beta = 0$, as presented in \cite{Steinigeweg2014,
Steinigeweg2015}: Crosses indicate DQT data while other symbols indicate ED
data. Similar ED data can be found in, e.g., \cite{Zotos1996,Heidrich-Meisner2003,Rabson2004,Herbrych2011,Karrasch2013}.}
\label{fig:dw_scaling}
\end{center}
\end{figure}

An example for ED data for the spin Drude weight of the XXZ chain is shown in Fig.\ \ref{fig:dw_scaling}; the data was obtained in a grand-canonical ensemble using periodic boundary conditions. These results will be discussed further in Sec.\ \ref{sec:xxz}.
Here, we note that for $\beta=0$ and, e.g., commensurate points $\Delta=\cos{(\pi/3)}=1/2$, the convergence seems fast and indeed yields agreement with other methods such as the lower bound from \textcite{Prosen2013} tDMRG  (see the discussion in Sec.~\ref{sec:xxz_numerical_approaches}) or  TBA \cite{Urichuk2018,Zotos1999}. \revision{A recent Bethe-ansatz-based  calculation \cite{Kluemper2019} of the spin Drude weight for commensurate values such as $m=3,4,5,6$  in $\Delta=\cos{(\pi/m)}$
observes increasingly large finite-size effects at lower temperatures. 
One should realize, though, that this calculation extracts the Drude weight from a set of rapidities, which is 
different from grand- or canonical ensemble used in exact diagonalization.
Therefore, no quantitative insight on the finite-size dependencies of other methods can be gained from \cite{Kluemper2019}.}

For thermal transport (or any transport channel for which the current is 
exactly conserved), the expression for the associated Drude weight can be 
further simplified from the form of Eq.~(\ref{eq:dw1}) resulting in 
\begin{equation}
\Dwe  = \frac{1}{2 T^2 L} \sum_n p_n \,\langle n | (\JE)^2 | n \rangle  \,. 
\end{equation} 
This quantity exhibits the same mild finite-size dependencies as, e.g., the 
specific heat \cite{Alvarez2002a}. For instance, for $L=20$, the ED data agree well with the exact 
solution for $\Dwe$ down to $ T\gtrsim 0.1 J$ 
\cite{Heidrich-Meisner2002}.

As an alternative to the aforementioned expressions, one can also extract the spin 
Drude weight from the average curvature of many-body eigenstates in systems 
with twisted boundary conditions parametrized via $\phi$ \cite{kohn64}:
\begin{equation} \label{eq:kohn_ed}
\Dws = \frac{1}{2L}\sum_n p_n \,\left. \left(\frac{\partial^2 E_n(\phi)}{ 
\partial \phi^2} \right) \right|_{\phi=0} \,.  
\end{equation}
This has the advantage that only eigenenergies need to be evaluated but a 
numerical differentiation is required. 

\subsubsection{Role of boundary conditions, symmetries and choice of ensemble}

The choice of the boundary conditions, symmetries and the ensemble can all  
affect the finite-size data and their convergence to the $L\to \infty$ limit.

For systems with periodic boundary conditions, one observes weight in 
$\sigma'(\omega)$ in a frequency window $\omega < 1/L$ for certain 
values of the anisotropy $\Delta$ \cite{Naef1998,Herbrych2012}. Similarly, for 
systems with open boundary conditions, the Drude weight is exactly zero for 
finite $L$, but there exist precursor peaks in $\sigma'(\omega)$ at small
frequencies that move towards $\omega=0$ as $L$ increases \cite{Rigol2008a, 
Brenes2018}. These observations suggest subtleties in extracting $\Dw^{(S)}$ 
from finite-size data at exactly zero frequency. A useful strategy is to work 
with twisted boundary conditions (also inspired by Kohn's expression 
\eqref{eq:kohn_ed}) and a finite nonzero twist angle. This reduces the 
symmetries of the problem (see the discussion below) and the convergence 
with respect to $L$ can be accelerated \cite{Sanchez2017}.
 
The choice of the ensemble for the computation of the Drude weight can matter 
as well. Specifically, states appearing in the sum over $n$ in, e.g., 
Eq.~\eqref{eq:dw1}, can either be chosen from a single subspace with a fixed  
$\Mtot$ (canonical approach) or an average over all $\Mtot$ (grandcanonical 
version). For concreteness, we focus on the case of a vanishing external 
magnetic field, corresponding to   a vanishing average $\langle \Mtot \rangle=0$. For 
very 
large systems, one expects these different ensembles to yield the same result, 
which is confirmed in numerical simulations \cite{Karrasch2013, Sanchez2017},
yet on finite systems, the differences can be significant. For instance, at 
$\Delta=0$, the grandcanonical version converges faster to the $L=\infty$ 
result, while close to $\Delta =1$, the convergence of canonical data seems to 
be faster \cite{Karrasch2013,Herbrych2012}.

Symmetry constraints on the matrix elements of $\langle n | \JS | m\rangle$ 
play another important role and are at the root of some of the aforementioned 
finite-size dependencies. For instance, in the $\Mtot=0$ subspace ($L$ even) 
that 
is symmetric under spin inversion $Z^\dagger \sz{r} Z =  - \sz{r} $, all diagonal matrix 
elements vanish identically, i.e.,  $\langle n | \JS | n\rangle =0$ since the 
spin current is antisymmetric under $Z$. One can extend this to show that there 
is no contribution from the $\Mtot=0$ subspace on finite systems with $L$ even 
and incommensurate values of $\Delta \not= \cos(\pi \ell/m)$ at all 
\cite{Sanchez2017}. Therefore, in a canonical evaluation of $\Dw^{(S)}$, the 
leading contribution in small $\Mtot$ comes from odd $L$ and $\Mtot=1/2$ 
\cite{Herbrych2012}. Interestingly, for commensurate $\Delta = \cos(\pi \ell/m)$, 
degeneracies appear for $L\geq L_{\rm min} = 2m$ \cite{Sanchez2017}, implying 
that for certain values of $\Delta$ and small $L$, essential contributions to 
$\Dw^{(S)}$ are missed. Because of the sum rule, these contributions must sit at 
small frequencies on smaller system sizes, and therefore, a rather intricate, 
size-dependent transfer of weight from low- to zero-frequency occurs (see 
\cite{Naef1998} for an early discussion). A comprehensive discussion of 
symmetry constraints on the matrix elements of the spin current and an analysis 
of contributions of degenerate and nondegenerate subspaces can be found in 
\cite{Mukerjee2008, Narozhny1998} and, in particular, in \cite{Sanchez2017}.
 
Obviously, a theory for the finite-size dependencies of the Drude weight would  
be highly desirable. An interpretation was put forward in 
\cite{Steinigeweg2013} [see also \cite{Prosen1999}]: the Drude weight [see Eq.~\eqref{eq:dw1}], up to 
degeneracies, measures the spread of diagonal matrix elements of current 
operators in eigenstates and is thus a measure of how closely this observable 
obeys the eigenstate thermalization hypothesis (ETH) \cite{dAlessio2016} 
already on finite systems. 
On general grounds, one therefore expects an 
exponential decrease with system size for nonintegrable models [consistent with many ED studies, see, e.g., \cite{Prosen1999,Heidrich-Meisner2004,Zotos1996}], which obey ETH,
and a power-law dependence for integrable models. These  qualitative expectations  
for the $L$-dependence of the Drude weight are supported in most cases
for system sizes larger than a crossover length scale \cite{Steinigeweg2013}.

The calculation of the regular part requires some strategy to deal with the 
$\delta$ functions such as broadening or binning procedures when working 
directly in frequency space. The finite system size sets a lower bound on the 
accessible frequency range below which finite-size effects dominate. At low 
temperatures, a conservative estimate is $\omega \gtrsim 1/L$ while at high 
temperatures, much lower frequencies can be accessed due to the dominant 
contributions from dense portions of the many-body spectrum.

\begin{figure}[t]
\begin{center}
\includegraphics[width=0.90\linewidth]{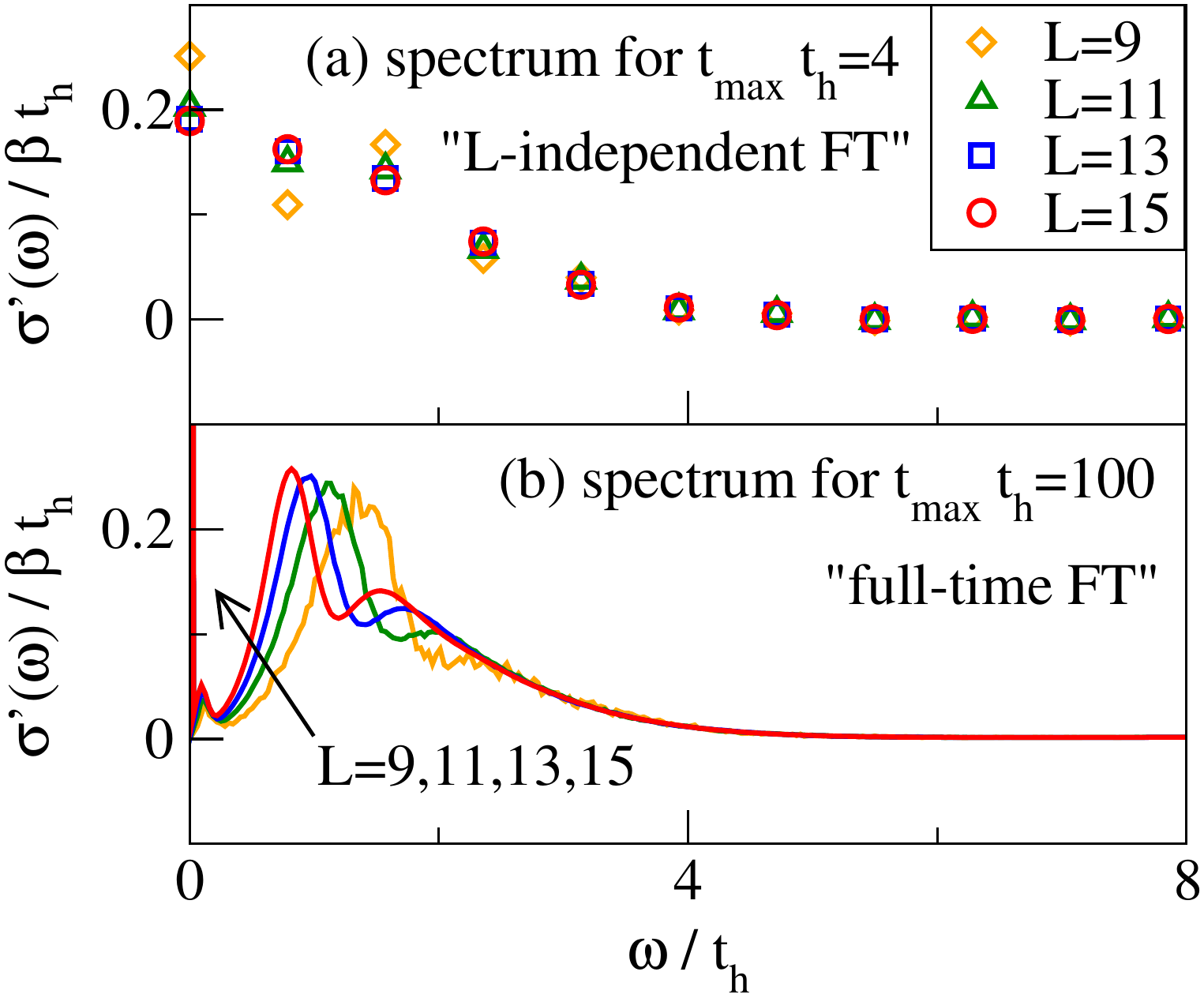}
\caption{(Color online) Frequency dependence of the charge conductivity in the 
Fermi-Hubbard chain at $U/t_\text{h} = 16$ and $\beta = 0$ and at half filling, as obtained from 
Fourier-transforming real-time data which uses (a) short times that are $L$-independent,
$t_\text{max} \, t_\text{h} = 4$, and (b) long times, $t_\text{max} \, t_\text{h} = 100$
\cite{Jin2015}. See Sec.~\ref{sec:TransportHubbard} and Eq.~\eqref{hub} for the definition of the Hamiltonian.
}
\label{fig:sigma_resolution} 
\end{center}
\end{figure}

\subsubsection{Pitfalls}

Let us now discuss the subtle point of the order of limits that was taken, 
i.e., $L\to \infty$ after $t\to \infty$, opposite to what is formally required. 
This is born out of the desire to operate with a closed expression for Drude 
weights rather than having to compute time-dependent quantities first and then 
carry out the limits. In fact, there is no known way of expressing the Drude
weight other than introducing a discrete set of eigenstates and hence going to infinite $t$ at 
finite $L$ first.

In ED, this approach is unavoidable, since system sizes are finite by 
definition. What could go wrong? One might be worried about mistaking a 
nonintegrable system for a ballistic conductor, since every finite system with 
discrete lattice translation invariance can have nonzero finite-$T$ Drude 
weights in the spin, charge or energy
channel. 
Thus, a careful finite-size analysis is required to deal with this. In those 
cases for which exact  or accepted results for the Drude weight are known (such as free 
systems or the energy Drude weight of the spin-1/2 XXZ chain) increasing 
system size in ED data leads to systematic convergence to the correct result. 
This observation lends confidence to the reliability of the analysis of 
finite-size trends.
Care must be taken in the vicinity of integrable points \revision{including limiting cases of free particles such as the spin-1/2 XX chain}, where the generic 
expectation is that microscopic physics will only unveil itself once very large 
systems are reached. Thus, the ED analysis of nonintegrable points better 
commences from points deep in the nonintegrable regime 
\cite{Heidrich-Meisner2004}.

Another pitfall can arise in the analysis of finite-frequency contributions, 
either from real-time data or directly in frequency space. A conservative 
approach is to only consider data that are $L$-independent, thus discarding the 
long times and low-frequency regime. An example is illustrated in the upper panel of
Fig.\ \ref{fig:sigma_resolution}; if the Fourier transformation is cut off at a short time scale, convergence in $L$ can be achieved. The Fourier transformation of long-time data (lower panel) shows significant finite-size effects at small frequencies
\cite{Prelovsek2004, Jin2015}.

\subsection{Dynamical quantum typicality}
\label{sec:typicality}

The concept of quantum typicality essentially states that a single pure state 
$| \psi \rangle$ can have the same properties as the ensemble density matrix 
$\rho$ \cite{gemmer2003, goldstein2006, popescu2006}. 
To be specific, here we look at the expectation
value of an observable $A$, i.e.,
\begin{equation}
\text{tr}[ \rho(t) A ] = \langle \psi(t) | A | \psi(t) \rangle + \varepsilon
\label{DQT}
\end{equation}
\cite{reimann2007, bartsch2009}, where $\varepsilon$ is a negligibly small 
correction (as discussed below in more detail). If $| \psi \rangle =  | n 
\rangle$ is a single eigenstate  with energy $E_n$ and $\rho = \rho_\text{mc}$ 
the microcanonical ensemble in an energy shell $E \approx E_n$, then Eq.\ 
(\ref{DQT}) becomes the diagonal part of the well-known eigenstate 
thermalization hypothesis (ETH)
\begin{equation}
\text{tr} [\rho_\text{mc} A] = \langle n | A | n \rangle + 
\varepsilon
\end{equation}
\cite{Deutsch1991, Srednicki1994, Rigol2008}. Even though the ETH is an 
assumption, there is solid evidence that it holds for local few-body observables in 
nonintegrable many-body systems \cite{Nandkishore2015, dAlessio2016}. However, 
in contrast to ETH, Eq.\ (\ref{DQT}) is a mathematically rigorous statement 
if $| \psi \rangle$ is essentially drawn at random from a sufficiently large 
Hilbert space \cite{reimann2007, bartsch2009}. In fact, the idea of using random states $| 
\psi \rangle$ has a long history \cite{alben1975, deraedt1989, jaklic1994} and
is at the basis of various numerical approaches to the density of states \cite{hams2000},
thermodynamic quantities \cite{devries1993, sugiura2012, sugiura2013, wietek2019},
equilibrium correlation functions \cite{itaka2003, elsayed2013, Steinigeweg2014,
Steinigeweg2016a, rousochatzakis2019}, non-equilibrium processes \cite{monnai2014,
endo2018, richter2019b}, as well as ETH \cite{steinigeweg2014b}. In this
review, we focus on the case of equilibrium correlation functions.

\begin{figure}[t]
\begin{center}
\includegraphics[width=0.90\linewidth]{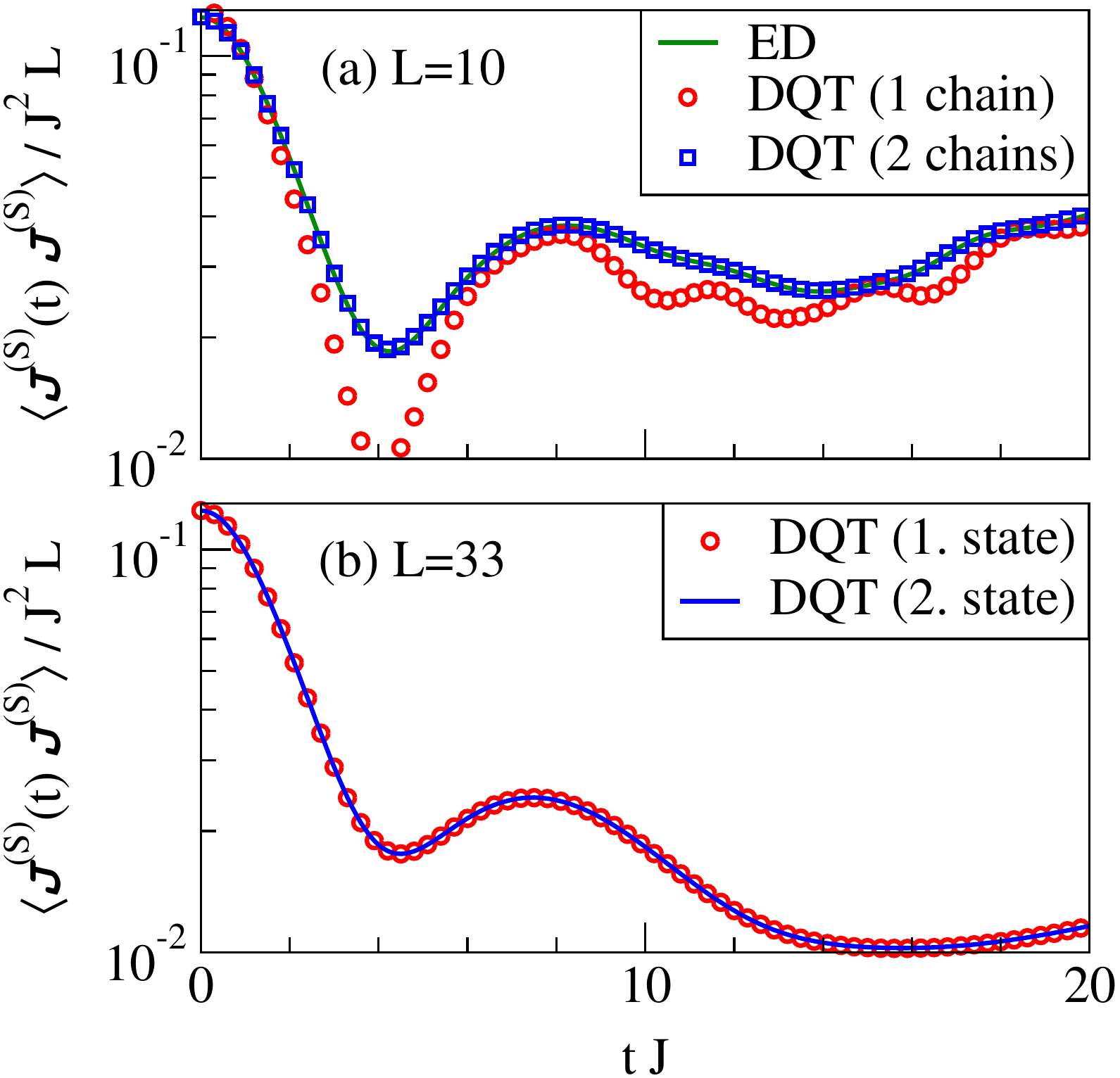}
\caption{(Color online) Accuracy of the DQT approximation, illustrated for the 
spin-current autocorrelation function in the spin-$1/2$ XXZ chain at 
the isotropic point $\Delta = 1$ and infinite temperatures $\beta = 0$ 
\cite{Steinigeweg2015}. (a) ED vs.\ DQT for a chain (with a total Hilbert-space dimension of $\text{dim} = 2^L$) and 
an uncoupled ladder ($\text{dim} = 4^L \gg 2^L$) with $L=10$. (b) DQT for 
$L=33$ and two randomly drawn pure states.
For the behavior of spin-spin correlations see, e.g., \cite{balz2018}.
}
\label{fig:DQT_error} 
\end{center}
\end{figure}

Using the idea of quantum typicality and considering, e.g., the canonical 
ensemble $\rho \propto e^{-\beta H}$, the equilibrium autocorrelation function
of an operator $A$ can be written as \cite{itaka2003, elsayed2013, Steinigeweg2014,
Steinigeweg2016a}
\begin{equation}
\text{Re} \, \langle A(t) A \rangle = \text{Re} \, \langle \psi | 
A(t) A | \psi \rangle + \varepsilon \label{DQT_correlation}
\end{equation}
with the pure state
\begin{equation}
| \psi \rangle = \frac{\sqrt{\rho} \, | \Phi \rangle}{\sqrt{\langle \Phi | \rho 
| \Phi \rangle}} \, , \,\, \rho \propto e^{-\beta H} \, , \label{DQT_rho}
\end{equation}
where the reference pure state $| \Phi \rangle$ reads
\begin{equation}
| \Phi \rangle = \sum_{k} c_k \, | k \rangle \, .
\end{equation}
Here, $| k \rangle$ can be any (orthonormal) basis, e.g., it can 
be the common eigenbasis of symmetries. In the  basis considered, the complex
coefficients $c_k$ must be chosen according to the unitary invariant Haar 
measure \cite{bartsch2009}, i.e., $\text{Re} \, c_k$ and $\text{Im} \, c_k$ 
have to be drawn at random from a Gaussian distribution with zero mean.\footnote{Note that
other types of randomness have been suggested as well \cite{alben1975, itaka2004}.} 
Assuming  $A$ to be a local operator \revision{in real space} (or a low-degree polynomial in 
$L$), the statistical error $\varepsilon$ in  Eq.\ (\ref{DQT_correlation}) is 
 bounded from above by $\varepsilon < {\cal O}(1/\sqrt{\text{dim}_\text{eff}})$, 
where $\text{dim}_\text{eff} = \text{tr}[e^{-\beta (H - E_0)}]$ is the 
partition function with the ground-state energy $E_0$. At $\beta = 0$, 
$\text{dim}_\text{eff} = \text{dim}$. Thus, $\epsilon$ decreases exponentially 
fast as $L$ is increased and eventually vanishes for $L \to \infty$. At $\beta 
\neq 0$, $\epsilon$ can still be expected to decrease exponentially but less 
quickly. The accuracy of the approximation (\ref{DQT_correlation}) for
finite $L$ is illustrated in Fig.\ \ref{fig:DQT_error} and can be 
checked in pratice by comparing to the exact correlation function or by 
comparing the results for two (or more) randomly drawn pure states. For a 
discussion of the full probability distribution of pure-state expectation values,
see \cite{reimann2019}.

The central advantage of the r.h.s.\ of Eq.\ (\ref{DQT_correlation}) is that 
its evaluation can be done without knowing eigenstates and eigenenergies. To 
this end, it is convenient to introduce the two auxiliary pure states
\begin{equation}
| \Phi_\beta(t) \rangle = e^{-\ii H t} \sqrt{\rho} \, | \phi \rangle \, , 
\,\, | \varphi_\beta(t) \rangle = e^{-\ii H t} A \, \sqrt{\rho} \, | 
\phi \rangle
\end{equation}
and to rewrite Eq.\ (\ref{DQT_correlation}) as
\begin{equation}
\text{Re} \, \langle A(t) A \rangle = \frac{\text{Re} \, 
\langle \Phi_\beta(t) | \, A \, | \varphi_\beta(t) \rangle}{\langle 
\Phi_\beta(0) | \Phi_\beta(0) \rangle} + \varepsilon
\end{equation}
\cite{itaka2003, elsayed2013, Steinigeweg2014, Steinigeweg2016a}. Then, the 
dependence on $t$ and $\beta$ occurs as a property of pure states only and can 
be obtained by solving the Schr\"odinger equation in real and imaginary time, 
respectively. For this purpose, any forward-iteration scheme can be used such as 
standard fourth-order Runge-Kutta \cite{elsayed2013} or more sophisticated 
Suzuki-Trotter decompositions \cite{devries1993} and Chebyshev polynomials
\cite{talezer1984, dobrovitski2003, weisse2006}. Since in these schemes, the
required matrix-vector multiplications can be performed  without storing (full)
matrices in computer memory, they can access long-time dynamics in large  Hilbert spaces. 
For instance, the spin Drude weight of the spin-$1/2$ XXZ
chain with $L = 33$ ($\text{dim} = 2^{33}$) \cite{Steinigeweg2014} [see Fig.\
\ref{fig:dw_scaling}] and the charge Drude weight of the Fermi-Hubbard chain
with $L=16$ ($\text{dim} = 2^{32}$) \cite{Jin2015} have been calculated. Similar
to tDMRG (see Sec.~\ref{sec:methods_dmrg}), real-time data can be Fourier-transformed
to  obtain also information in frequency space \cite{itaka2003}, e.g., the optical 
conductivity \cite{Steinigeweg2016a}.

Note that recently, dynamical quantum typicality has  been combined with
numerical linked cluster expansions \cite{Tang2013} to obtain current 
autocorrelations in the thermodynamic limit \cite{Richter2019}.

\subsection{Microcanonical Lanczos method}

The microcanonical Lanczos method (MCLM) \cite{Long2003} also works with single  
pure states drawn at random. Yet, in contrast to the last section, these states 
are constructed so as to give an accurate approximation to equilibrium 
expectation values in the microcanonical ensemble, i.e., Eq.\ (\ref{DQT_rho}) 
becomes
\begin{equation}
| \psi \rangle = \frac{\sqrt{\rho} \, | \Phi \rangle}{\sqrt{\langle \Phi | \rho 
| \Phi \rangle}} \, , \,\, \rho = \rho_\text{mc} \propto \sum_{n=1}^N | n 
\rangle \langle n | \, ,
\end{equation}
where $\rho_\text{mc}$ is a projector onto an energy shell which (i) is narrow 
but at the same time (ii) contains sufficiently many energy eigenstates $N \gg 
1$. Therefore, due to (ii), ETH is not required and typicality arguments 
can still be applied  \cite{steinigeweg2014b}. Moreover, MCLM has been designed 
to work directly in frequency space (instead of the time domain discussed 
before). 
See \cite{Long2003} for an extensive discussion of the method. 

In the algorithm presented in \cite{Long2003}, a pure state $| \psi \rangle$ 
is prepared around a desired energy $E$ by performing a Lanczos procedure on 
$K = (H- E)^2$. Then, the dynamical susceptibility is obtained from 
\begin{equation}
\sigma'(\omega) = - \lim_{\eta \to 0} \frac{\mbox{Im} \left \langle \psi |  
\JS \frac{1}{z-H+E} \JS | \psi \right \rangle}{\pi \langle \psi | (\JS)^2 | 
\psi \rangle } \, ,
\end{equation} 
where $z = \omega + i \eta$, by using, e.g., a  continued fraction expansion.
The quality of the corresponding results was demonstrated for $\sigma'(\omega)$ 
of spin-1/2 XXZ chains \cite{Long2003}. For the extraction of the Drude weight, 
which cannot be directly resolved in this approach but appears as a contribution 
at small frequencies $\omega < 1/L$, an ad-hoc integration over a low-frequency 
regime needs to be employed.

Since MCLM is a pure-state, Lanczos-based approach, it can access systems of
similar size, e.g., for spin-1/2  chains as long as $L \lesssim 32$ sites are
feasible. The approach has been 
applied to various physical situations, including spin-1/2 chains \cite{Long2003, 
Mierzejewski2011, Herbrych2012, Okamoto2018}, ladders \cite{Zotos2004, 
Steinigeweg2016a}, spin-1 chains \cite{Karadamoglou2004}, spin-full fermions 
\cite{Prelovsek2004}, and disordered spin systems \cite{karahalios09, Barisic2016}. Although 
MCLM has been originally formulated in the frequency domain, carrying out 
microcanonical calculations in the time domain is also possible 
\cite{steinigeweg2014b}. Moreover, other energy filters than $K =  
(H- E)^2$ can be chosen \cite{Yamaji2018}. For reviews on MLCM and 
other methods in the context of Lanczos diagonalization, see 
\cite{Prelovsek-review, jaklic2000}.

\subsection{Finite-temperature matrix product state methods}
\label{sec:methods_dmrg}
The density-matrix renormalization group (DMRG) method was originally devised as a tool to accurately determine static ground-state properties of one-dimensional systems \cite{white1992}. Later on, the method was extended in various directions, e.g., to access spectral functions, real-time evolutions, or thermodynamics \cite{Schollwoeck2005}. From a modern perspective, all DMRG algorithms can be formulated elegantly if one introduces the concept of matrix product states (MPS) \cite{Schollwoeck2011},
\begin{equation}\label{eq:mps}
|\psi\rangle = \sum_{\{\sigma_{r}\}} \textnormal{tr}\big[M^{\sigma_1}\cdot M^{\sigma_2} \cdots
M^{\sigma_L}\big] |\sigma_1\sigma_2\ldots\sigma_L\rangle,
\end{equation}
where $\sigma_r$ denote single-site quantum numbers at the $r$-th site. The (so-called bond) dimension $\chi$ of the matrices $M^{\sigma_l}$ grows exponentially with the amount of entanglement in the state $|\psi\rangle$. The idea of a ground-state DMRG calculation is to determine $M^{\sigma_l}$ variationally for a fixed, small $\chi$, which is a well-suited tactic for 1d systems obeying the area law \cite{Eisert2010}.

The above language allows one to deal with \textit{pure} states and is thus not directly applicable at finite temperatures. In order to access $T>0$, one can introduce the notion of matrix-product operators, or -- equivalently -- one can purify the thermal density matrix $\rho=e^{-\beta H}/Z$ by expressing it as a partial trace over a pure state living in an enlarged Hilbert space,
\begin{equation}\label{eq:purification}
\rho = \textnormal{tr}_{\cal Q} |\Psi_\beta\rangle\langle\Psi_\beta|,
\end{equation}
where auxiliary degrees of freedom ${\cal Q}$ encode the thermal bath 
\cite{Verstraete2004,Feiguin2005,Barthel2009,Barthel2013}. 
\revision{This purification step is not unique, and a simple choice is for the bath to be a copy of the
system's degrees of freedom, yet without any unitary dynamics of its own.

The key point is that a purification of the infinite-temperature state   $\rho=1/Z$ can be written down analytically.
Again, the representation of this state is not unique and a common choice is to put each physical degree of freedom into a
maximally entangled state with its copy in the bath by putting both into a singlet state.
A subsequent imaginary time evolution where $H$ acts only on the physical degrees of freedom that is carried out
using standard DMRG time-evolution methods can then (in principle) provide a purified version of the thermal state  $\rho$
at any finite temperature. The final thermal expectation values are obtained by taking the trace over the auxiliary degrees of 
freedom. Note that imaginary-, real-time evolution as well as the trace operation are all linear, and therefore, we exploit that they can be
applied in arbitrary order.}

For instance, 
correlation functions can be obtained via
[similar to Eq.\ \eqref{DQT_correlation}]
\begin{equation}\label{eq:dmrg1}
\langle A(t)B\rangle = \langle\Psi_0| e^{-\beta H/2} U(t)^\dagger A U(t)Be^{-\beta H/2}|\Psi_0\rangle,
\end{equation}
where the (matrix product) state $|\Psi_0\rangle$ purifies $\rho=1/Z$ at 
$\beta=0$. If the Hamiltonian at hand contains only 
short-ranged interactions, both the real and the imaginary time evolutions 
appearing in Eq.~(\ref{eq:dmrg1}) can be computed straightforwardly 
\cite{Vidal2004,Daley2004,White2004}, e.g., by splitting them up into small 
steps, $U(t)=\exp(-iHt)=\exp(-iH\delta t)\exp(-iH\delta t)\cdots$. One can then 
Trotter-decompose the exponentials $\exp(-iH\delta t)$ into mutually commuting local terms, which can be applied straightforwardly to a MPS 
\cite{Vidal2004,Schollwoeck2011,Paeckel2020}. Other ways to incorporate finite temperatures 
within DMRG include a Lindbladian superoperator approach \cite{Zwolak2004}, a 
transfer-matrix formulation \cite{Sirker2005}, or a probabilistic sampling over 
pure states \cite{White2009}.

\begin{figure}
\begin{center}
\includegraphics[width=0.95\linewidth,clip]{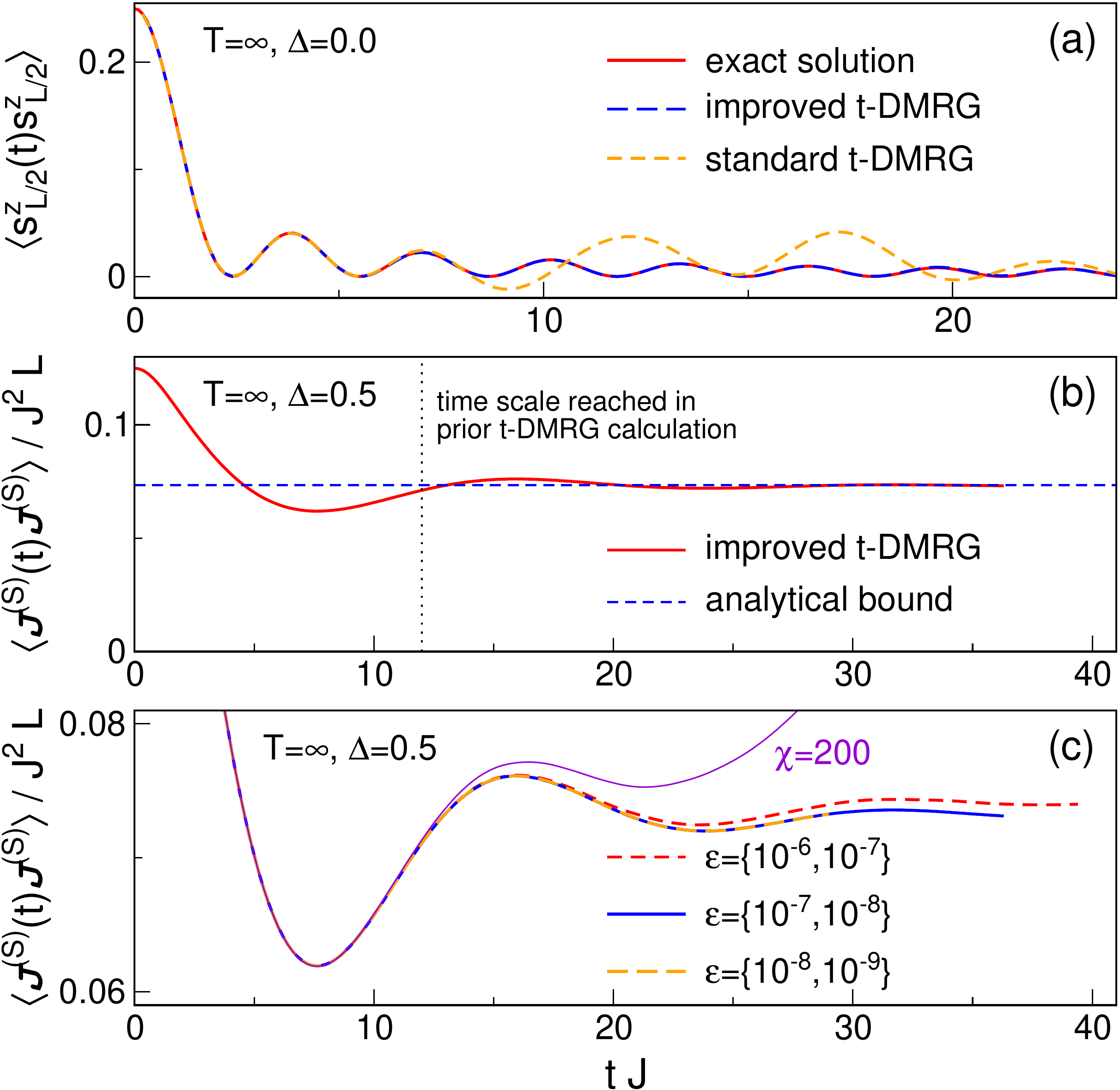}
\caption{(Color online) Benchmark of the improved finite-$T$ t-DMRG algorithm
for the spin-1/2 XXZ chain (\ref{eq:XXZ-intro}). (a) Spin autocorrelation function for
$\Delta=0$ computed using both the standard algorithm (\ref{eq:dmrg1}) and the
improved version ($U$ replaced by $\tilde U$) with a fixed bond dimension of
$\chi=60$. The exact solution is shown as a reference. The data is taken from
\cite{Karrasch2012}. (b) Spin current autocorrelation function at
$\Delta=0.5$ calculated using Eq.~(\ref{eq:dmrg2}) with a fixed discarded
weight. The data is taken from \onlinecite{Karrasch2015}. Both the analytical
result of \onlinecite{Prosen2013} and the time scale reached in the t-DMRG
calculation of \onlinecite{Sirker2009} are shown for comparison. (c) Same as in (b) but for different discarded weights which each differ by one order of magnitude (the two values denote the discarded weight during the two different time evolutions in Eq.\ (\ref{eq:dmrg2}), see \cite{Kennes2016}). Data obtained using a fixed bond dimension of $\chi=200$ is shown for comparison.}
\label{fig:dmrg}
\end{center}
\end{figure}

The crucial step when applying $e^{-i H \delta t}$ to a given MPS is to truncate the bond dimension by neglecting singular values below a certain threshold. 
This is the best approximation in the 2-norm of the wavefunction. The discarded weight is the key numerical control parameter; fixing it means fixing the error of the calculation. One usually runs calculations for several different values until physical observables have converged up to a desired accuracy [an example for this is shown in Fig.\ \ref{fig:dmrg}(c)].

If entanglement 
builds up linearly with time, the bond dimension $\chi$ grows exponentially and so does the 
computational effort. This severely limits the accessible time scales 
\cite{Barthel2009}. The strengths of time-dependent DMRG (t-DMRG) are that 
the system size can easily be chosen large enough to be effectively in the 
thermodynamic limit (due to a finite effective speed of information propagation \cite{liebrobinson}) and that it is not limited to 
integrable models or translationally invariant cases.

At finite temperatures, one can exploit the fact that some of the entanglement 
growth is ``unphysically'' taking place in ${\cal Q}$ and can thus be removed 
\cite{Karrasch2012,Karrasch2013p}. Mathematically, the state 
$|\Psi_\beta\rangle$ appearing in Eq.~(\ref{eq:purification}) is not unique but 
only determined up to an arbitrary unitary rotation, which can be chosen such 
that the entanglement is minimized. If this unitary is taken as a backward time 
evolution in ${\cal Q}$ with an operator $H_{\cal Q}$ that has the same form as $H$ (but acts 
in ${\cal Q}$), which amounts to replacing $U(t)$ by $\tilde U(t) = \exp\{-i(H-H_{\cal Q})t\}$ 
in Eq.~(\ref{eq:dmrg1}), then the entanglement growth is slowed significantly and 
larger time scales become accessible. This is illustrated in Fig.~\ref{fig:dmrg}(a) and 
\cite{Karrasch2012,Karrasch2013p}. It was later shown that the backward 
time evolution in ${\cal Q}$ appears naturally in an operator-space language 
\cite{Barthel2013,Tiegel2014}. Its form can also be motivated from  the fact that 
$|\Psi_\beta\rangle$ is an eigenstate of $H-H_{\cal Q}$ but not of $H$ 
\cite{Kennes2016}. Further optimization schemes were discussed in 
\cite{Barthel2013,Karrasch2013p}. A method which in practice 
allows one to find the minimally entangled representation
by iteratively minimizing the second Renyi entropy was presented in \cite{Leviatan2017b}.

Moreover, it was suggested \cite{Barthel2013} to rewrite $\langle A(2t)B\rangle=\langle A(t)B(-t)\rangle$ as 
\begin{equation}\label{eq:dmrg2}
\langle A(2t)B\rangle = \underbrace{\langle\Psi_0| e^{-\beta H/2} A \tilde U(t)}_{\langle\phi_1|} \underbrace{\tilde U(t) Be^{-\beta H/2}|\Psi_0\rangle}_{|\phi_2\rangle},
\end{equation}
and to determine the states $|\phi_{1,2}\rangle$ via separate time evolutions. This again gives access to larger time scales by about a factor of two or less  [see Fig.~\ref{fig:dmrg}(b)].

The so-improved finite-$T$ t-DMRG algorithm can be used to determine Drude weights and diffusion constants by looking at the long-time limit of, e.g., the current correlation function \cite{Karrasch2012,Karrasch2013,Karrasch2014,Karrasch2014a,Karrasch2015} or from local quenches \cite{Karrasch2014,Karrasch2017a} as well as in the bipartitioning protocol \cite{Vasseur2015,Karrasch2017}. The strength of such quenches can be tuned in order to reduce the build-up of entanglement and thus extend the simulation time, and it is observed that 
certain bipartitioning protocols  (see also Sec.~\ref{sec:bipartitioning}) are the best suited route to determine Drude weights \cite{Karrasch2017}. Frequency-resolved quantities can be determined from Fourier transformations \cite{Karrasch2014a,Karrasch2015,Karrasch2016}, which can be improved by using so-called linear prediction methods \cite{Barthel2009}.

Another possibility to obtain transport properties on longer time scales is to 
employ the time-dependent variational principle approach \cite{Haegemann2011}, 
but the approach has its own advantages \cite{Leviatan2017} and shortcomings 
\cite{Kloss2018}. Descendants of `Lanczos DMRG' methods, which directly yield 
frequency-dependent quantities \cite{Holzner2011,Tiegel2014}, are another 
promising avenue but have not been pursued yet in transport setups.
\revision{A very promising direction has recently been pursued by \cite{Rakovszky2020}. Operators with a local support
are evolved in the presence of a bath with a coupling strength $\Gamma$ that controls dissipation. The diffusion constant is recovered in the limit of
$\Gamma \to 0$ and agreement with previous studies has been observed \cite{Steinigeweg2014a}. }

\subsection{Quantum Monte Carlo} \label{sec:qmc}

For all spin systems on non-frustrated lattices, quantum Monte Carlo methods,
such as the stochastic series expansion (SSE) \cite{Syljuasen2002,Sandvik2010}, 
or the cluster methods using loop updates \cite{Evertz1993}  
 provide essentially exact results for the thermodynamics and static
correlations on large systems. Computing frequency-resolved quantities, though, 
is notoriously difficult due to the ill-defined problem of the analytic 
continuation from imaginary time to real time. One can avoid the problem by 
directly computing the response on the imaginary axis and comparing
to theoretical predictions expressed in imaginary rather than real time.
This method works best at low temperatures, where the set of available Matsubara 
frequencies $\omega_\beta = 2\pi n /\beta$ ($n \in \mathbb{Z}$) is more dense. Therefore, QMC studies of 
transport in 1d spin systems \cite{Louis2003,Alvarez2002, Alvarez2002a,Alvarez2002c, 
Heidarian2007, Grossjohann2010} and Fermi-Hubbard models \cite{Kirchner1999} are complementary to finite-temperature DMRG, 
ED, and dynamical typicality. The claims  of some of these QMC studies 
conflict with the bulk of the literature. For instance, both \cite{Kirchner1999} 
and \cite{Heidarian2007} claim evidence of ballistic transport in gapless 
nonintegrable models.
\revision{While there has not been any systematic comparison between QMC data and other numerical methods (which is hampered by
the different temperature regimes that these methods work in), a generic issue related to the analytical continuation
from the imaginary axis to the real--frequency axis arises at low temperatures. Since Matsubara frequencies $\omega\propto T$, there is a poor
resolution whenever the width of a peak in the spectral feature is smaller than $k_B T$.}

The statistical errors in QMC calculations are typically larger for higher-order 
correlation functions and it is therefore preferable 
\cite{Grossjohann2010,Alvarez2002,Alvarez2002a} to work with
two-site
correlation functions instead of directly evaluating current-current 
correlations \cite{Heidarian2007}.
At finite momentum $q$ and frequency 
$\omega_n$, one can relate the dynamical conductivity $\sigma_q(\omega_n) $ 
given by
\cite{Alvarez2002}
\begin{equation}
\sigma_q(\omega_n) = \frac{\langle -T_{\rm kin} \rangle - J_q^{(\text S)}(\omega_n)}{\omega_n}
\end{equation}
to the dynamical spin susceptibility $S_q(\omega_n)$ via
\begin{equation}
\label{eq:qmc_sq}
\langle -T_{\rm kin} \rangle - J_q^{(\text S)}(\omega_n) = \frac{\omega_n^2}{\tilde{q}^2} \, S_q(\omega_n)\,.\end{equation}
Note that, compared to Eq.\ (\ref{eq:sigma_chi}), there is a minus sign,
due to imaginary time. The expressions entering here are:
\begin{eqnarray}
J_q^{(\text S)}(\omega_n) &=& \frac{1}{L} \int_0^\beta e^{i\omega_n \tau}  
\langle\JS_q(\tau) \JS_{-q}(0) \rangle \, \text{d}\tau \, , \\
S_q(\omega_n) &=& \frac{1}{L} \int_0^\beta e^{i\omega_n \tau} \langle 
S_q^z(\tau) S^z_{-q}(0) \rangle \, \text{d}\tau \, ,  
\end{eqnarray}
where $\tau$ is imaginary time.

The strategy pursued in \cite{Alvarez2002,Alvarez2002a} is to fit the numerical 
data to a phenomenological ansatz (see \cite{Alvarez2002} 
for details). One notable result of \cite{Alvarez2002, Alvarez2002a} is a
Drude weight $\Dws(T) = \text{const.}$ at low temperatures for commensurate points 
$\Delta = \cos(\pi/m)$ ($m=1,2,\dots$) in the gapless phase of the spin-1/2 XXZ chain, in 
contradiction to the TBA results for the temperature dependence \cite{Zotos1999}.
It thus remains open whether the specific ansatz of \cite{Alvarez2002} is  
justified and whether finite temperatures were actually resolved in these QMC 
studies, which reproduce the zero-temperature Drude weight away from $\Delta=1$ 
with excellent accuracy.

Another QMC work \cite{Grossjohann2010} focused on the spin-1/2 XXX chain 
and aimed at verifying the field-theoretical prediction of 
\cite{Sirker2009,Sirker2011} for the dynamical spin susceptibility 
$S_q(\omega_n)$. Qualitatively, a diffusive form at small wavelength is 
expected based on the perturbative bosonization analysis of \cite{Sirker2009}, 
cf.\ Sec.\ \ref{sec:fieldtheory}. This is consistent with QMC data, yet 
quantitative deviations for the decay rate $\gamma$ were reported.

\section{Open quantum systems}
\label{sec:open_systems}

In this section, we describe methods that use an explicit external driving, \revision{such that a system evolves to} a nonequilibrium steady state (NESS)~\cite{Zia95,Dickman99}. \revision{The NESS describes a time-averaged system's density operator from which one can then evaluate expectation values of observables}. A particular emphasis will be put on the boundary-driven Lindblad setting as the most frequently used framework to obtain the NESS.

We note that open quantum systems are sometimes also studied numerically with a unitary time evolution, i.e., the leads are treated on the Hamiltonian level
and as finite system. We will not further discuss this approach here, but mention studies that looked at spin chains  \cite{Branschaedel2010,Lange2019,Lange2018} or electronic systems \cite{Heidrich-Meisner2010,Einhellinger2010,Kirino2010,Knap2011} sandwiched between leads. \revision{An alternative formulation used in studies of mesoscopic systems, particularly in the absence of interactions, is to describe the system's properties by a scattering matrix and the leads by occupation numbers, leading to Landauer-B\" uttiker type formulas~\cite{nazarov09}.} \revision{Finally, we mention that there exist some settings that are able to produce a NESS within the unitary dynamics. One is the bipartitioning protocol where one prepares two semi-infinite chains in different initial states and then evolves unitary in time [see Sec.~\ref{sec:bipartitioning}]. Another is to use a Lagrange multiplier to add a current operator to the Hamiltonian, see, e.g.,~\cite{antal97}.}

\subsection{Non-equilibrium steady-state driving}

\label{Lindblad}

A canonical way of studying nonequilibrium properties is to induce a NESS using some kind of reservoirs and to measure its properties. In studies of classical systems~\cite{lepri2003,dhar2008}, where many different types of reservoirs are available, this is, in fact, a method of choice to study transport~\cite{Zia95,Dickman99,derrida07}. Compared to linear-response calculations, no extra care is needed when treating anomalous transport often observed in classical nonintegrable 1D systems, such as, for instance, in the celebrated Fermi-Pasta-Ulam-Tsingou model~\cite{FPU,FPUPT}. 
Quantum NESS studies are fewer, one reason being that it is not so easy to construct quantum reservoirs that one can efficiently simulate. 
As we shall see in this section,  the situation has been changing in recent years, with increased research into quantum master equations.

In a one-dimensional system it suffices to use one reservoir at each chain end and, provided they are different, the system will, after a long time, evolve into a NESS $\rho_\infty$. Once one gets the NESS, the main quantity used to assess the transport is the NESS current $\jS{r}$, which is just the expectation value of the current operator. The current is always defined such that the continuity equation holds, and therefore,  at sites $r$, on which the reservoirs act, it must account also for the bath action. In the bulk, though, where the evolution is unitary, the current operator is the standard $\jS{r}$ obtained from the commutator between $h_r$ and the local density $\sz{r}$, see, e.g.,  Eq.~\eqref{eq:jS}, and therefore, the NESS current is $j^{(\rm S)} =\tr{(\rho_\infty \jS{r})}$. Due to the continuity equation, $j^{(\rm S)}$ is independent of the lattice site $r$. Provided one has diffusion the current 
will scale as $j^{(\rm S)} =-\DS \frac{\Delta \mu}{L}$ (i.e., Fourier's, Fick's, or Ohm's 
law), where $\Delta \mu$ is the difference in driving potentials,\footnote{For spin transport $\Delta\mu$ will be equal to a magnetization difference between chain ends and should not be confused with the chemical potential.} and $\DS$ 
a diffusion constant. If the system is not diffusive one will 
instead have a more general scaling, namely, keeping $\Delta \mu$ fixed \revision{the current will scale with system length $L$ as}
\begin{equation}
j^{(\rm S)} \sim \frac{1}{L^\gamma}.
\label{eq:NESSJ}
\end{equation}
Depending on $\gamma$ one has (i) diffusive transport for $\gamma=1$, (ii) 
ballistic transport for $\gamma=0$, (iii) superdiffusive transport for 
$0<\gamma<1$, and (iv) subdiffusive transport for $\gamma>1$. Localization 
corresponds to $\gamma \to \infty$. \revision{See \cite{Dhar19} for a review of anomalous transport in classical systems.} The type of transport can also be 
recognized from the NESS profile of a conserved density. Similarly as in 
classical systems, one expects some finite boundary ``jumps'' close to the 
location of driving, i.e., an  impedance mismatch. Disregarding those, in the bulk, one 
will have a linear profile for a diffusive system, a flat profile for a 
ballistic system, and a domain-wall-like profile for an insulator. In short, in 
order to keep $j^{(\rm S)}$ constant, local areas with higher resistivity will support 
higher density gradients, and vice versa. Heuristic profile shapes can also be 
associated to anomalous $\gamma$~\cite{znidaric11a,Scardicchio16}, though it is 
not clear how universal they are. Assuming a single-exponent scaling 
\cite{li2003}, these $\gamma$ are connected to the corresponding dynamical 
exponents in the context of linear response functions, see Eq. 
(\ref{eq:alpha2}),
\begin{equation}
\alpha' = \frac{2}{\gamma +1} \, .
\end{equation}

A crucial  question is how to efficiently implement reservoirs. One possibility is to describe the system and the (infinite) reservoirs as one large Hamiltonian system. Then one can derive the evolution equation of the system alone by tracing out the reservoir degrees of freedom. A problem with this approach is that the obtained equations are in general very complicated.
 For instance, the resulting master equation is nonlocal in time with a complicated memory kernel and in general,  is no easier to treat than the original problem~\cite{Breuer2002}. Depending on further approximations one gets a so-called Redfield master equation~\cite{redfield}, which  we shall not discuss, or a simpler Lindblad master equation. An exception are quadratic systems (i.e., non-interacting) where the physics is rather simple since quadratic translationally invariant systems display ballistic transport.

A more pragmatic approach is to simply seek an evolution equation for the system's density matrix that is able to describe the NESS situation and which is as simple as possible -- meaning that it still obeys all the rules of quantum mechanics. After all, in the thermodynamic limit, the bulk conductivity or transport type should not depend on the details of the driving, provided the dynamics is sufficiently ergodic. 
While this is seemingly natural, this assumption has to be checked in each individual system, especially in integrable systems. Next, we shall elaborate on such a setting.

\subsection{Lindblad master equation}
\label{sec:methods_open}

Let us argue for the simplest master equation governing the evolution of the system's density operator. Quantum mechanics is linear and therefore, we require that the evolution of $\rho(t)$ is also linear, and furthermore, that it maps density operators to density operators. Namely, if $\rho(0) \ge 0$,  $\rho(t)\ge 0$ should also hold. Requiring also that a map that is trivially extended to a larger space (i.e., one that acts as an identity on added degrees of freedom) also maps any positive semidefinite operator on that larger space to a positive semidefinite operator, means that such a map should be completely positive and not just positive. Such maps are known as completely positive trace-preserving (CPTP) maps~\cite{AlickiLendi07}. The class of CPTP maps is still too broad a set and therefore,  one requires an additional property, namely, that the action of reservoirs is as ``random'' as possible. In other words, the maps have no memory, i.e., they correspond to a Markovian evolution. Formally, this means that the evolution generated by the linear (super)operator ${\cal L}$ should form a dynamical semigroup: the evolution can be split into smaller steps, $\rho(t_1+t_2)=\e^{{\cal L}\cdot(t_1+t_2)}\rho(0)=\e^{{\cal L}\cdot t_1} \e^{{\cal L}\cdot t_2}\rho(0)$. It has been shown that any such evolution in a finite Hilbert space~\cite{GKS76} as well as in an infinite one~\cite{Lindblad76} can be written in the form of the Lindblad master equation (also LGKS -- Lindblad, Gorini, Kossakowski, Sudarshan),
\begin{eqnarray}
\label{eq:Lin}
\frac{\di}{\di t}{\rho(t)}={\cal L}(\rho(t))=\ii [ \rho(t),H ]+ {\cal L}^{\rm diss}(\rho(t)),\\
{\cal L}^{\rm diss}(\rho)=\sum_j [ L_j\, \rho,L_j^\dagger ]+[ L_j,\rho\, L_j^{\dagger} ], \nonumber
\end{eqnarray}
where $L_j$ are Lindblad operators that describe the action of reservoirs. Note that  the $L_j$ can be any operators, also non-Hermitian ones. Conversely, a Lindblad master equation with given $L_j$ and $H$ generates a CPTP map. For a historical account and earlier uses and occurrences of such an equation, see~\cite{Dariusz07}. In a finite-dimensional Hilbert space, Brouwer's fixed point theorem~\cite{Brouwer} guarantees the existence of at least one fixed point. Namely, a continuous map, $\e^{{\cal L}t}$ in our case, of a compact convex set (a set of density matrices) on itself has a fixed point, ${\cal L}\rho_\infty=0$. 
Typically and under certain algebraic conditions on $L_j$ and $H$~\cite{spohn77,frigerio77,evans77}, there is exactly one steady state  and, therefore, any initial state converges after long time to that unique NESS, $\lim_{t\to\infty} \e^{{\cal L}\cdot t}\rho(0)=\rho_\infty$. Systems described by the Lindblad equation are often called open systems~\cite{Breuer2002,AlickiLendi07}, as opposed to closed systems where the evolution is unitary.

 Depending on the driving type, one can distinguish the case of global $L_j$, see 
e.g.,~\cite{saito00,saito02,Saito2003}, or that of local $L_j$, 
e.g.,~\cite{michel03,Mejia2007,steinigeweg09}. A somewhat related scheme is that of a stochastic heat bath in which one measures and stochastically resets the boundary spin~\cite{Mejia2005,Mejia2007}. Another hybrid way to model a bath is by describing it as a lead (with a 
certain number of lattice sites) that is in addition coupled to a Lindbladian 
dissipation. For noninteracting leads, one can construct dissipators that 
thermalize such free systems~\cite{Dzhioev11,Ajisaka12,Landi16}, or model nontrivial spectral properties of the bath~\cite{Arrigoni13,Arrigoni16,Goold19}. For a discussion of thermalization properties of such baths, see~\cite{podolsky18}.

\begin{figure}[tb]
\centering
\includegraphics[width=0.6\linewidth]{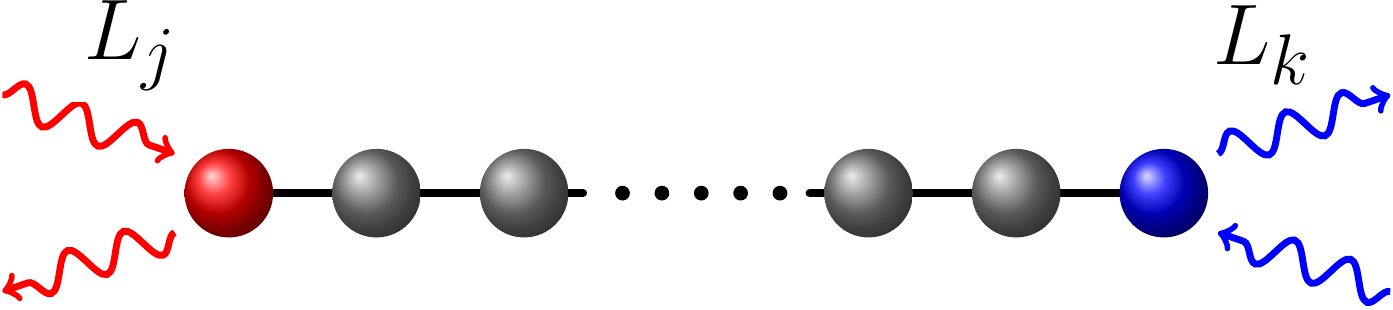}
\vskip2mm
\includegraphics[width=0.8\linewidth]{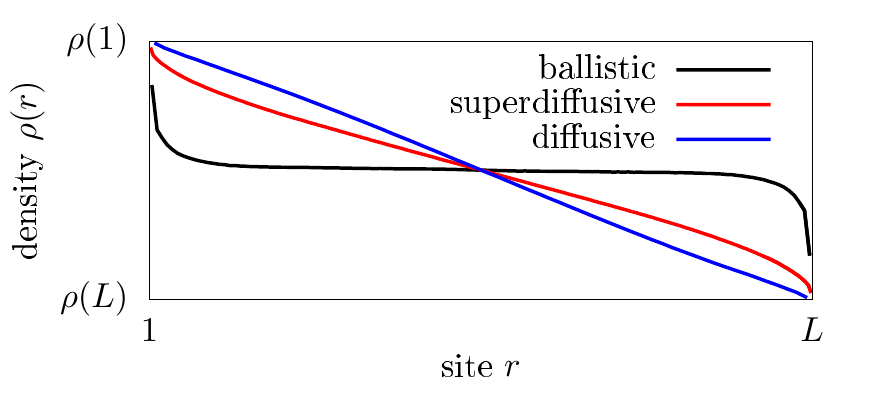}
\caption{(Color online) Top: NESS boundary Lindblad driving. Bottom: typical NESS density profiles for diffusive (blue), superdiffusive (red), and ballistic transport (black).}
\label{fig:Lin}
\end{figure}
One of the simplest choices are local $L_j$ that act only on the edges of the chain, such that the bulk dynamics is still fully coherent and determined by $H$ (see Fig.~\ref{fig:Lin}). This is similar to the way classical nonequilibrium lattice models are driven~\cite{Zia95,Dickman99,derrida07}, where the bath acts only on the boundary. The resulting locally-driven Lindblad equation is a mathematically sound NESS setting, without any shortcomings such as  the violation of positivity at short times encountered in the Redfield equation. Moreover, this setting often allows for  the  simulation of very large systems (hundreds of spins), and, sometimes, even permits  exact solutions. Justifying local $L_j$ on microscopic grounds is not easy; the standard weak-coupling microscopic derivation~\cite{Breuer2002} will typically result in nonlocal $L_j$. In particular, requiring an exact thermal steady state for equilibrium driving demands nonlocal $L_j$ (so-called Davies generators~\cite{davies74}) that have to be constructed by diagonalizing each particular $H$. This is neither practical nor in the spirit of having an effective bath description that is system-independent. From a practical point of view, demanding exact thermal states is anyway too strong as  it suffices that one is sufficiently close. For a system possessing good thermalization properties, it should not matter how one drives such a system in the thermodynamic limit.
The reason it that,  far away from the boundaries, a generic system will anyway self-thermalize and therefore, boundary effects, protruding a finite distance into the system, are expected to cause only subleading corrections.
This behavior, however,  is not guaranteed in an integrable system~\cite{znidaric10pre,Mendoza2015}.

Note that things are different if one studies small systems -- there one should pay close attention to thermodynamic details of local Lindblad driving~\cite{barra15} as well as to the fact that quantities such as, e.g., the temperature, might not have a well defined thermodynamic meaning~\cite{hartmann04,eisert14}. We remark that one can nevertheless provide a kind of ``microscopic'' picture also to local $L_j$. Hermitian $L_j$, such as the dephasing $L_j \sim \sz{j}$, can be obtained via Gaussian noise~\cite{Gardiner}, while general $L_j$ can be, somewhat more artificially, obtained by a continuous non-ideal measurement~\cite{Breuer2002}, or by an instantaneous repeated interaction with a bath~\cite{Attal06,Karevski09}. 

\subsubsection{Infinite-temperature magnetization driving}

Let us have a closer look at one of the simplest cases of Lindblad driving, where the Lindblad operators act on a single site and induce infinite-temperature spin transport. A one-site driving is given by two Lindblad operators that flip a spin up or down with different probabilities, thereby trying to induce a net magnetization at that site. They are given by 
\begin{equation}
L_1=\sqrt{\Gamma(1+\mu)}\splus{r},\quad L_2=\sqrt{\Gamma(1-\mu)}\sminus{r},
\label{eq:Lspin}
\end{equation}
where $\Gamma$ is the coupling strength, $\mu$ the driving strength, and 
$\spm{r}=(\sx{r}\pm \ii \sy{r})$. In the absence of any Hamiltonian, \revision{that is, driving a single-site system}, they have a unique 1-site steady state $\rho \sim \mathbbm{1}+\mu 2 \sz{r}$, and therefore, they model a bath that tries to induce a magnetization $+\mu$ at site $r$, i.e., $2\tr{(\sz{r} 
\rho)}=\mu$.

To induce a NESS in a long chain, one uses one such pair of $L$s at each 
chain end. For instance, using $+\mu$ driving at the left end and $-\mu$ at the 
right end results in a NESS with a position-dependent magnetization along the 
chain and a nonzero spin current (see Fig.~\ref{fig:Lin}). Similar Lindblad 
driving has already been used in early 
studies~\cite{michel03,michel04,Wichterich2007} and numerous subsequent ones, 
e.g.,~\cite{Prosen2009,Popkov13a,mendoza-arenas13,Landi14,Poletti18}. For $\mu=0$, one has a trivial steady state 
$\rho \sim \mathbbm{1}$, i.e., an infinite-temperature state, and one can 
interpret (\ref{eq:Lspin}) as spin driving at infinite temperature. For non-zero $\mu$, the NESS 
current $\jS{r}$ is nonzero and is the main observable. 

\revision{As described in Sec.~\ref{Lindblad}, the  transport type can then be extracted by evaluating the expectation value of the current $\jS{r}$ and of the magnetization $\sz{r=1,L}$. Due to a ``boundary resistance'' associated to a particular driving one will typically have boundary jumps in magnetization -- the expectation value of $\sz{r=1,L}$ will not be exactly $\pm \mu$. 
However, the size of such jump is proportional to $\jS{r}$ and therefore goes to zero in the thermodynamic limit provided the current goes to zero, which is true for  sub-ballistic transport. 
In the thermodynamic limit,  the difference in driving potential is therefore simply $\Delta \sz{}=\mu$, and furthermore, the current expectation value in the NESS can be evaluated at any site $r$. From its dependence on $L$ given in Eq.~(\ref{eq:NESSJ}), one can therefore extract the transport type, and in the case of diffusion, also the diffusion constant from  $\js=-\DS \frac{\mu}{L}$.}

\revision{We note that the Lindblad driving parameters are, in general, not simply related or equal to thermodynamic parameters. For instance, for a 1-site spin driving (\ref{eq:Lspin}) and $H=0$ one gets a 1-site steady state density operator $\rho \sim \mathbbm{1}+\mu \sigma^{\rm z}$ for which the ratio of probabilities of finding the spin in up and down states is $\frac{1+\mu}{1-\mu}$. Arguing that this ratio can be equated to $\e^{-\Delta E/T}$ given by the equilibrium distribution, where $\Delta E$ is the energy difference between the up and down states, would incorrectly associate a particular finite $T$ to a given $\mu$. At the boundary where the $L$ act, there will be a nontrivial interplay between driving and a nonzero $H$ (boundary effects), causing the state there, in general, not to be thermal. However, far away from the boundaries, one does expect thermalization (at least in non-integrable systems) and therefore, thermodynamic parameters describing local equilibrium can be determined from  local observables~\cite{znidaric10pre,Mendoza2015} [see also~\cite{Mori2018} for an alternative].}

An important question is whether the linear-response Green-Kubo type calculation and the NESS one in the limit of small $\mu$ give the same transport coefficient. This is, in general, a difficult question with no rigorous mathematical connection between the two transport coefficients known in general, either for quantum or for classical systems~\cite{Lebowitz00}. We shall outline one specific result for the spin driving given in Eq.~(\ref{eq:Lspin}).

\begin{figure}[tb]
\includegraphics[width=0.9\linewidth]{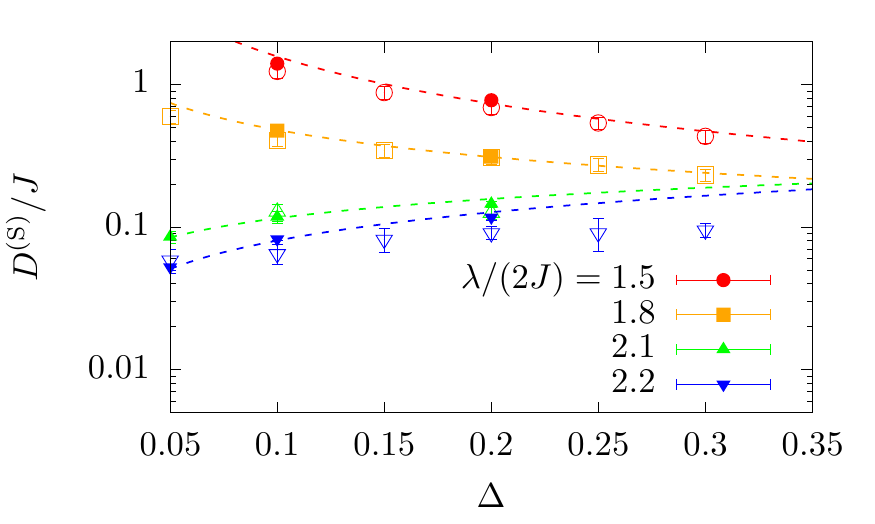}
\caption{(Color online) Comparison of results for the spin-diffusion constant
$\DS$ obtained from NESS simulations (empty symbols) and unitary domain-wall
spreading (full symbols) in the spin-$1/2$ XXZ chain with a quasiperiodic
magnetic field of amplitude $\lambda$. Details can be found in~\cite{Znidaric18}. \revision{Unitary domain-wall spreading is a particular case of a bipartitioning protocol, see Sec.~\ref{sec:bipartitioning} for details.}}
\label{fig:Dcmp}
\end{figure}

In the limit of weak driving $\mu\ll1$, one can use perturbation theory to get 
the NESS linear-response correction to the infinite-temperature equilibrium 
state $\sim \mathbbm{1}$. Similar to  classical systems~\cite{dhar09}, one can 
derive a NESS Kubo-like expression~\cite{Kamiya13,Znidaric2019} for the 
diffusion constant $\DS$  at infinite temperature, 
\begin{equation}
\DS = \lim_{L \to \infty} 4 L \int_0^\infty \frac{\tr{ (\jS{r} \jS{r'}(t)) 
}}{2^L} \di t
\label{eq:NESSKubo}
\end{equation}
for any $r$ and $r'$, where $\jS{r}(t) := \e^{{\cal L}_0 t} 
\jS{r}$, defined with ${\cal L}_0$ being the equilibrium Liouvillian propagator 
(i.e., ${\cal L}$ with $\mu=0$). Although looking similar to the equilibrium 
Kubo formula Eq.~(\ref{eq:KuboJJ}), the content is quite different.  For instance, the time 
integral is, due to the dissipative nature of ${\cal L}_0$, always well 
defined, 
even for finite $L$. Alternatively, the expression can be 
rewritten as~\cite{Znidaric2019} 
\begin{equation}
  \DS = L (8 \Gamma)^2 \int_0^\infty \frac{\tr{(\sz{L} \sz{1}(t))}}{2^L} \di t,
  \label{eq:NESSD}
\end{equation}
where $\sz{1}(t):=\e^{{\cal L}_0 t} \sz{1}$. The NESS diffusion constant $\DS$ 
is equal to the transfer 
probability accross the chain under evolution by ${\cal L}_0$ that is unitary in 
the bulk and dissipative at the edges. Even though it looks as if 
$\DS$ depends on $\Gamma$, this is not the case. One can show that in the thermodynamic limit, provided the unitary dynamics is perfectly diffusive and all parameters are held fixed (including $\Gamma$), this dependence is exactly compensated 
by a dissipative decay of $\sz{1}(t)$, resulting in exactly the same diffusion 
constant as the unitary Green-Kubo approach~\cite{Znidaric2019}. Quantitative 
agreement between the Lindblad and the unitary linear response Green-Kubo calculations of the 
diffusion constant has been verified in chaotic models, for 
instance, the spin-1/2 XX ladder~\cite{Znidaric2013a,Steinigeweg2014a}. 
\revision{Similarly, one can compare the Lindblad approach with the unitary dynamics in an out-of-equilibrium setting. Once again, to have a meaningful comparison one should focus on quantities accessible by both methods, such as the diffusion constant. So far, an extensive comparison has not been performed, however, an example is shown in Fig.~\ref{fig:Dcmp} for a spin-1/2 XXZ model 
in the presence of a quasiperiodic potential. Specifically, the figure shows a comparison between the diffusion constant obtained in the Lindblad evolution using a driving as specified in Eq.~(\ref{eq:Lspin}) and that extracted from the domain-wall spreading in a  bipartitioning protocol (see Sec.~\ref{sec:bipartitioning}).}

For non-diffusive systems (i.e., for superdiffusive or subdiffusive) the relationship between unitary and NESS approaches is less clear, although usually the same scaling exponent is observed (both for superdiffusive systems, such as the spin-$1/2$ Heisenberg chain~\cite{Ljubotina2017}, as well as for subdiffusive dynamics in a quasiperiodic potential~\cite{Znidaric2019Fibo}). \revision{It has been also demonstrated~\cite{michele20} that for ballistic systems the above Lindblad driving (\ref{eq:Lspin}) gives the same result as the Landauer-B\" uttiker formula at infinite temperature.} One case, believed to be special, where NESS and unitary dynamics do not agree, is a noninteracting critical model displaying multifractality~\cite{Varma17,Kulkarni18}. \revision{One should be also aware that in the non-linear response regime, i.e., at large $\mu$, one can get a different behavior. An explicit example is the spin-1/2 XXZ chain  at maximal driving $\mu=1$~\cite{ProsenExact11} or close-to-maximal driving~\cite{benenti09,benenti09b}.} It remains to be explored if and how a boundary-driven Lindblad setting can be used to extract the Drude weight or a frequency-dependent conductivity. Using simply a time-periodic driving $\mu$ in a Markovian Lindblad equation (Floquet Lindblad), see, e.g.~\cite{Bojan}, likely does not give the same information as $\sigma'(\omega)$.

Note that the coupling strength $\Gamma$ introduces an ``energy'' scale into the system and therefore, the limits of $\Gamma \to 0$ (or $\Gamma \to \infty$) will typically not commute with either the thermodynamic limit or, for instance, the limit of $\Delta \to \infty$ in the Heisenberg model. The limit of weak boundary coupling $\Gamma \to 0$ causes a decoupling of the bulk from the boundary, resulting in a different scaling of current and density with $\Gamma$~\cite{Prosen2011}. This means that weak boundary coupling $\Gamma \ll 1$ cannot be used to probe bulk transport. Similar caution is required also in the limit of $\Gamma \to \infty$, especially if there is any other diverging energy scale with which a scale introduced by $\Gamma$ can ``compete''. As an example, if one takes the limit of $\Delta \to \infty$ in the spin-1/2 XXZ spin chain, then a different behavior of the diffusion constant might be obtained depending on how one scales $\Gamma$~\cite{Znidaric2011}. This is a likely cause of a discrepancy in the value of the diffusion constant at large $\Delta$ between closed-system Kubo formula calculations~\cite{steinigeweg09,Karrasch2014} and the NESS result~\cite{Znidaric2011} obtained for a particular coupling-strength scaling $\Gamma \sim \Delta$. 
Namely, in the studies \cite{steinigeweg09,Karrasch2014}, the infinite-temperature limit is taken first, and therefore, even at large $\Delta$, one has a coupling between all states. The Lindblad setting, by contrast, with its finite (but large $\Gamma$), is closer to the case when one takes the limit $\Delta \to \infty$ first, i.e., at finite $T$ in the limit $\Delta \to \infty$ one decouples states with differing number of domain-walls.

\subsubsection{Solving the Lindblad equation}

How does one solve a many-body Lindblad equation? Provided the whole Liouvillian 
is quadratic (in, e.g., fermionic operators) one can use the so-called 3rd 
quantization method~\cite{3rdQuant}, simplifying diagonalization of the full 
$4^L \times 4^L$ Liouvillian to a diagonalization~\cite{ProsenSpectral} of a $2L 
\times 2L$ matrix describing decay modes. In some exceptional 
quadratic~\cite{znidaricJPA10} as well as non-quadratic 
systems [see, e.g.,~\cite{ProsenExact11,Karevski13,Popkov13,Prosen14,TPreview15,enej17,lenart18}], one can even get a closed matrix-product-operator (MPO) NESS solution.

Numerically, one can use full diagonalization (allowing $L \approx 10$ for spin-$1/2$), \revision{or an explicit integratation of an exponential set of differential equations [see~\cite{Orus19} for an overview]. An alternative approach, often used in atomic, molecular and optical system, is the quantum-trajectory method~\cite{Breuer2002} that evolves $\ket{\psi(t)}$ and averages over stochastic jumps to get $\rho$. Writing down $\ket{\psi(t)}$ in full one is again limited to small systems $L \approx 16$~\cite{Nori12}, however, using the MPS ansatz one can extend the available system sizes~\cite{Daley14}. For large systems, as needed in transport studies, the method of choice is a version of time-dependent DMRG (tDMRG)~\cite{Zwolak2004,Verstraete2004,Daley2004,White2004}, also called time-evolved block decimation (TEBD) method, used to evolve in time $\rho(t)$ until the NESS is reached. One could also try to avoid time evolution by directly targeting the NESS (i.e., the ``ground state'' of a non-Hermitian ${\cal L}$) using Lanczos-type methods (Arnoldi, $L \approx 15 $) or again employ the DMRG~\cite{Banuls15,Casagrande2020}. Invariably though, as in time evolution, the bottleneck will be a small gap of ${\cal L}$~\cite{ZnidaricPRE15gap}.}

The tDMRG for Lindblad equations works by writing the state $\rho$ 
in terms of a matrix-product operator  ansatz, exactly the same as for pure states (\ref{eq:mps}), 
the only difference being that the local Hilbert-space dimension in the 
operator space is the square of the pure-state dimension. For instance, for spin-$1/2$, 
it is 4, spanned, for instance, by Pauli matrices and the identity, which are orthogonal with respect to the Hilbert-Schmidt inner product. By discretizing  the time 
evolution into small time steps $\delta t$ and by using a Trotter-Suzuki decomposition of the time-evolution operator resulting in $\e^{{\cal L}\delta t}$, one propagates  some initial density operator, such as $\rho(0) \sim \mathbbm{1}$, in time until it converges to the NESS. The basic ingredient 
is a two-site nearest-neighbor transformation, similar to the time evolution of marix-product states 
~\cite{Schollwoeck2011}. 
Because non-unitary evolution eventually spoils the 
optimal truncation via Schmidt decomposition, it is worthwhile to occasionally 
reorthogonalize the state [see, e.g.,~\cite{znidaricnjp10}]. For further implementations by various groups, including open-source codes, \revision{see~\cite{tntonline,ITensor,Anto18,Brenes2018,Kollath18,Denisov19}.}

In unitary tDMRG simulations of MPS or MPOs (see Sec.~\ref{sec:methods_dmrg}), 
where one needs to account for the unavoidable entanglement growth, one fixes 
the discarded weight to a set number. As a consequence, the bond dimension  
necessary to maintain the same truncation per time step grows as a function of 
time, generically in an exponential way \cite{Schollwoeck2011}. As a 
consequence, the accessible time scales are of the order of several 
$\mathcal{O}(10/J)$. The tDMRG simulations for solving Lindblad master equations 
are, on the other hand, often carried out with a fixed bond dimension. This 
methodological choice (fixed bond dimension) is motivated by two arguments: 
First, dissipative dynamics is expected to exhibit a much milder entanglement 
growth than pure-state simulations, albeit still present [see the discussion in, 
e.g., \cite{Prosen2009}]. Second, one is not interested in the time evolution 
of, e.g., currents as such, but only in the NESS. Since in the cases of 
interest, the NESS is unique, different initial states should lead to 
the same NESS. Thus, numerical errors in accounting for the real-time evolution 
due to working at a fixed bond dimension should, to a certain degree, not 
prevent the system from converging to the correct NESS. 
A detailed analysis regarding the role of the discarded weight
in tDMRG simulations of Lindblad equations has not
been reported in the literature.
For instance, it is unclear whether 
significant truncation errors during the time evolution can possibly spoil the 
approach to the correct NESS. In the practical analysis of tDMRG simulations of Lindblad systems, 
one  checks the convergence of the NESS currrent
with the bond dimension, as will be illustrated next.

There are two main quantities that determine the efficiency and accuracy of the tDMRG simulations for Lindblad equations. 
One is the truncation error due to representing the NESS with a finite-bond MPO, the other is the required convergence time to the NESS that is given by the inverse gap of the Lindbladian superoperator ${\cal L}$ from Eq.~(\ref{eq:Lin}). Note that the Trotter time step should be chosen small enough such that it does not dominate over the truncation error. The size of the truncation error is connected to the operator ``entanglement''~\cite{Pizorn07} of the resulting NESS $\rho_\infty=\sum_k \sqrt{\lambda_k} A_k \otimes B_k$ given by the Shannon entropy of the non-negative $\lambda_k$ ($\sum_k \lambda_k=1$) obtained via the operator Schmidt decomposition, for instance, for a symmetric bipartition. If one starts from an identity initial state, which has zero operator entanglement, the operator entanglement will typically monotonically grow with time until it reaches its maximum value once the NESS is reached. Provided the operator entanglement of the NESS is low, the method is efficient as a small bond dimension suffices. For a small magnetization driving $\mu$ (\ref{eq:Lspin}), one typically observes the asymptotic scaling
\be 
\lambda_k \sim \mu^2 L^r/k^p
\ee 
for large $k$, with some model-dependent power-law exponents $r$ and $p$. The powers $r$ and $p$ crucially determine the size of the truncation error, and therefore, the required bond dimension $\chi$.

\begin{figure}[tb]
\includegraphics[width=0.9\linewidth]{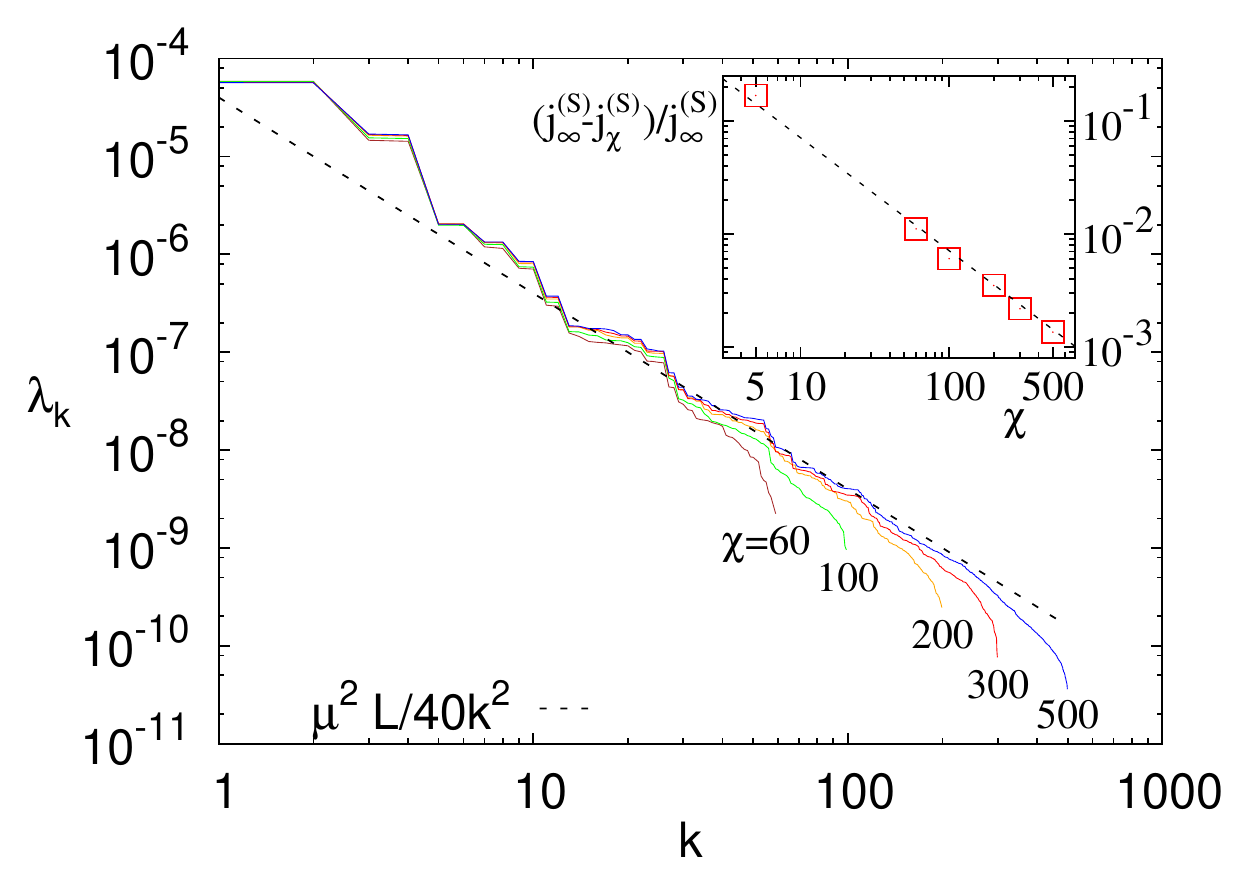}
\caption{(Color online) Schmidt spectrum for the NESS of a boundary driven spin-1/2 Heisenberg chain with \revision{$\Delta=1$ and} $L=64$ sites, $\mu=0.005$, and MPO bond dimensions $\chi=60-500$. The dashed line is the best-fitting asymptotic decay. Inset: Convergence of the NESS current $j_\chi$ obtained from a fixed-$\chi$ calculation, with the dashed line being the predicted error from the main plot $\frac{j_\infty-j_\chi}{j_\infty} \approx \frac{L}{90\chi}$ (no fitting parameters).}
\label{fig:XXX_schmidt}
\end{figure}

An example of the truncation-error analysis for the \revision{isotropic} spin-1/2 Heisenberg model is shown in Fig.~\ref{fig:XXX_schmidt}. Note that in this case, the NESS spin current (\ref{eq:jS}) scales superdiffusively with $\js \approx \mu\frac{0.39}{\sqrt{L}}$~\cite{Znidaric2011}. We evolve with a fixed bond dimension $\chi$ until the NESS is reached. The spectrum  $\lambda_k$ in the NESS is plotted in Fig.~\ref{fig:XXX_schmidt}. Analyzing its dependence on $L$ and $k$ one gets that the two exponents characterizing the NESS are $r\approx 1$ and $p \approx 2$. At fixed bond dimension $\chi$, the discarded probability weight is given by all dropped eigenvalues, $\sum_{k=\chi}^\infty \lambda_k$, and therefore scales as $\approx \mu^2 \frac{L}{40 \chi}$, where $\frac{1}{40}$ is an empirically fitted parameter obtained in the main plot of Fig.~\ref{fig:XXX_schmidt}. Since for small $\mu$, one has $\rho_\infty \sim \mathbbm{1}+{\cal O}(\mu)$, the largest eigenvalue is trivially $\lambda_0 \approx 1$. For relative precision, what matters is the ratio of the discarded weight and the first non-trivial eigenvalue $\lambda_1$, which is $\lambda_1 \approx 2.3 \mu^2$ for the data shown. The relative error of the NESS calculated using a finite $\chi$ can therefore be estimated as $\approx \frac{L}{90\chi}$ ($\frac{1}{90}\approx \frac{1}{2.3\cdot 40}$). Even though the error of a particular observable, such as the current, could involve some extra factors due to overlaps of Schmidt eigenvectors with the observable, we see in the inset of Fig.~\ref{fig:XXX_schmidt} that the agreement of the error estimate based only on the Schmidt spectrum with the actual error of the NESS current without any additional fitting parameters is very good.\footnote{The current $\js_\infty=\js_{\chi \to \infty}$ has been estimated using linear extrapolation in $1/\chi$.} For the boundary-driven Heisenberg chain, one therefore has to increase the bond dimension as $\chi \propto L^{\frac{r}{p-1}} \sim L$ if one wants to keep the error constant. For instance, $\chi \sim 100$ results in a relative error of 
about $1\%$  at $L=100$. If a slightly larger error of a few percent suffices, and one uses larger $\chi$, even systems with close to $L=10^3$ sites can be simulated. Such simulations, though,  take weeks of CPU time.

The other important parameter is the relaxation time required to converge to the NESS. For the spin-1/2 Heisenberg model, it scales as $\sim L^3$~\cite{ZnidaricPRE15gap}, and therefore, the complexity of finding the NESS with a fixed precision ($\chi \sim L$) scales as $L^3\cdot L\cdot \chi^3 \sim L^7$. However, in the spin-1/2 Heisenberg chain, it turns out that the spin current actually relaxes on a shorter time-scale $\sim L^{1.5}$~\cite{znidaric11a} and therefore, a fixed-precision NESS current can be obtained in ${\cal O}(L^{5.5})$ steps.

We note that if the Schmidt spectrum $\lambda_k$ decays slower than $1/k^2$ (which often happens) the required bond-dimension  scaling will be worse, see, e.g.,~\cite{Prosen2009} for some examples. Nevertheless, at infinite (or sufficiently high) temperature, when the NESS is close to $\mathbbm{1}$ (which is a product operator), the method typically works well since high temperatures are expected to decrease entanglement, especially compared to unitary evolution for which the complexity of $\rho(t)$ will typically grow exponentially with 
time, regardless of whether dynamics is chaotic or integrable~\cite{Prosen07}. For very slow transport (e.g., strongly subdiffusive dynamics), a long convergence time to the steady state can become a problem, rendering a boundary-driven Lindblad setting less suitable.

The single-site driving described above can be generalized to many sites. That is, 
one can choose Lindblad operators such that in the absence of $H$,
the steady state on those sites is a thermal state (or any other 
$\rho$)~\cite{Prosen2009,znidaric10pre}. Such many-site driving is, for instance, 
required in order to have an efficient coupling to the energy density (being at least a 2-site operator) and therefore, is used to  study energy transport~\cite{Prosen2009,znidaric11a,Mendoza2015,dario19}. 
An exception are   
weakly-coupled  systems~\cite{michel03,michel08,steinigeweg09}, for  
which  energy transport is essentially the same as spin transport.

A comparison of Lindblad and other master equations, such as 
the Redfield one, can be found in, e.g.,~\cite{Wichterich2007,bojan10,kulkarni16,Poletti19}, \revision{and with the Landauer-B\" uttiker formalism in~\cite{michele20}.}

\section{Transport in the spin-1/2 XXZ chain}
\label{sec:xxz}

Recently, significant progress has been made in the computation of
finite-temperature linear-response transport properties of one of the seemingly 
simplest interacting one-dimensional lattice models, the spin-$1/2$ XXZ chain. 

In this section, we give an overview of the current understanding, more than 
accounting for its historical development. We are presenting the results in 
the following order and discuss the relation between them: (i) exact 
statements, (ii) analytical results involving assumptions, (iii) results from 
numerical methods. We first discuss results for $\kappa(\omega)$ and 
$\sigma(\omega)$ in the linear-response regime in  Secs.\ 
\ref{sec:thermal_transport} -- \ref{sec:conductivity} and then 
cover insights from numerical open-quantum system simulations in  
Sec.\ \ref{sec:xxz.open}.

In the evolution of the field, \cite{Zotos1997} plays a seminal role, as it 
established the first exact lower bounds to the energy and spin Drude weight
of the spin-1/2 XXZ chain. Numerous early exact-diagonalization studies laid out 
the foundations for much of the future research (see, e.g., \cite{Fabricius1998, 
Zotos1996, Narozhny1998, Naef1998}). A next milestone is the work by Kl\"umper 
and Sakai \cite{Kluemper2002, Sakai2003} who computed the full temperature 
dependence of the energy Drude weight in the whole parameter range of the model 
from the quantum transfer-matrix method at zero magnetization. 
Finite-temperature Bethe-ansatz calculations of the spin Drude weight were 
carried out in \cite{Zotos1999,Benz2005} using different assumptions (see the 
discussion in Sec.~\ref{sec:DrudeTBA}). The significance of the work by Prosen and 
collaborators \cite{Prosen2011, Prosen2013} in proving the existence of nonzero 
finite-temperature Drude weights at vanishing magnetization was highlighted in 
Sec.~\ref{sec:charges}. GHD has very  recently emerged as a quite complete 
theoretical framework for the description of transport in integrable lattice 
models \cite{Bertini2016,Castro-Alvaredo2016} and is thus frequently referred 
to in the next sections. From the field-theory side, the work by Sirker {\it et 
al.} \cite{Sirker2009, Sirker2011} played a particularly inspiring role as it 
established the generic expectation for a gapless model after accounting for 
umklapp scattering. This can be considered the currently most advanced effective 
theory for the low-temperature transport of nonintegrable models 
\cite{Sirker2020}. For the integrable spin-1/2 XXZ chain, the theory predicts a diffusive 
form of spin autocorrelations at low $T$ \cite{Karrasch2015c}. For a recent 
review on transport in this model from the field-theory and Bethe-ansatz 
perspectives, see \cite{Sirker2020}.

Many results from numerical methods such as ED, tDMRG, and dynamical typicality 
were covered in Sec.~\ref{sec:methods} and will be mentioned in the context of 
the spin-1/2 XXZ chain in this section.

In the context of transport in the spin-1/2 XXZ chain and related models, open 
quantum systems were studied already in \cite{Saito1996, michel03, Saito2003, 
michel08}, yet acquired a much larger weight and higher reception after the 
first studies using tDMRG as the solver of the underlying Lindblad equations 
\cite{Prosen2009,Znidaric2011}. Notably,  numerical tDMRG solutions of Lindblad equations 
for transport through spin-1/2 XXZ chains were first in making  predictions for 
superdiffusion in the spin-1/2 Heisenberg chain and provided strong support 
for diffusive dynamics in the regime of $\Delta >1$ \cite{Znidaric2011}.

\subsection{The model}

The Hamiltonian governing the spin-1/2 XXZ chain is given in Eq.\ (\ref{eq:XXZ-intro}). For 
our choice of units ($J>0$), $\Delta > 0$ and $\Delta < 0$ correspond to the 
antiferromagnetic and ferromagnetic regimes, respectively. By using a 
Jordan-Wigner transformation \cite{Giamarchi},
\begin{equation}
\splus{r} = c_r^\dagger \, e^{i\pi\sum_{k=-\infty}^{r-1} n_k},~ \sz{r} = n_r - 
\frac{1}{2},
\end{equation}
the spin-$1/2$ XXZ chain can be mapped to a system of spinless lattice fermions 
$c_r^{(\dagger)}$:
\begin{equation} \label{eq:xxz_fermions}
\begin{split}
h_{r,r+1} = &\frac{J}{2} \, c_r^\dagger c_{r+1} + \textnormal{h.c.} + 
 J \Delta \left ( n_r-\frac{1}{2} \right) \left ( n_{r+1} -\frac{1}{2} \right).
\end{split}\end{equation}
The limit $\Delta=0$ corresponds to free fermions and can thus be solved 
analytically by a simple Fourier transform from real to (quasi)momentum space.

In this section, we will focus mainly on $\magdens=2\langle \sz{r} 
\rangle=0$,\footnote{We assume translational invariance.} which corresponds to zero magnetization (half filling) in the 
spin (fermion) language. The system is then gapless for $|\Delta|\leq1$ and at 
low energies falls within the Tomonaga-Luttinger-liquid universality class 
\cite{Giamarchi}. A gap opens for $|\Delta| > 1$. There, the ground state is two-fold 
degenerate and exhibits staggered spin order in the antiferromagnetic regime 
$\Delta > 1$, while in the ferromagnetic case $\Delta < -1$, the system 
exhibits phase separation. For finite $0<|\magdens|<1$, the system is a gapless 
Tomonaga-Luttinger liquid at any $\Delta$.
\subsection{Thermal transport} \label{sec:thermal_transport}

The energy-current operator, which is given in Eq.\ (\ref{eq:jE}), is exactly 
conserved for systems with periodic boundary conditions, $[H, \JE]=0$ 
\cite{Huber1969, Niemeijer1971, Zotos1997}. Thus, the zero-frequency thermal  
conductivity is  divergent at all temperatures and as a conseqence, the Drude weight is nonzero,
\begin{equation}
\Dwe(T>0) > 0 \, , \quad \kappar = 0 \, .
\end{equation}
At $\Delta=0$, the XXZ chain maps to free fermions via 
Eq.~(\ref{eq:xxz_fermions}) and $\Dwe$ can be obtained analytically for any 
$T \geq 0$:
\begin{equation} \label{eq:xxz_dwe_free}
\Dwe = \frac{1}{4\pi T^2}\int_{-\pi}^\pi \left[ \epsilon_k v_k f(\epsilon_k) 
\right]^2 e^{\epsilon_k/T} dk \, ,
\end{equation}
where $\epsilon_k=J\cos(k)$ denotes the single-particle dispersion, $v_k = 
\partial_k \epsilon_k$, and $f(\epsilon)= 1/(1 + e^{\epsilon/T})$.

The energy Drude weight of the XXZ chain has been calculated analytically for 
$\Delta \neq 0$ and arbitrary temperatures by exploiting the integrability of 
the system \cite{Kluemper2002, Sakai2003}. Since $\JE$ is a conserved quantity, 
$\Dwe \sim \langle (\JE)^2 \rangle$. This (time-independent) expectation value 
can be computed from a modified partition function $\rho \propto \tr [e^{-\beta H + \lambda 
\JE}]$ which serves as a generating functional and which can be determined 
within a quantum transfer-matrix formalism. One ultimately obtains an 
expression for $\Dwe$ in terms of a set of nonlinear integral equations. For 
high and low temperatures, these equations were solved analytically, and the 
result in the gapless phase $|\Delta| \leq 1$ reads 
\begin{equation}
\Dwe=\begin{cases}
\frac{\pi}{6} v T & T \to 0 \\ 
\frac{1}{128 \pi} \left[3 + \frac{\sin(3\eta)}{\sin(\eta)} 
\right] \frac{J^4}{T^2} & T \to \infty \, ,
\end{cases}
\end{equation}
with $v$ defined in Eq.\ (\ref{eq:xxz_luttingerpara}), and $\Delta = 
\cos(\eta)$. The low-$T$ limit agrees with the expression 
(\ref{eq:dw_luttinger}) obtained using bosonization 
\cite{Heidrich-Meisner2002}. In the gapped regime, one finds $\Dwe \sim 1/T^2$ 
at high $T$ as well as $\Dwe \sim e^{-\delta/T}/\sqrt{T}$ at low $T$, where 
$\delta$ is the one-spinon gap in the antiferromagnetic regime $\Delta > 1$ 
and the one-magnon gap in the ferromagnetic regime $\Delta < -1$. For arbitrary 
$T$, the set of nonlinear integral equations can be solved numerically.
Results for $0 \leq \Delta \leq 1$ as well as for $|\Delta|>1$ were presented 
in \cite{Kluemper2002} and in \cite{Sakai2003}, respectively. The temperature  
dependence of $\Dwe$ is shown in Fig.\ \ref{fig:drude_bethe} for three values 
of $\Delta$.

It was subsequently shown that for $|\Delta|<1$, the energy Drude weight can 
also be computed using the thermodynamic Bethe ansatz \cite{Zotos2016}. On the 
numerical side, $\Dwe$ was calculated via an exact diagonalization of  
small systems \cite{Alvarez2002a, Heidrich-Meisner2002}. At sufficiently high temperatures, the ED results are in 
agreement with the exact ones (obtained in the thermodynamic 
limit).

The energy Drude weight away from zero magnetization was obtained exactly using the 
quantum-transfer matrix approach \cite{Sakai2005}, TBA \cite{Zotos2016} as well 
as via exact diagonalization \cite{Heidrich-Meisner2005,Louis2003} and quantum 
Monte Carlo simulations \cite{Louis2003}. Magnetothermal corrections to the 
energy Drude weight due to the coupling of the energy to the spin current
were addressed in \cite{Louis2003,Heidrich-Meisner2005,Sakai2005,Psaroudaki2016,Zotos2016}.

For a discussion of the energy Drude weight in other integrable spin chains, 
see, e.g., \cite{Ribeiro2010}.

\begin{figure}[t]
\begin{center}
\includegraphics[width=0.90\linewidth]{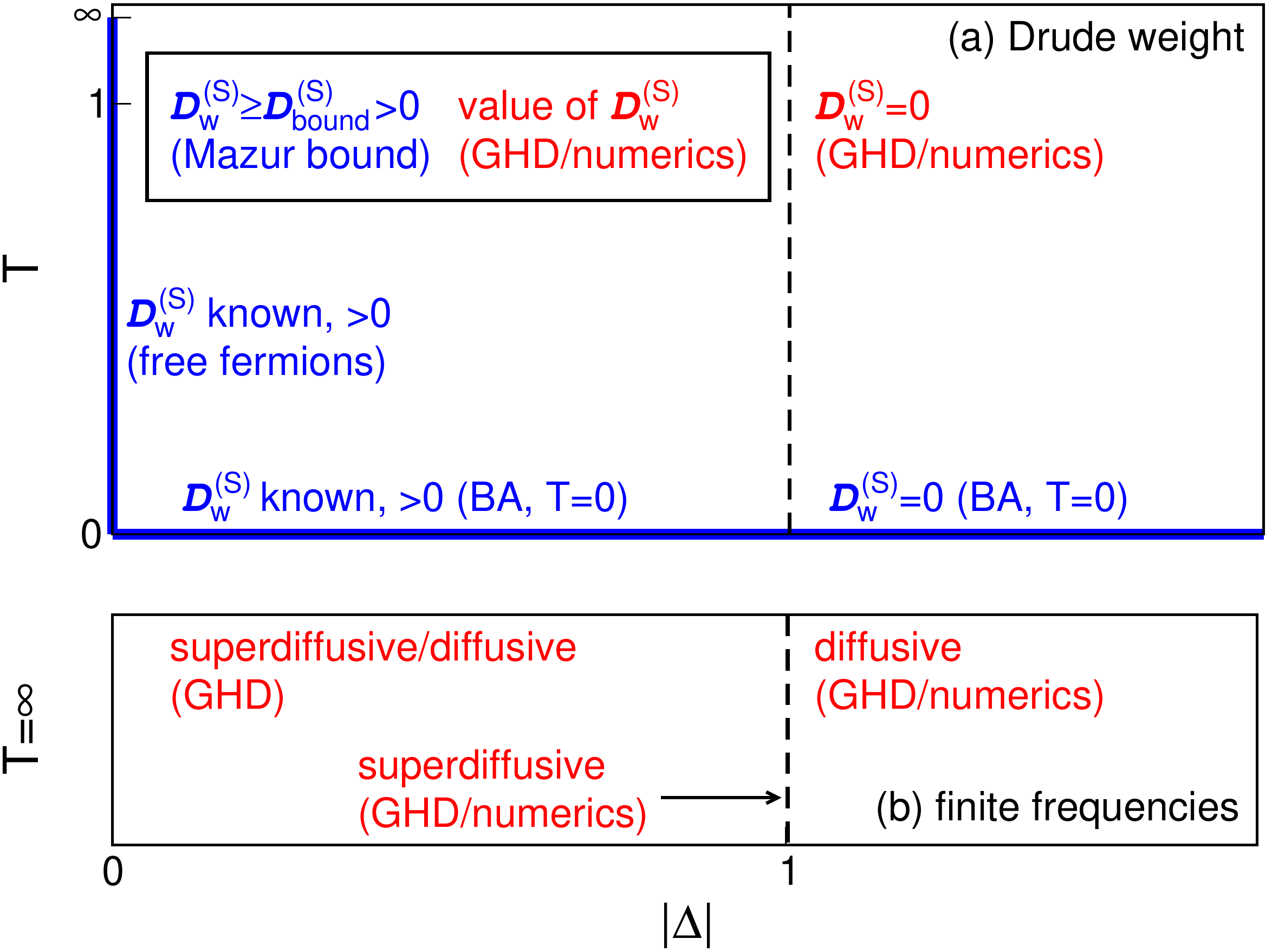}
\caption{(Color online) (a) Overview of all known \textit{exact results} (free  
fermions, Bethe ansatz (BA) at $T=0$, and Mazur bounds) as well as 
\textit{predictions} obtained using GHD and numerics for the spin Drude weight 
of the spin-$1/2$ XXZ chain at zero magnetization. \revision{(b) Overview of the 
high-temperature results for the leading contribution at low but finite 
frequencies. In the regime $\Delta<1$, this low-frequency contribution is either superdiffusive or diffusive.}}
\label{fig:drude_overview} 
\end{center}
\end{figure}

\subsection{Spin transport: Drude weight}
\label{sec:xxz_drudespin}

For the spin Drude weight, the following picture has been established at 
zero magnetization. $\Dws$ is known exactly in the limit $T=0$ 
via the Bethe ansatz \cite{Shastry90} as well as in the limit $\Delta = 0$ at 
any $T$ via a mapping to free fermions. At $T=0$, $\Dws>0$ for $|\Delta|\leq 1$ 
and $\Dws=0$ otherwise. Using the Mazur inequality, one can prove 
\revision{for a dense set of commensurate anisotropies covering the range $|\Delta|<1$}
that $\Dws$ is nonzero for any temperature $T>0$ \cite{Prosen2011, 
Prosen2013}. These are the only exact statements available at $\magdens=0$; they 
are complemented by various analytical and numerical results. The spin Drude 
weight can be computed using GHD at $T>0$ for any $\Delta$ 
\cite{Ilievski2017a,Bulchandani2017a}. For $|\Delta|<1$, the GHD prediction 
coincides with the lower Mazur bound at infinite temperature \cite{Prosen2013} 
as well as with the result obtained using the thermodynamic Bethe ansatz 
\cite{Zotos1999} at any $T>0$ \cite{Urichuk2018}. At $|\Delta|>1$, GHD predicts 
that the Drude weight vanishes. In addition to these analytical statements, 
numerical data for the Drude weight is provided by ED, tDMRG, and dynamical 
typicality [see Sec.\ \ref{sec:xxz_numerical_approaches}].

Away from zero magnetization ($\magdens \neq0$), the Drude weight is finite for any 
value of $T$ and $\Delta$. This follows from the exact Bethe-ansatz calculation 
at $T=0$ \cite{Shastry90} as well as from the exact lower Mazur bound 
\cite{Zotos1997}, respectively. The Drude weight was also computed numerically 
\cite{Heidrich-Meisner2005}.

We will now discuss the above results in more detail (see also 
Fig.~\ref{fig:drude_overview} for a summary).

\subsubsection{Free fermions, Bethe ansatz at $T=0$}

At $\Delta=0$, the spin Drude weight $\Dws$ can be obtained analytically for any 
$T\geq0$ by using the mapping to free fermions. The result is given by  
\cite{Giamarchi}:
\begin{equation}
\Dws = \frac{1}{4\pi T} \int_{-\pi}^\pi \left[v_k f(\epsilon_k) \right]^2 
e^{\epsilon_k/T} dk\,.
\end{equation}

In the zero-temperature limit, the spin Drude weight can be computed exactly for 
any $\Delta$ by evaluating Kohn's formula (\ref{eq:Kohn}) \cite{kohn64} via the 
Bethe ansatz \cite{Shastry90}. The result for $|\Delta| \leq 1$  reads
\begin{equation} \label{eq:xxz_dws_bethe0}
\Dws = \frac{Kv}{2\pi} \, ,
\end{equation}
with $K$ and $v$ given in Eq.\ (\ref{eq:xxz_luttingerpara}) for $\magdens=0$.  
This agrees with the expression (\ref{eq:dw_luttinger}) obtained using 
bosonization. For $|\Delta|>1$, the Drude weight vanishes. Note that Eq.\ 
(\ref{eq:xxz_dws_bethe0}) does not approach zero for $\Delta \nearrow 1$, and 
$\Dws$ thus shows a discontinuity at $\Delta = 1$ for $T=0$.

\subsubsection{Mazur bounds}

Away from zero magnetization (i.e., for $\magdens \neq 0$), the spin Drude weight is 
finite for any value of $T$ and 
$\Delta$. This can be shown by evaluating the Mazur inequality (\ref{eq:MazurD}) 
with the energy-current operator as a single conserved local charge $Q_2 = \JE$ 
that has a nonzero overlap with $\JS$ in the thermodynamic limit. At high $T$, 
the bound can be evaluated analytically \cite{Zotos1997}:
\begin{equation} \label{eq:xxz_mazur1} 
\Dws \geq \frac{J^2}{T}\frac{\Delta^2\magdens^2}{4}\frac{(1-\magdens^2)}{1+\Delta^2(2+2\magdens ^2)} > 0 \, ,
\end{equation}
which is valid for any value of $\Delta$. In the gapless phase at low $T$ and 
close to $\magdens=0$, one can add a Zeeman term $b\sum_r \sz{r}$ to the Hamiltonian 
($b$ is the magnetic field) and then use bosonization to obtain 
\cite{Sirker2011}
\begin{equation}
\Dws \geq \frac{Kv}{2\pi} \frac{1}{1 + \frac{\pi^2}{3K} \left( \frac{T}{b} 
\right)^2} > 0 \, .
\end{equation}
Directly at zero magnetization, the energy-current operator has a vanishing 
overlap with the spin-current operator and thus does not yield a nonzero 
contribution to the Mazur inequality (\ref{eq:MazurD}). This follows from 
symmetry arguments and can also be seen by setting \revision{$\magdens = 0$} in Eq.\ 
(\ref{eq:xxz_mazur1}): $Q_2$ is even under the transformation $\sz{r} \to 
-\sz{r}$, $s^{\pm}_r \to s^{\mp}_r$ while $\JS$ is odd, and hence $\langle Q_2 
\JS \rangle = 0$. The same holds true for all other strictly local conserved quantities 
associated with the integrability of the system. Note that the vanishing of 
$\langle Q_2 \JS \rangle$ also implies the absence of a magnetothermal 
correction in the zero-magnetization sector \cite{Louis2003}, while this term 
generally contributes at finite magnetizations.

As discussed in Sec.\ \ref{sec:charges} in detail, an exact lower bound can be 
derived for zero magnetization as well by using quasilocal conserved charges 
that do have a nonzero overlap with the spin-current operator \cite{Prosen2011, 
Prosen2013}. This bound is given in Eq.\ (\ref{eq:prosen_bound}) and is  
visualized in Fig.\ \ref{fig:prosen_bound} (see also Fig.~\ref{fig:dw_comparison} below); it generally features a fractal 
dependence on $\Delta$. An important question concerns the completeness of this 
set of charges: a numerical study of finite systems suggests that there are no 
additional  charges beyond the known ones \cite{Mierzejewski2015}.

\subsubsection{Bethe ansatz at $T>0$, GHD}

An exact Bethe-ansatz calculation at finite $T$ using Kohn's formula is hindered by 
the fact that excited states can only be treated approximately. In 
\cite{Zotos1999}, the calculation has been carried out by means of the 
thermodynamic Bethe ansatz, which, as discussed in Sec.~\ref{sec:tba}, involves 
the string hypothesis for bound states of magnons. An alternative calculation 
based on a spinon-antispinon basis was presented in \cite{Benz2005}. The 
thermodynamic Bethe-ansatz approach predicts that $\Dws(T)$ is finite in the 
gapless phase and decreases monotonically with $T$ except at the isotropic 
point $\Delta=1$ where $\Dws(T>0) = 0$. At low $T$, one obtains a nontrivial 
power-law dependence for commensurate values of $\Delta= \cos(\pi/m)$, 
$m=1,2,\dots$:
\begin{equation}\label{eq:xxz.betheTdep}
\Dws(T) = \Dws(T=0) - \mbox{const.} \times T^{\alpha}\quad \alpha = 
\frac{2}{m-1}\,.
\end{equation}
For $\Delta=1$, a second Bethe-ansatz-based calculation \cite{Carmelo2015}  
concludes in favor of $\Dws=0$, in agreement with GHD \cite{Ilievski2018}.

One can show that for commensurate values of $\Delta=\cos(\ell\pi/m)$ ($\ell$, $m$ coprime), the TBA 
result of \cite{Zotos1999} coincides with the GHD prediction 
\cite{Ilievski2017a,Bulchandani2017a} at any temperature \cite{Urichuk2018}. Moreover, it 
also coincides with the exact lower bound of \cite{Prosen2013} at infinite 
temperature, which is given in Eq.\ (\ref{eq:prosen_bound}) and shown in 
Fig.~\ref{fig:prosen_bound}. In a nutshell,
\begin{equation}
\frac{\Dws |_\textnormal{TBA}}{\beta}
=  \frac{\Dws \big|_\textnormal{GHD}}{\beta} 
\stackrel{T\to\infty}{=}\tilde{\mathcal{D}}_{\rm w}^{\rm 
(S)}|_\textnormal{bound}.
\end{equation}
For $\Delta > 1$, both GHD and a TBA-based analytical calculation  
\cite{peres99} suggest that the Drude weight vanishes in this regime. 

\subsubsection{Numerical approaches} \label{sec:xxz_numerical_approaches}

A variety of numerical methods has been used to compute the spin Drude weight 
$\Dws$ in the thermodynamic limit, such as (i) exact diagonalization, 
which is limited to small systems $L\leq 20$ \cite{Zotos1996, Rabson2004,Narozhny1998, 
Heidrich-Meisner2003, Herbrych2011,Karrasch2013,Sanchez2017}, (ii) the  
microcanonical Lanczos method, which is also limited to small $L\leq 28$ 
\cite{Long2003} (iii) quantum Monte Carlo, which requires an analytical 
continuation to extract $\Dws$ \cite{Alvarez2002, Heidarian2007}, (iv) the 
time-dependent DMRG, where the accessible time scales are bounded by 
the entanglement growth \cite{Karrasch2012, Karrasch2013, Karrasch2015, 
Karrasch2017,Karrasch2014a}, and (v) dynamical typicality \cite{Steinigeweg2014, 
Steinigeweg2015}, which is also limited in terms of the system size $L\leq 36$.

\begin{figure}[t]
\begin{center}
\includegraphics[width=0.90\linewidth]{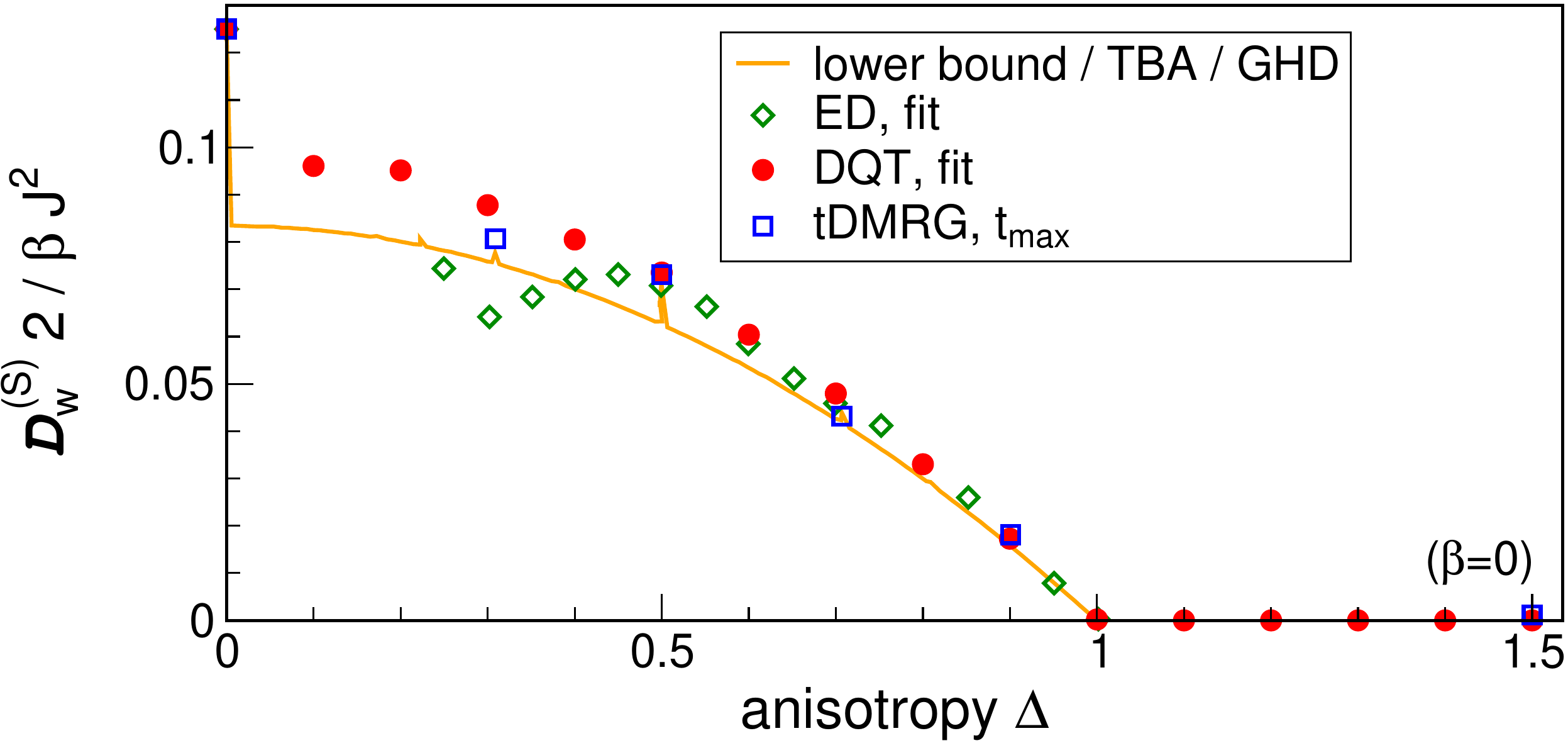}
\caption{(Color online) Comparison of different results for the spin Drude 
weight $\Dws$ in the high-temperature limit $\beta = 0$ at zero magnetization: the  
lower bound \cite{Prosen2013}, which is given by Eq.\ 
(\ref{eq:prosen_bound}) and which at infinite temperature coincides with the 
TBA result \cite{Zotos1999,Urichuk2018,Pavlis2019} and the GHD prediction 
\cite{Ilievski2017,Bulchandani2017a}. Moreover, we show  ED 
\cite{Herbrych2011}, DQT \cite{Steinigeweg2014, Steinigeweg2015}, and 
tDMRG data \cite{Karrasch2014a, Karrasch2015}. For the latter, the 
Drude weight is taken as the value of $\langle \JS (t) \JS \rangle/L$ 
at the largest accessible time without any further extrapolation. Note that  
while the lower bound was only computed for commensurate 
$\Delta=\cos(\ell\pi/m)$, numerical data are also shown away from these 
points.}
\label{fig:dw_comparison} 
\end{center}
\end{figure}

Figure \ref{fig:dw_comparison} shows a comparison of tDMRG \cite{Karrasch2012, 
Karrasch2013, Karrasch2014a, Karrasch2015, Karrasch2017}, exact diagonalization  
\cite{Herbrych2011, Karrasch2013}, and dynamical typicality data \cite{Steinigeweg2014, 
Steinigeweg2015} with the lower bound \cite{Prosen2013} at infinite temperature. Note that  
the numerical results are also shown away from commensurate $\Delta$.
At certain commensurate points such as $\Delta=1/2$, the numerical results and the bound agree
very well. For generic values at $\Delta>1/2$, all methods result in larger values than the bound.
For $0<\Delta <1/2$ and in particular close to $\Delta=0$, the diasgreement is  evident
and has not been resolved yet \revision{(see Secs.~\ref{sec:methods_ed}, \ref{sec:typicality}, and \ref{sec:methods_dmrg} for a discussion of the 
strengths and limitations of the numerical approaches and Sec.~\ref{sec:xxz.open} for a discussion of open questions concerning the spin-1/2 XXZ chain).}

For $\Delta > 1$, there are no exact results available for $T>0$ at zero 
magnetization. Both GHD and a TBA-based analytical calculation \cite{peres99} 
suggest that the Drude weight vanishes in the regime. Numerical studies also 
point in this direction \cite{Zotos1996, Heidrich-Meisner2003, Karrasch2012, 
Steinigeweg2014}. In particular, one observes a clean scaling of the Drude 
weight as $\Dws \propto 1/L$ in systems of finite size. As an example, Fig.\ 
\ref{fig:dw_scaling} shows the scaling found in \cite{Steinigeweg2014, 
Steinigeweg2015}.

At $\Delta=1$, exact-diagonalization results indicate a vanishing 
\cite{Herbrych2011} or at best very small Drude weight 
\cite{Heidrich-Meisner2003}, with the actual numbers depending on details of 
the finite-size extrapolation \cite{Karrasch2013,Sanchez2017} [see Sec.\  
\ref{sec:methods_ed}]. Both dynamical typicality \cite{Steinigeweg2014, 
Steinigeweg2015} and tDMRG \cite{Karrasch2012,Kennes2016,Sirker2009} yield a 
current correlation function $\langle\JS(t)\JS\rangle$ which decays slowly. The 
DQT data was interpreted in terms of a zero (finite) Drude weight at infinite 
(finite) temperature; the tDMRG results were interpreted in terms of a finite 
Drude weight. 

The previous discussion focused on infinite temperature, yet the temperature  
dependence of the Drude weight is also of interest 
\cite{Zotos1999,Fujimoto2003,Benz2005,Alvarez2002,Heidrich-Meisner2003,
Karrasch2013}. The verification of the TBA result for the low-$T$ 
behavior (\ref{eq:xxz.betheTdep}) in a numerical calculation has not been 
accomplished yet \cite{Alvarez2002,Karrasch2013}.

\subsection{Spin transport: Finite frequencies} \label{sec:conductivity}

We recall that it is now established \textit{exactly} that at zero magnetization, the  
Drude weight $\Dws$ is finite for $\Delta<1$ at any temperature $T\geq0$. For 
the regime $\Delta > 1$, the current understanding is that $\Dws$ vanishes. 
In recent years, there has been substantial progress in understanding the 
spin transport in the XXZ chain beyond the mere existence of the spin Drude 
weight. In this section, we summarize results for the regular part of the spin  
conductivity in the three different regimes $\Delta > 1$, $\Delta < 1$, and 
$\Delta = 1$. We focus exclusively on zero magnetization.

As outlined in Sec.~\ref{sec:theory.ballistic}, one can envision three  
different scenarios for the low-frequency behavior: The spin conductivity is (a) 
diffusive, $\sigma_\textnormal{reg}(\omega\to0)=\sigma_\textnormal{dc}>0$, (b) 
superdiffusive, $\sigma_\textnormal{reg}(\omega)\sim\omega^{\alpha}$ with 
$\alpha<0$, or (c) subdiffusive, $\sigma_\textnormal{reg}(\omega) 
\sim\omega^{\alpha}$ with $\alpha>0$. This is illustrated in 
Fig.~\ref{sketch_sigma}.

\begin{figure}[t]
\begin{center}
\includegraphics[width=0.9\linewidth]{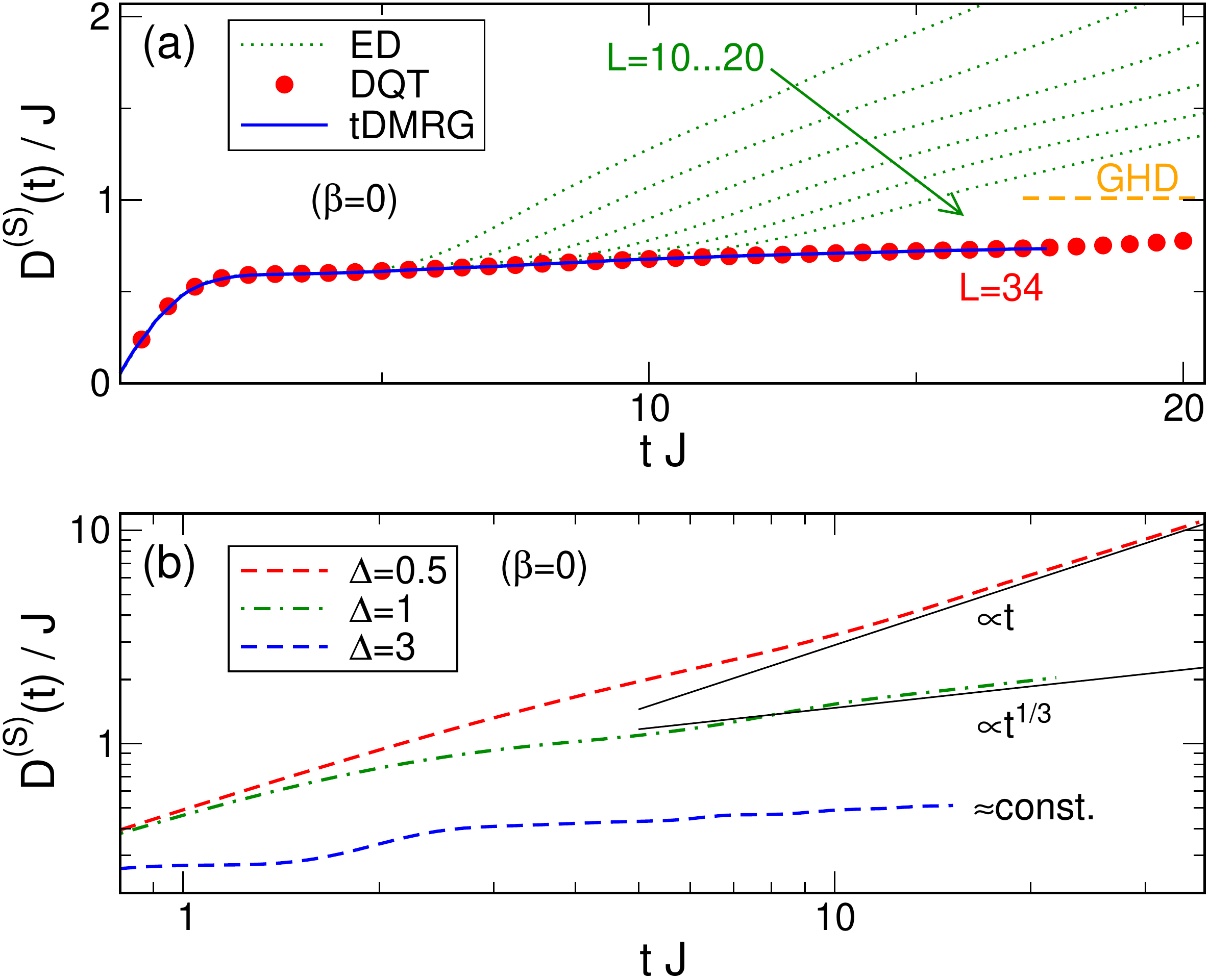}
\caption{(Color online) (a) Comparison of different results 
for the time-dependent diffusion constant $\DS(t)$ of the spin-1/2 XXZ chain at $\Delta=1.5$ in the 
infinite-temperature limit $\beta = 0$: ED \cite{Steinigeweg2009}, DQT 
\cite{Steinigeweg2015}, and tDMRG \cite{Karrasch2014}. (b) tDMRG data for 
$\DS(t)$ at various $\Delta$ (obtained from integrating the current 
autocorrelation function published in \cite{Karrasch2014a, Karrasch2015}).} 
\label{fig:diffusion_comparison}
\end{center}
\end{figure}

\subsubsection{$\Delta > 1$}

A lower bound for the diffusion constant, which is related to the DC  
conductivity via Eq.\ (\ref{eq:Ds}), was established in 
\cite{Medenjak2017,Ilievski2018} (see Sec.~\ref{sec:lowerbound_diff}). The 
diffusion constant is expressed in terms of the magnetization-dependence (i.e., 
curvature) of the Drude weight, which can be bounded from below using conserved 
charges. It was shown that the bound is finite for 
$\Delta>1$, which rules out a subdiffusive form of 
$\sigma_\textnormal{reg}(\omega)$ in this regime 
\cite{Medenjak2017,Ilievski2018}. This is an exact result.

One can define a time-dependent diffusion constant $\DS(t)$, $\DS(t\to\infty) 
= \DS$, using the generalized Einstein relation (\ref{Einstein_relation}) via a 
time integral of the current autocorrelation function \cite{bohm1992, 
Steinigeweg2009, yan2015}. This quantity was evaluated using GHD at infinite 
temperature and arbitrary $\Delta$ \cite{DeNardis2018, Gopalakrishnan2019}. In 
the limit of large $\Delta$, the result takes the form
\begin{equation}\label{eq:diffconst_ghd}
\lim_{\Delta \to \infty} \lim_{t \to \infty} \DS(t) = \lim_{\Delta \to \infty}  
\DS \approx 0.42 \, J \,,
\end{equation}
which is consistent with the non-vanishing lower bound for the diffusion  
constant from  \cite{Ilievski2018} and moreover predicts that transport is 
diffusive and not superdiffusive. Within GHD, the finite-time corrections 
have the form $\DS(t) = a + b/\sqrt{t}$.

\begin{figure}[t]
\begin{center}
\includegraphics[width=0.9\linewidth]{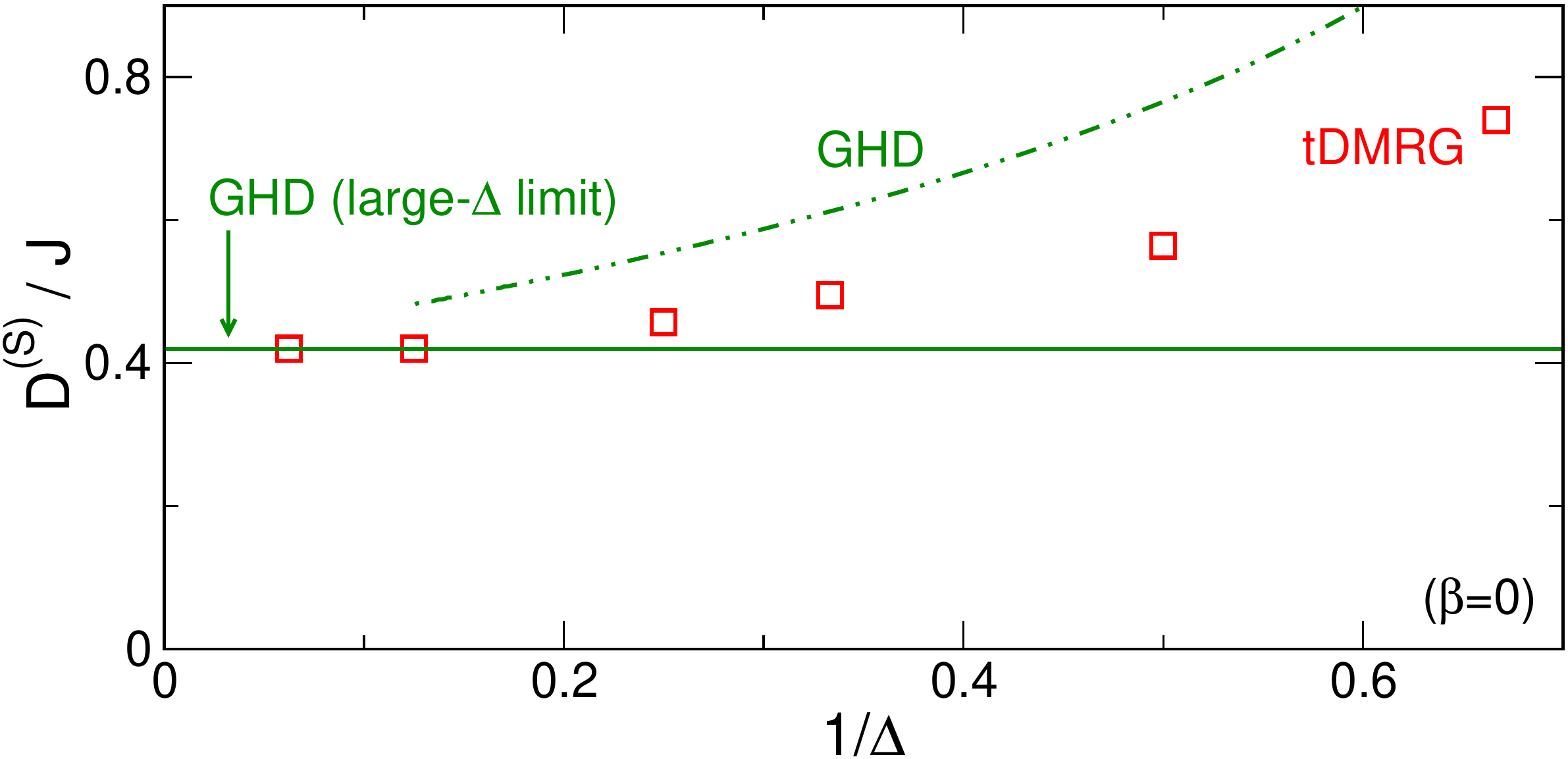}
\caption{(Color online) Diffusion constant of the spin-1/2 XXZ chain at infinite temperature as a function 
of $1/\Delta$ for $\Delta >1$. tDMRG results \cite{Karrasch2014} are compared to the GHD 
prediction for $\Delta \to \infty$, cf.\ Eq.\ (\ref{eq:diffconst_ghd}), and 
$\Delta < \infty$ \cite{DeNardis2018}. The tDMRG results close to $\Delta=1$ 
only give a lower bound to the true diffusion constant.
}
\label{fig:diffusion_comparison2}
\end{center}
\end{figure}

The time-dependent diffusion constant has also been calculated numerically via ED  
and DQT \cite{Steinigeweg2009, Steinigeweg2015} as well as using tDMRG 
\cite{Karrasch2014}. We show $\DS(t)$ at infinite temperature for $\Delta=1.5$ 
and $\Delta=3$ in Fig.\ \ref{fig:diffusion_comparison}(a) and (b), respectively 
(the curves with $\Delta=0.5,1$ will be discussed below). The system size can be 
chosen large enough, both within DQT and tDMRG, such that the results are 
effectively in the thermodynamic limit at the largest time depicted in the 
figure, which is illustrated explicitly in the case of DQT. This data was 
interpreted in terms of a finite diffusion constant $\DS=\DS(t\to\infty)$ and 
thus regular diffusive transport. In Fig.\ \ref{fig:diffusion_comparison2}, we 
show a comparison between the GHD prediction for large $\Delta$ and the value of 
$\DS$ extracted from the tDMRG data \cite{Karrasch2014}; both results agree in 
the limit $\Delta\to\infty$. The tDMRG results close to $\Delta=1$ only give a 
lower bound to the true diffusion constant.

The time-dependent diffusion constant was also studied via perturbation theory 
in powers of $\Delta$ under the assumption that the current autocorrelation 
function decays monotonically in time to zero 
\cite{steinigeweg10, Steinigeweg2011c}. A good agreement with numerics was found 
at short and intermediate time scales, where this assumption is well satisfied.

\begin{figure}[t]
\begin{center}
\includegraphics[width=0.90\linewidth]{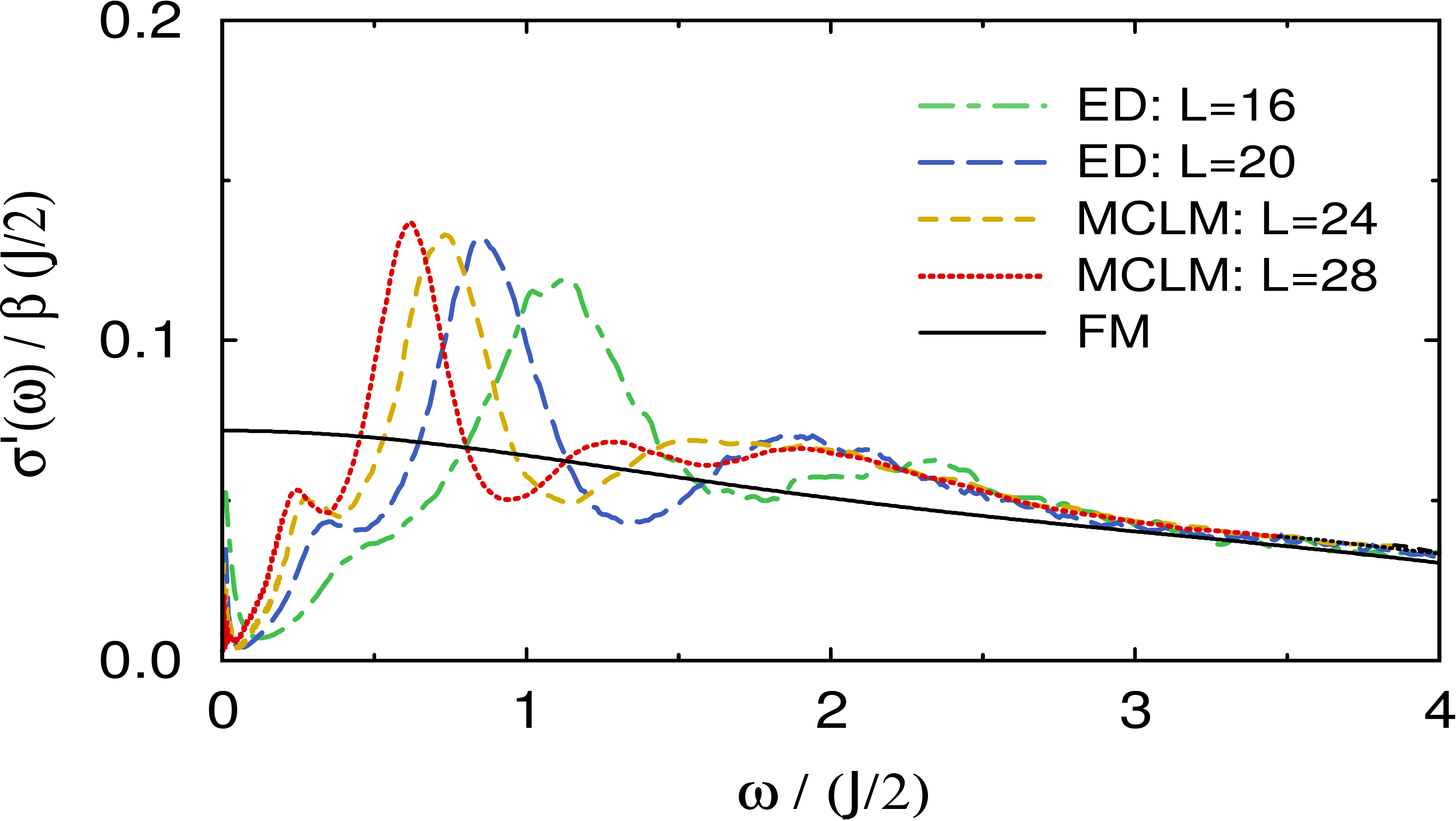}
\caption{(Color online) Real part of the optical conductivity $\sigma(\omega)$ of the spin-1/2 XXZ chain, 
as obtained in  \cite{Prelovsek2004} from ED and MCLM for $\Delta = 2$ and infinite 
temperature $\beta = 0$. The anomalous scaling with system size is still 
consistent with a well-behaved low-frequency part and a finite dc conductivity 
in the thermodynamic limit.}
\label{fig:prelovsek_sigma} 
\end{center}
\end{figure}

The full frequency-dependent conductivity $\sigma_\textnormal{reg}(\omega)$ was 
investigated for $\Delta>1$  using ED and MCLM \cite{Prelovsek2004}. As 
illustrated in Fig.\ \ref{fig:prelovsek_sigma}, for system sizes accessible to 
those methods, $\sigma_\textnormal{reg}(\omega)$ typically has an anomalous 
form, with strongly reduced spectral weight at small $\omega$ in the vicinity 
of the finite-size spin Drude weight $\Dws$ and a pronounced shoulder at larger 
$\omega$. As argued in \cite{Prelovsek2004}, however, the position of this 
shoulder moves with increasing system size to smaller $\omega$ as $1/L$, and 
might eventually take on a simple form with a well-behaved low-frequency part 
and a finite dc conductivity in the thermodynamic limit. Whether this anomalous 
form is indeed a spurious effect of small systems or whether it persists 
for large systems is still an open problem (see the discussion in  
Sec.~\ref{sec:methods_ed}). Yet, it is clear that the degree of 
anomalous behavior substantially depends on the frequency scale or, 
equivalently, the time scale considered \cite{Jin2015, Steinigeweg2016c}. 

The full frequency-dependent conductivity $\sigma_\textnormal{reg}(\omega)$ was  
also calculated for $\Delta>1$ via a Fourier transform of finite-time tDMRG data 
\cite{Karrasch2014a}. It shows a regular diffusive peak at small frequencies as 
well as contributions for frequencies above the spectral gap.

In addition to the numerous works on current-current correlations, a significant
body of works studied density-density correlations as well, either in momentum  
or real space \cite{Huber1969, Fabricius1997, Fabricius1998, Narozhny1998}. This 
allows one to also study the momentum dependence of the diffusion coefficient. 
In the context of diffusion, a result from exact and Lanczos diagonalization 
\cite{Steinigeweg2011, steinigeweg12a} 
is that the time-dependent susceptibility $\chi_q(t)$ defined in Eq.\ (\ref{eq:chi_time}) decays at small $\beta$ according to  
\begin{equation}
\frac{\text{d} \chi_q(t)}{\text{d} t} = - \tilde{q}^2 \, \DS_q(t) \, \chi_q(t) \, .
\end{equation}
Here, the decay rate $\DS_q(t)$ becomes independent of $q$ for small momenta $q > 
0$ in a finite lattice and coincides with the time-dependent diffusion 
coefficient $\DS(t)$ in the Einstein relation (\ref{Einstein_relation}). The 
number of diffusive momenta was shown to decrease with decreasing 
temperature, while the diffusion constant increases, as long as temperature 
is large compared to the gap.

\subsubsection{$\Delta < 1$}

The Drude weight has been shown to be  finite for any commensurate  value of $\Delta<1$ with $\Delta = \cos(\ell \pi/m )$ 
and is thus conjectured to be finite everywhere.
An exact lower bound for 
the diffusion constant was also obtained in this regime \cite{Ilievski2018}. It 
was shown analytically that the bound is finite for commensurate 
$\Delta=\cos(\pi/m)$, which rules out a subdiffusive form of 
$\sigma_\textnormal{reg}(\omega)$ for these parameters. For incommensurate 
values of $\Delta$ (i.e., almost everywhere), the lower bound diverges, and 
transport  cannot be diffusive but must be faster than diffusive.
Combined with the expectation that the frequency-integrated conductivity should be 
continuous everywhere due to sum rules, this hints at the possibility of
superdiffusive corrections away from the commensurate points. This conjecture 
was put onto firmer grounds in \cite{Agrawal2019}. 
Let us consider a value 
$\Delta=\cos(\pi \lambda_\infty)$ where $\lambda_\infty$ is a generic  irrational number.
The reasoning uses that $\lambda_\infty$ can be approximated by a series
of rational values $\lambda_m = \ell/m$ with growing denominators $m$.
Using fairly general 
arguments, one can show that the dc conductivity  
at infinite temperature can be approximated as 
\begin{equation}
\label{eq:agrawal}
\sigma_{\rm dc}(\lambda_m) \sim m^{2\alpha/(1-\alpha)} \,.\end{equation} 
Equation~\eqref{eq:agrawal} describes a subleading correction that, as a function of time, is 
greater than $1/t$ and hence diverges logarithmically.  

GHD allows one to obtain the exponents associated with the superdiffusive 
correction \cite{Agrawal2019}: The low-frequency conductivity behaves as 
$\sigma(\omega) \propto \omega^{-\alpha}$ with $\alpha=1/2$ for generic values 
of $\Delta$. This divergence is cut-off at the rational points, leading to a 
diffusive correction. Furthermore, a qualitative picture emerges from GHD: the 
subleading  correction arises from scattering of charged quasiparticles off 
neutral quasiparticles and an  interpretation in terms of   a L{\'e}vy flight 
has been put forward \cite{Agrawal2019,Gopalakrishnan2019a}.

At low temperatures, a field-theory calculation which incorporates the leading  
irrelevant umklapp term and accounts for conserved charges via the 
memory-matrix formalism suggests a diffusive form of the (subleading) 
$\sigma_\textnormal{reg}(\omega)$ \cite{Sirker2011} (see \cite{Sirker2006} for 
earlier work). This is consistent with earlier results for the generic behavior 
of a Tomonaga-Luttinger liquid in the presence of umklapp scattering 
\cite{giamarchi91}, but it is an open question whether field theory away from 
commensurate values of $\Delta$ is consistent with the GHD result that subleading correction 
cannot be diffusive there. The field theory was used to compute the 
density-density correlation function and a diffusive behavior was found in 
agreement with tDMRG data \cite{Karrasch2015c}.

The spin conductivity has been computed numerically via various approaches.  
Using Lanczos diagonalization, it was concluded that 
$\sigma_\textnormal{reg}(\omega)\sim\omega^2$ at low frequencies 
\cite{Herbrych2012}, which is at odds with the lower bound established in 
\cite{Ilievski2018}. The Fourier transform of finite-time tDMRG data is 
consistent with a finite 
$\sigma_\textnormal{reg}(\omega\to0)=\sigma_\textnormal{dc}>0$, and for certain 
values of $\Delta$ suggests an additional peak structure at larger frequencies 
\cite{Karrasch2015}. For completeness, Fig.\ \ref{fig:diffusion_comparison}(b) 
depicts tDMRG results for $\DS(t)$ at $\Delta=0.5$. One finds that $\DS(t)\sim 
t$ for $tJ\gtrsim10$ due to the finite Drude weight. A convincing numerical 
confirmation of the GHD prediction for the power-law decay of $C(t)$ at generic 
values of $\Delta$ towards the Drude weight is still missing.

\subsubsection{$\Delta = 1$}
\label{sec:xxz_delta1}

While no exact results are available, the current belief is that the Drude  
weight vanishes at the isotropic point $\Delta = 1$ in the thermodynamic limit. 
There is, in principle, the possibility of diffusive transport to occur 
\cite{Sirker2009,Sirker2011}. However, this scenario has been discussed 
controversially in the literature and, in contrast to the regime $\Delta > 1$, 
there is mounting evidence that diffusion is not realized.

The exact lower bound on the diffusion constant diverges in the limit $\Delta 
\to 1$ at infinite temperature \cite{Ilievski2018}, which is indicative of 
superdiffusion at this point. A divergence was also obtained within the GHD 
approach, and for $\Delta\to1^+$ it was found that \cite{DeNardis2018}
\begin{equation}
\DS = \lim_{t \to \infty} \DS(t) \propto \frac{1}{\sqrt{\Delta -1}}.
\end{equation}
The same result has been derived via a GHD-based kinetic picture  
\cite{Gopalakrishnan2019}, and the time-dependent diffusion constant was 
predicted to scale as $\DS(t) \propto t^{1/3}$. This was confirmed in another 
GHD study \cite{Agrawal2019}, yet without invoking 
the Kardar-Parisi-Zhang (KPZ) scaling mechanism 
\cite{Kardar1986,Ljubotina2017,Ljubotina2019,Weiner2020,Spohn2019}.

This superdiffusive behavior is consistent with finite-time tDMRG data for 
$\DS(t)$ at $\Delta=1$ [see Fig.\ \ref{fig:diffusion_comparison}(b)] as well as 
with with numerical linked-cluster expansions \cite{Richter2019, Richter2019e}. 
Signatures of superdiffusion at $\Delta=1$ were found in the unitary evolution 
of inhomogeneous initial states \cite{Ljubotina2017}. In particular, for initial 
states with a magnetization profile of domain-wall type, the profiles at later 
times collapse after a rescaling of space with the power law $t^{3/2}$, which 
corresponds to $\DS(t)\sim t^{1/3}$ \cite{Ljubotina2017}.

The field-theory calculation of \cite{Sirker2009,Sirker2011} was also carried  
out directly at $\Delta=1$. It predicts diffusive dynamics of 
density-density correlations in the hydrodynamic regime of small momenta $q \to 
0$ and low frequencies $\omega \to 0$, where the diffusion constant scales with 
temperature as
\begin{equation}
\DS \propto \frac{\ln T}{T} \, .
\end{equation}
The possibility of diffusion in the hydrodynamic regime was also scrutinized in 
quantum Monte-Carlo simulations \cite{Grossjohann2010}, where the bosonization 
prediction was transformed to imaginary time in order to avoid transformations 
of Monte-Carlo data to real times.  
\revision{A fit of these QMC data to the bosonization result supports a finite
diffusion constant $\DS$. It is presently not fully understood how this
can be reconciled with the fact that the lower bound as well as GHD
predict superdiffusion at $\Delta  = 1$.}

\subsection{Open quantum systems}
\label{sec:xxz.open}

Complementary to works on Kubo response functions in closed  systems, one can 
study open quantum systems, in particular, using the Lindblad equation with 
a driving at the two boundaries of the XXZ chain [cf.\ Sec.\ \ref{sec:methods_open}]. 
For $\Delta > 1$ and in the case of weak driving, i.e., in the linear-response 
regime, a linear magnetization profile as well as a spin current scaling as 
$\JS \propto 1/L$ were observed, which corresponds to diffusion 
\cite{Znidaric2011, michel08, Prosen2009}. In the limit of large $\Delta$, the 
Lindblad approach yields a scaling of the diffusion constant as $\DS \propto 
1/\Delta$ \cite{Znidaric2011}, which is different from $\DS = \text{const.}$ as 
resulting from tDMRG calculations \cite{Karrasch2014} and generalized 
hydrodynamics \cite{DeNardis2018, Gopalakrishnan2019} [see Fig.\ 
\ref{fig:diffusion_comparison2}]. However, an equivalence of the linear-response 
and the open-system results is not expected for the particular choice of the 
system-bath coupling made in \cite{Znidaric2011}, where $\Gamma\sim\Delta$.

\begin{figure}[t]
\begin{center}
\includegraphics[width=0.90\linewidth]{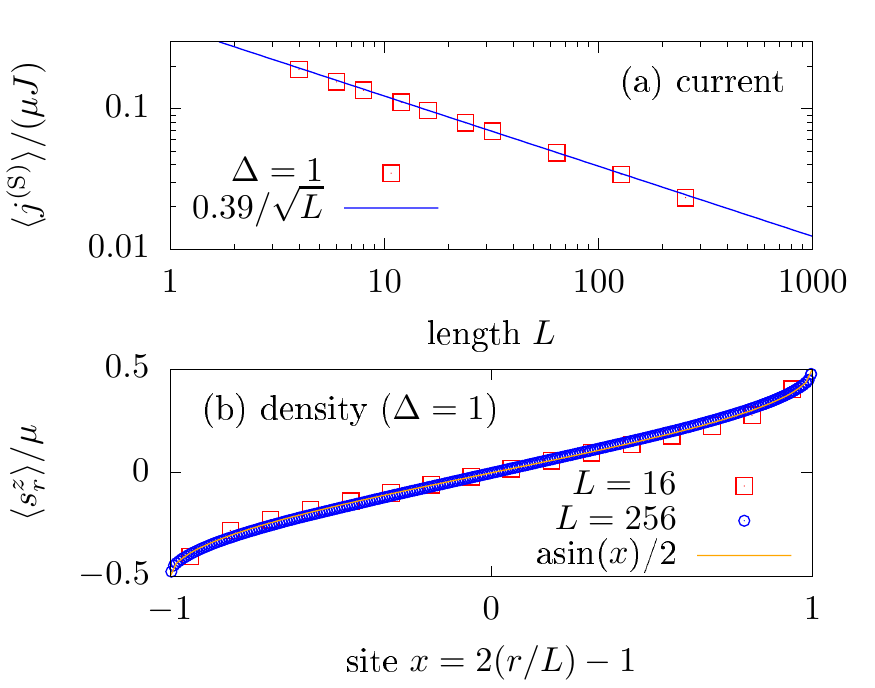}
\caption{(Color online) Results from the Lindblad quantum-master equation for simulating spin transport in the spin-1/2 XXZ chain, as
obtained in \cite{Znidaric2011} for $\Delta = 1$. (a) Superdiffusive scaling of
the spin current as $j^{\mathrm (S)} \propto 1/\sqrt{L}$. (b):
$L$-independent magnetization profile.
}
\label{fig:superdiffusion1}
\end{center}
\end{figure}

At $\Delta=1$ and high $T$, the open-system spin current does not scale as $\JS 
\propto 1/L$ anymore, but is instead found to scale slower according to the 
power law $\JS \propto 1/\sqrt{L}$, see Fig.\ \ref{fig:superdiffusion1}. Since 
the magnetization profile does not show a significant dependence on $L$, this 
scaling of the spin current shows the emergence of superdiffusion in the 
Lindblad approach \cite{Znidaric2011,znidaric11a}, which was the first 
observation of this behavior in the spin-1/2 Heisenberg chain. This is in 
agreement with numerics for unitary time evolution \cite{Ljubotina2017} and with 
the fact that the lower bound to the diffusion constant diverges for 
$\Delta\to1$ \cite{Ilievski2018}.

\subsection{Open questions}
\label{sec:open_questions}

As discussed in the previous sections, our theoretical understanding of 
transport in the spin-1/2 XXZ chain at nonzero temperatures has seen substantial 
progress during the past decade, due to the combination of various analytical 
and numerical techniques. Yet, there are certainly many open questions. A few of 
these questions are summarized in the following.

One important question is whether or not the exact lower bound for the spin 
Drude weight $\Dws$ in Eq.\ (\ref{eq:prosen_bound}) is exhaustive for all 
commensurate values of $|\Delta|<1$. It is now established that at infinite 
temperature, this lower bound coincides with the TBA \cite{Zotos1999} and GHD 
\cite{Ilievski2017a, Bulchandani2017a} results, which are also identical at any $T$ 
for $|\Delta|<1$ \cite{Urichuk2018}. However, a central assumption invoked 
within the TBA and therefore GHD, which is formulated in the TBA language, is 
the string hypothesis. It is an interesting open question whether or not GHD can 
be formulated without resorting to the string hypothesis. 
Similarly, the question remains whether the spin Drude weight can be computed 
in a Bethe-ansatz approach without resorting to the string hypothesis.
A recent study by \textcite{Kluemper2019} carried out such an alternative calculation without using the string hypothesis.
Using a numerical solution of the resulting nonlinear equations for $\Delta = \cos{(\pi/m)}$,
results for $\Dws(T)$ were obtained. These exhibit finite-size effects that become more significant
as temperature is lowered.
Eventually, the data converge to the TBA results by \textcite{Zotos1999} in the thermodynamic
limit. These results indicate the potential limitations of finite-size based numerical simulations in accessing
the low-temperature dependence of the spin Drude weight.

Moreover, while the possibility of a 
fractal-like dependence of $\Dws$ on $\Delta$ is intriguing, no numerical 
method \revision{will likely be capable to confirm the fractal structure.
 The sudden drop of $\Dws$ when going from $\Delta = 0$ to $\Delta > 0$ has not been 
verified numerically so far.} Particularly useful would be a lower bound with finite-size 
corrections, which would allow for more reliable extrapolations to the limit of 
large system sizes as well.

Another important and closely related issue concerns subleading corrections 
to the spin Drude weight $\Dws$ in the regime $|\Delta|<1$. It is now 
established from exact lower bounds and GHD that the diffusion constant  is finite for 
commensurate values of $\Delta$ but diverges away from these points 
\cite{Ilievski2018,Agrawal2019}.
This rapid change is explained by a significant weight 
transfer in the low-frequency window as one goes from \revision{commensurate} values of 
$\Delta$ to \revision{incommensurate} ones; concrete exponents for the divergence of 
$\sigma_\textnormal{reg} (\omega)$ at generic values of $\Delta <1$ and $\Delta 
=1$ were obtained from GHD \cite{Agrawal2019}.
This results in a more appealing picture as it satisfies the physical 
expectation of a smooth parameter dependence of at least the integral over the 
low-frequency part of $\sigma_\textnormal{reg}(\omega)$. Nevertheless, at least 
within GHD, the distinction between rational and irrational values again relates 
to properties of the quasiparticles, and thus the existence of both diffusive 
and superdiffusive corrections may also rely on Takahashi's string hypothesis. 
This leads to the same question again: can the string hypothesis be replaced in 
GHD and what would be the results? 
A convincing numerical 
confirmation of the exponents for the subleading correction for generic values of $\Delta<1$ is also missing.

Several open questions remain in the regime $\Delta > 1$ as well. While  
analytical calculation based on certain assumptions \cite{Carmelo2015} conclude 
in favor of $\Dws = 0$ at $T>0$, a strict proof is still missing. An exact lower 
bound to the diffusion constant was obtained and shown to be finite, ruling out 
subdiffusion \cite{Ilievski2018}. While substantial evidence has been provided 
that spin dynamics is diffusive, it still needs to be qualitatively explained why diffusion 
can occur in integrable systems, where concepts such as chaos, ergodicity, 
etc.\ do not apply. It would be interesting to obtain a better numerical 
estimate for the diffusion constant in the long-time limit, since the deviation 
from the GHD data in Fig.\ \ref{fig:diffusion_comparison2} is most 
likely related to the finite times reached in the simulations.

Another puzzling issue concerns the notion of a mean-free path. While the 
quantitative values for the diffusion constant suggest a mean-free path on the 
order of one lattice site, strong finite-size effects and anomalous scaling to 
the thermodynamic limit appear anyhow \cite{Prelovsek2004}, which suggests that 
the mean-free path is not the only length scale involved for a finite system 
\cite{steinigeweg12a}. It would be interesting to investigate the behavior of  
higher-order current correlation functions \cite{steinigeweg2013c}.

At the isotropic point $\Delta = 1$, transport at infinite temperature is faster 
than diffusive since the exact lower bound to the diffusion constant diverges 
for $\Delta\to1$ \cite{Ilievski2018}; such a divergence is also observed in GHD 
calculations \cite{Agrawal2019,DeNardis2018}. Numerical simulations 
\cite{Znidaric2011, Prosen13b} point to the emergence of superdiffusion. 
However, the origin and nature of this non-diffusive process is not 
fully understood yet. First attempts have been undertaken, and there is mounting support for \revision{the dynamical exponent of $z=3/2$ \cite{Ljubotina2017}, consistent with  
KPZ-scaling \cite{Ljubotina2019} further corroborated and discussed in
 \cite{Gopalakrishnan2019,
Bulchandani2019,Spohn2019,Weiner2020,DeNardis2020a,bulch2019kardarparisizhang}}.  
Whether the KPZ-like 
scaling persists in other isotropic spin models with or without integrability is
currently the object of intense scrutiny (see, e.g., \cite{DeNardis2019,Dupont2019}).
\revision{A recent study concludes that superdiffusion with an exponent of $z=3/2$ is generally realized in
all integrable, Heisenberg-like magnets that are invariant under global non-Abelian continuous symmetry \cite{Ilievski2020,Krajnik2020}.
A first-principle derivation of KPZ scaling for these integrable models (besides predicting the exponent) is still lacking and the possibility
of other types of superdiffusion in integrable spin chains cannot be fully ruled out either \cite{Znidaric2013b}.}

From a methodological point of view, it is unclear how the field-theory  
prediction of diffusion \cite{Sirker2009} and the associated low-$T$ QMC data 
\cite{Grossjohann2010} for $\Delta=1$ can be reconciled with the exact 
statement that the diffusion constant diverges at high $T$. 
\revision{Note that GHD also predicts superdiffusion at the isotropic point \cite{DeNardis2018,Agrawal2019}
and includes more types of excitations such as bound states than what is 
captured by field theory.}

Numerical methods 
such as ED, DQT, and tDMRG become less useful at low $T$ since the relevant time 
scales and finite-size effects are known to increase substantially as 
temperature is reduced from infinity (see the discussion in 
Secs.~\ref{sec:methods_ed} and \ref{sec:methods_dmrg}).

So far, there is no example for the spin-1/2 XXZ chain for which open system simulations  
and linear response agree for the actual values of the diffusion constants. It 
would be interesting to further investigate whether or not agreement between the 
open-system and the linear-response calculation can be achieved. This is also 
of fundamental interest, and respective studies may shed light on the 
differences between the dynamics in isolated and open systems in a much broader 
context. 

A phenomenological picture of transport in the spin-1/2 XXZ chain was developed  
in \cite{Huber2012,Sanchez2018}. The rich phenomenology of transport in the XXZ 
chain (ballistic with (super)diffusive corrections for $0\leq \Delta \leq 1$, 
superdiffusive at $\Delta=1$ and diffusive for $\Delta>1$) partially carries 
over to other integrable spin models [see, e.g., \cite{Piroli2016,Dupont2019}].
For instance, the $S = 1$ Zamolodchikov-Fateev (ZF) model  
\cite{Zamolodchikov1980} exhibits a similar transport behavior \cite{Dupont2019} 
with the exception of extra superdiffusion at $\Delta=0$. The exact general 
necessary and sufficient criteria for superdiffusion to occur are not fully understood, e.g., the role of 
SU(N) symmetry.

Generally, it is a crucial question if and in how far the rich dynamical 
behavior of the spin-1/2 XXZ chain is stable against weak integrability-breaking 
perturbations \cite{Zotos2004, Jung2006,Huang2013,Jung2007, Steinigeweg2016a} [see Sec.\  
\ref{sec:nonint}]. From a theoretical point of view, this question is 
challenging, for instance, because  conventional perturbation theory starts from a 
noninteracting problem. From an experimental point of view, this question is 
vital, because the coupling to environments or other degrees of freedom can 
never be suppressed completely [see also Sec.\ 
\ref{sec:experiments}].

\section{Transport in the Hubbard chain}
\label{sec:TransportHubbard}

The 1d fermionic Hubbard model $H=\sum_l h_l$ with 
\begin{eqnarray}\label{hub}
h_r&=&-t_{\rm h}  \sum_\sigma \left(c_{r\sigma}^\dagger c_{r+1\sigma}^{\phantom{\dagger}}
 + {\rm h.c.} \right) \\
	&& +
 U \left(n_{r\uparrow}-\frac{1}{2} \right)\left(n_{r\downarrow}-\frac{1}{2}\right)
\nonumber\end{eqnarray}
is a more general integrable model than the spin-1/2 XXZ chain as it also includes charge fluctuations.
Much less attention has been devoted to computing its finite-temperature transport properties, for either charge, spin, or thermal transport.
The main thermodynamic parameters, characterizing transport properties in the Hubbard chain are, besides temperature $T$, the chemical
potential and the magnetic field. These control the filling $\rho=(N_\uparrow + N_\downarrow)/(2L)$ and the magnetization 
density $\magdens=(N_\uparrow-N_\downarrow)/L$. Here, we will mostly assume a canonical situation, where $\rho$ and $m$ are fixed.

The fermionic Hubbard model possesses a pair of global $SU(2)$ symmetries \cite{esslerbook}, where one of them,  the spin symmetry, is related to transport of magnetization, while the other, the so-called $\eta-$spin symmetry, 
is related to charge conservation and transport of charge.
While the integrability of the Hubbard model was shown by coordinate Bethe ansatz already in \cite{LiebWu1968}, it was only in 1986 when Shastry proposed the Lax operator and the toolbox of 
algebraic integrability which allowed to explicitly construct an infinite sequence of local conservation laws \cite{shastry1986infinite}. These conservation laws allow one to obtain some rigorous results for transport properties.

\revision{GHD has also been applied to investigate Drude weights \cite{Ilievski2017,Ilievski2018,Fava2020}, the emergence of diffusion and superdiffusion as well as KPZ behavior \cite{Fava2020} in the Hubbard chain.
A comprehensive overview is given in \cite{Fava2020}.}
\subsection{Thermal conductivity}

The energy-current operator $\JE$ is given by \cite{Zotos1997}:
\begin{eqnarray}
\JE    &=&  \sum_{r,\sigma}t_{\rm h}^2\Big[\big(i c_{r+1\sigma}^\dagger c_{r-1\sigma}^{\phantom{\dagger}}+{\rm h.c.}\big)\Big. \nonumber \\
&& \Big. -\frac{U}{2}\big(j_{r-1\sigma}^\mathrm{(C)} 
+ j_{r\sigma}^\mathrm{(C)} \big)\big(n_{r\bar\sigma} -\frac { 1 } { 2 } 
\big)\Big]\,,\, \label{eq:JEhubbard}
\end{eqnarray}
where $j_{r,\sigma}^\mathrm{(C)}$ is the charge current and $\bar \sigma 
=\uparrow(\downarrow)$ for  $\sigma=\downarrow(\uparrow)$.
Here, we will restrict the discussion to the case of a vanishing chemical potential and magnetic field and therefore, half filling and zero magnetization. 
Under these conditions, the energy current
does not couple to the charge or spin current and offdiagonal matrix elements in the Onsager matrix of transport coefficients can be ignored \cite{Mahan}. 
This is equivalent to saying that the Seebeck coefficient, which is proportional 
to $\langle \JE(t) \J^\mathrm{(C)} \rangle$, vanishes identically at half 
filling at all temperatures \cite{Beni1975},
where $\J^\mathrm{(C)}$ is the particle current.

Similar to the spin-1/2 XXZ chain, the Hubbard model is a ballistic thermal conductor \cite{Zotos1997}, which is a \textit{rigorous} statement: While $\JE$ is not conserved, it still has a nonzero overlap
with a local conserved quantity $Q_2$.
This $Q_2$ is the first nontrivial conserved charge in the Hubbard chain beyond 
energy $E$, particle number $N$ and the $z$-component $\Mtot$ of the total 
spin, and it happens to be quite
similar to $\JE$ in structure: $Q_2$ results from $\JE$ by $U/2 \to U$.
Consequently, the Mazur inequality \eqref{eq:MazurD} provides a nonzero lower bound.

This  lower bound to the energy Drude weight was evaluated analytically
in \cite{Zotos1997} for $T=\infty$. This 
expression reads:
\begin{eqnarray}
\Dwe &\geq& \frac{\beta^2}{2L}\sum_\sigma 2 \rho_\sigma (1-\rho_\sigma)+ \frac{U^4}{4}  \\
    &\times &\-\-\-\-\-\-   \frac{\lbrack  \sum_\sigma 2 \rho_\sigma (1-\rho_\sigma)(2 \rho^2_{-\sigma} -2 \rho_{-\sigma} +1) \rbrack^2 }{ \sum_\sigma 2 \rho_\sigma (1-\rho_\sigma) \lbrack 1+ U^2 (2 \rho_{-\sigma}^2 -2 \rho_{-\sigma} +1 \rbrack}
\nonumber \end{eqnarray}
where $\rho_\sigma$ is the density of electrons with spin $\sigma=\uparrow,\downarrow$.
A tDMRG study showed that contributions from other conserved charges $Q_{n >2}$ 
with a nonzero overlap with $\JE$ are fairly small for all $U/t_{\rm h}$ 
\cite{Karrasch2016}
at infinite temperature and half filling. 

\begin{figure}[t]
\includegraphics[width=0.9\columnwidth]{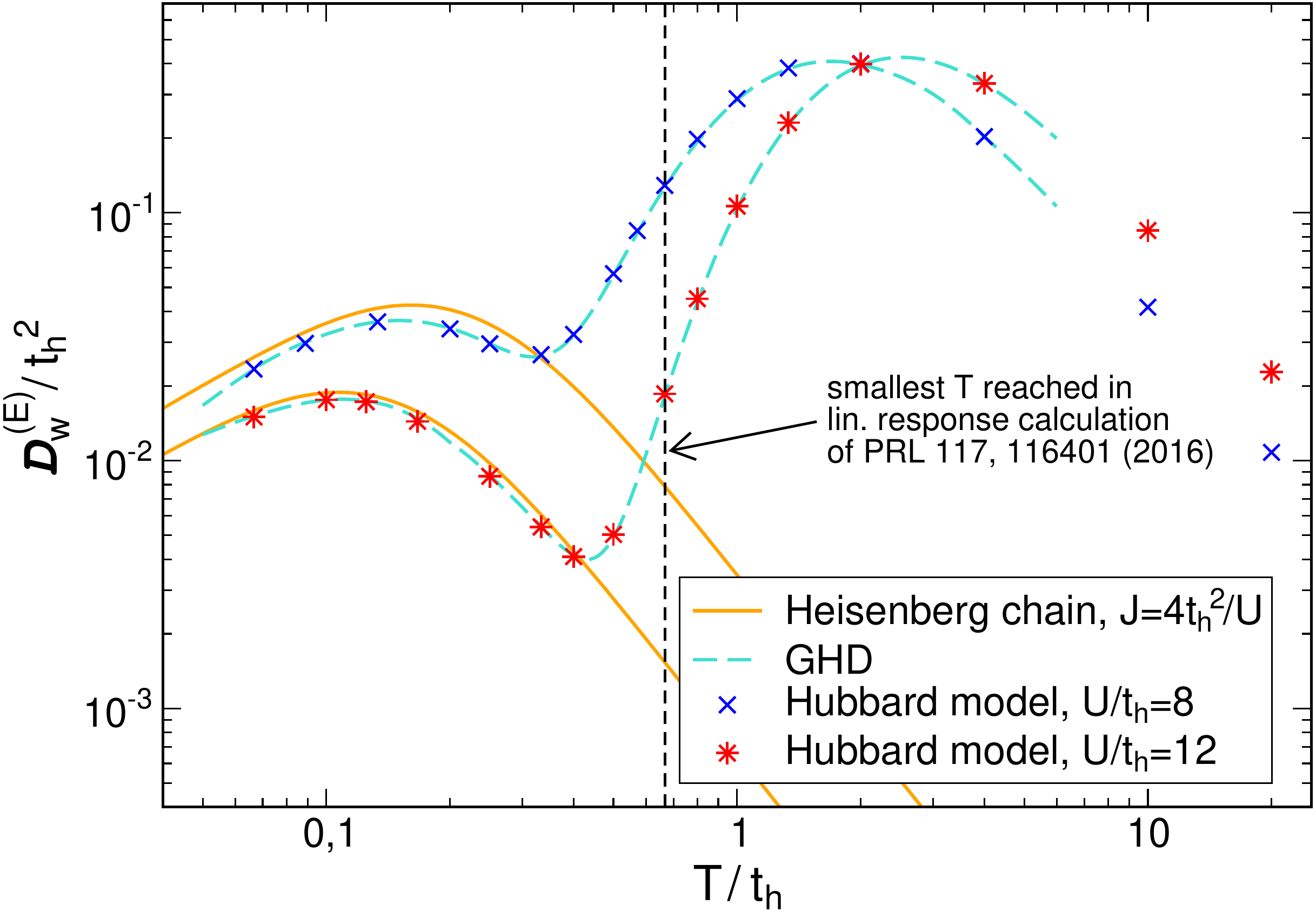}
\caption{(Color online)  Energy Drude weight \revision{of the Fermi-Hubbard chain} as a function of 
temperature computed from finite-$T$ tDMRG \cite{Karrasch2016,Karrasch2017} and GHD 
\cite{Ilievski2017}.
 }
\label{fig:DE_hubbard}
\end{figure}

The full temperature dependence of the energy Drude weight was computed only 
recently from both finite-$T$ tDMRG \cite{Karrasch2016, Karrasch2017} and GHD 
\cite{Ilievski2017}, which are in quantitative agreement.
This Drude weight has, for $U\gg t_{\rm h}$, two maxima: the low-temperature regime $T\lesssim \Delta_{\rm Mott}$ is dominated by spin excitations. This part of ${\cal D}_{\rm w}^{\rm (E)}(T)$ agrees well with 
the results for the spin-1/2 Heisenberg chain from \cite{Kluemper2002}. At high temperatures, charge contributions are activated and dominate the thermal transport.
This behavior is illustrated in Fig.~\ref{fig:DE_hubbard}.  \revision{A more complete picture of the various temperature regimes and the relevant contributing excitations is described in \cite{Fava2020}}.

Since the energy current is not exactly conserved, there are finite-frequency contributions that were studied in \cite{Karrasch2016}, but a conclusion about the nature of the
subleading correction at low frequencies could not be drawn. 
Other related directions include the thermolelectric response of the model \cite{Peterson2007,Zemljic2005}.

\subsection{Charge conductivity}

The local conserved charges have a nonzero overlap with the charge current away from half filling $\rho\neq 1/2$. Using the Mazur inequality (\ref{eq:MazurD}), 
one can thus show that charge transport is ballistic, i.e., the  charge Drude weight is positive ${\cal D}_{\rm w}^{\rm (C)} > 0$ \cite{Zotos1997,Garst2001}. This is a rigorous statement. At half filling, which corresponds to the most symmetric and thermodynamically dominant sector, the Mazur bound based on the known local charges vanishes. However, this does not imply that the Drude weight necessarily has to be zero.

Nevertheless, some rigorous results have been obtained at half filling. It has been shown in \cite{Carmelo2018} that for any $U>0$ and any positive temperature $T>0$ and within the canonical ensemble where $N=N_\uparrow+N_\downarrow-L$ is held fixed (while
$L\to\infty$), one has a strict upper bound on the charge Drude weight
\begin{equation}\label{eq:hub.rig1}
{\cal D}^{\rm (C)}_{\rm w}|_{\rm canonical} \le \frac{c'(U) t_{\rm h}^2}{T} L (2\rho-1)^2,
\end{equation}
which scales as $1/L$ at half filling $\rho=1/2$ since, by including leading 
finite-size corrections, $ 2 \rho - 1 = 0 + {\cal O}(1/L)$. Therefore, the 
r.h.s.\ of Eq.\ (\ref{eq:hub.rig1}) vanishes in the thermodynamic limit.

One can argue that within the grand-canonical ensemble where the number of electrons fluctuates according to the law of large numbers, 
\be
\langle (2(N/L)-1)^2\rangle_{\rm grand-can} \propto 1/L\,,  
\ee this bound is no longer useful.
There, instead, one can derive an improved bound which, however, only holds within leading order in $1/T$ but for  any value and sign of $U$
\begin{equation}\label{eq:hub.rig2}
{\cal D}^{\rm (C)}_{\rm w}|_{\rm grand-can} \le \frac{c(U) t_{\rm h}^2}{T} (2\rho-1)^2 \,.
\end{equation}
This bound indicates that ${\cal D}^{(C)}_{\rm w} =0$ if $\rho=1/2$, consistent with the GHD result \cite{Ilievski2018}.
The full temperature dependence of the charge Drude weight ${\cal D}^{\rm (C)}_{\rm w}$ at $0<\rho< 1/2$ was computed in a recent GHD study \cite{Ilievski2017}.

\begin{figure}[t]
\includegraphics[width=0.9\columnwidth]{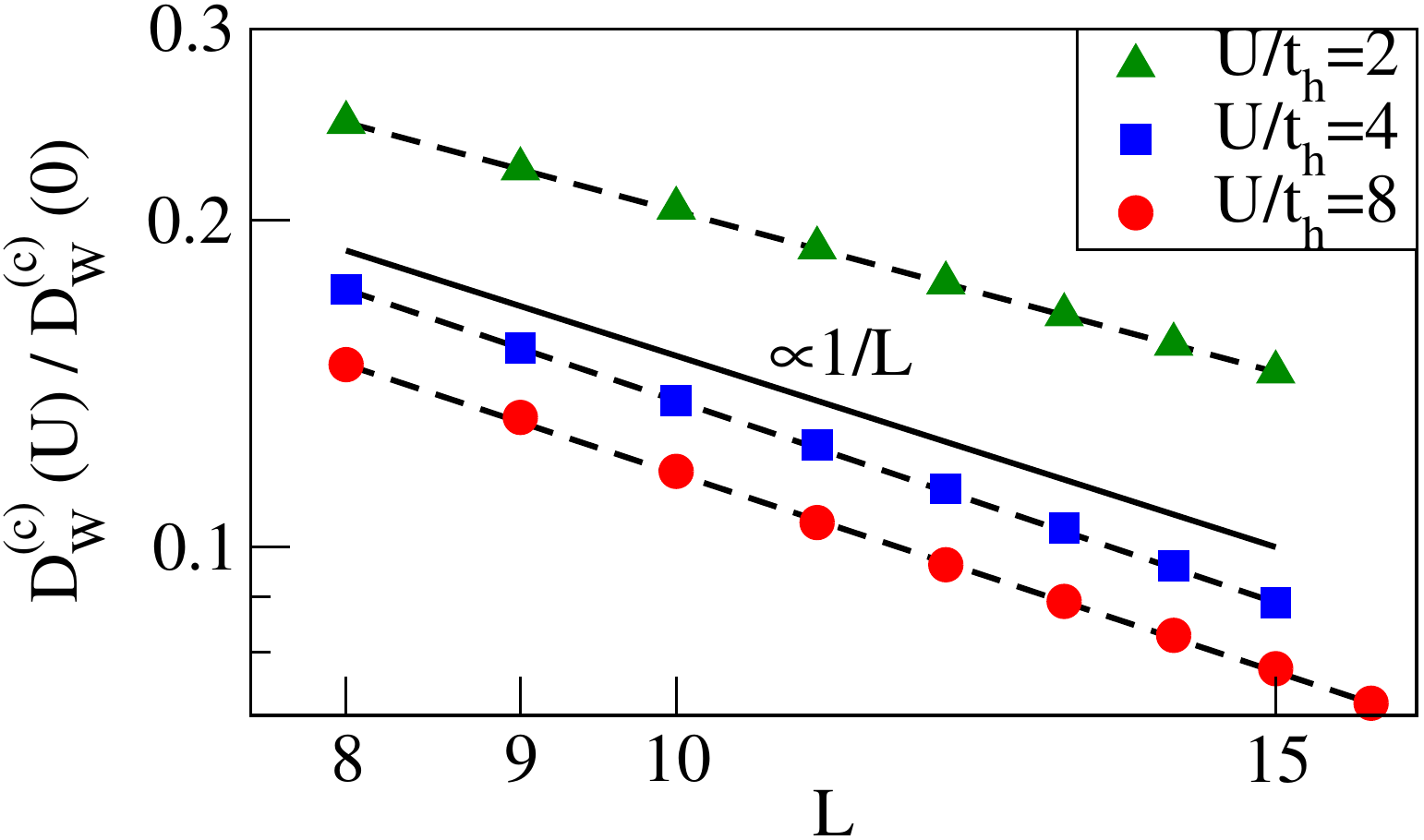}
\caption{(Color olnline)   
Charge  Drude weight \revision{of the Fermi-Hubbard chain} at half filling and infinite temperature versus system size obtained from dynamical typicality \cite{Jin2015}, plotted using a logarithmic scale on both axis.}
\label{fig:Dc_hubbard}
\end{figure}

The question of whether or not the charge Drude weight in the half-filled Fermi-Hubbard chain is zero has historically  been a controversial topic. 
Several early studies reported evidence for a finite Drude weight \cite{Kirchner1999,Fujimoto1998}.
This result was later challenged by Bethe-ansatz studies that emphasized 
symmetry constraints on the diagonal matrix elements of 
the charge-current operator  \cite{Carmelo2013, Carmelo2018}. Numerically, charge transport was studied using exact diagonalization and MCLM \cite{Prelovsek2004}, finite-$T$ tDMRG \cite{Karrasch2014a,Karrasch2016}, dynamical typicality \cite{Jin2015}
and tDMRG simulations of open quantum systems \cite{Prosen2012}. All these studies agree in so far as they find no evidence for a ballistic contribution.
As an example, we show the infinite-temperature Drude weight computed from dynamical typicality in Fig.~\ref{fig:Dc_hubbard} as a function of system size for several values of $U/t_{\rm h}$.
The Drude weight decays with a power-law in $1/L$, consistent with the observation for other integrable models [see, e.g., the large $\Delta$ phase of the spin-1/2 XXZ chain \cite{Heidrich-Meisner2003,Steinigeweg2014}].
 
A rigorous lower bound using the method of \cite{Medenjak2017} for the charge-diffusion constant was recently obtained \cite{Ilievski2018}. This bound diverges at half filling, which shows that transport cannot be diffusive. Therefore, the charge transport is similar to the spin transport in the spin-1/2 Heisenberg chain, with presumably no Drude weight in both cases and superdiffusion. Still, the spreading of density perturbations at finite times and in finite systems is indicative of diffusion \cite{Steinigeweg2017b} (see also Sec.\ \ref{sec:spreading}), leaving reconciling these two observations as an open problem.

Using finite-$T$ tDMRG \cite{Karrasch2014}, an attempt was made to extract the temperature dependence of the dc-conductivity 
at low temperatures in order to verify field-theoretical predictions from \cite{Damle1998,Sachdev1997}. The presence of anomalous finite-size effects in $\sigma'(\omega)$ was pointed out in the MCLM study of \cite{Prelovsek2004}. Both these numerical works and \cite{Jin2015} argued for a diffusive form of the conductivity, which is at odds with the rigorous results of \cite{Ilievski2018}.

In \cite{Prosen2012}, the steady-state master equation with boundary Lindblad reservoirs was used to investigate transport in the Hubbard model. It was argued that transport is diffusive in the TDL (results were reported for $L\sim 100 $). Indications for superdiffusion were presented for short systems and large $U/t_{\rm h}$, leading to the speculation that the two limit $U/t_{\rm h}\to \infty$ and $L\to \infty$ may not commute \cite{Prosen2012}.

\subsection{Spin conductivity}

For a non-zero magnetization $\magdens\neq 0$, the Mazur inequality (\ref{eq:MazurD}) shows that the spin Drude weight is finite \cite{Zotos1997}.

One can make rigorous statements similar to Eqs.~(\ref{eq:hub.rig1}) and (\ref{eq:hub.rig2}) about spin transport in the attractive Hubbard model $U<0$. Specifically, under a partial particle-hole transformation, where 
$(c_{l\uparrow},c^\dagger_{l\uparrow},c_{l\downarrow},c^\dagger_{l\downarrow})\to (c_{l\uparrow},c^\dagger_{l\uparrow},c^\dagger_{l\downarrow},c_{l\downarrow})$, the sign of $U$ changes $U\to-U$, while the
spin current(spin Drude weight) maps to charge current(charge Drude weight)and vice versa. At  asymptotically high temperatures, the sign of $U$ becomes irrelevant, and we then have a full symmetry between spin and charge transport.
For example, at zero magnetization,  the leading term in a high-$T$ expansion of the  spin Drude weight vanishes. It is believed that ${\cal D}^{\rm (S)}_{\rm w}=0$ at $\magdens=0$.

The full temperature dependence of the spin Drude weight ${\cal D}^{\rm (S)}_{\rm w}$ at $0<\magdens<1$ was computed in a recent GHD study \cite{Ilievski2017}. At zero magnetization, GHD predicts  ${\cal D}^{\rm (S)}_{\rm w}=0$  \cite{Ilievski2018}, and, in the same regime, the spin diffusion constant was shown to diverge \cite{Ilievski2018}. Ballistic spin transport away from $\magdens=0$ was observed numerically in a tDMRG study of the spreading of density wave packets \cite{Karrasch2017}.

Clarifying the nature of the deviations from diffusion and investigating how exactly the Heisenberg regime is recovered out of the transport properties of the Hubbard chain in the low-temperature regime are open. For instance, the most recent developments
of finite-$T$ tDMRG methods have not been exploited yet in order to address these questions again \cite{Karrasch2017}.

\revision{In a recent GHD study, several aspects of spin transport have been explored, including the crossover from the spin-coherent to the spin-incoherent regime and the emergence of superdiffusion 
at points with non-Abelian symmetry (vanishing chemical potential and/or magnetic field) \cite{Fava2020}.}

\section{Beyond integrable systems}
\label{sec:nonint}

While integrability is particularly appealing because it allows for exact solutions, most systems of relevance for condensed matter physics (experimental or theoretical) do not share this property. In particular, even though the spin-1/2 XXZ chain describes many features of real materials (such as thermodynamics or spectral functions), it cannot describe generic transport. Indeed, the latter is governed by relaxation mechanisms and external scattering off impurities or phonons is unavoidable.     

In this section, we assume that the only relevant conserved quantities are energy,
particle number and magnetization. Hence we exclude Floquet systems, where energy is not conserved,
and unusual or specifically engineered nonintegrable systems that  possess a finite number of nontrivial conserved quantities. 

Theoretically, there is much interest in the stability of properties of integrable models
against adding integrability-breaking perturbations. In classical systems of few particles, the Kolmogorov-Arnold-Moser (KAM) theorem~\cite{Gutzwiller} 
makes a precise statement on this stability whereas there is no such result for quantum systems.
Within the wider field of nonequilibrium dynamics in closed quantum systems, the accepted view
is that in most cases, any arbitrarily small strength of an integrability breaking term
leads to thermalization \cite{Vidmar2016,dAlessio2016} and diffusive transport.
This, however, may not be easy to see
in actual numerical simulations. For finite time scales, the perturbed system may very well still
remember the existence of now only approximately conserved quantities and exhibit prethermalization
behavior \cite{Moeckel2008,Kollar2011,Essler2014,Bertini2015}. For transport, there are so far 
\revision{only few studies that explicitly made a connection between prethermalization and a respective behavior in a transport coefficient
(see, e.g.,  the discussion in \cite{Nessi2015}), while a number of studies touched upon the topic (see, e.g., \cite{Jung2006,Jung2007}).} For sure, the observation  of slow dynamics in finite-size simulations 
is ascribed to weak violations of conservation laws or quasi-localization physics in translational invariant systems [see, e.g.,  \cite{Yao2016,Schiulaz2014,Michailidis2018}]. 

\begin{figure}[t]
\centerline{\includegraphics[width=0.6\linewidth]{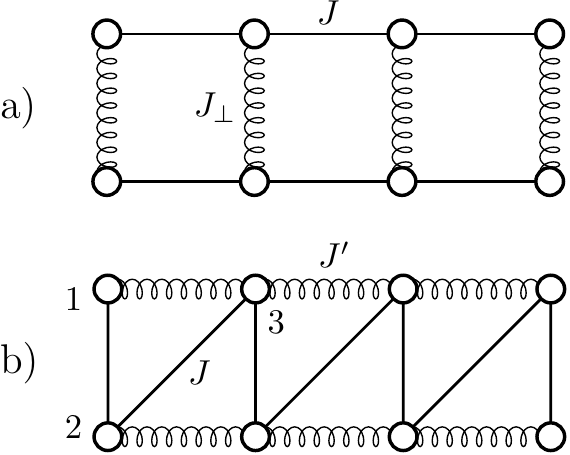}}
\caption{(a) A spin-1/2 ladder  with a coupling strength $J$ and $J_\perp$ along legs 
and rungs, respectively, and (b) a frustrated chain with a next-nearest-neighbor coupling 
of strength $J'$.}
\label{fig:ladders}
\end{figure}

We will concentrate the discussion on nonintegrable models that result from perturbing the spin-1/2
XXZ chain.  The choice of the integrability breaking term is  motivated
by either a particular relevance for experiments (e.g., spin-1/2 Heisenberg ladders), the possibility of obtaining analytical or exact results (e.g., for 
the spin-1/2 XXZ chain with a staggered magnetic field),
or by the desire to obtain the simplest possible cases (e.g., spin-1/2 XX ladders or simple types of short-range interactions).
See Fig.~\ref{fig:ladders} for an illustration of ladders and frustrated chains.

In this section, we will present those statements that describe the majority of those models and will cover selected examples
for which there are either particularly convincing numerical or analytical results. Relevant and important results were certainly obtained for many other models
that are not covered in detail here.
These include spin-$S$ XXZ chains with $S>1/2$ \cite{Karadamoglou2004,Dupont2019,Richter2019a}, Kitaev-Heisenberg chains and ladders \cite{steinigeweg2016b,Metavitsiadis2017,Pidatella2019,Metavitsiadis2019}, and Hubbard models with integrability-breaking terms \cite{Znidaric2013a,Znidaric2013b,Karrasch2016}. 
We will here focus on results obtained via the Kubo formula for closed quantum systems; studies of open quantum systems can be found in \cite{Znidaric2013b,Znidaric2013a,Mendoza2015}.

\subsection{Universal description of the low-energy behavior}

We first turn to the predictions from field theory for the low-temperature behavior.
In a generic gapless system, the field theory developed in \cite{Sirker2011} provides the generic behavior: a Tomonaga-Luttinger-liquid
becomes a diffusive conductor after including sufficiently many umklapp terms \cite{Rosch2000}.
As an example for spin transport in a gapless system,  we consider a spin-1/2 XXZ chain with a staggered magnetic field of strength $h$
that breaks the integrability. For small values of $h$, the system is in the Tomonaga-Luttinger-liquid phase and by applying the field theory of  \cite{Sirker2011}
one obtains \cite{Huang2013}
\begin{equation}
\sigma_{\rm dc} \propto  h^{-2}T^{3-2K}\,,
\end{equation}
where $K$ is the Luttinger-liquid parameter. Further related studies can be found in \cite{Bulchandani2019b,Szasz2017}.

\revision{Another generic insight can be drawn from the fact that at low energy scales, (regular) momentum emerges as an additonal
approximate conserved quantity due to the mapping to a continuum model.
This does not give rise to a finite Drude weight at small but finite temperatures, but causes the dc conductivities of those
currents with a finite overlap to momentum to be exponentially large as a function of decreasing temperature \cite{Rosch2000,Rosch2006}. These predictions 
are based on a memory-matrix formalism.}

For gapped systems, the \revision{semiclassical} theory of \revision{ \cite{Damle1998} }
leads to
\begin{equation}
\sigma_{\rm dc} \propto \frac{1}{\sqrt{T}}\,,
\end{equation}
\revision{see Sec.~\ref{sec:fieldtheory}}. This divergence (as $T\to 0$) can be understood from the fact that on the one hand, the density
of carriers is exponentially suppressed but on the other hand, this dilution leads to an exponential
suppression of scattering as well.
The available tDMRG results \cite{Karrasch2014} for the Hubbard chain with a nearest-neighbor repulsion and the gapped phase of the integrable spin-1/2 XXZ chain
seem more consistent with a $1/T$ dependence. An outstanding question is to compute $\sigma_{\rm dc}(T)$ for a spin-1 chain or a spin-1/2 ladder, for which the predictions of
\cite{Damle2005} were developed.

\subsection{Absence of Drude weights}
\label{sec:nonint_drude}

Within our working definition of nonintegrable models given above, it is clear that there is 
no nonzero Mazur bound for Drude weights. Hence, the expectation is that Drude weights vanish at
any finite temperature. Note that at zero temperature, any metallic phase has a nonzero Drude weight
as long as the system preserves translational invariance \cite{Scalapino1993,Mastropietro2013}.
This nonzero spin and charge Drude weight results from the fact that 
the low-energy theory is a gapless Tomonaga-Luttinger liquid with one or two modes and is thus a consequence
of the conservation of momentum in the continuum limit.
Yet, these zero-temperature Drude weights are not
related to the  integrability of the microscopic models. A concrete example is the frustrated spin-1/2 Heisenberg chain, which in its
gapless phase has a nonzero spin Drude weight at $T=0$ \cite{Bonca1994}.

Most numerical studies confirm the expectation that spin, charge and energy 
Drude weights vanish
in nonintegrable models at any $T>0$, including, for example, spin-1/2 Heisenberg ladders \cite{Zotos2004,Heidrich-Meisner2003,rezania13},
frustrated spin-1/2 Heisenberg chains \cite{Heidrich-Meisner2003,Heidrich-Meisner2004,hm04c,hm04d}, dimerized spin-1/2 Heisenberg chains \cite{Heidrich-Meisner2003,hm04d},
spin-1/2 XXZ chains with additional nearest-neighbor interactions $S^z_\ell S^z_{\ell+2}$ (equivalent to spinless fermions with density-density interactions) \cite{Zotos1996}, and spin-1/2 XXZ chains with staggered magnetic fields \cite{Huang2013,Steinigeweg2015}.

\begin{figure}[t]
\begin{center}
\includegraphics[width=0.90\linewidth]{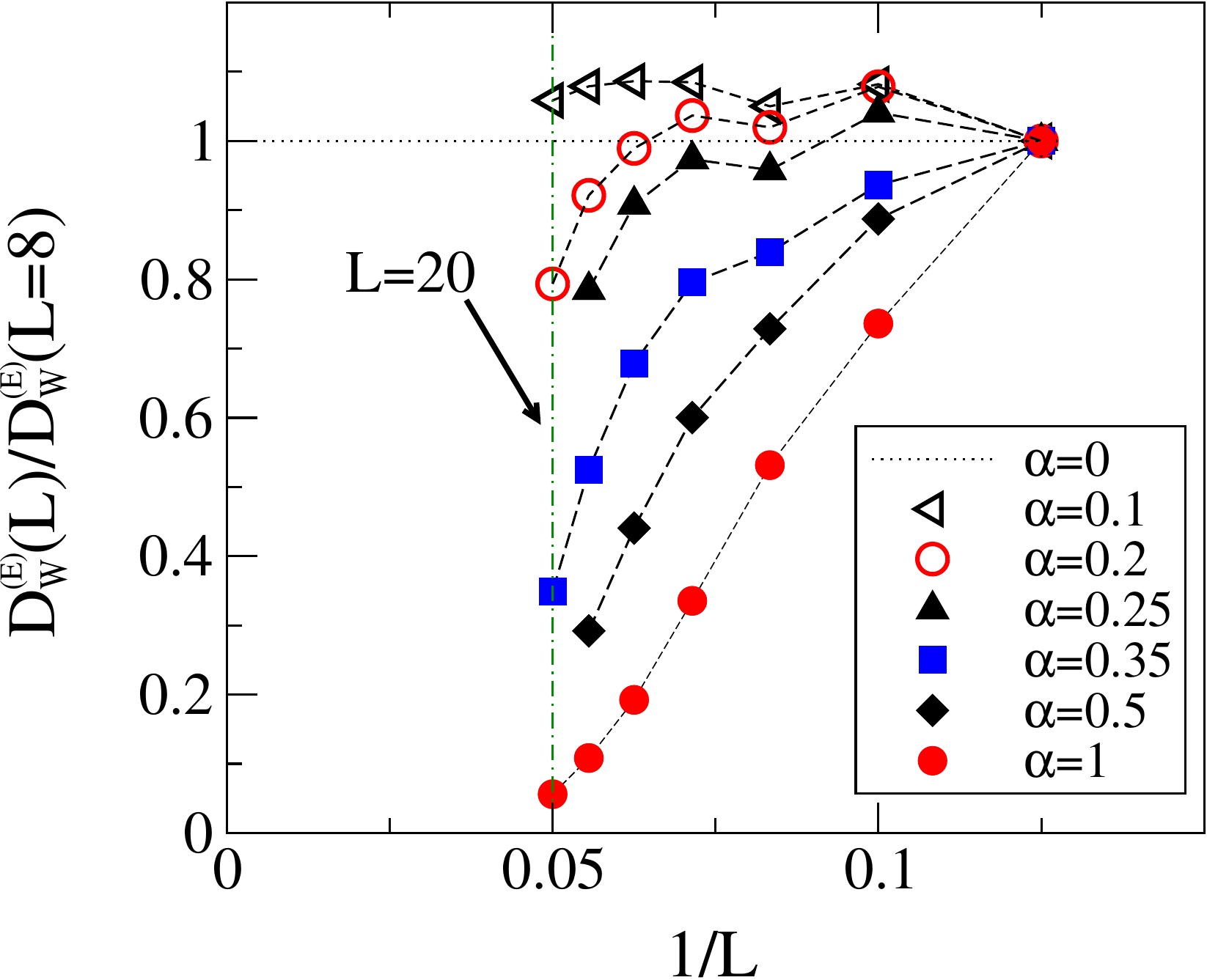}
\caption{(Color online) 
Infinite-temperature energy Drude weight of frustrated spin-1/2 Heisenberg 
chains
computed with exact diagonalization  \cite{Heidrich-Meisner2003,Heidrich-Meisner2004}.
}
\label{fig:dw_frust}
\end{center}
\end{figure}

In the vicinity of integrable models and on \textit{finite} systems, the Drude weight may still account for most of the weight in the conductivity $\sigma'(\omega)$
with only a very slow transfer of weight to finite frequencies. 
A particularly interesting example is the spin-1/2 frustrated Heisenberg chain 
where the relevant parameter is the ratio $\alpha=J'/J$ 
and  $J(J')$ are the nearest(next-to-nearest) neighbor exchange couplings [see Fig.~\ref{fig:ladders}(b)]. The Hamiltonian reads
\begin{equation}
H=J\sum_{r =1}^{L} h_{r,r+1}+J'\sum_{r=1}^{L} h_{r,r+2}\,,
\label{eq:Hnnnchain}
\end{equation}
where we assume periodic boundary conditions. As an example, we show 
exact-diagonalization data for the energy Drude weight in 
Fig.~\ref{fig:dw_frust};
specifically, for its leading coefficient $\DwEbeta$ in a $1/T$ expansion:
\begin{equation}
{\cal D}^{\rm (E)}_{\rm w} = \frac{\DwEbeta}{T^2} + \dots \,.
\end{equation}
At small values of $\alpha \lesssim 0.3$ and for the accessible system sizes, $\DwEbeta$ decays only mildy compared to the integrable case
and seems to saturate \cite{Alvarez2002a,Heidrich-Meisner2004}. Upon increasing $\alpha$, though, the decrease of $\DwEbeta$ with $L$ becomes faster and is consistent with an 
exponential decay or at least faster than any power law. The latter is expected from ETH arguments \cite{Steinigeweg2013} and numerically observed in nonintegrable
models far away from integrable limits \cite{Prosen1999,Zotos1996,Heidrich-Meisner2004,Rabson2004,Jin2015}. 

For the frustrated spin-1/2 chain, there is a theoretical argument 
that explains why on small system sizes and for small values of $\alpha \lesssim 0.3$, the thermal Drude weight still amounts to a substantial fraction
of the total spectral weight.
It turns out that the energy-current conservation is only violated
in next-to-leading order in $\alpha$ \cite{Jung2006}.
Within the memory-matrix formalism, one can then show that current lifetimes are enhanced in the small-$\alpha$ regime.

Other cases in which the proximity to integrable limits can lead to a slow decay of Drude weights
on finite systems (or to a slow temporal decay of current-autocorrelations computed with t-DMRG) are certain spin-1/2 dimerized XXZ chains \cite{Karrasch2013}
and gapped quantum models in large magnetic fields \cite{Langer2010,Psaroudaki2014,Stolpp2019}. 
In the former case, the  existence of several integrable limits (vanishing dimerization, zero exchange anisotropy $\Delta=0$, decoupled dimers) has been 
speculated to give rise to a slow decay of current correlation functions.
In the latter case, the application of 
a longitudinal magnetic field induces a transition into a gapless phase. For spin-1 chains  \cite{Psaroudaki2014}, that field-induced phase 
can be approximately described by an effective spin-1/2 XXZ chain Hamiltonian, explaining the numerically observed large finite-size Drude 
weights. 
There is, in none of these examples, any theoretical evidence to believe that the finite-size Drude weights remain nonzero for $L\to \infty$.
Other claims of nonzero Drude weights in generic spin ladders, frustrated spin chains or dimerized spin chains 
were either based on a mapping to noninteracting effective theories \cite{Orignac2003,Saito2003} or due to the difficulties involved with interpreting 
finite-size exact diagonalization  \cite{Alvarez2002a} or QMC data \cite{Kirchner1999}.

\subsection{Frequency-dependence of the conductivity}
The simplest picture for the frequency dependence was already given in Sec.~\ref{sec:theory}
and is based on  the Drude model: a Lorentzian whose width is controlled by a single
relaxation time. One may wonder whether such a simple structure is possible at all
in strongly correlated models in one dimension where there is no Landau quasi-particle picture in the
first place.

\begin{figure}[t]
\begin{center}
\includegraphics[width=0.90\linewidth]{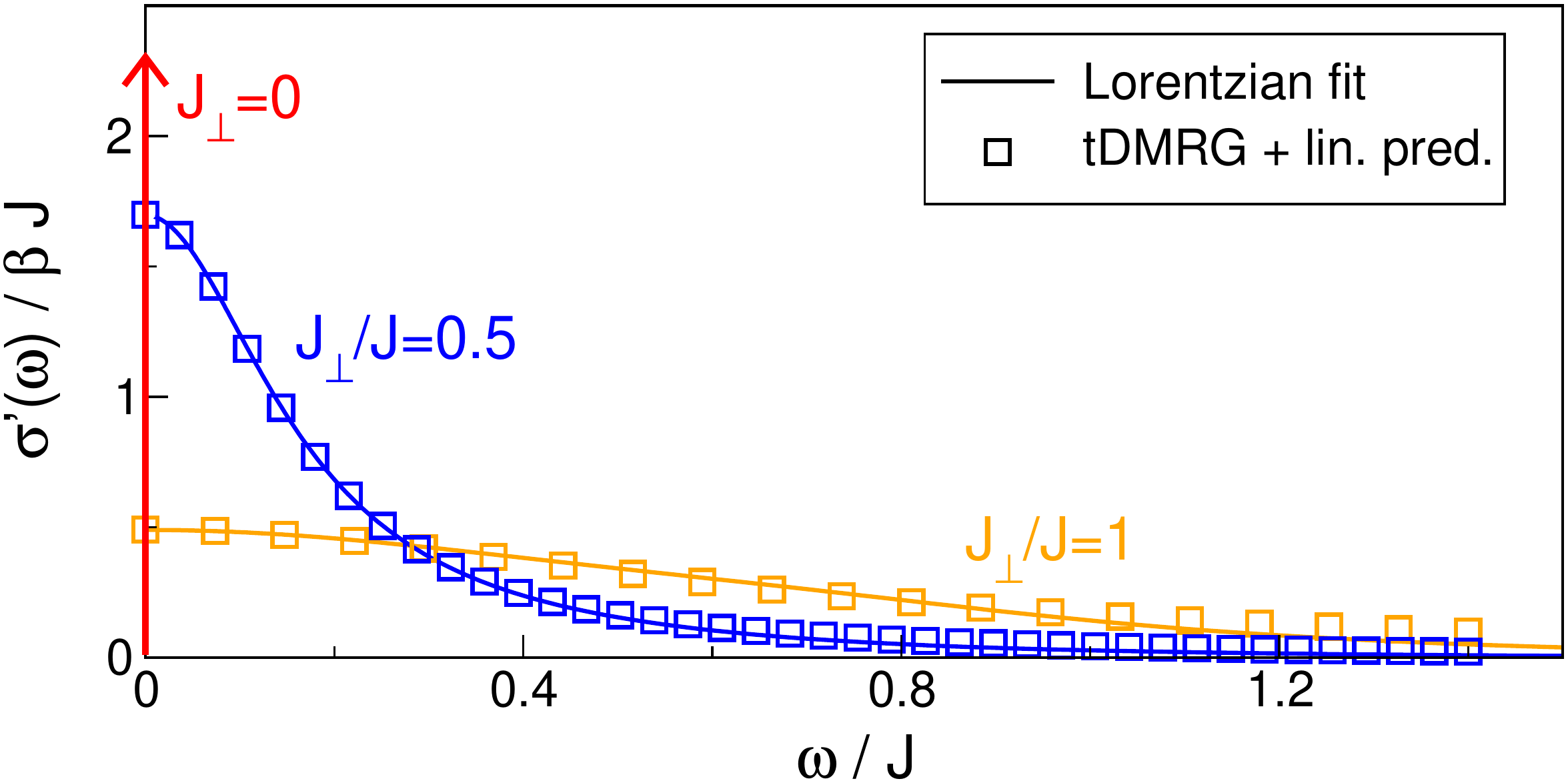}
\caption{(Color online) Spin conductivity of a spin-1/2 XX ladder for various rung
couplings $J_\perp$ \cite{Karrasch2015}.
}
\label{fig:ladder_sigma}
\end{center}
\end{figure}

For infinite temperature, there are many numerical results available. A particularly clear picture
emerges for spin-1/2 XX ladders. In that case, the integrable limits are two chains that have only a Drude weight, i.e., $\sigma'(\omega)= 2 \pi \, \Dws \delta(\omega)$.
The Hamiltonian reads (with $\Delta =0$ in the $h_r^{\parallel,\perp}$ terms):
\begin{equation}
H=H^{||}+H^{\perp}=J\sum_{\ell=1,2}\sum_{r=1}^{L} h^{||}_{\ell;r,r+1}+J_\perp\sum_{r=1}^{L}
h^{\perp}_{r}\,.
\label{eq:Hladder}
\end{equation}
$\ell=1,2$ labels the two legs of the ladder.
Upon coupling the chains with a nonzero $J_\perp$, 
the Drude peak is indeed broadened into a Lorentzian. This is illustrated in Fig.~\ref{fig:ladder_sigma}, obtained from tDMRG \cite{Karrasch2015},
which agrees with dynamical typicality and perturbation theory \cite{Steinigeweg2014a,Richter2019d, Richter2019e}.
For more complicated models, the situation is less clear based on the numerical data. For instance, in spin-1/2 Heisenberg ladders, there is presumably
superdiffusion for $J_\perp=0$ [see Sec.~\ref{sec:xxz_delta1}] and it is not obvious that there exists a single Lorentzian at low frequencies. Numerical results for $\sigma'(\omega)$ of
nonintegrable models are available for spin-1/2 XXZ ladders \cite{Karrasch2015}, spin-1/2 XXZ chains with a staggered magnetic field \cite{Huang2013},
 dimerized spin-1/2 Heisenberg chains \cite{Langer2011}, and interacting spinless fermions with next-to and nearest-neighbor hopping \cite{Mukerjee2007}. The thermal conductivity $\kappa(\omega)$ was computed numerically for spin-1/2 XXZ chains with a staggered magnetic field \cite{Huang2013,Steinigeweg2015}. 

\begin{figure}[t]
\begin{center}
\includegraphics[width=0.8\linewidth]{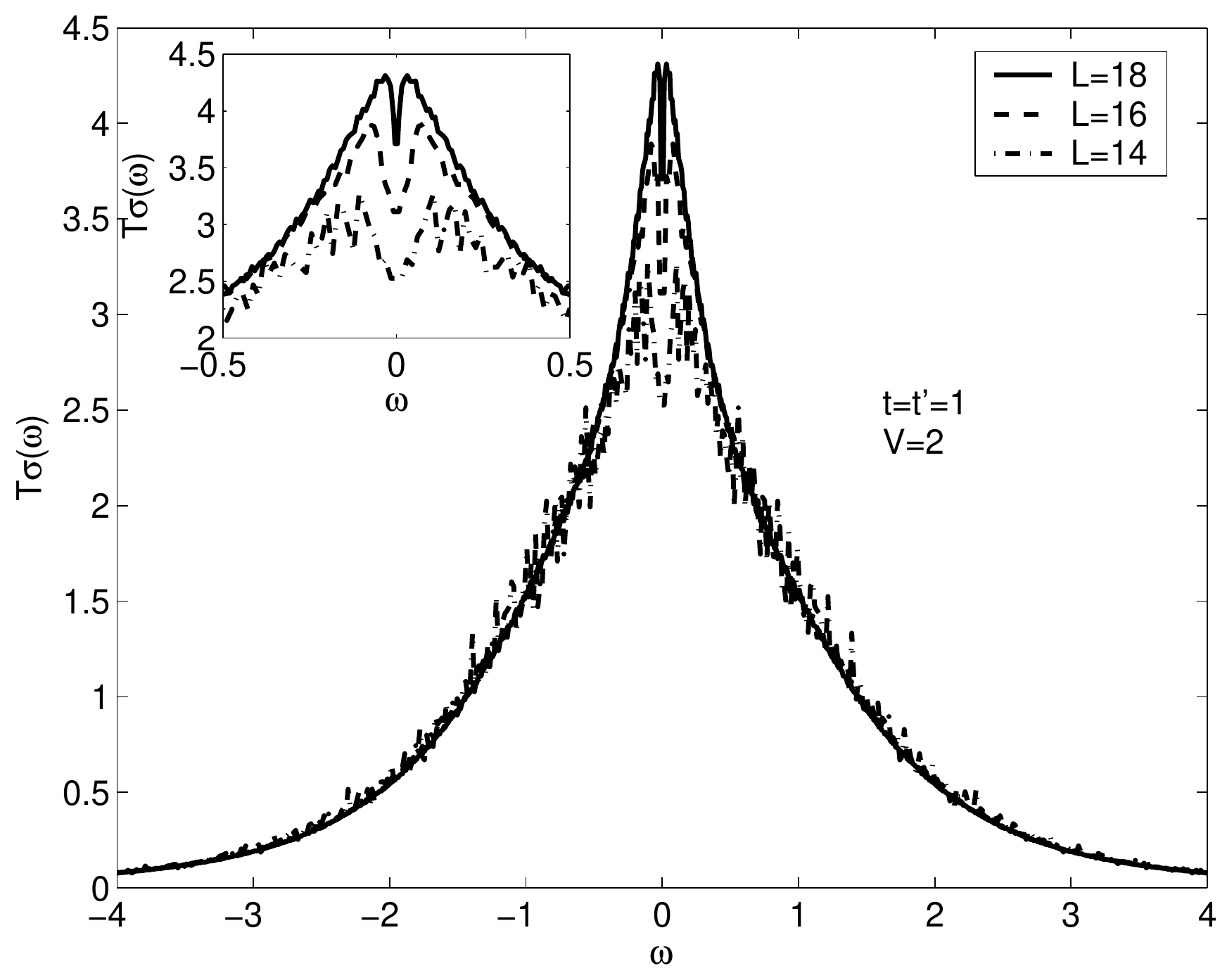}
\caption{Real part of the conductivity of a nonintegrable model versus $\omega$ (in units of $t$), namely, spinless fermions with a nearest-neighbor repulsion of strength $V=2t$ and an additional next-to-nearest neighbor hopping $t'=t$, where
$t$ is the nearest-neighbor hopping matrix element. The exact-diagonalization data indicates that the low-frequency behavior is incompatible with a simple Lorentzian (see \textcite{Mukerjee2007} for a discussion). Other examples of a similar
shape were reported in \textcite{Zotos2004,hm04c}.
The figures is taken from  \cite{Mukerjee2007}.
}
\label{fig:mukerjee}
\end{center}
\end{figure}

While a finite dc-limit is enough to classify the system as a normal conductor, there remains the possibility of potential anomalous low-frequency behaviors [see, e.g., \cite{Garst2001}
for a discussion of the mass-imbalanced Hubbard model].
For instance, evidence for such a situation was reported for a nonintegrable model of spinless fermions  \cite{Mukerjee2007}, where $\sigma'(\omega)= a- b \sqrt{\omega}+\dots$ was observed in numerical data and explained as a 
hydrodynamic tail. The corresponding $\sigma'(\omega)$ is shown in Fig.~\ref{fig:mukerjee}. A systematic study of such tails in nonintegrable models for larger systems
and a broader class of models remains to be done [see also \cite{Zotos2004}], in particular, by making 
more quantitative contact the predictions of hydrodynamics.

\subsection{DC conductivity and diffusion constant}

We next turn to the available results for the temperature dependence of dc conductivities and diffusion constants
and their dependence on model parameters. The latter dependency is relevant to understand the effect of integrability 
breaking terms (parametrized by a coupling constant $J_{\text{pert}}$). The crossover from GHD describing integrable models to regular
hydrodynamics in nonintegrable models has been discussed in \cite{Friedman2019}. Based on Fermi's Golden rule, one generically expects
$\D^{(Q)} \propto 1/J_{\text{pert}}^2$ and thus a similar scaling for the conductivity \cite{Steinigeweg2016a,Jung2007,Zotos2004}.

\begin{figure}[t]
\begin{center}
\includegraphics[width=0.90\linewidth]{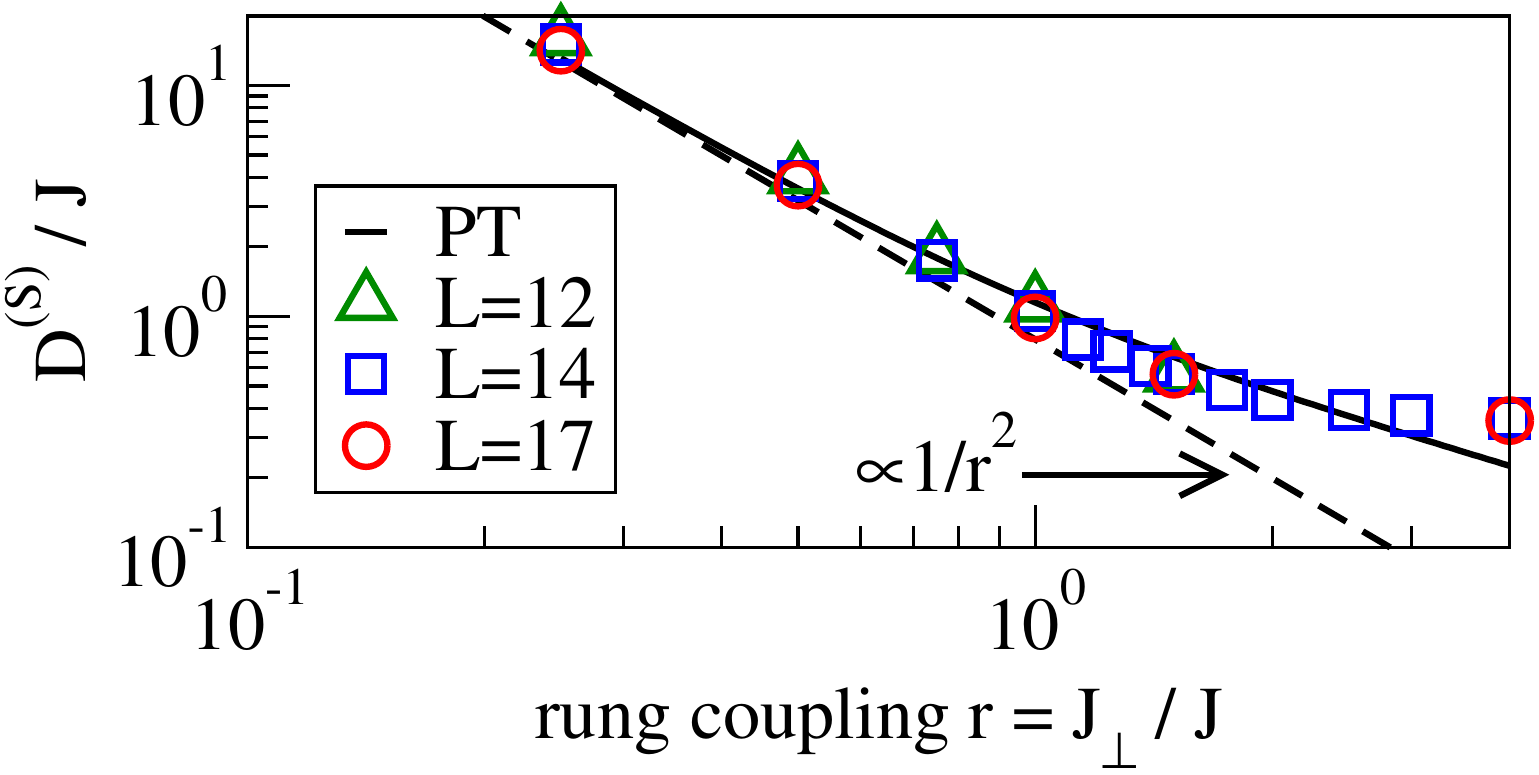}
\caption{(Color online) Spin-diffusion constant of the spin-1/2 XX ladder as a
function of $r=J_{\perp}/J$ at infinite temperature, as obtained from dynamical typicality
for $L=12,14,17$ rungs (i.e., $2L$ sites in the ladder)  \cite{Steinigeweg2014} and perturbation theory  (PT) \cite{Richter2019e},
with $\DS / J = 1/(2 \gamma r^2)$ and $\gamma \approx 0.63$ in the limit of small $r$ with no free
parameter.}
\label{fig:D_XX}
\end{center}
\end{figure}

We first discuss the infinite-temperature limit and then move on to cover predictions and results for finite temperatures.
Numerical results for the diffusion constant - and via Einstein relations - the dc conductivity are available for spin transport in 
spin-1/2 XX ladders \cite{Steinigeweg2014a,Znidaric2013a} and thermal transport in spin-1/2 XXZ ladders \cite{Alvarez2002,Saito2003,Orignac2003,Heidrich-Meisner2003,Steinigeweg2016a,Karrasch2015,Zotos2004}
and spin-1/2 chains with staggered magnetic fields \cite{Huang2013,Steinigeweg2015} as well as for charge transport in the mass-imbalanced
Fermi-Hubbard chain \revision{\cite{Jin2015,Heitmann2020}} [see also \cite{Garst2001}].

Figure~\ref{fig:D_XX} shows the dependence of the spin-diffusion constant $\DS$ on $J_\perp$ for spin-1/2 XX ladders: at small $J_\perp$, $\DS \propto 1/J_\perp^2$
in agreement with perturbation theory \cite{Steinigeweg2014a,Richter2019e} with 
a crossover to $\DS =\textnormal{const.}$ at large $J_\perp \gg J$. The latter 
behavior is typical for systems with a bandstructure
(here controlled by $J_\perp$ in the strong dimer limit) at $T=\infty$ and is also seen in the large $U/t_{\rm h}$ regime of the Fermi-Hubbard chain \cite{Jin2015}.
A perturbative dependence of diffusion constants on an integrability-breaking parameter was reported for thermal transport in spin ladders as well \cite{Steinigeweg2016a,Zotos2004}.

\begin{figure}[t]
\begin{center}
\includegraphics[width=0.90\linewidth]{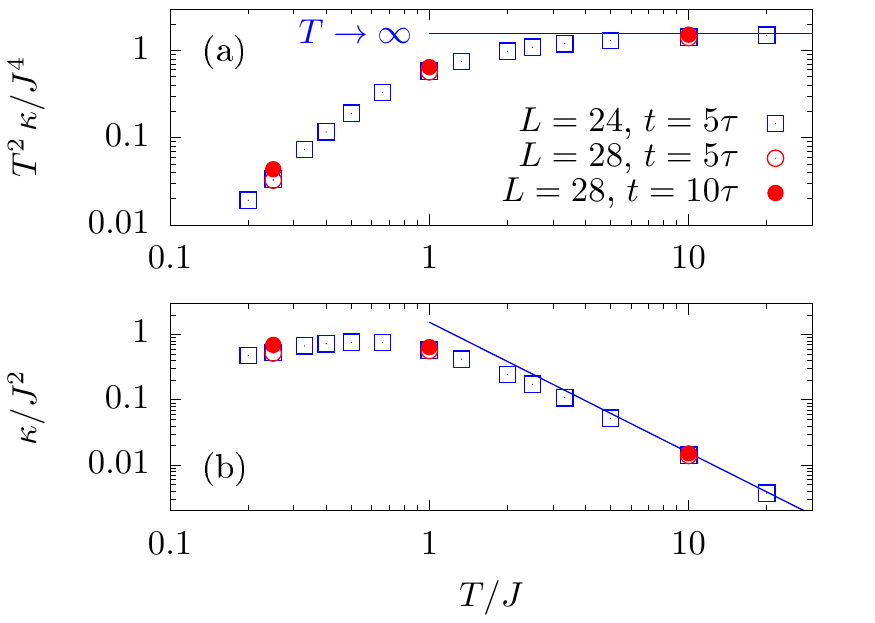}
\caption{(Color online) Temperature dependence of the thermal conductivity $\kappa$
in a Heisenberg chain with a staggered field, as shown in
\cite{Steinigeweg2015} [see also \cite{Huang2013}]. $\tau$ is the relaxation time, defined as the time at which the current correlation has decayed to
a fraction of $1/e$ \cite{Steinigeweg2015}. }
\label{fig:kappa_staggered}
\end{center}
\end{figure}

Numerical methods now also give access to a wider temperature range. As an example, we show the thermal conductivity of spin-1/2 XXZ chains with
a staggered magnetic field in Fig.~\ref{fig:kappa_staggered} \cite{Steinigeweg2015,Huang2013} [see also \cite{Prosen2009}].
The maximum at a temperature $T \sim \mathcal{O}(J)$ is  resolved, while the data indicate a power-law dependence at low $T$.

\subsection{Special cases and outlook}
We conclude this section by giving a brief account of special cases and ongoing directions.

Local perturbations that act only on one or few sites can behave completely
differently from global perturbations that were covered so far. Having in mind that integrable systems
possess infinitely long-lived excitations, this  is not very surprising:
looking at the transmission from one end to the other, an excitation 
will scatter only once, regardless of the system's length, and therefore, one
will have zero bulk resistance and ballistic transport.

\begin{figure}[t]
\centerline{\includegraphics[width=0.9\linewidth]{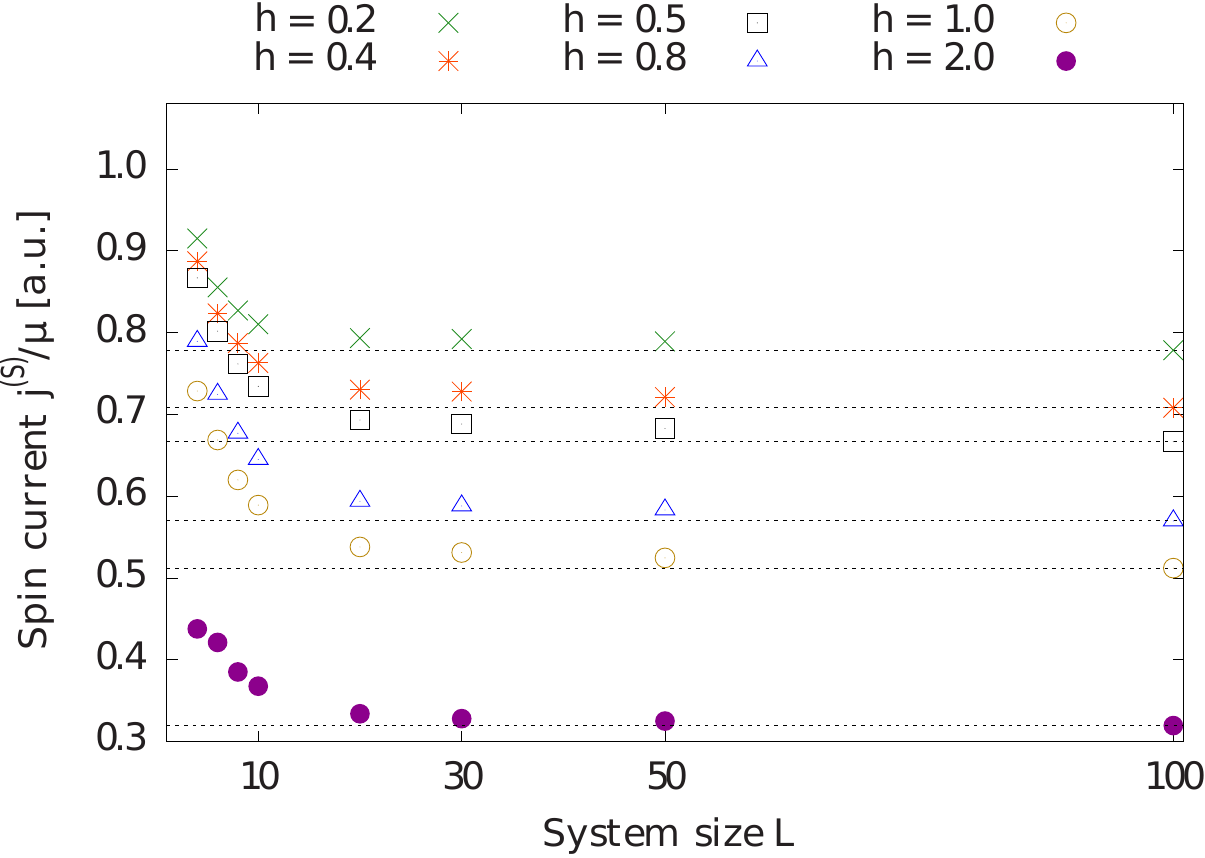}}
\caption{Dependence of the NESS spin current on system size $L$ for a spin-1/2  XXZ chain ($\Delta = 0.5$) with a single-impurity of strength $h$. Adapted from~\cite{Brenes2018}.
}
\label{fig:brenes18}
\end{figure}

Let us
take a concrete model, the spin-1/2 XXZ Heisenberg chain, Eq.~(\ref{eq:XXZ-intro}), 
and a single impurity at the middle of the chain described by
$\frac{h}{2} \sz{L/2}$, where $h$ is a local magnetic field. Analyzing the distribution of energy spacings between
nearest-eigenenergy levels~\cite{Santos2004,Santos2008,barisic09,Fagotti:2017aa,Torres-Herrera2014,Brenes2018}, one observes
level repulsion and agreement with random-matrix theory for a large range of $h$
(in the thermodynamic limit likely for any finite nonzero $h$), typical of
quantum chaotic systems. Studies of spin transport with  a boundary-driven Lindblad
setting as well as with a linear-response calculation of $\sigma'(\omega)$ 
suggest 
\revision{ballistic transport~\cite{Brenes2018,Brenes2020}} [see Fig.~\ref{fig:brenes18}] (\revision{see also \cite{Brenes2020a}}). We note
that in order to identify a nonzero Drude weight, one has to carry out a careful finite-size
scaling analysis of $\sigma'(\omega)$ because for open boundary conditions, 
$\Dws$
gets ``transferred'' to finite frequencies \cite{Rigol2008}, getting a Lorentzian (Cauchy)
representation of a Dirac delta function, whose width decreases $\sim 1/L$ while
its height increases as $\sim L$. For the case of a spin-$S$ impurity with $S>1/2$, this was interpreted in
\cite{barisic09} and in \cite{metavitsiadis10,metavitsiadis11} 
 as an ``anomalous incoherent'' energy and spin transport. One therefore 
has a ``quantum chaotic'' system according to the level-spacing statistics but
ballistic transport, typically associated with integrability and conserved
quantities. 

How can one reconcile these two seemingly contradicting findings? In a
many-body system, the level spacing is exponentially small in $L$ in the
thermodynamic limit. Therefore, starting with  eigenstates
of the integrable spin-1/2 XXZ chain, even a small local perturbation can, in the
thermodynamic limit, cause mixing of close eigenenergies, leading to level
repulsion. However, level spacing measures properties on an exponentially small
energy scale that can be potentially irrelevant for local physics. For transport,
timescales that are polynomial in $L$ are what matters.

A currently intensely investigated question 
concerns the precise time scale and conditions for hydrodynamics to set in \revision{(see, e.g., \cite{Lopez-Piqueres2020,Glorioso2020})}. 
This question is not new,
yet numerical methods are now in a position to simulate this while novel theoretical concepts 
from quantum information theory  such as entanglement spreading or out-of-time-ordered correlators provide for 
a novel view onto this problem.
 In that regime, the system should behave
``classically'' and be subject to the 
laws of hydrodynamics
\cite{Lux2014,Leviatan2017,Bohrdt2017,Karrasch2014,Rakovszky2018}. 
\revision{Recently, a generalized relaxation-time approximation framework has been proposed to study the crossover from generalized hydrodynamics, applicable to integrable systems,
to hydrodynamics in a generic model \cite{Lopez-Piqueres2020}.}
Related efforts address the emergence of hydrodynamics in random unitary 
circuits [see, e.g., \cite{Nahum2018, Rakovszky2018}].
 
Earlier work studied the emergence of diffusion in Hamiltonian systems with random couplings \cite{Steinigeweg2006,Steinigeweg2007}.
\revision{In addition to a hydrodynamical description as a generic framework and numerical approaches, a semiclassical method based on the
truncated Wigner approximations has recently been developed to study diffusion in spin systems \cite{Wurtz2020}. 
Finally, the possibility of anomalous transport in nonintegrable models is still of interest and an example of subdiffusion has been reported
in systems that conserved dipole and/or higher moments \cite{Feldmeier2020}.}

\section{Far-from-equilibrium transport}
\label{sec:noneq}

There is a growing interest in the nonequilibrium dynamics induced by  initial states with inhomogeneous densities
across various branches of theoretical physics.
The ensuing dynamics when starting from  such initial states is in fact of interest across several disciplines in physics, including condensed matter theory \cite{Liu2014}, quantum field theory \cite{BernardDoyon_2016}, AdS/CFT correspondence \cite{Bhaeseen2015}, statistical physics \cite{Antal1997}, and ultracold quantum gases \cite{Schneider2012,Ronzheimer2013,Vidmar2015,Vidmar2017}.

\subsection{Spreading of density perturbations}
\label{sec:spreading}

A prototypical nonequilibrium setup is to prepare a local energy-, charge- or spin-density perturbation in an otherwise equilibrated background. Such a ``wave packet'' can, for instance, be realized via an initial density matrix of the form $\rho_L(T)\otimes\rho_C\otimes\rho_R(T)$, where the density matrices $\rho_{L,R}$ associated with the left and right regions have the standard equilibrium form. The center region can, e.g., be chosen as a thermal density matrix with a different temperature $T+\Delta T$ in order to model an energy density perturbation.

In Sec.~\ref{sec:transport_inhom}, it was discussed that if this initial local perturbation is small ($\Delta T\to0$ and the size of the center region $C$ being small in the above example), the time evolution of its variance $\Sigma(t)$ is strictly related to the time-dependent diffusion constant via Eq.~(\ref{Einstein_relation}) \cite{bohm1992, Steinigeweg2009a, yan2015}. It implies that at long times, $\Sigma\sim t$ for ballistic transport and $\Sigma\sim\sqrt{t}$ for diffusive transport, with the prefactors given by the Drude weight and the diffusion constant, respectively. In the context of this review, the validity of the time-dependent Einstein relation was confirmed numerically for spin, charge, and energy transport within the spin-1/2 XXZ chain and the Fermi-Hubbard model, by tDMRG \cite{Karrasch2017a} and dynamical dynamical typicality calculations \cite{Steinigeweg2017, Steinigeweg2017b}.

One still expects that the long-time behavior of the variance is of the above-mentioned form ($\Sigma\sim t$ and $\Sigma\sim\sqrt{t}$, respectively), even if one considers the spreading of local perturbations which are not necessarily small. This was first shown for the integrable spin-1/2 XXZ chain as well as for nonintegrable systems at zero temperature using tDMRG \cite{Langer2009,Jesenko2011,Langer2011}. For instance, it was illustrated that spin propagates ballistically for $\Delta<1$ and diffusively for $\Delta>1$ in agreement with the zero-temperature behavior of the Drude weight which is finite in the former but vanishes in the latter case \cite{Langer2009}; while energy always propagates ballistically at any $\Delta$ \cite{Langer2011}.

\begin{figure}[tb]
\includegraphics[width=0.90\columnwidth]{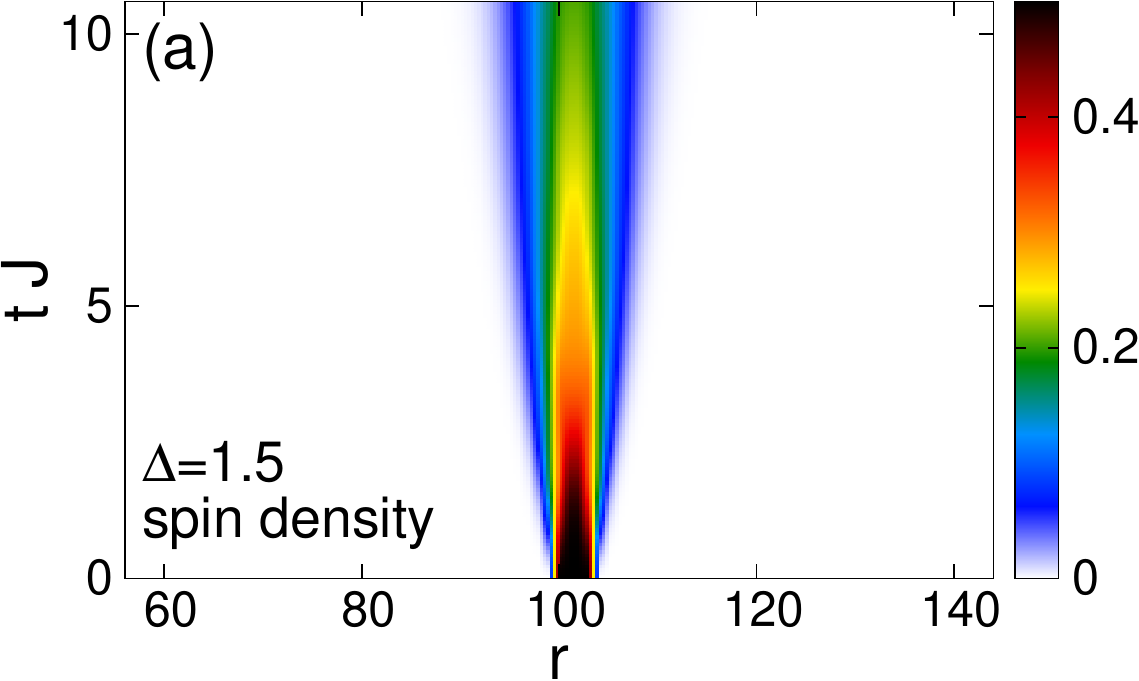}
\vspace{0.2cm}
\includegraphics[width=0.90\columnwidth]{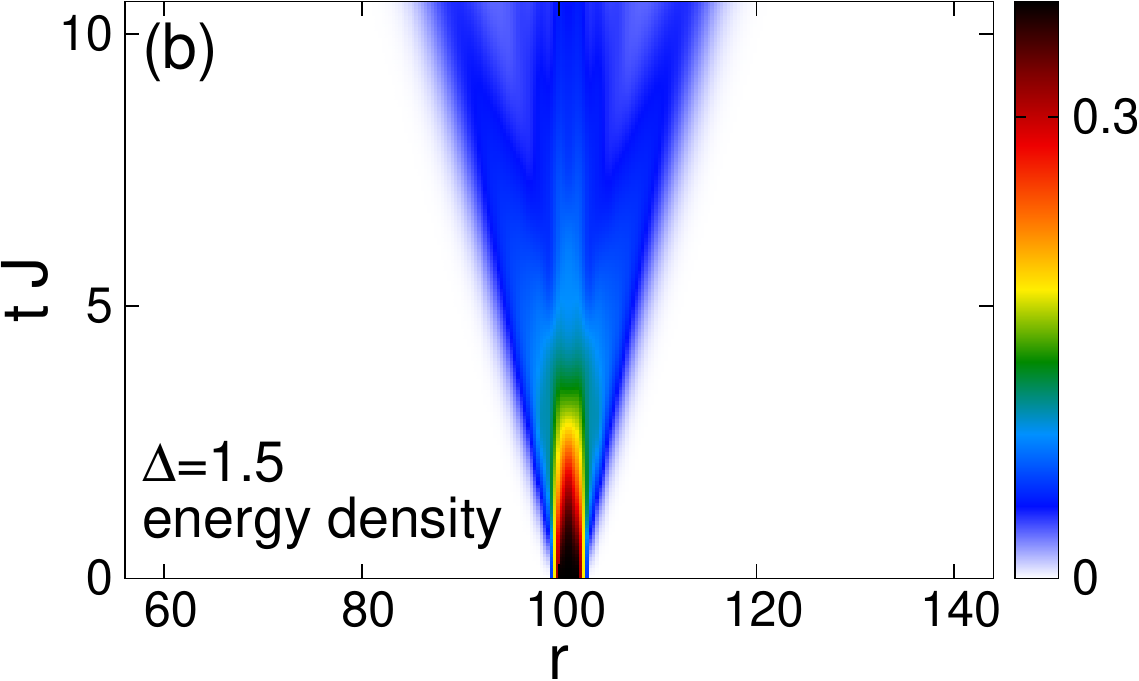}
\caption{Densities as a function of time $t$ and position $r$ for a local quench inducing (a) spin dynamics and (b) energy dynamics
in the spin-1/2 XXZ chain at $\Delta =1.5$ at $T=\infty$. The dynamics is induced by introducing a local perturbation in the initial
state. Adapted from \cite{Karrasch2014}.}
\label{fig:local_xxz}
\end{figure}

These studies were  extended to finite temperatures and to pure-state dynamics, and the spreading of spin and energy wave packets were studied for the spin-1/2 XXZ chain, for spin ladders, and for the Fermi-Hubbard model \cite{Karrasch2014,Karrasch2017a,Steinigeweg2017,Richter2018a,Richter2019d,Foster2011,Foster2010}, \revision{including the mass-imbalancec case \cite{Heitmann2020}}. For instance, one can prepare a spin-polarized central region in a $T=\infty$ background within the XXZ chain, which leads to a simultaneous propagation of spin and energy densities. In this setup, spin propagates diffusively and energy propagates ballistically for $\Delta>1$, and both quantities propagate ballistically for $\Delta<1$.
The typical behavior of the spin and energy density after local quenches of this type is illustrated in Fig.~\ref{fig:local_xxz} for $\Delta=1.5$. 
On the time scales accessed in these tDMRG simulations, the variance behaves as $\Sigma^2 \propto t^{1.2}$ for the spin density and $\Sigma^2 \propto t^2$ for the energy density \cite{Karrasch2014}.
Similar initial states can be prepared within the Fermi-Hubbard model and could in principle be realized in a cold-atom experiment \cite{Karrasch2017a}.

The generic behavior of a diffusive spreading of a local perturbation in 
nonintegrable models was investigated in 
\cite{Langer2009,Kim2013,Karrasch2016,Leviatan2017}.
We mention that solving the problem of the real-time evolution from a state with a few spins flipped compared to a background of full polarization is also of interest in the integrability community as some aspects of the finite-time dynamics can be understood exactly in this case [see, e.g., \cite{Ganahl2012,Liu2014}].

\begin{figure}[t]
\begin{center}
\includegraphics[width=0.90\linewidth]{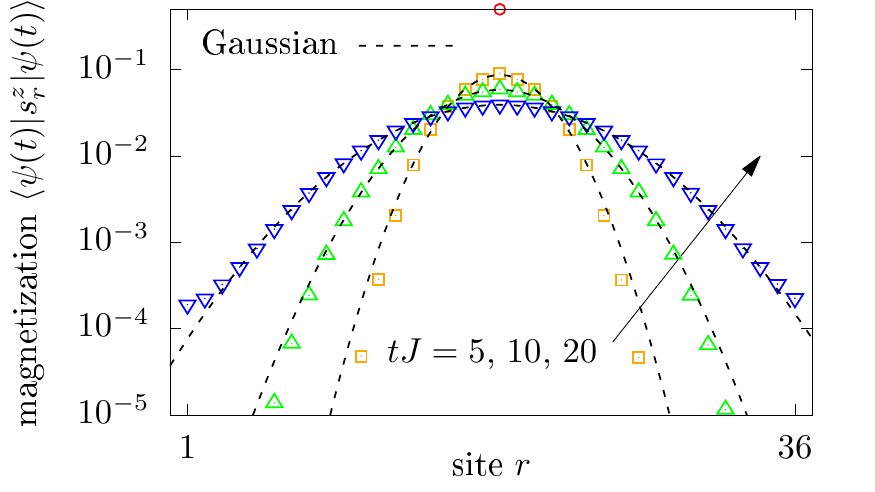}
\caption{(Color online) Spatial dependence of magnetization profiles at 
different times, as obtained in \cite{Steinigeweg2017} for a spin-1/2 XXZ chain at $\Delta = 1.5$, the 
full Hilbert space of $L = 36$ sites, and a randomly chosen initial pure state 
with a $\delta$ peak on top of a many-particle background at high temperatures. 
These profiles are remarkably well described by Gaussian fits over several 
orders of magnitude. Similar Gaussian profiles have been found in 
\cite{Ljubotina2017}.}
\label{fig:diffusion_Gaussian} 
\end{center}
\end{figure}

Recently, the analysis of the time- and space-dependent densities has been extended beyond just the 
spatial variance [see, e.g.,  \cite{Ljubotina2017,Steinigeweg2017}].
For $\Delta > 1$, as illustrated in Fig.\ 
\ref{fig:diffusion_Gaussian}, clean Gaussian profiles can be observed and 
provide another strong evidence of diffusion. 

\subsection{Bipartitioning Protocols}
\label{sec:bipartitioning}

\begin{figure}[t]
\begin{center}
\includegraphics[width=0.95\linewidth]{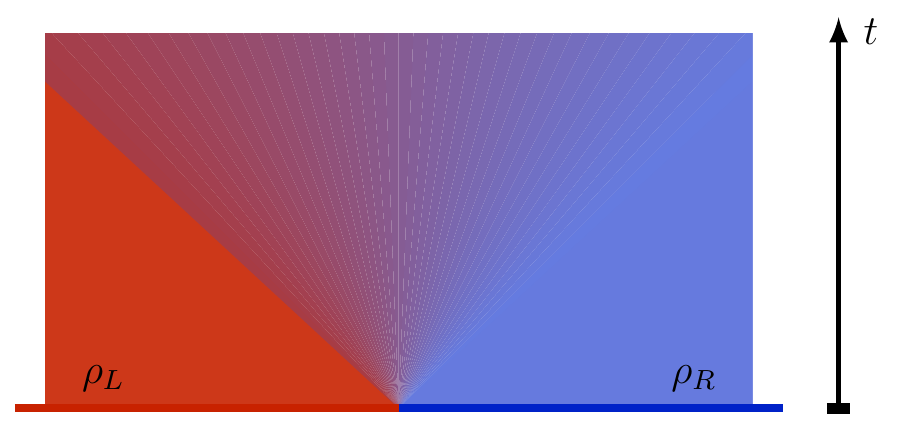}
\caption{(Color online) Pictorial representation of a generic bipartitioning protocol. The two halves of the chain are prepared in two different homogeneous states at time $t=0$. A nonequilibrium region emerges from the junction at the middle and expands at the maximal allowed speed.}
\label{fig:lightcone}
\end{center}
\end{figure}

Bipartitioning protocols emerged in the last two decades as a paradigmatic setting to study far-from-equilibrium transport in the context of isolated quantum many-body systems.\footnote{For a recent and more extended  discussion, see also the reviews \cite{BernardDoyon_2016, VasseurMoore_2016} dedicated to the subject.} These protocols are very simple: one prepares the two halves of the system in different homogeneous states, then joins them and lets the entire system evolve under the dynamics of  a spatially homogeneous Hamiltonian. In formulae, the state of the system at time $t$ is represented as 
\be
\rho(t)= e^{i H t}(\rho_{\rm L}\otimes \rho_{\rm R})e^{-i H t}\,,
\ee
where $H$ is the homogeneous Hamiltonian of the entire system and $\rho_{\rm R/L}$ are the two initial homogeneous states of the two halves. We refer the reader to  Fig.~\ref{fig:lightcone} for a pictorial illustration of the setting. 
Relevant examples, extensively studied in the literature, include the sudden junction of two half chains prepared at different temperatures \revision{[see, e.g.,~\cite{Nozawa2020,Ogata2002, Aschbacher2003,aschbacher2006out, Deluca2013, Karrasch2013a, Eisler2014, ColluraKarevski, ColluraMartelloni2014, DeLucaVitiMartelloni, Bhaeseen2015, DoyonBaseen2015, DOYON2015190, Biella2016, Castro2014, DeLucaVitiMazzaRossini, Castro-Alvaredo2016, Bertini2016, Zotos2016, Kormos2017, BertiniPiroli2018, Mazza2018, Mestyan2018, Bertini2019, Karevski2018}] or at different averaged magnetizations (or filling) [see, e.g.,~\cite{Misguich2017,Ljubotina2017,Santos2008,Santos2009,Santos2011,Antal2008,Antal1998,Antal99, ColluraAnalytic, Lancaster2010, Lancaster2010a,Calabrese_2008, Vidmar2017, eisler13, Alba2014, Vidmar2015,Sabetta2013, Viti_2016, Bertini2016, Piroli:2017aa, Eisler2016, DeLucaspintransport, Hauschild2015, Gobert2005, collura2020domainwall}]. We note that the latter kind of bipartitioning protocols, also referred to as ``geometric quenches" in the literature \cite{Mossel2010}, can be  realized in experiments on the sudden expansion of quantum gases in optical lattices (cf. Sec.~\ref{sec:experiments_optical_lattices}).    
}

In the two examples above, the two halves are prepared in homogeneous \emph{stationary} states. This means that a nontrivial time evolution is observed only in a region, ``the light cone", expanding from the junction at the maximal allowed speed. In locally-interacting lattice models with a finite-dimensional Hilbert space, this velocity is finite~\cite{liebrobinson}.
The light-cone region  contains information about the ``inhomogeneous nature" of the system (see Fig.~\ref{fig:lightcone}). In general, one can also prepare the two halves in homogeneous, \emph{non-stationary} states generating nontrivial dynamics also away from the junction. Importantly, however, the information about the inhomogeneous nature of the system is still contained in a light-cone region expanding from the junction at the maximal speed.  

Bipartitioning protocols are appealing because they give a minimal setting in which a genuine NESS, i.e., steady states supporting
nontrivial currents,  can be observed at infinite times. 
This has first been analytically observed  in 
noninteracting systems~\cite{Antal99}, then in conformal field 
theories~\cite{Bernard2012, bernard2015non}, and finally --- with the 
introduction of GHD --- in interacting integrable 
models~\cite{Bertini2016,Castro-Alvaredo2016}. On the contrary, for generic 
systems --- at least for those with Hamiltonians invariant either under 
space-inversion $P$ or time-reversal $\cal T$ ---  currents are seen to vanish 
in numerical studies~\cite{Karrasch2013a, Biella2016, Biella_2019}.

This fact can be explained using the ``hydrodynamic picture" discussed in Sec.~\ref{sec:GHD}.  Assuming that, at very large times, the 
expectation values of local observables
can be computed in 
 a locally quasi-stationary state, we generically have 
\be
\lim_{t\to\infty}{\rm tr }\left[{j}_x^\mathrm{(Q)}  e^{-i { H} t}  
{\rho}_0 e^{i { H} t}\right]={\rm tr }\left[{j}_0^\mathrm{(Q)} {\rho}_{\rm 
st}({x,\infty})\right]
\label{eq:currentEV}
\ee
for any current ${j}^{(Q)}_x$. For generic systems, we can assume that \revision{at the leading order in time}  
${\rho}_{\rm st}({x,t})$ is a Gibbs ensemble with a space-time dependent inverse 
temperature (and chemical potential if the system enjoys some U(1) symmetry). 
Generic lattice systems with a $P$-invariant Hamiltonian  have no $P$-odd charge 
 since  momentum is not conserved. This means that if the Hamiltonian is 
$P$-symmetric so is the Gibbs state. Noting that ${j}^{(Q)}_x$ is $P$-odd, we 
then conclude that \eqref{eq:currentEV} vanishes. The same reasoning applies for 
$\cal T$-symmetric systems. On the contrary, for integrable models, the state 
${\rho}_{\rm st}({x,t})$ is a GGE at each fixed $(x,t)$, and it generically 
includes parity-odd and time-reversal-odd charges. In this case, the 
expectation values of the currents are generically nonzero. \revision{Note that the above reasoning applies only in the infinite-time limit. At finite times, the quasi-stationary state of a non-integrable system is not exactly a space-time dependent Gibbs ensemble: it includes corrections (proportional to gradients of temperature and chemical potentials) that produce non-zero expectation values of the currents. These corrections, however, vanish in the infinite-time limit.}

\begin{figure}[t]
\begin{center}
\begin{tabular}{l}
\centerline{\includegraphics[width=0.72\linewidth]{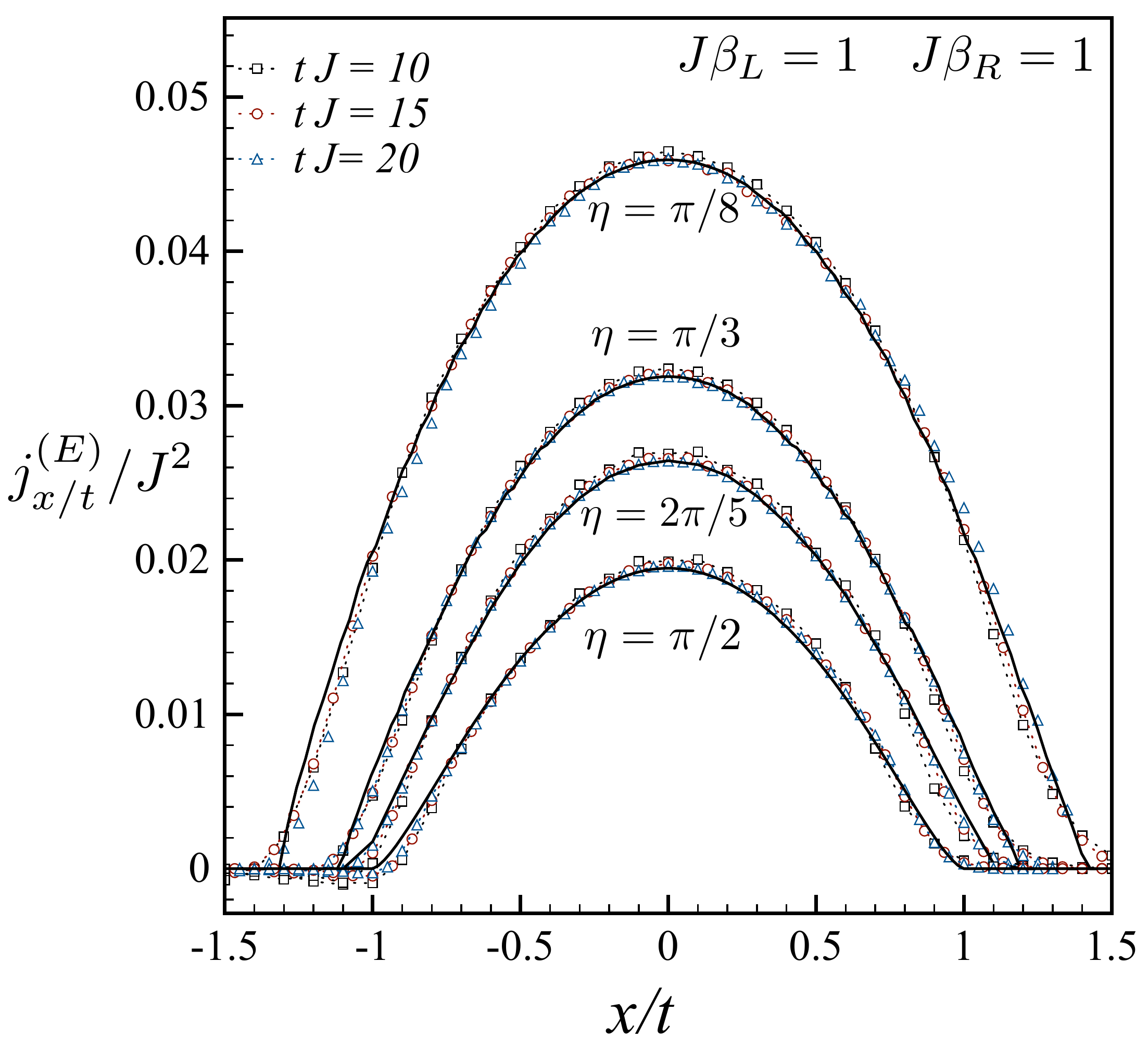}}
\\
\centerline{\includegraphics[width=0.99\linewidth]{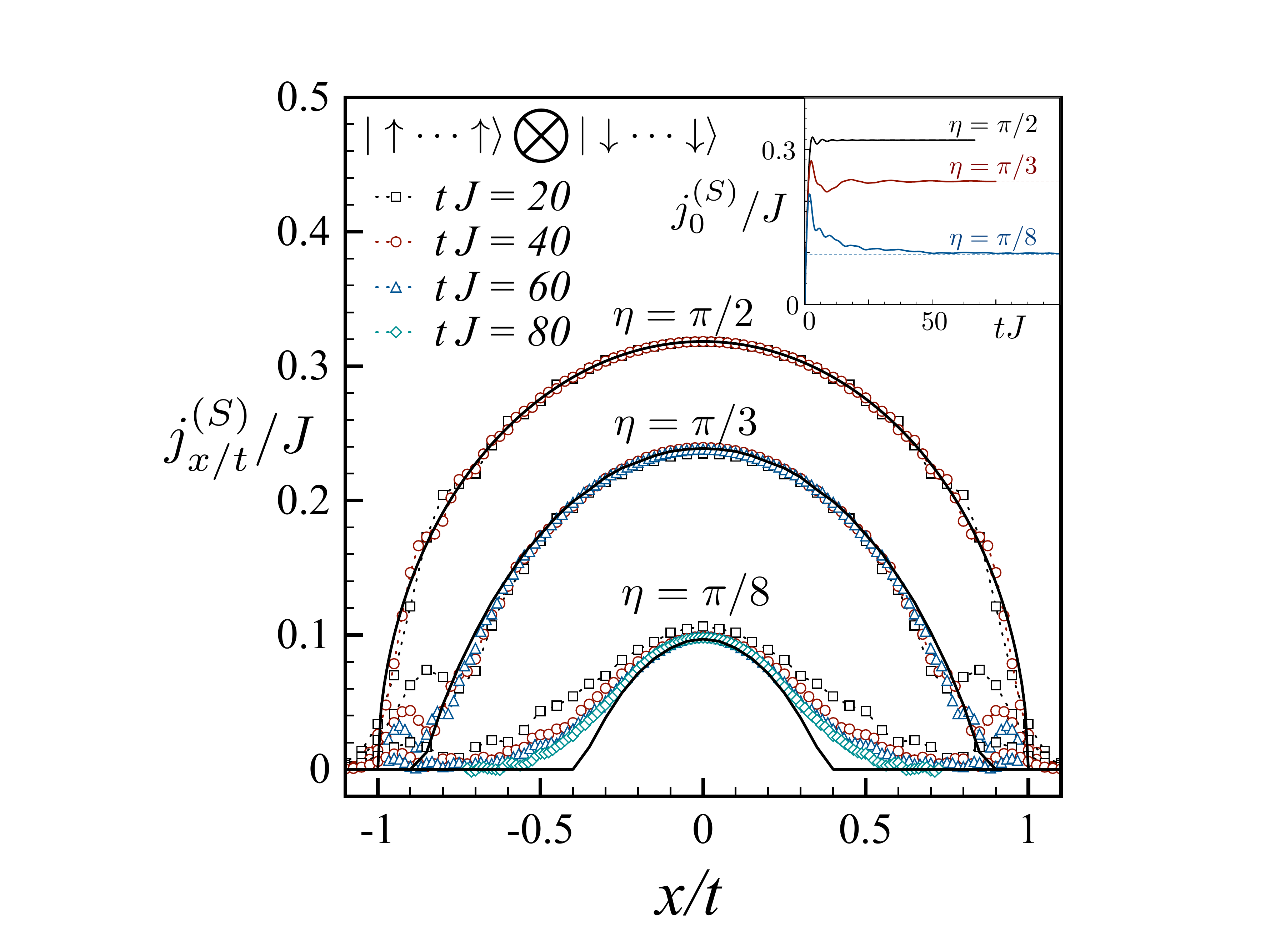}}
\end{tabular}
\caption{(Color online)
Profiles of the local currents in  the spin-1/2 XXZ chain for different values of $\Delta=\cos(\eta)$ as a function of rescaled position $x/t$. Symbols denote time-evolving block-decimation data for a chain of length $L=60$ (top) $L=120$ (bottom) and different times; full black lines are the GHD predictions. Top: Energy current after the two halves of the system have been initially prepared at inverse temperatures $\beta_{\rm L}=1$ and $\beta_{\rm R}=2$. Bottom: Spin current after the two halves have been prepared in two ferromagnetic states with opposite magnetization. The inset shows the time-dependent approach of $j^{(S)}_x$ (full colored lines) to the prediction (dashed lines). Figure adapted from~\cite{Bertini2016}.}
\label{fig:current_profile}
\end{center}
\end{figure}

For integrable systems, this argument can be checked by comparing the GHD solution (cf.~Eq.~\eqref{eq:continuitythetasol}) with tDMRG. In particular, note that for bipartitioning protocols, GHD predicts ${\rho}_{\rm st}({x,t})$ to become a function of the scaling variable $\zeta=x/t$ for large times, in agreement with previous observations in the context of noninteracting systems~\cite{Antal99}. This can be understood intuitively noting that an observer moving away from the junction at velocity $\zeta$ measures quasiparticles coming from the left (right) state if their velocity is larger (smaller) than $\zeta$. Since quasiparticles from the left and right state carry different physical information it is natural to expect the result of the measurement to depend on $\zeta$. Therefore, when studying bipartitioning protocols, it is customary to view expectation values of physical quantities for large times as functions of $\zeta$. As a representative example, in Fig.~\ref{fig:current_profile}, we report the comparison between GHD and tDMRG for profiles of energy and spin currents as a function of $\zeta$ for the spin-1/2 XXZ chain for different values of $\Delta\in[0,1]$ taken from~\cite{Bertini2016}. The upper panel displays the profile of the energy current at infinite times after joining together two chains prepared at different temperature, while the lower panel describes the profile of the spin current at infinite times after connecting  two chains prepared in two ferromagnetic states with opposite magnetization. This state is also known as the domain-wall state. From Fig.~\ref{fig:current_profile}, we clearly see that the current is finite within a light cone propagating from the junction, with a velocity that generically depends on the interaction strength. 

The emergence of a nonzero current at infinite times in integrable models 
signals ballistic transport of the related charge by the stable quasiparticles 
and corresponds to a finite Drude weight in the linear-response regime. In 
accordance with the linear-response physics, also when studying  bipartitioning 
protocols, there can be cases where certain currents vanish at infinite times, 
signalling subballistic transport. Such an inhibition of the transport of 
specific charges is typically caused by discrete symmetries. For instance, this 
happens for the transport of spin in the  spin-1/2 XXZ chain with 
$|\Delta|\geq1$, where all local conserved charges are invariant under a $Z_2$ 
spin-reversal symmetry except for the total magnetization~\cite{Piroli:2017aa}. In 
this case, considering a bipartitioning protocol that connects together a chain 
in a certain state with one in its spin-reversed copy (for example, two thermal 
states at the same temperature yet with opposite magnetization), one finds a 
vanishing  spin current in the infinite-time limit. In particular, the transport 
of spin has been observed to be diffusive for $|\Delta|>1$ and 
superdiffusive for $\Delta=1$~\cite{Ljubotina2017, Ljubotina2019}. The 
former case is described by GHD with diffusive corrections~\cite{DeNardis2018} 
[see Sec.~\ref{sec:GHDDiffusion}], while a complete theoretical description of 
the latter is still missing and the problem is currently subject of active 
research~\cite{Gopalakrishnan2019, DeNardis2019, bulch2019kardarparisizhang, 
DeNardis2020, Weiner2020, Agrawal2019, Medenjak2019b}.

Finally, we note that, even though in generic spin chains no nontrivial NESS is 
believed to emerge at infinite times, NESS-like physics can emerge in some 
intermediate-time window. This is the case of gapless systems subject to 
``low-temperature" bipartitioning protocols. Namely, these are bipartitioning 
protocols connecting two thermal states at different temperatures that are both 
small~\cite{BernardDoyon_2016}. In this regime, for large intermediate times, 
the behavior of energy density and current is well described by the 
Tomonaga-Luttinger liquid theory. The energy current is nonzero in the 
light-cone region and takes a conformal form~\cite{Bernard2012, bernard2015non}. 
On the other hand, for describing the profiles of generic observables, such as, e.g., 
the spin current in the gapless phase of the spin-1/2 XXZ chain, it is 
necessary to keep track of the non-linearities in the dispersion of low-energy 
modes. One can make progress in this direction by using the framework of 
nonlinear Tomonaga-Luttinger liquids~\cite{BertiniPRL2018}. For gapless 
integrable models at low temperatures, this approach reproduces the low-$T$ 
expansion of the GHD solution~\cite{BertiniPiroli2018, Mestyan2018}.

\section{Overview over experiments} \label{sec:experiments}

In this final section, we give an account of some of the experimental
efforts devoted to investigating transport in either quantum magnets or
with ultracold quantum gases. We stress that the survey of the literature cannot
be complete and refer the reader to recent reviews, where available \cite{Hess2019}.

   \subsection{Quantum magnets}

\begin{figure}[tb]
\begin{center}
\includegraphics[width=0.9\linewidth]{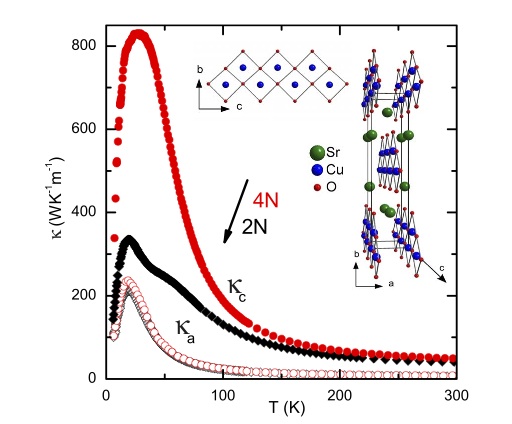}\\
\includegraphics[width=0.75\linewidth]{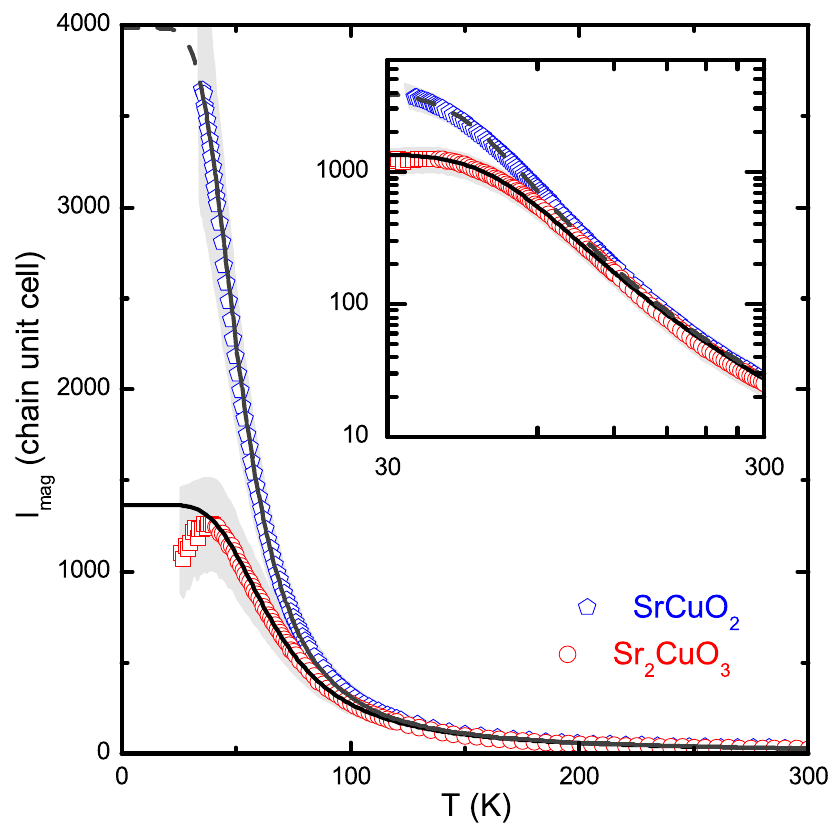}
\caption{(Color online) Top: experimental data for the thermal conductivity of SrCuO$2$ for two sample purities [2N (black curves) and 4N (red curves)]. The high-quality samples (4N) show a remarkably high thermal conductivity $\kappa_c$ in the crystal direction that is parallel to the spin chains, which is attributed to spin excitations. In the transverse directions, presumably only phonons contribute. From~\cite{Hlubek2010}. Bottom: Extracted magnetic mean free paths of spinons, from~\cite{Hlubek2012}.}
\label{fig:exp_hlubek}
\end{center}
\end{figure}

While this review focusses on the theoretical developments and results, the field was also strongly driven by
experimental results. Most notably, many cuprate-based low-dimensional magnets exhibit a contribution from magnetic
excitations to the thermal conductivity (see \cite{Hess2007,Sologubenko2007,Hess2019} for a review).
The values of the thermal conductivity $\kappa$ can be extremely large, given that these materials are electrical insulators and that they have
typically quite complicated structures. Originally, the largest thermal conductivities were reported for spin-ladder materials \cite{Hess2001,Sologubenko2000},
yet later on, much purer samples 
of the spin-chain materials SrCuO$_2$~\cite{Hlubek2010} and Sr$_2$CuO$_3$~\cite{Kawamata2008,Hlubek2012} became available
that show higher thermal conductivities
(see \cite{Sologubenko2000a,Sologubenko2001} for earlier experimental results). Those compounds are good realizations of the isotropic Heisenberg spin-chain model formed by Cu-O-Cu bonds with the exchange coupling being $J/k_{\rm B} \sim 2000-3000\,$K and the coupling between the chains $|\frac{J_\perp}{J}|\sim 10^{-1}$. The results for $\kappa(T)$, probing energy transport at low temperatures ($k_{\rm B}T \ll J$), are shown in Fig.~\ref{fig:exp_hlubek}. 
Considering the complicated struture of these materials, the
  conductivities are surprisingly large.
Other examples of one-dimensional materials that show a similar phenomenology are  copper pyrazine dinitrate \cite{Sologubenko2007a}, CaCu$_2$O$_3$ \cite{Hess2007a} and Haldane chains \cite{Sologubenko2008}.

While it is tempting to relate these large conductivities to the integrability of the underlying spin-chain Hamiltonians,
a rigorous experimental or theoretical verification of such a connection is very difficult: measuring thermal transport necessarily
requires a coupling of phonons to spins and thus a complete theory of thermal transport in such material requires the incorporation
of phonons (see \cite{Narozhny1996,Louis2005,Louis2006,Bartsch2013,Gangadharaiah2010,Chernyshev2015,Chernyshev2005,Rozhkov2005a,Shimshoni2003,Boulat2007,Chernyshev2016}. 

Assuming simple additivity of different contributions to conductivity, one can subtract the phononic contribution by measuring $\kappa$ in the direction orthogonal to the orientation of spin chains (where only phonons contribute, and whose contribution can be well described~\cite{Kawamata2008} by the Debye model). The resulting magnetic $\kappa_{\rm mag}$ contribution is then finite despite the ballistic energy transport in the Heisenberg model. This is caused by residual scattering on few magnetic impurities (due to  residual impurity of solvents used in the crystal growth), a nonzero interchain coupling  and/or due to spinon-phonon scattering. One can even deliberately introduce impurity doping~\cite{Kawamata2008} and  study how such disorder reduces transport~\cite{Hlubek2011,Mohan2014}. Precisely accounting for different scattering effects is not easy~\cite{Hlubek2012}, however, a picture that seems to account for most experimentally measured features seems to be compatible with a dominant impurity scattering at low temperatures ($T<50\,$K) while spinon-phonon scattering is the leading term at higher $T$. One can in fact infer~\cite{Sologubenko2000a} the mean-free path $l_{\rm mag}$ of magnetic excitaions (spinons) by using a simple kinetic expression for the conductivity of spinons, $\kappa_{\rm mag}=C v l_{\rm mag}$, where $C$ and $v$ are the heat capacity and the velocity of spinons, respectively. The heat capacity of the spin-$1/2$ Heisenberg model at low $T$ is proportional to $T$~\cite{Takahashi73}, leading to $ l_{\rm mag} \propto \kappa_{\rm mag}/k_{\rm B}T$ (see Fig.~\ref{fig:exp_hlubek} bottom).

The quasi-2d parent compounds of the HT$_C$s also exhibit a magnon contribution to the thermal conductivity, for instance, in La$_2$CuO$_4$ \cite{Hess2003},
Sr$_2$CuO$_2$Cl$_2$ \cite{Hofmann2003}, Ba$_2$Cu$_3$O$_4$Cl$_2$ \cite{Ohno2019}, or Nd$_2$CuO$_4$ \cite{Jin2003}. The values are smaller
than in their quasi-one dimensional relatives (where $l_{\rm mag} \sim 1\,\mu$m, see Fig.~\ref{fig:exp_hlubek}), yet this can partly be ascribed to the dependence of the specific heat on dimensionality. 

Thermal transport in quantum magnets cannot only be measured in the steady state, but also using time-resolved methods or at specific finite frequencies.
In the context of spin ladders, both the time-domain thermoreflectance  method \cite{Hohensee2014} and fluorescent microthermal imaging technique \cite{Otter2009,Otter2012}
were used.
 Moreover, one can induce a heat pulse on one end of a macroscopically large sample and then measure the time-resolved evolution of temperature at its other end \cite{Montagnese2013}. Such techniques can be used to extract electron-phonon coupling strength.

Measuring spin transport is much more difficult: until very recently, the only 
experiments were indirect ones using NMR \cite{Thurber2001,Kuehne2009} or 
muon-spin resonance ($\mu$SR) techniques to obtain a relaxation rate (of a 
nuclear spin in NMR, or muon in $\mu$SR) which is given by the spin 
autocorrelation function. The frequency dependence of the latter can be probed 
by the magnetic field dependence of the relaxation rate, allowing one to 
distinguish, e.g., diffusive from ballistic behavior from the tail of the spin 
autocorrelation function. NMR studies on SrCuO$_2$ found 
diffusive relaxation~\cite{Thurber2001}, while $\mu$SR experiments on high-purity samples 
in found ballistic relaxation~\cite{Maeter2013} (both studies probe $k_{\rm 
B}T \ll J$). $\mu$SR measurements on an organic salt~\cite{Pratt2006} or on 
Cu(C$_4$H$_4$N$_2$)(NO$_3$)$_2$~\cite{Xiao2015}
 were interpreted in 
terms of  diffusion, \revision{while a more recent $\mu$SR experiment \cite{Huddart2020} on [pym-Cu(NO$_3$)$_2$(H$_2$O)$_2$] and [Cu(pym)(H$_2$O)$_4$]SiF$_6$$\cdot$H$_2$O reports ballistic and diffusive dynamics, respectively.}\footnote{Both materials are rather perfect
realizations of the antiferromagnetic isotropic Heisenberg model with $J/k_{\rm
B} \sim 10-50\,$K and $|\frac{J_\perp}{J}|\lesssim 10^{-3}$.} Recently, the spin-Seebeck effect was exploited to directly 
induce and measure spin currents in a quasi-1d cuprate material 
\cite{Hirobe2017}.

We  mention that within solid-state NMR, experimental schemes were developed to study  spin transport in quasi-1d spin-chain systems 
after initializing the system in a state with an inhomogeneous magnetization. An example is an apatite crystal in which flourine atoms form chains that can be under an appropriate pulse sequence described by a nearest-neighbor dipolar Hamiltonian (being related to the XX Hamiltonian by a unitary transformation), and with an inter-chain coupling being as small as $|\frac{J_\perp}{J}|\sim 0.02$. A mixed initial state with a boundary imbalance of magnetization can be prepared (exploiting different energy scales of bulk and boundary spins) whose time evolution can then be studied~\cite{Ramanathan2011,Kaur2012}.

Besides experiments with bulk materials, there are novel synthetic one-dimensional structures that may in the future
be used to study transport in correlated one-dimensional systems. These include arrays of atoms arranged on various surfaces
(metallic, insulating or superconducting), whose properties are in some realizations believed to be related to the physics of
spin systems \cite{Khajetoorians2013,Toskovic2015}.

\revision{The prediction of superdiffusive dynamics of the Kardar-Parisi-Zhang type for the spin-1/2 Heisenberg chain has stimulated
a recent neutron-scattering study using the well-known quasi-one-dimensional material  KCuF$_3$ \cite{Scheie2020}.
By studying the regime of high temperatures $\hbar \omega \ll k_B T$, the authors report evidence that the data are  more consistent 
with KPZ behavior than diffusve or ballistic dynamics}

As a future challenge for theory, the development of efficient numerical methods for the description of transport in electron-phonon systems
is desirable.  An open  question is the applicability of wave-function based 
methods to the study of
transport in spin-phonon systems. Recent advances with DMRG methods using optimized local phonon basis \cite{Zhang1998} already give access to real-time dynamics
in electron-phonon systems 
\cite{Brockt2015,Dorfner2015,Stolpp2020,Guo2012,Kloss2019}, calling for 
extensions to finite temperatures and spin-phonon systems.

 \subsection{Ultracold quantum gases in optical lattices}
 \label{sec:experiments_optical_lattices}
Ultracold quantum gases provide another promising route to experimentally study the transport properties of
low-dimensional many-body systems. In optical lattices, both Fermi- and Bose-Hubbard models can be rather routinely realized \cite{Bloch2008,Bloch2017}.
A direct emulation of Heisenberg models or even more generally, spin-1/2 XXZ systems is more difficult: starting from single-bands and contact interactions,
these models arise only in the strong-coupling regime of Hubbard models and the degree to which they can be realized with high fidelity
depends  on the quality of loading processes and state preparation. The fact that here we are interested in finite-temperature properties
implies that no particular cooling schemes are needed, unlike in the ongoing efforts to reach the regime of long-range antiferromagnetic correlations
in the Fermi-Hubbard model \cite{Mazurenko2017,Solomon2019,Hilker2017,Schneider2008,Joerdens2008,Hart2015,Cheuk2015,Edge2015,Omran2015,Haller2015,Cheuk2016a,Greif2016,Boll2016,Cocchi2016,Cheuk2016,Parsons2015,Parsons2016}.

Besides working with the Fermi-Hubbard model, one can also emulate the Heisenberg model in two-component Bose-Hubbard models. Using this route, the decay
of a spin-spiral initial state was studied in 1d and 2d Heisenberg systems with a ferromagnetic exchange coupling \cite{Hild2014}.
\revision{For a 1d system with an isotropic exchange interaction, a diffusice decay of the spin spiral was found. 
A very recent studied succeded to extends this to the entire range of exchange anisotropies by working with a different atomic species (namely the bosonic isotope $^7$Li) and by exploiting a specific
Feshbach resonance \cite{Jepsen2020}. As a main result, the transition from a ballistic decay at $\Delta=0$ to a variety of transport behaviors is reported: superdiffusion for a
range of $0<\Delta<1 $, diffusion at $\Delta=1$, and subdiffusive dynamics for $\Delta>1$. These observations are quite different from the linear-response predictions 
discussed in Sec.~\ref{sec:xxz_drudespin}, yet the initial spiral state may lead to genuinely nonequilibrium dynamics.
}

It thus appears that studying the role of integrability directly with Fermi-Hubbard models is the more promising route. Given the rapid
emergence of many fermionic quantum-gas microscopes \cite{Guardado-Sanchez2020,Cheuk2015,Edge2015,Omran2015,Haller2015,Greif2016,Boll2016,Cocchi2016,Cheuk2016,Parsons2015,Parsons2016,Mazurenko2017,Solomon2019,Hilker2017,Brown2019,Nichols2019}, which all work with the two-dimensional Fermi-Hubbard model and which do allow
to chop such 2d systems into individual 1d systems \cite{Boll2016,Vijayan2020,Solomon2019}, the finite-temperature transport properties of the 1d Fermi-Hubbard model might be the easiest
accessible integrable lattice model. Ultracold quantum gases have some drawbacks: particle numbers (or system sizes) cannot be made arbitrarily large, the systems have a finite life-time
and they realize closed quantum systems, i.e., it is not straightforward to couple such a gas to leads (see \cite{Brantut2012,Stadler2012,Brantut2013,Esslinger2017}, though).
Nevertheless, one could exploit the single-site manipulation and resolution capabilities of quantum-gas microscopes to investigate the spreading of perturbations in
the particle or spin density, as suggested in \cite{Karrasch2017}.
 Numerical simulations show that it is fairly well possible to resolve the difference between
(presumably) diffusive and ballistic dynamics at high temperatures $T\gg J$ on time scales of less than $4/t_{\rm h}$, where $t_{\rm h}$ is the hopping-matrix element, and thus within the time window of coherent many-body dynamics in such systems \cite{Trotzky2012}. Such an experiment could directly probe the linear-response regime. 
A recent experiment addressing spin-charge separation in the 1d Hubbard model utilizes a quite similar protocol to induce spin- and charge dynamics \cite{Vijayan2020}.

\begin{figure}[tb]
\begin{center}
\includegraphics[width=0.90\linewidth]{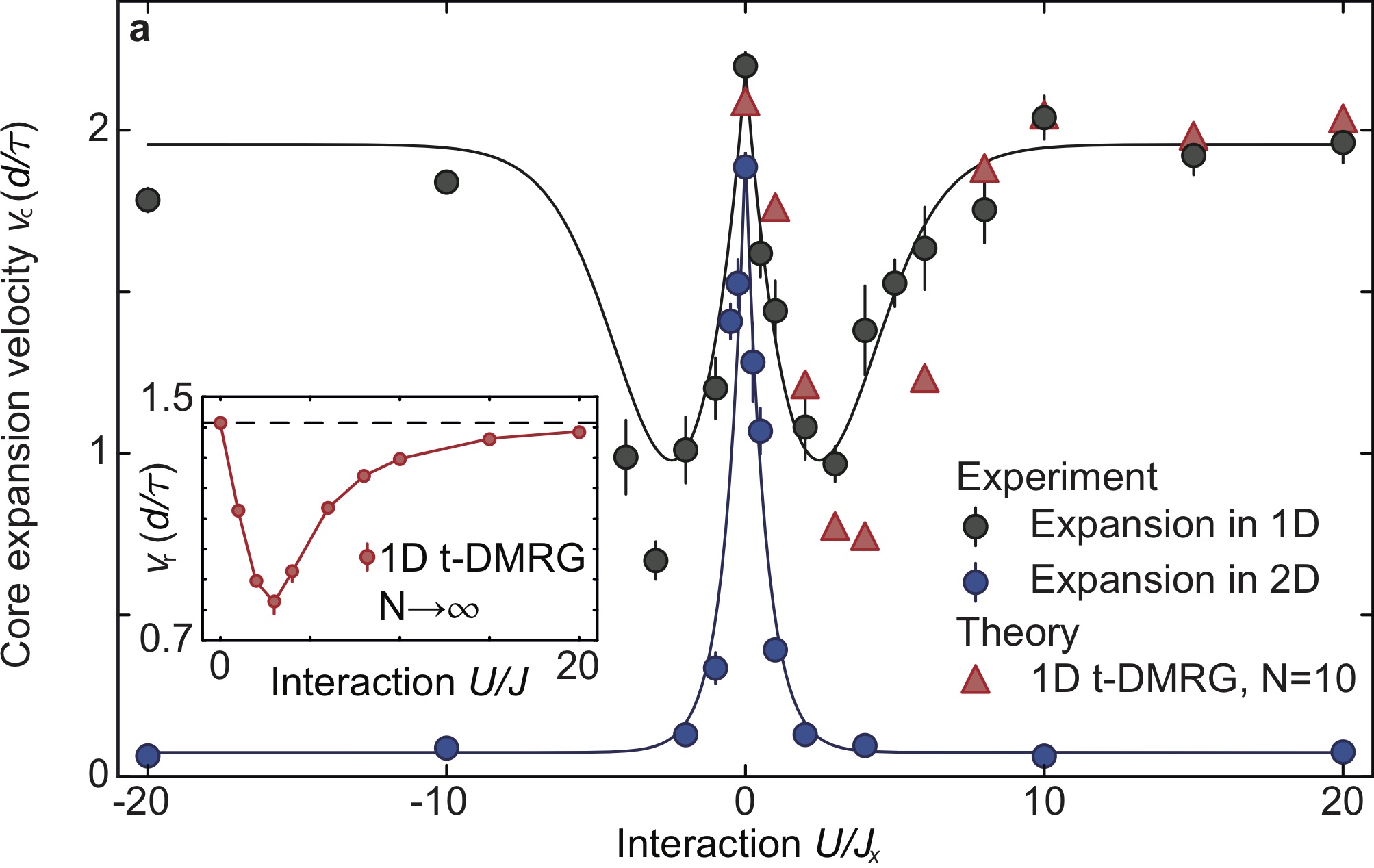}
\caption{(Color online)
Expansion velocity of a cloud of bosons that are released from a trap into an empty optical lattice.
The main panel shows experimental and DMRG data for the expansion velocity as a function of interaction strength $U/J_x$ extracted from the half-width-half-maximum. $J_x$ is the hopping matrix element
along the $x$-direction of the two-dimensional lattice and $J_x=J$ for the one-dimensional case. $U$ is the onsite interaction strength in the Bose-Hubbard model.
The inset shows DMRG data for the radial velocity as a function of $U/J$. Both noninteracting and strongly interacting bosons expand ballistically with the same
expansion velocity \cite{Ronzheimer2013}.
\label{fig:exp_ronzheimer}}
\end{center}
\end{figure}

Nonequilibrium mass transport can be investigated in a much more straightforward fashion using optical lattices. In the so-called sudden expansion, an originally
trapped quantum gas is released from its confining potential and allowed to expand in a homogeneous and flat optical lattice. This method was used
to study the nonequilibrium transport of the 2d Fermi-Hubbard model \cite{Schneider2012}, the Fermi-Hubbard chain \cite{Scherg2018} as well as of bosons in 1d and  2d lattices \cite{Ronzheimer2013}.
In the latter experiment with bosons, an impressive difference between the dynamics of strongly-interacting bosons in 1d versus 2d lattices was observed:
in 1d, the sudden expansion is as fast as for noninteracting bosons (assuming the same initial conditions), while in 2d, the cloud expands much slower, more consistent
with the notion that interactions should induce scattering and degrade currents (see Fig.~\ref{fig:exp_ronzheimer}). The reason for the behavior of such strongly-interacting 1d gases lies in their exact mapping
onto spinless noninteracting fermions via a Jordan-Wigner transformation \cite{Cazalilla2011}. Therefore, strongly interacting bosons with densities not exceeding unity
realize an integrable model in 1d, equivalent to the spin-1/2 XX chain.
Experimentally, integrability can be broken in three ways: (i) coupling 1d systems to a 2d system, (ii) inducing double occupancies in the initial state and (iii)
by going to finite interaction strength $0<U/t_{\rm h}< \infty$, where the Bose-Hubbard model is nonintegrable. All three cases show clear deviations from the fast and
ballistic expansion of hardcore bosons. In cases (i) and (ii), this can be traced back to the breaking of integrability \cite{Vidmar2013,Steinigeweg2014a}.
The dynamics in the 1d Bose-Hubbard model at $U/t_{\rm h}< \infty$ is more involved in this particular experiment as it also involves a quantum quench in the interaction and thus probes the dynamics
at different energy densities, depending on $U/t_{\rm h}$ \cite{Vidmar2013}. The experiment \cite{Ronzheimer2013} is therefore a clear realization of integrability-protected ballistic
mass transport in the spirit of this review, albeit in the nonequilibrium regime (see Sec.~\ref{sec:bipartitioning}).
Extensions of this approach are possible using quantum-gas microscopes as well, where so far, only the expansion dynamics of two bosons was investigated \cite{Preiss2015,Tai2016}.
More recent experiments studied transport in the two-dimensional Fermi-Hubbard model using the capabilities of quantum-gas microscopes \cite{Vijayan2020,Brown2019,Nichols2019}. 
All these studies investigate the interplay of spin- and charge in transport, with \cite{Vijayan2020} focussing on spin-charge separation in one dimension, while \cite{Brown2019,Nichols2019}
observe diffusion in  two-dimensional systems.

We note that experiments with ultracold bosons in optical lattices in the strongly interacting regime thus offer a unique and controlled way to study integrability breaking by perturbing around the limit of the spin-1/2 XX chain, resulting then in the 2d XX model or ladders \cite{Vidmar2013,Steinigeweg2014a,Hauschild2015}.
Besides measuring densities, one can further study one-body correlations in such sudden expansions, for which theory predicts a dynamical quasi-condensation phenomenon
\cite{Rigol2004,Vidmar2017} as a result of the emergent eigenstate solution for this nonequilibrium problem \cite{Vidmar2017}.
Remarkably, even this effect, another consequence of integrability in nonequilibrium transport, has been observed experimentally \cite{Vidmar2015}.

\section{Summary and Outlook}

This article reviewed the state-of-the-art of the theoretical understanding of transport in translationally invariant one-dimensional quantum lattice models at finite temperatures from the theoretical physics perspective.
We discussed, in particular, the important role of integrability and its breaking, focusing primarily on the paradigmatic  spin-1/2 XXZ  and the Fermi-Hubbard chain as minimal models for  spin, charge and energy transport. The progress that has been achieved in recent years for these systems and their theoretical description in general is due to methodological breakthroughs, both in fundamental concepts, such as establishing the existence of quasi-local conservation laws \cite{Prosen2011,Prosen2013,Pereira2014,Prosen2014,Ilievski15a} and their connection to a complete hydrodynamic description [the so-called generalized hydrodynamics \cite{Bertini2016,Castro-Alvaredo2016}], as well as in numerical methods such as, e.g., matrix-product-based techniques \cite{Karrasch2012,Karrasch2013p,Karrasch2017} or dynamical typicality \cite{Steinigeweg2014,Steinigeweg2016a} for utilizing time-evolution methods at finite temperatures for the calculation of transport properties.
Establishing time-dependent DMRG as a solver of Lindblad master equations 
\cite{Prosen2009} opened up possibilities for complementary qualitative and 
quantitative insights from studying open quantum systems \cite{Znidaric2011}.

We may say that the understanding of ballistic  transport at high temperatures or even in nonequilibrium states have by now matured.  
The thermal Drude weight in both the spin-1/2 XXZ chain and the 1d Fermi-Hubbard model were computed as a function of model parameters and temperature \cite{Kluemper2002,Sakai2003,Ilievski2017,Karrasch2017b}.
The exact and complete calculation of magnetothermal corrections involving off-diagonal coefficients and the spin Drude weight at finite magnetizations remains as an open task \cite{Louis2003,Heidrich-Meisner2005,Sakai2005,Zotos2016},
in particular, for the Fermi-Hubbard model where in principle, three currents can couple.
\revision{The calculation of all cross-coefficients could be accomplished using the methodology of GHD.}

For spin transport in the spin-1/2 XXZ chain, the existence of a 
finite-temperature Drude weight at both finite and zero magnetization and any 
value of $\Delta $ \cite{Zotos1997} and for zero magnetization at $|\Delta|<1$
is now well established  and accepted \cite{Zotos1999,Prosen2011,Prosen2013,Pereira2014,Urichuk2018}. Its full temperature dependence is accessible as well \cite{Zotos1999,Urichuk2018,Ilievski2017a}, yet has not been convincingly computed with numerical methods.
The agreement between TBA, GHD, and the lower bound supports the notion of  a 
fractal structure as a function of $\Delta$, yet neither approach is rigorous, 
involving either Takahashi's string hypothesis or
relies on the assumption of knowing all relevant charges.
For the spin-1/2 Heisenberg chain, the overwhelming evidence suggests that the 
spin Drude weight vanishes at finite temperature. 
The same goes for the regime of $\Delta>1$, while in both cases, a rigorous proof is missing.
 
For those cases that prohibit ballistic transport channels, or when studying subleading corrections, the situation is still much less clear, yet actively studied. 
Although normal diffusion is the most commonly observed type of non-ballistic transport, both in integrable \cite{DeNardisDiffusion} and nonintegrable quantum lattice systems \cite{Sirker2011,Sirker2009}, one often encounters also other types of transport, including, in particular, superdiffusive dynamics \cite{Znidaric2011,Ljubotina2017}. Notably, the conjectured KPZ scaling \cite{Ljubotina2019} \revision{[see also \cite{Dupont2019,Weiner2020,Krajnik2019,Das2019,bulch2019kardarparisizhang,Fava2020,DeNardis2020a,Ilievski2020}]} of spin-correlation functions and spin-transport in the isotropic Heisenberg chain and other integrable models of magnetism with non-Abelian symmetries is a particularly pressing question, on which much work is expected in the near future.  Another universal option suggested by recent studies is the one of marginally superdiffusive transport characterized by a diffusive exponent and a  logarithmic correction \cite{DeNardis2020}.
The exact nature of subleading corrections in the ballistic regimes of the 1d Fermi-Hubbard model or the exact nature of spin- and charge transport at zero magnetization and filling
is much less well understood. In general, a complete qualitative understanding of the emergence of diffusion in integrable models is still lacking.

\revision{Both the now solidly established aspects of spin transport in the spin-1/2 XXZ chain and the open questions on, e.g., superdiffusion and the connection between linear-response
behavior and transport in specific far-from-equilibrium settings have stimulated additional recent experiments using both quasi-one-dimensional materials \cite{Scheie2020} and ultracold atoms \cite{Jepsen2020}. The neutron-scattering study \cite{Scheie2020} reports consistency of their data with the KPZ scenario.}

For nonintegrable models, we presented examples where the notion of diffusion is soundly supported from approximate analytical methods as well as numerically exact techniques. 
These include the dimerized spin-1/2 XX ladder \cite{Steinigeweg2014a}, spin-1/2 chains with a staggered magnetic field \cite{Huang2013,Steinigeweg2015}, 
as well as general spin ladders and frustrated chains \cite{Karrasch2015,Zotos2004,Steinigeweg2016a}. 
In the latter cases, the long-time dynamics is usually more complex and diffusion is harder to establish.
At low energies, field-theoretical studies are strongly suggestive of diffusive dynamics as well [see, in particular, \cite{Sirker2011}].

Although the work presented here considers only quantum lattice systems, it is not clear if the transport phenomena are in any fundamental way affected by the quantum nature of the microscopic equation of motions as compared to classical deterministic Hamiltonian dynamics governing classical lattice systems. So far, we see no argument against the conclusion that all the 
emerging transport phenomena at finite temperatures have analogous counterparts in classical lattice models. \revision{An exception might be the putative many-body localization \cite{Abanin2019,Nandkishore2015}, where temperature is ill-defined.} However, both a systematic semiclassical analysis and elucidating the quantum-classical correspondence for transport in many-body lattice systems  would be desirable for the  future.

The transition between diffusion to  types of non-diffusive transport can be expected to be a manifestation of a form of ergodicity breaking. The latter is currently being intensely studied even in translationally invariant, disorder-free settings, prominent examples being the so-called quantum scars in models with constrained dynamics [see, e.g., \cite{Bernien2017,Lan2018,Turner2018,Bernevig2018}].
However, a possible connection to finite-temperature transport in such models has not been investigated. Another interesting set of open questions in relation to ergodicity breaking concerns the connection between spectral statistics, described by  random-matrix theory, and transport properties. Since spectral statistics contain information on different time scales, it may happen \cite{Brenes2018} 
that models with a 
local integrability breaking are ergodic on the Heisenberg time scale, i.e.,  on 
time scales controlled by the inverse mean level spacing, while transport is 
ballistic on shorter time scales.

\acknowledgements
We thank H. De Raedt, J. Gemmer, C. Hess, P. Maass, V. Meden, J. Moore, P. 
Prelov\v{s}ek, U. Schneider, H. Spohn, and X. Zotos for discussions and for 
their comments on a previous version of the manuscript.
We thank M. Fagotti for valuable discussions and his contributions to this project in its early stages.
Furthermore, we acknowledge helpful comments from S. Gopalakrishnan, D. Huse, E. Ilievski,  A. Kl\"umper, and L. Santos on a previous version of the manuscript.
This work was supported in parts by the Starting Grant No.~679722 and Advanced Grant OMNES No.~694544 of European Research Council (ERC). 
F.H.-M. and M.Z. are grateful for the hospitality at KITP, where part of this work was performed. 
This research was supported in part by the National Science Foundation under Grant No. NSF PHY-1748958.
C.K. acknowledges support by the Deutsche Forschungsgemeinschaft through the Emmy Noether program (KA 3360/2-1) as well as by `Niedersächsisches Vorab' through `Quantum- and Nano-Metrology (QUANOMET)' initiative within the project NL-2. R.S. acknowledges support by the DFG - 397067869 (STE 2243/3-1) - within the DFG Research Unit FOR 2692 - 355031190. 
F. H.-M. was supported by the Deutsche Forschungsgemeinschaft  (DFG, German Science Foundation) -  217133147 - via CRC 1073 (project B09).
T.P. and M.Z. acknowledge support by Program P1-0402 of Slovenian Research Agency (ARRS). M.Z. acknowledges support by the Grants J1-7279 and J1-1698 of the Slovenian Research Agency (ARRS).

\bibliographystyle{apsrmp4-1}

\bibliography{references}

\end{document}